\documentstyle[11pt,epsf,rotate]{article}

\newcommand{\RE}{{\rm Re}}
\newcommand{\IM}{{\rm Im}}
\newcommand{\vcb}{|V_{cb}|}
\newcommand{\vtd}{|V_{td}|}
\newcommand{\vub}{|V_{ub}/V_{cb}|}
\newcommand{\vts}{|V_{ts}|}

\def\eps{\varepsilon}
\def\epe{\varepsilon'/\varepsilon}
\def\as{\alpha_s}
\newcommand{\eqn}{\ref}
\def\Heff{{\cal H}_{\rm eff}}
\newcommand{\nn}{\nonumber}
\newcommand{\mt}{m_{\rm t}}
\newcommand{\mtb}{\overline{m}_{\rm t}}
\newcommand{\mcb}{\overline{m}_{\rm c}}
\newcommand{\mc}{m_{\rm c}}
\newcommand{\ms}{m_{\rm s}}
\newcommand{\md}{m_{\rm d}}
\newcommand{\mb}{m_{\rm b}}
\newcommand{\mw}{M_{\rm W}}
\newcommand{\mz}{M_{\rm Z}}
\newcommand{\gev}{\, {\rm GeV}}
\newcommand{\mev}{\, {\rm MeV}}

\newcommand{\Lms}{\Lambda_{\overline{\rm MS}}}

\newcommand{\Bsg}{$B \to X_s \gamma$ }

\newcommand{\bea}{\begin{eqnarray}}
\newcommand{\eea}{\end{eqnarray}}
\newcommand{\bd}{\begin{displaymath}}
\newcommand{\ed}{\end{displaymath}}
\newcommand{\aem}{\alpha}

\newcommand{\beq}{\begin{equation}}
\newcommand{\eeq}{\end{equation}}
\newcommand{\be}{\begin{equation}}
\newcommand{\ee}{\end{equation}}
\newcommand{\ord}{{\cal O}}
\newcommand{\order}{{\cal O}}
\newcommand{\f}{\frac}
\newcommand{\Ctilde}{\tilde{C}}
\def\kpnn{$K^+\rightarrow\pi^+\nu\bar\nu$}
\def\kpn{K^+\rightarrow\pi^+\nu\bar\nu}
\def\klpn{K_{\rm L}\rightarrow\pi^0\nu\bar\nu}
\def\klpnn{$K_{\rm L}\rightarrow\pi^0\nu\bar\nu$}
\def\klm{K_{\rm L} \to \mu^+\mu^-}
\def\aspi{\frac{\as}{4\pi}}
\newcommand{\imlt}{\IM\lambda_t}
\newcommand{\relt}{\RE\lambda_t}
\newcommand{\relc}{\RE\lambda_c}

\begin{document}


\thispagestyle{empty}

\rightline{TUM-HEP-275/97}
\rightline{TTP97-15}
\rightline{hep-ph/9704376}
\rightline{April 1997}
\vspace*{1.2truecm}
\bigskip

\centerline{\LARGE\bf  Quark Mixing, CP Violation and Rare Decays}
\vspace{0.3truecm}
\centerline{\LARGE\bf  After the Top Quark Discovery}
 \vskip1truecm
\centerline{\large\bf Andrzej J. Buras${}^1$ and Robert Fleischer${}^2$}
\bigskip
\centerline{\sl ${}^1$ Technische Universit{\"a}t M{\"u}nchen, Physik 
Department}
\centerline{\sl D-85748 Garching, Germany}
\vskip0.6truecm
\centerline{\sl ${}^2$ Institut f{\"u}r Theoretische Teilchenphysik}
\centerline{\sl Universit{\"a}t Karlsruhe}
 \centerline{\sl D-76128 Karlsruhe, Germany}
\vskip1.5truecm
\centerline{\bf Abstract}

We review the highlights of quark mixing, particle--antiparticle mixing,
CP violation and rare $K$- and $B$-decays in the Standard Model. The top 
quark discovery, the precise measurement of its mass, the improved knowledge
of the couplings $V_{cb}$ and $V_{ub}$, and the calculations of NLO
short distance QCD corrections improved considerably the predictions
for various decay rates, the determination of the couplings $V_{td}$
and $V_{ts}$ and of the complex phase in the Cabibbo-Kobayashi-Maskawa 
matrix. After presenting the general theoretical framework for weak decays,
we discuss the following topics in detail: i) the CKM matrix, its
most convenient parametrizations and the unitarity triangle,
ii) the CP-violating parameter $\varepsilon_K$ and $B^0_{d,s}-\bar B^0_{d,s}$
mixings, iii) the ratio $\varepsilon'/\varepsilon$, iv) the rare $K$-decays 
$K_{\rm L}\to\pi^0e^+e^-$, $K^+\to\pi^+\nu\bar\nu$, $K_{\rm L}\to\pi^0\nu
\bar\nu$ and $K_{\rm L} \to \mu^+\mu^-$, v) the radiative decays $B \to X_s
\gamma$ and $B\to X_s l^+l^-$, vi) the rare $B$-decays $B\to X_{s,d}\nu
\bar\nu$ and $B_{d,s}\to l^+l^-$, vii) CP violation in neutral and charged 
$B$-decays putting emphasis on clean determinations of the angles of the 
unitarity triangle, and viii) the role of electroweak penguins in 
$B$-decays. We present several future visions demonstrating very clearly 
the great potential of CP asymmetries in $B$-decays and of clean $K$-decays 
such as $K^+\to\pi^+\nu\bar\nu$ and $K_{\rm L}\to\pi^0\nu\bar\nu$ in the 
determination of the CKM parameters and in decisive testing of the Standard 
Model. An outlook for the coming years ends our review.

\vspace*{1.0cm}

\begin{center}
{\small To appear in {\it Heavy Flavours II}, World Scientific (1997), Eds.\
A.J. Buras and M. Lindner.}
\end{center}


\newpage

\thispagestyle{empty}

\mbox{}

\newpage

\pagenumbering{roman}

\tableofcontents

\newpage

\pagenumbering{arabic}

\setcounter{page}{1}

\centerline{\LARGE\bf  Quark Mixing, CP Violation and Rare Decays}
\vspace{0.3truecm}
\centerline{\LARGE\bf  After the Top Quark Discovery}
 \vskip1truecm
\centerline{\large\bf Andrzej J. Buras${}^1$ and Robert Fleischer${}^2$}
\bigskip
\centerline{\sl ${}^1$ Technische Universit{\"a}t M{\"u}nchen, Physik 
Department}
\centerline{\sl D-85748 Garching, Germany}
\vskip0.6truecm
\centerline{\sl ${}^2$ Institut f{\"u}r Theoretische Teilchenphysik}
\centerline{\sl Universit{\"a}t Karlsruhe}
  \centerline{\sl D-76128 Karlsruhe, Germany}
\vskip1truecm
\centerline{\bf Abstract}

We review the highlights of quark mixing, particle--antiparticle mixing,
CP violation and rare $K$- and $B$-decays in the Standard Model. The top 
quark discovery, the precise measurement of its mass, the improved knowledge
of the couplings $V_{cb}$ and $V_{ub}$, and the calculations of NLO
short distance QCD corrections improved considerably the predictions
for various decay rates, the determination of the couplings $V_{td}$
and $V_{ts}$ and of the complex phase in the Cabibbo-Kobayashi-Maskawa 
matrix. After presenting the general theoretical framework for weak decays,
we discuss the following topics in detail: i) the CKM matrix, its
most convenient parametrizations and the unitarity triangle,
ii) the CP-violating parameter $\varepsilon_K$ and $B^0_{d,s}-\bar B^0_{d,s}$
mixings, iii) the ratio $\varepsilon'/\varepsilon$, iv) the rare $K$-decays 
$K_{\rm L}\to\pi^0e^+e^-$, $K^+\to\pi^+\nu\bar\nu$, $K_{\rm L}\to\pi^0\nu
\bar\nu$ and $K_{\rm L} \to \mu^+\mu^-$, v) the radiative decays $B \to X_s
\gamma$ and $B\to X_s l^+l^-$, vi) the rare $B$-decays $B\to X_{s,d}\nu
\bar\nu$ and $B_{d,s}\to l^+l^-$, vii) CP violation in neutral and charged 
$B$-decays putting emphasis on clean determinations of the angles of the 
unitarity triangle, and viii) the role of electroweak penguins in 
$B$-decays. We present several future visions demonstrating very clearly 
the great potential of CP asymmetries in $B$-decays and of clean $K$-decays 
such as $K^+\to\pi^+\nu\bar\nu$ and $K_{\rm L}\to\pi^0\nu\bar\nu$ in the 
determination of the CKM parameters and in decisive testing of the Standard 
Model. An outlook for the coming years ends our review.

\section{Introduction}
\setcounter{equation}{0}
Quark mixing, CP violation and rare decays of $K$ and $B$ mesons
constitute an important
part of the Standard Model and particle physics in general. There are
several reasons for this:
\begin{itemize}
\item
This sector probes in addition to weak and electromagnetic
interactions also the strong interactions at short and long distance
scales. As such it involves essentially the dominant part of the
dynamics present in the Standard Model.
\item
It contains most of the free parameters of the Standard Model
such as the quark masses and the Cabibbo-Kobayashi-Maskawa
parameters \cite{CAB,KM}.
\item
The presence of a large class of processes, which take place only as
loop effects, tests automatically the quantum structure of the theory and
offers the means to probe (albeit indirectly) the physics at very short
distance scales which may possibly imply modifications and/or extensions
of the Standard Model.
\item
The renormalization group effects play here an important role  in
view of the vast difference between the weak interaction ${\cal{O}}(\mw)$ and
strong interaction ${\cal{O}}(1\gev)$ scales.
\item
The nature of CP, T and CPT violations can be investigated.
\end{itemize}

\noindent
The processes in this sector originate in weak interactions and
can be divided naturally into two distinct classes:
\begin{itemize}
\item
Tree level decays
\item
One-loop induced decays and transitions known as {\it flavour-changing
neutral current processes} (FCNC).
\end{itemize}

\noindent
The predictions for these two classes can be obtained from the
Lagrangian of the Standard Model by means of the usual techniques of
quantum field theory, in particular  the {\it operator product expansion}
and the {\it renormalization group.} In deriving and subsequently 
testing these
predictions one encounters, however, several difficulties:
\begin{itemize}
\item
There are many free parameters.
\item
The strong interaction effects at long distances must be evaluated
outside the perturbative framework which results in large theoretical
uncertainties.
\item
The experimental data are often not sufficiently accurate to allow for
firm conclusions.
\end{itemize} 

\noindent
Yet it is evident that the field of quark mixing, CP violation and
rare decays of $K$ and $B$ mesons played a very important role in
particle physics and there is no doubt that it will play
this role in the future in the continuing tests of the Standard
Model and in searches for physics beyond it.

\noindent
The main purpose of this chapter is to review the present status
of this field and to provide an outlook for the future. In 1992
a review of this type was presented in the first edition
of {\it Heavy Flavours} under the title: {\it A Top Quark Story} \cite{BH}.
During the last five years several things happened which forced
us to rewrite this chapter to a large extent.

{\bf On the experimental side:}
\begin{itemize}
\item
The top quark has been discovered and its mass considerably
constrained.
\item
The $B$-meson life-times and $B^0_d-\bar B^0_d$-mixing have
been measured with improved accuracy.
\item
The uncertainty in the element $V_{cb}$ of the CKM matrix has
been substantially decreased both due to improved data
and theory. 
\item
The values for $\vub$ have improved and
decreased by almost a factor of two.
\item
The radiative transition $b \to s\gamma$ has been observed for the first
time and several upper bounds on rare decays have been lowered.
\end{itemize}

{\bf On the theoretical side:}
\begin{itemize}
\item
The next-to-leading (NLO) QCD corrections to the most interesting decays
have been calculated thereby considerably reducing  the theoretical
uncertainties and modifying the previous predictions.
\item
The application of the Heavy Quark Effective Theory (HQET) and Heavy Quark
Expansions (HQE) improved considerably the theoretical status of 
$B$-decays and, as stated above, allowed an improved 
determination of $V_{cb}$.
\item
Considerable progress has been made in analyzing CP asymmetries in $B$ decays,
in designing new methods for extracting CP violating phases, and in
understanding the role of electroweak penguins in $B$ decays.
\item
The intensive studies of certain rare $K$ and $B$ decays show that these
decays, when combined with the future measurements of CP asymmetries,
should allow the determination of the CKM matrix and tests of
the Standard Model without any hadronic uncertainties.
\item
Some progress has also been achieved in calculating relevant 
non-perturbative parameters such as $B_K$ and $F_B$, and in
extracting some hadronic matrix elements entering the theoretical
estimate of $\varepsilon'/\varepsilon$ from experimental data.
\end{itemize}
\par
\noindent
Finally another change relative to the {\it Top Quark Story}
took place: the second author has been changed.

\noindent
All these reasons motivated us to rewrite the previous review to
a large extent.
Consequently the present review is not really an update of the
{\it Top Quark Story} but rather an independent article even if there
are some similarities, in particular in the first part of section 2. 
The main new ingredients are the
inclusion of NLO QCD corrections to all decays for which these
corrections have been calculated and a considerably extended discussion
of CP violation in $B$ decays. Moreover the full numerical analysis
presented in the {\it Top Quark Story} had to be changed in view of the
top quark discovery and the changes listed above. In preparing
this review we benefited enormously from a recent review on NLO
corrections by Gerhard Buchalla, Markus Lautenbacher and the first
author \cite{BBL} as well as from 
the Ph.D.\ thesis on CP violation in $B$ decays
completed by the second author in February 1995 \cite{ROBF}.
Also our recent reviews \cite{AJBW}-\cite{F97} were helpful in this respect.
\par
\noindent
In {\bf section 2} we present the general theoretical framework 
for analyzing tree
level decays and flavour-changing neutral current processes (FCNC).
Beginning with a simple classification of basic Feynman diagrams
and effective FCNC vertices \cite{BH}, we discuss briefly a more
formal and more complete approach 
based on the operator product expansion (OPE) and the renormalization
group.  We give the classification of all the operators relevant
for subsequent sections as well as Feynman diagrams from which
they originate. We give a list of seven universal
$\mt$ dependent functions $F_r(x_t)$ which result from various penguin
and box diagrams and constitute an important ingredient in Feynman
rules for the effective
FCNC vertices. In the formal approach based on the OPE 
these functions enter the initial conditions for the renormalization
group evolution of the Wilson coefficients. It is, however, possible
to rewrite the OPE in the form of the so-called
penguin--box expansion (PBE) \cite{PBE0} in which the decay
amplitudes are given directly in terms of $F_r(x_t)$.
This offers a systematic way of
exhibiting the $\mt$ dependence of FCNC processes and is useful
for phenomenological applications. 
In this section we also
summarize briefly the present status of higher order QCD corrections 
to weak decays. These are discussed in great detail in \cite{BBL} and
will be taken into account in subsequent sections.
Finally we will make a few comments on Heavy Quark Effective Theory
(HQET) and Heavy Quark Expansions (HQE) which are discussed in great
detail by Neubert in another chapter of this book and in his review
\cite{NEU}.

\noindent
In {\bf section 3} we discuss the
Cabibbo-Kobayashi-Maskawa matrix \cite{CAB,KM}, 
its two most convenient parametrizations and
its geometrical representation given by the {\it main} unitarity triangle.
The properties of this triangle are listed. 
Next the present status of the CKM matrix based on tree
level decays is summarized. Only a part of this
matrix can be determined this way. Using finally 
the unitarity of this matrix 
we estimate the top quark couplings. This analysis is
refined in later sections with the help of other processes.

\noindent
In {\bf section 4} we use the existing experimental information on one--loop
decays in order to complete the determination of the CKM matrix.
The two quantities at our disposal are the
parameters $\varepsilon_K$ describing {\it indirect CP violation} in
K--meson decays, and the mass difference $\Delta M_d$ 
(or the parameter $x_d$) which measures the size of 
$B^0_d-\bar B^0_d$ mixing. The present theoretical and experimental status of
these two quantities will be given with particular emphasis on QCD effects at
both short and long distances. The latter introduce considerable
uncertainties in the phenomenological analysis and consequently do not allow 
for firm conclusions.
Yet, as we will see, some general implications on the structure of the
CKM matrix and on the shape of the unitarity triangle can be found this way.
It should be stressed that significant progress relative to the
situation at the time of \cite{BH} has been made in this field. We discuss
here also $B^0_s-\bar B^0_s$ mixing which when measured should offer
an improved determination of the unitarity triangle. 
This section contains also a few messages which should be
useful for the unitarity triangle practitioners.
The information on CKM parameters obtained in this section is essential
for the material of the subsequent 
sections which deal exclusively with the weak decays of the late nineties
and of the next decade: the rare
$K$ and $B$ decays, and the CP asymmetries in the $B$--meson system. 
This section ends
with present ranges for various parameters which
one can find on the basis of $\varepsilon_K$ and $\Delta M_d$ alone.

\noindent
In {\bf section 5} the ratio $\varepsilon'/\varepsilon$ is discussed in some
detail including the implications of a rather low value of the
strange quark mass found in most recent lattice calculations.

\noindent
In {\bf section 6} the decays 
$K_{\rm L}\to\pi^0 e^+ e^-$, $B \to X_s\gamma$ and
$B\to X_s \mu^+ \mu^-$ are analyzed. We discuss these three decays in one
section because they have a similar theoretical structure.

\noindent
In {\bf section 7} we discuss the rare decays $K^+\to\pi^+\nu\bar\nu$,
$K_{\rm L}\to\pi^0\nu\bar\nu$, $K_{\rm L}\to\mu\bar\mu$, $B\to X_s \nu\bar\nu$
and $B \to l\bar l$ which also have a similar theoretical structure.
Except for $K_{\rm L} \to \mu\bar\mu$ all these decays are theoretically
very clean offering this way excellent means for the determination
of CKM parameters and tests of the Standard Model.

\noindent
In {\bf section 8} CP violation in non-leptonic $B$-meson decays and various 
strategies for the determination of the angles of the unitarity triangle 
at future $B$ meson facilities are reviewed. We discuss in detail 
general aspects, the ``benchmark modes'' to determine $\alpha$, $\beta$
and $\gamma$, some recent developments including CP-violating asymmetries
in $B_d$ decays, the $B_s$ system in light of a possible width difference
$\Delta\Gamma_s$, charged $B$ decays, and relations among certain
non-leptonic $B$ decay amplitudes.  

\noindent
{\bf Section 9} is devoted to the role 
of electroweak penguins in non-leptonic
$B$-decays. Because of the large 
top-quark mass electroweak penguins may become important and may even 
compete with QCD penguins. These effects led to considerable interest in 
the recent literature. We will see in section 9 that some non-leptonic 
$B$ decays are affected significantly by electroweak penguins and that a 
few of them should even be dominated by these contributions. The question 
to what extent the strategies for extracting the angles of the unitarity
triangle reviewed in section 8 are affected by the presence of electroweak 
penguins is also addressed and methods for obtaining experimental insights 
into the world of electroweak penguins are discussed.

\noindent
{\bf Section 10} is an attempt to classify $K$- and $B$- decays 
from the point
of view of theoretical cleanliness.

\noindent
{\bf Section 11} offers some future visions.
In particular
we illustrate here how  future measurements of  CP asymmetries 
in $B$ decays and the measurements of the very clean rare decays
$\kpn$ and $\klpn$
may offer precise determinations of the CKM matrix. 

\noindent
Finally in {\bf section 12} we close this review by giving a 
shopping list for the
late nineties and the next decade.

\noindent
In this article we did not have space and energy to review {\it all} aspects
of the fascinating field of quark mixing, CP violation and rare decays.
Rather we concentrated on a series of
selected topics which we expect to play an important role in the future.
Certain interesting topics have, however, not been covered by us. These are: 
electric dipole moments \cite{HE,BERSU,BARR}, 
CP violation in hyperon decays \cite{Hyperon}, 
CP violation and mixing in the $D$-system \cite{Burdman}
and long distance dominated $K$-decays \cite{KLONG}.

\section{Theoretical Framework}
\setcounter{equation}{0}
\subsection{The Basic Theory}
Throughout this review we will work in the context of the three
generation model of quarks and leptons based on the gauge group
$SU(3)\otimes SU(2)_L\otimes U(1)_Y $ spontaneously broken 
to $SU(3)\otimes U(1)_Q$.
Here $Y$ and $Q$ denote the weak hypercharge and the electric charge
generators, respectively. $SU(3)$ stands for $QCD$ which describes the
strong interactions mediated by eight gluons $G_a$.

Concerning electroweak interactions,
the left-handed leptons and quarks are put in $ SU(2)_L $ doublets
\begin{equation}\label{2.31}
\left(\begin{array}{c}
\nu_e \\
e^-
\end{array}\right)_L\qquad
\left(\begin{array}{c}
\nu_\mu \\
\mu^-
\end{array}\right)_L\qquad
\left(\begin{array}{c}
\nu_\tau \\
\tau^-
\end{array}\right)_L
\end{equation}
\begin{equation}\label{2.66}
\left(\begin{array}{c}
u \\
d^\prime
\end{array}\right)_L\qquad
\left(\begin{array}{c}
c \\
s^\prime
\end{array}\right)_L\qquad
\left(\begin{array}{c}
t \\
b^\prime
\end{array}\right)_L       
\end{equation}
with the corresponding right-handed fields transforming as singlets
under $ SU(2)_L $. The primes in (\ref{2.66}) are discussed below.

The electroweak interactions of quarks and leptons are mediated by
the massive weak gauge bosons $W^\pm$ and $Z^0$ and by the photon $A$.
The physical neutral Higgs has no impact on our review. The effects of charged 
Higgs particles present in the extensions of the Standard Model are discussed
in a separate chapter of this book.

The dynamics of this theory is described by the fundamental Lagrangian
\begin{equation}\label{LL}
{\cal L} = {\cal L}({\rm QCD}) + {\cal L}({\rm SU(2)_L\otimes U(1)_Y})+
{\cal L}({\rm Higgs})
\end{equation}
from which -- after quantization and spontaneous symmetry breaking -- the 
Feynman rules can be derived. Before discussing these rules let us
say a few more things about the fermion--gauge--boson electroweak 
interactions 
resulting from (\ref{LL}). They play a crucial role in this review. 

These interactions are summarized by the Lagrangian
\begin{equation}\label{3}
{\cal L}_{\rm int}={\cal L}_{\rm CC}+{\cal L}_{\rm NC}\,,
\end{equation}
where
\begin{equation}\label{4}
{\cal L}_{\rm CC}=\frac{g_2}{2 \sqrt{2}}(J^+_\mu W^{+\mu}+J^-_\mu W^{-\mu})
\end{equation}
describes the {\it charged current} interactions and
\begin{equation}\label{5}
{\cal L}_{\rm NC}=
- e J^{\rm em}_\mu A^{\mu}+ \frac{g_2}{2 \cos \Theta_{\rm W}} J^0_\mu Z^\mu
\end{equation}
the {\it neutral current} interactions. Here $e$ is the QED coupling constant,
$g_2$ is the $SU(2)_L$ coupling constant and $\Theta_{\rm W}$ is the Weinberg
angle. The currents are given as follows
\begin{equation}\label{6}
J^+_\mu=
(\bar{u} d')_{V-A} +
(\bar{c} s')_{V-A} +
(\bar{t} b')_{V-A} +
(\bar{\nu}_e e)_{V-A} +
(\bar{\nu}_\mu \mu)_{V-A} +
(\bar{\nu}_\tau \tau)_{V-A}
\end{equation}

\begin{equation}\label{7}
J^{\rm em}_\mu=\sum_f {Q_f \bar f \gamma_\mu f}
\end{equation}
\begin{equation}\label{8}
J^0_\mu=\sum_f \bar f \gamma_\mu (v_f-a_f\gamma_5) f
\end{equation}
\begin{equation}\label{9}
v_f=T^f_3-2 Q_f \sin^2\Theta_{\rm W},
\qquad
a_f=T^f_3,
\end{equation}
where $Q_f$ and $T^f_3$ denote the charge and the third component of the
weak isospin of the left-handed fermion $f_L$, respectively. 
The relevant electroweak
charges are given in table \ref{tab:ewcharges}.

\begin{table}[htb]
\begin{center}
\begin{tabular}{|c||c|c|c|c|c|c|c|}
\hline
 & $\nu^e_L$ & $ e^-_L$ & $ e^-_R$ & $u_L$ & $d_L$ & $u_R$ & $d_R$ \\
\hline
$Q$ & 0 & $-1$ & $-1$ & 2/3 & $-1/3$ & 2/3 & $-1/3$ \\
\hline
$T_3$ &1/2 & $-1/2$ & 0 & 1/2 & $-1/2$ & 0 & 0 \\
\hline
$Y$ & $-1$ & $-1$ & $-2$ & 1/3 & 1/3 & 4/3 & $-2/3$ \\
\hline
\end{tabular}
\end{center}
\caption[]{Electroweak charges $Q$, $Y$ and the third component of
the weak isospin $T_3$ for quarks and leptons in the Standard Model.
\label{tab:ewcharges}}
\end{table}

\subsection{Elementary Vertices}
Let us next recall those
elementary interaction vertices which govern the physics of quark
mixing, CP violation and rare decays.
They are given in fig.\ 1.

\begin{figure}[hbt]
\vspace{0.10in}
\centerline{
\epsfysize=3in
\epsffile{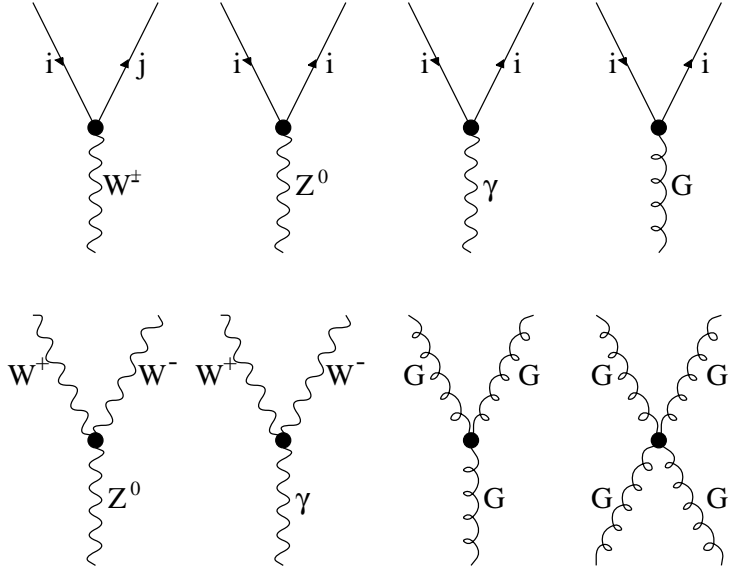}
}
\vspace{0.08in}
\caption[]{
Elementary Vertices
\label{fig:1}}
\end{figure}

The following comments should be made:
\begin{itemize}
\item
The indices $i,j$ denote flavour: $i,j = u,d,c,t,\ldots$
\item
In non--physical gauges also vertices involving fictitious
Higgs particles in place of $W^{\pm}$, $Z^0$ have to be included in
this list.
\item
In the processes considered, the triple and quartic gluon
couplings enter only through the running of the QCD coupling constant
and in higher order QCD corrections to weak decays. The quartic
electroweak couplings do not enter our discussion at the level of
approximations considered.
\item
The striking property of the interactions listed above is
the flavour conservation in vertices involving neutral gauge bosons
$Z^0$, $\gamma$ and $G$. This fact implies the absence of flavour
changing neutral current (FCNC) transitions at the tree level. This is
the GIM mechanism \cite{GIM1} which has a crucial impact on the dynamics 
of weak decays in the Standard Model. However, in the generalizations of
this model tree level FCNC transitions are possible. GIM mechanism will be 
discussed in more detail below.
\item
The charged current processes mediated by $W^{\pm}$ are
obviously flavour violating with the strength of violation given by
the gauge coupling $g_2$  and effectively at low energies 
by the Fermi constant 
\begin{equation}\label{2.100}
\frac{G_{\rm F}}{\sqrt{2}}=\frac{g^2_2}{8 \mw^2}
\end{equation}
and a unitary $3\times3$
CKM matrix \cite{CAB,KM}. This matrix connects the weak
eigenstates $(d^\prime,s^\prime,b^\prime)$ and the corresponding mass 
eigenstates $d,s,b$ through
\begin{equation}\label{2.67}
\left(\begin{array}{c}
d^\prime \\ s^\prime \\ b^\prime
\end{array}\right)=
\left(\begin{array}{ccc}
V_{ud}&V_{us}&V_{ub}\\
V_{cd}&V_{cs}&V_{cb}\\
V_{td}&V_{ts}&V_{tb}
\end{array}\right)
\left(\begin{array}{c}
d \\ s \\ b
\end{array}\right)=\hat V_{\rm CKM}\left(\begin{array}{c}
d \\ s \\ b
\end{array}\right),
\end{equation}
so that for instance
\begin{equation} 
(d\buildrel{W^+}\over\longrightarrow t)=
  i{{g_2}\over{2\sqrt2}}V_{td}\;\gamma_{\mu}(1-\gamma_5),
  \qquad\quad
  (t\buildrel{W^-}\over\longrightarrow d)=
  i{{g_2}\over{2\sqrt2}}V_{td}^*\;
  \gamma_{\mu}(1-\gamma_5).
\end{equation}
In the leptonic sector the analogous mixing matrix is a unit matrix
due to the masslessness of neutrinos in the Standard Model.
The fact that the CKM matrix is unitary assures the absence of
elementary FCNC vertices. Consequently the
unitarity of $\hat V_{\rm CKM}$ is at the basis of the GIM mechanism. On
the other hand, the fact that the $V_{ij}$'s can a priori be complex
numbers allows the introduction of CP violation in the Standard Model.
The structure
and the experimental status of $\hat V_{\rm CKM}$ is discussed in
sections 3 and 4.
\item
The strength of the neutral current vertices is described by
the gauge couplings $g_3,g_2,e$ and the relevant strong and
electroweak charges. For completeness we give in figs.\ 2 and 3 the most
important Feynman rules in the Standard Model. 
\item
It should be stressed that the photonic and gluonic
vertices are vectorlike (V), the $W^{\pm}$ vertices are purely $V-A$,
whereas, as can be seen in fig.\ 3, 
the $Z^0$ vertices involve both $V-A$ and
$V+A$ structures.
\end{itemize}

\noindent
With the help of the elementary vertices of fig.\ 1,
the propagators and  Feynman rules
at hand, one can build physically interesting
processes and subsequently evaluate them. The simplest of such
processes, which forms the basis for subsequent considerations, is the
$W^{\pm}$ exchange between two fermion lines shown in fig.\ 11a.
Neglecting the momentum of the W-propagator relative to $\mw$,
this process gives the following tree level effective Hamiltonian
describing the charged weak interactions of quarks and leptons:
\begin{equation}\label{WE}
{\cal H}^{\rm tree}_{\rm eff} = {{G_{\rm F}}\over{\sqrt2}}\;{\cal J}^+_\mu
{\cal J}^{-\mu}
\end{equation}
with ${\cal J}^+_\mu$  given in (\ref{6}).

\subsection{Effective FCNC Vertices}

Next one--loop effects have to be considered. At the one--loop level
in addition to corrections to the vertices of fig.\ 1 new structures
appear which were absent at  tree level. These are the flavour
changing neutral current (FCNC) transitions which can be summarized by a set
of basic triple and quartic effective vertices. In the literature they
appear under the names of penguin and box diagrams, respectively.

\begin{figure}[thb]
\centerline{
\epsfysize=9in
\epsffile{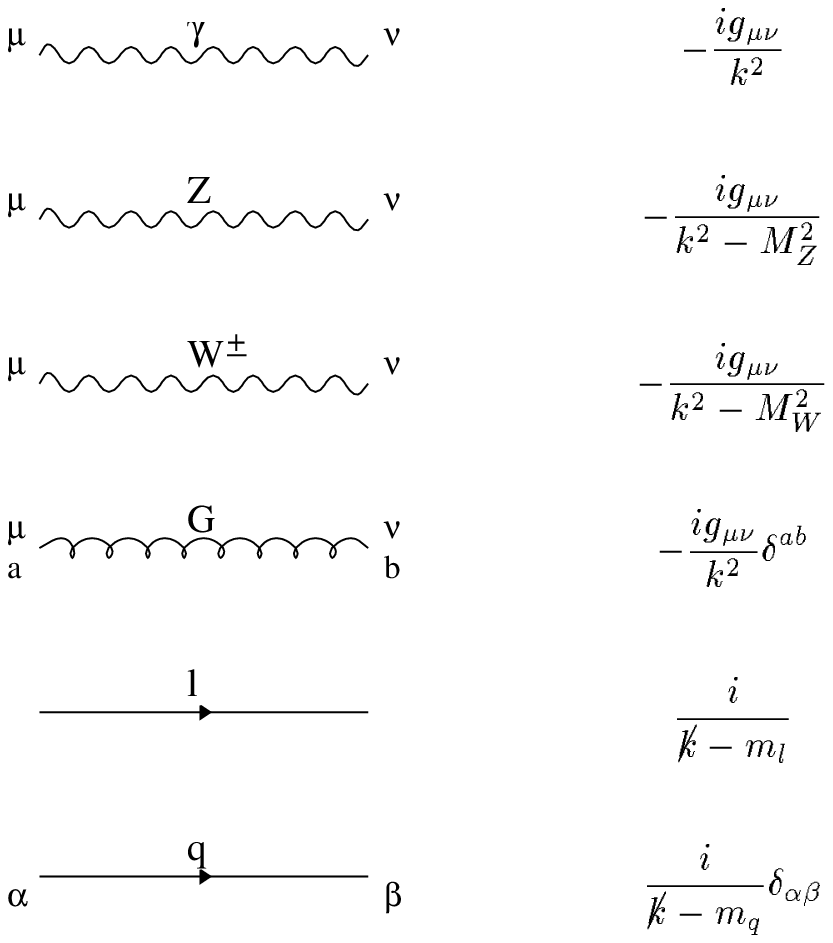}
}
\vspace{-5.25in}
\caption[]{
Feynman Rules (Propagators)
\label{fig:2}}
\end{figure}


\subsubsection{Penguin vertices}

\noindent
These vertices involve only quarks and can be depicted as in fig.\ 4
where $i$ and $j$ have the same charge but different flavour and $k$
denotes the internal quark whose charge is different from that of $i$
and $j$. 
These effective vertices 
can be calculated by using the elementary vertices and 
propagators of figs.\ 2 and 3.
Important examples are given in fig.\ 5.
The diagrams with fictitious Higgs exchanges in place of $W$ have not been
shown. Strictly
speaking, also self--energy corrections on external lines have to be
included to make the effective vertices finite.

\begin{figure}[thb]
\centerline{
\epsfysize=9.5in
\epsffile{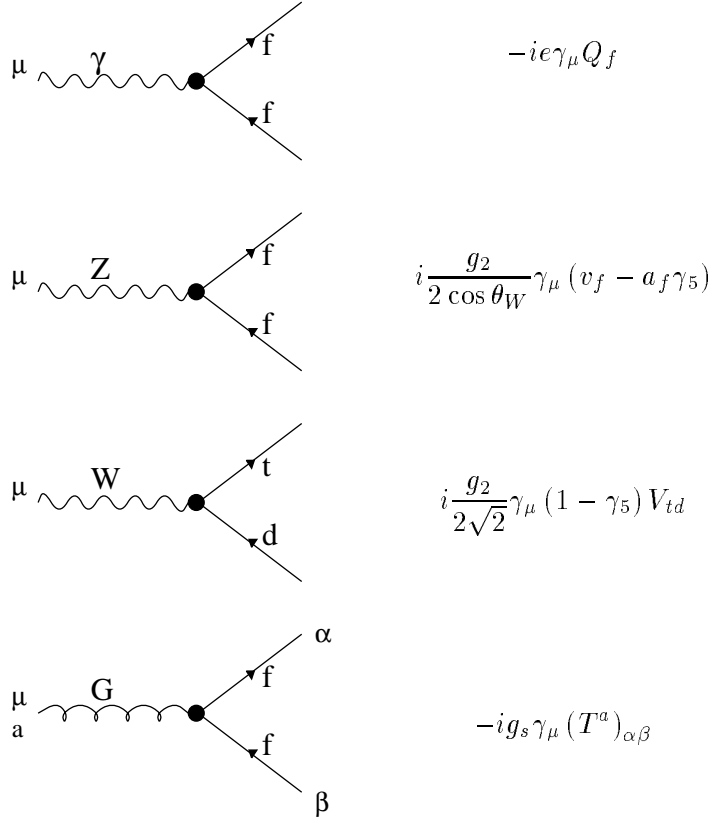}
}
\vspace{-4.95in}
\caption[]{
Feynman Rules (Vertices)
\label{fig:3}}
\end{figure}

\subsubsection{Box vertices}
These vertices involve in general both quarks and leptons and can be
depicted as in fig.\ 6,
where again $i,j,m,n$ stand for external quarks or leptons and $k$ and
$l$ denote the internal quarks and leptons. In the vertex (a) the
flavour violation takes place on both sides (left and right) of the
box, whereas in (b) the right--hand side is flavour conserving. These
effective quartic vertices can also be calculated using the elementary
vertices and propagators of figs.\ 2 and 3. We have
for instance the vertices in fig.\ 7
which contribute to $B^0-\bar B^0$ mixings and $K^+\to\pi^+\nu\bar\nu$, 
respectively.
The fictitious Higgs exchanges have not been shown.
Other interesting examples will be discussed in the course of this
review. \par

\begin{figure}[hbt]
\centerline{
\epsfysize=1.6in
\epsffile{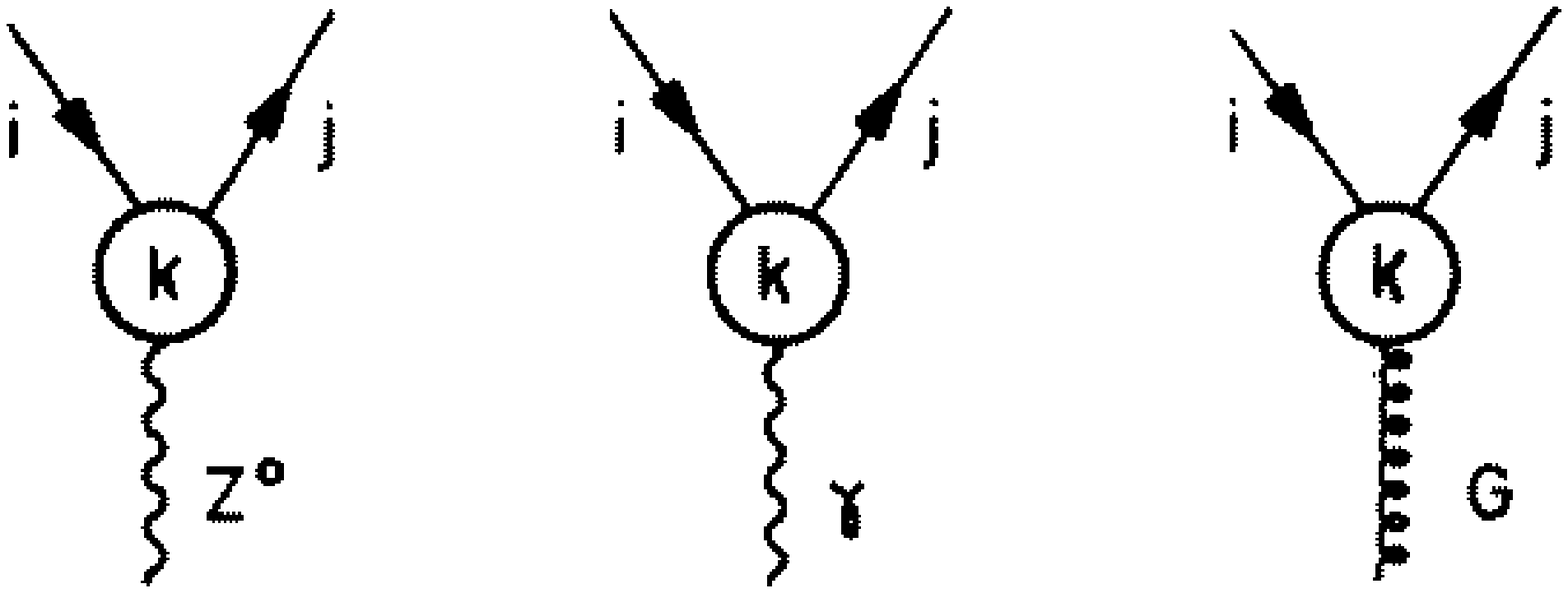}
}
\vspace{0.08in}
\caption[]{
Penguin vertices
\label{fig:4}}
\end{figure}

\begin{figure}[htb]
\centerline{
\epsfysize=3.2in
\epsffile{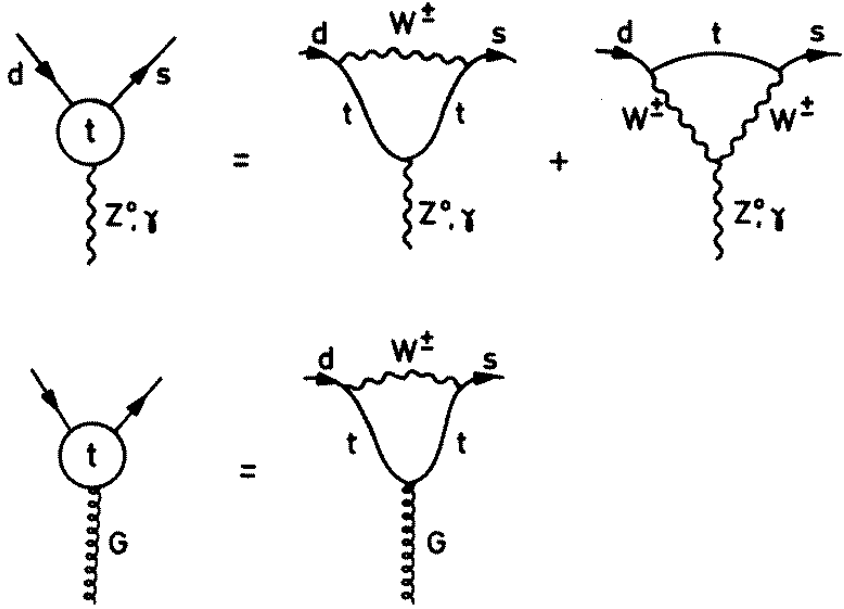}
}
\caption[]{
Penguin vertices resolved in terms of basic vertices
\label{fig:5}}
\end{figure}

\subsubsection{Effective Feynman Rules}
With the help of the elementary vertices and propagators 
shown in figs.\ 2 and 3, 
one can now derive
``Feynman rules'' for the effective vertices discussed above by
calculating simply the diagrams on the r.h.s. of the equations in
figs.\ 5 and 7. These rules are given  
in the \\
't Hooft--Feynman gauge as follows:
\begin{equation}\label{FR}
  {\rm Box} (\Delta S = 2)
~=~ \lambda^2_i {{G^2_{\rm F}}\over{16\pi^2}} \mw^2S_0(x_i) 
   (\bar s d)_{V-A} (\bar s d)_{V-A} 
\end{equation}
\begin{equation}
{\rm Box}(T_3= -1/2)~ =~ \lambda_i {{G_{\rm F}}\over{\sqrt 2}}
   {{\alpha}\over{2\pi \sin^2\Theta_{\rm W}}} B_0(x_i) (\bar s d)_{V-A} 
   (\bar\mu\mu)_{V-A}
\end{equation}
\begin{equation}
 {\rm Box}(T_3= 1/2)~ =~ \lambda_i {{G_{\rm F}}\over{\sqrt 2}}
   {{\alpha}\over{2\pi \sin^2\Theta_{\rm W}}} \lbrack -4 B_0(x_i)\rbrack 
   (\bar s d)_{V-A} (\bar\nu\nu)_{V-A}
\end{equation}
\begin{equation}
 \bar s Z d~ =~i \lambda_i {{G_{\rm F}}\over{\sqrt 2}} {{e}\over{2\pi^2}} 
M^2_Z
   {{\cos\Theta_{\rm W}}\over{\sin\Theta_{\rm W}}} C_0(x_i) \bar s \gamma_\mu 
   (1-\gamma_5)d
\end{equation}
\begin{equation}
 \bar s\gamma d~ =~- i\lambda_i {{G_{\rm F}}\over{\sqrt 2}} {{e}\over{8\pi^2}}
   D_0(x_i) \bar s (q^2\gamma_\mu - q_\mu \not\!q)(1-\gamma_5)d 
\end{equation}
\begin{equation}
 \bar s G^a d~ =~ -i\lambda_i{{G_{\rm F}}\over{\sqrt 
2}} {{g_s}\over{8\pi^2}}
   E_0(x_i) \bar s_{\alpha}(q^2\gamma_\mu - q_\mu \not\!q)
(1-\gamma_5)T^a_{\alpha\beta}d_\beta 
\end{equation}
\begin{equation}\label{MGP}
 \bar s \gamma' b~ =~i\bar\lambda_i {{G_{\rm F}}\over{\sqrt 2}} {{e}\over
   {8\pi^2}} D'_0(x_i) \bar s \lbrack i\sigma_{\mu\lambda} q^\lambda
   \lbrack m_b (1+\gamma_5) \rbrack\rbrack b
\end{equation}
\begin{equation}\label{FRF}
 \bar s G'^a b~ =~ 
i\bar\lambda_i{{G_{\rm F}}\over{\sqrt 2}}{{g_s}\over{8\pi^2}}
   E'_0(x_i)\bar s_{\alpha} \lbrack i\sigma_{\mu\lambda} q^\lambda
   \lbrack m_b (1+\gamma_5) \rbrack\rbrack T^a_{\alpha\beta} b_\beta \,,
\end{equation}
where $\lambda_i=V^*_{is}V_{id}$ and $\bar\lambda_i=V^*_{is}V_{ib}$.
Here $q_\mu$ is the {\it outgoing} gluon or photon momentum.
Moreover we have set $m_s=0$ in the last two rules.
The rules in (\ref{FR})-(\ref{FRF}) correct for the correspoding rules 
given in \cite{BH}
which contained unfortunately some misprints.
Together with the rules of figs.\ 2 and 3 they allow the calculation of
the effective Hamiltonians for FCNC processes without the inclusion
of QCD corrections. To this end some care
is needed. The penguin vertices should be used in the same way as the
elementary vertices of fig.\ 3 and 
which follow from $i {\cal L}$. Once a
mathematical expression corresponding to a given diagram has been found,
the contribution of this diagram to the relevant effective Hamiltonian
is obtained by multiplying this mathematical expression by ``i". On the
other hand our conventions for the box vertices are such that they directly 
give the contributions to the effective Hamiltonians. We will give an
example below by calculating the internal top contributions to 
$K^+\to \pi^+\nu\bar\nu$.
First, however, we would like to make
general remarks emphasizing the new features of these effective
vertices as compared to the ones of fig.\ 1.:

\begin{figure}[hbt]
\centerline{
\epsfysize=1.6in
\epsffile{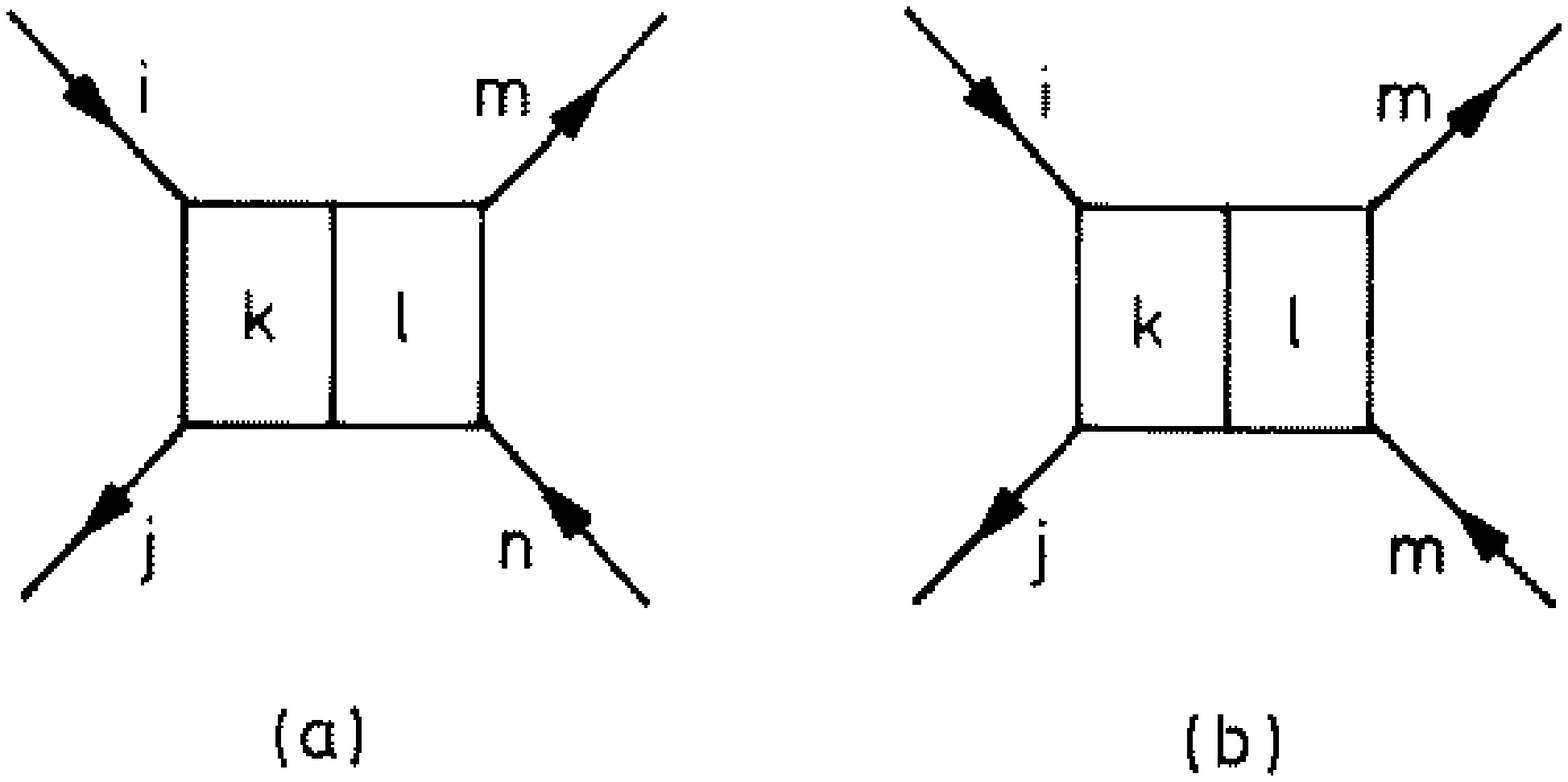}
}
\caption[]{
Box vertices 
\label{fig:6}}
\end{figure}   

\begin{figure}[hbt]
\centerline{
\epsfysize=3.2in
\epsffile{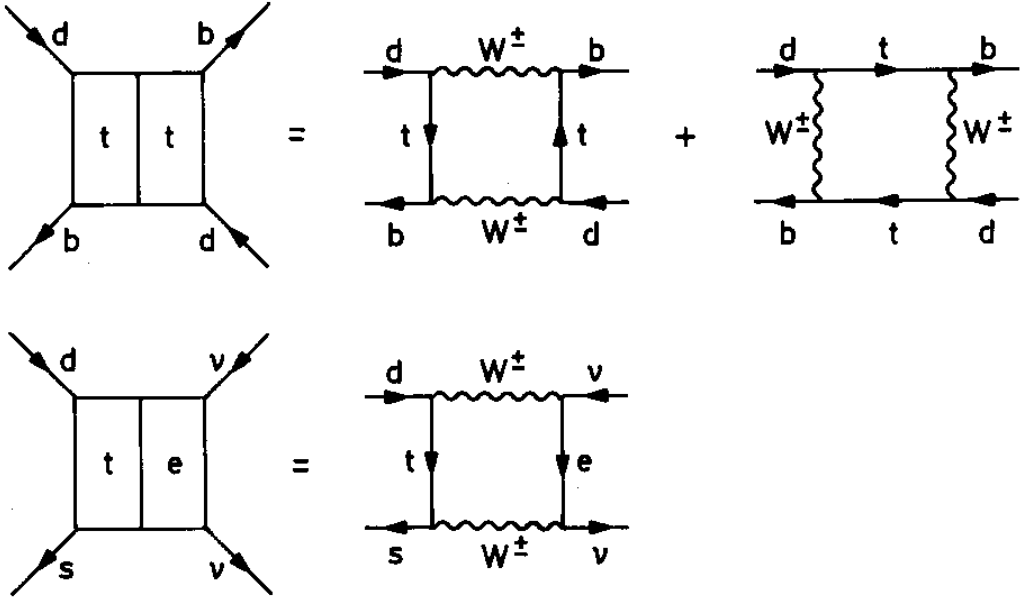}
}
\caption[]{
Box vertices resolved in terms of elementary vertices
\label{fig:7}}
\end{figure}

\begin{itemize}
\item
They are higher order in the gauge couplings and consequently
suppressed relative to elementary transitions.
\item
Because of the internal $W^{\pm}$ exchanges all penguin
vertices in fig.\ 5 are purely $V-A$, i.e.\ the effective vertices
involving $\gamma$ and $G$ are parity violating as opposed to their
elementary interactions in fig.\ 1! Also the structure of the $Z^0$ coupling
changes since now only $V-A$ couplings are involved. The box vertices
are of the $(V-A)\otimes(V-A)$ type.
\item
The effective vertices depend on the masses of internal
quarks or leptons and consequently are calculable functions of
\begin{equation}
x_i = {{m_i^2}\over{M_W^2}},\qquad i=u,c,t.
\end{equation}
A set of basic universal functions can be found. These functions
govern the physics of all FCNC processes. They are given below.
The masses of internal leptons except for the $\tau$ contribution
to $K^+\to\pi^+\nu\bar\nu$ can be set to zero.
\item
The effective vertices depend on elements of the CKM
matrix and this dependence can be found directly from the diagrams of
figs.\ 5 and 7.
\item
The dependence on external fermions manifests itself in two
ways. First the CKM factors and the type of internal fermions depend
on the external fermions considered. This in turn has an impact on the
argument $x_i$ of the basic function and consequently on the strength of
the vertex in question.
\item
The second dependence enters when one considers mass
effects of external fermions. Since generally these masses are
substantially smaller than $M_W$, it suffices to include this
dependence to first order in $m_{\rm ext}/M_W$. 
In this case one can summarize the
effects of $m_{\rm ext}$ by introducing new effective vertices without
changing the structure of the vertices of figs.\ 5
and 7 which have been obtained by setting $m_{\rm ext}=0$. For all
practical purposes only external mass effects in penguin diagrams need
to be considered. The new vertices are then described as in fig.\ 8
where the cross indicates which external mass has been taken into
account. These vertices are proportional to $m_{\rm ext}$, introduce new
$x_i$ dependent functions and have different Dirac structure as seen in 
the last two rules of (\ref{FR})-(\ref{FRF}). They have, 
however, the same dependence on the CKM parameters as the
corresponding vertices with $m_{\rm ext}=0$. It turns out that only the
external mass effects in photonic and gluonic vertices are relevant.
\item
Another new feature of the effective vertices of figs.\ 5, 7
and 8 as compared with the elementary vertices is their dependence on
the gauge used for the $W^{\pm}$ propagator. We will return to this
point below.
\end{itemize}

\begin{figure}[hbt]
\centerline{
\epsfysize=1.6in
\epsffile{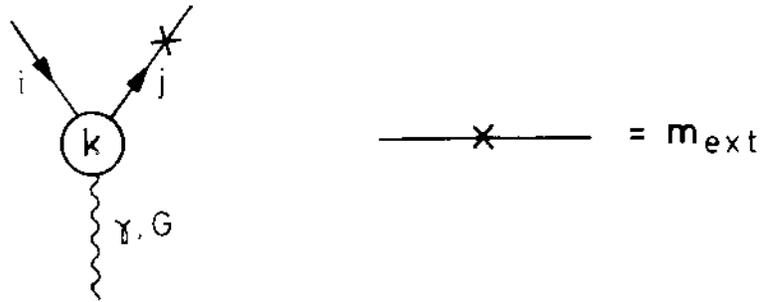}
}
 \caption[]{
External mass in an effective penguin vertex (``magnetic
penguin'')
\label{fig:8}}
\end{figure}

\subsubsection{Basic Functions}
The basic functions present in (\ref{FR})-(\ref{FRF}) 
were calculated by various authors, in
particular by Inami and Lim \cite{IL}.  
They are given explicitly as follows:
\begin{equation}\label{BF}
B_0(x_t)={1\over
4}\left[{x_t\over{1-x_t}}+{{x_t\ln x_t}\over{(x_t-1)^2}}\right]
\end{equation}
\begin{equation}\label{C0xt}
C_0(x_t)={x_t\over 8}\left[{{x_t-6}\over{x_t-1}}+{{3x_t+2}
\over{(x_t-1)^2}}\;\ln x_t\right] 
\end{equation}
\begin{equation}
D_0(x_t)=-{4\over9}\ln x_t+{{-19x_t^3+25x_t^2}\over{36(x_t-1)^3}}
+{{x_t^2(5x_t^2-2x_t-6)} \over{18(x_t-1)^4}}\ln x_t
\end{equation}
\begin{equation}\label{E0}
E_0(x_t)=-{2\over 3}\ln x_t+{{x_t^2(15-16x_t+4x_t^2)}\over{6(1-x_t)^4}}
\ln x_t+{{x_t(18-11x_t-x_t^2)} \over{12(1-x_t)^3}}
\end{equation}
\begin{equation}
D'_0(x_t)= -{{(8x_t^3 + 5x_t^2 - 7x_t)}\over{12(1-x_t)^3}}+ 
          {{x_t^2(2-3x_t)}\over{2(1-x_t)^4}}\ln x_t
\end{equation}
\begin{equation}
E'_0(x_t)=-{{x_t(x_t^2-5x_t-2)}\over{4(1-x_t)^3}} + {3\over2}
{{x_t^2}\over{(1 - x_t)^4}} \ln x_t
\end{equation}
\begin{equation}\label{S0}
S_0(x_t)=\frac{4x_t-11x^2_t+x^3_t}{4(1-x_t)^2}-
 \frac{3x^3_t \ln x_t}{2(1-x_t)^3}
\end{equation}
\begin{equation}\label{BFF}
S_0(x_c, x_t)=x_c\left[\ln\frac{x_t}{x_c}-\frac{3x_t}{4(1-x_t)}-
 \frac{3 x^2_t\ln x_t}{4(1-x_t)^2}\right]\,,
\end{equation}
where in the last expression we keep only linear terms in $x_c\ll 1$, but of 
course all orders in $x_t$. The subscript ``$0$'' indicates that 
these functions
do not include QCD corrections to the relevant penguin and box diagrams.
These corrections will be discussed in subsequent sections. 

The functions $D_0$ and $D'_0$ given here are valid for internal 
up--type
quarks. Denoting by $\tilde D_0$ and $\tilde D'_0$ the corresponding 
functions
involving internal down--type quarks, one has
\begin{equation}
\tilde D_0(x_b)=D_0(x_b)-E_0(x_b);\quad \tilde D'_0(x_b)=D'_0(x_b)-E'_0(x_b).
\end{equation}
In writing the expressions in (\ref{BF})-(\ref{BFF}) we have 
omitted $x_t$--independent 
terms which
do not contribute to decays due to the GIM mechanism. Moreover 
\begin{equation}
S_0(x_t)\equiv
F(x_t,x_t) + F(x_u,x_u) - 2 F(x_t,x_u)
\end{equation}
 and 
\begin{equation}
S_0(x_i,x_j)=F(x_i,x_j) + F(x_u,x_u)
- F(x_i,x_u) - F(x_j,x_u),
\end{equation}
where $F(x_i,x_j)$ is the true function
  corresponding to the box diagram. In this way the effective Hamiltonian for
  $\Delta S=2$ transitions as given in section 4 can be directly obtained in 
the usual form by  summing only over $t$ and $c$ quarks.

\subsection{Effective Hamiltonians for FCNC Transitions and GIM Mechanism}

With the help of the Feynman rules given in figs.\ 2 and 3 and in 
(\ref{FR})-(\ref{FRF}) 
it is an easy
matter to construct an effective Hamiltonian for any FCNC process. As
an example consider the decay $K^+\to \pi^+\bar\nu_e\nu_e$ to which
the diagrams in fig.\ 9 contribute.

\begin{figure}[hbt]
\centerline{
\epsfysize=1.6in
\epsffile{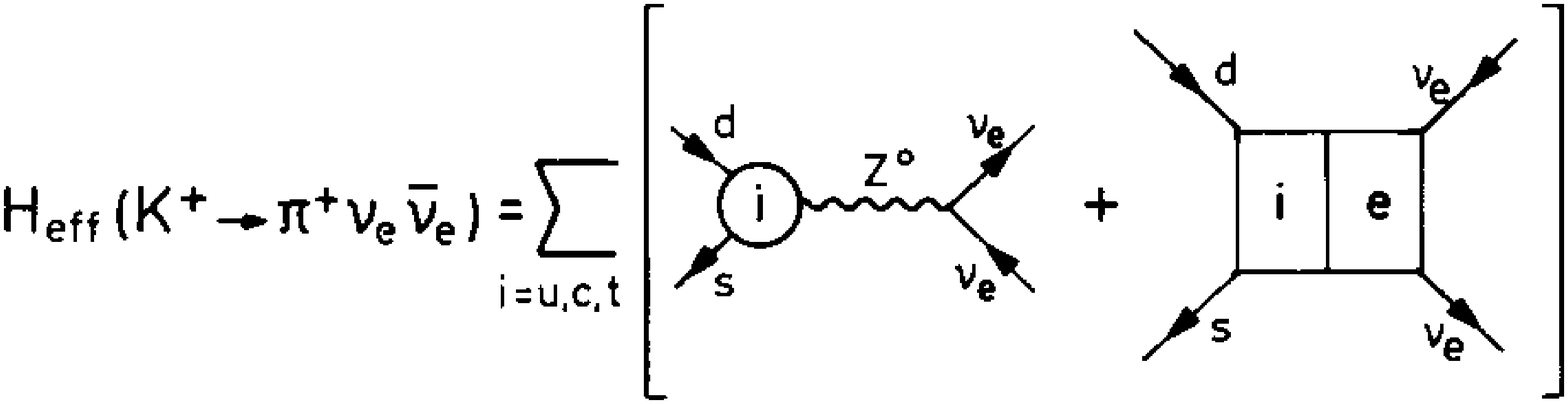}
}
\caption[]{
Calculation of ${\cal H}_{\rm eff}(K^+\to\pi^+\nu_e\bar\nu_e)$
\label{fig:9}}
\end{figure}

Replacing the $Z^0$ propagator by $i g_{\mu\nu}/M_Z^2$, using the rules
of figs.\ 2 and 3 and 
(\ref{FR})-(\ref{FRF}), and multiplying the first diagram 
by ``i", we
find the well-known result for the top contribution to this decay:
\begin{equation}
{\cal H}_{\rm eff}(K^+\to\pi^+\nu_e\bar\nu_e) = {{G_{\rm F}}\over{\sqrt2}}\;
{{\alpha}\over{2\pi\sin^2\Theta_{\rm W}}}\; V_{ts}^* V_{td}\; 
[C_0(x_t)-4 B_0(x_t)]\;
(\bar
s d)_{V-A} (\bar\nu_e\nu_e)_{V-A}.
\end{equation}

For decays involving photonic and/or gluonic penguin vertices, the
$1/q^2$ in the propagator cancels the $q^2$ in the vertex and the
resulting effective Hamiltonian can again be written in terms of local
four--fermion operators. Thus generally an effective Hamiltonian for
any decay considered can be written in the absence of QCD corrections
as
\begin{equation}\label{exp}
{\cal H}_{\rm eff}^{\rm FCNC}=\sum_k C_k O_k,
\end{equation}
where $O_k$ denote local operators such as $(\bar s d)_{V-A}(\bar s
d)_{V-A}$, $(\bar s d)_{V-A}(\bar uu)_{V-A}$ etc. The coefficients
$C_k$ of these operators are simply linear combinations of the
functions of eq. (\ref{BF})-(\ref{BFF}) times the corresponding 
CKM factors which can
be read off from our rules. Later we will exhibit these CKM factors. 
The fact that the coefficients $C_k$ for
any process considered can be expressed in terms of universal
functions (\ref{BF})-(\ref{BFF}) demonstrates the usefulness 
of the formulation
of FCNC decays in terms of effective vertices. We will encounter many
examples of the expansion (\ref{exp}) in the course of this review.

At this stage it is useful to return to the GIM mechanism which did
not allow tree level FCNC transitions. This mechanism is also felt in
the Hamiltonian of (\ref{exp}) and in fact it is fully effective when
the masses of internal quarks of a given charge  are set to be equal, e.g.\
$m_u=m_c=m_t$. Indeed the CKM factors in any FCNC process enter in the
combinations
\begin{equation}\label{CK}
C_k \propto \sum_{i=u,c,t}\lambda_i\; R(x_i)~~~~{\rm or}~~~~\sum_{i,j=u,c,t}
\lambda_i\lambda_j\; \tilde R(x_i,x_j),
\end{equation}
where $R,\tilde R$ denote any of the functions of (\ref{BF})-(\ref{BFF}), and
the $\lambda_i$ are given in the case of $K$ and $B$ meson decays and 
particle--antiparticle
mixing as follows:
\begin{equation}
\lambda_i=\cases{V_{is}^*V_{id}&~~~$K$--decays, ~~$K^0-\bar K^0$\cr
                   V_{ib}^*V_{id}&~~~$B$--decays, ~~$B^0_d-\bar B^0_d$\cr
                   V_{ib}^*V_{is}&~~~$B$--decays, ~~$B^0_s-\bar B^0_s$\cr} 
\end{equation}
They satisfy the unitarity relation
\begin{equation}
\lambda_u + \lambda_c + \lambda_t =0,
\end{equation}
which implies vanishing  coefficients $C_k$ in (\ref{CK}) if
$x_u=x_c=x_t$. For this reason the mass--independent terms in the calculation
of the basic functions in (\ref{BF})-(\ref{BFF})  can always be omitted.
In this limit, FCNC decays and transitions are absent.
Thus beyond tree level the conditions for a complete GIM
cancellation of FCNC processes are:
\begin{itemize}
\item
Unitarity of the CKM matrix
\item
Exact horizontal flavour symmetry which assures the equality
of quark masses of a given charge.
\end{itemize}

\noindent
It should be emphasized that such a horizontal symmetry is very
natural, as the quantum numbers of all fermions of a given charge are
equal in the Standard Model and so these fermions can be naturally put
into multiplets of some horizontal symmetry group. 
Now in nature such a horizontal symmetry, even if it exists
at very short distance scales, is certainly broken at low energies by
the disparity of masses of quarks of a given charge. This in fact is
the origin of the breakdown of the GIM mechanism at the one--loop
level and the appearance of FCNC transitions. 
The size of this breakdown, and consequently the size of 
FCNC transitions, depends on the disparity of masses,
on the behaviour of the basic functions of (\ref{BF})-(\ref{BFF}), and can be
affected by QCD corrections as we will see below. Let us make two
observations: 
\begin{itemize}
\item
For small $x_i\ll 1$, relevant for $i\not= t$, 
the functions (\ref{BF})-(\ref{BFF})
behave as follows:
\begin{equation}
S_0(x_i)\propto x_i, \quad 
  B_0(x_i)\propto x_i \ln x_i, \quad
  C_0(x_i)\propto x_i \ln x_i 
\end{equation}
\begin{equation}
D_0(x_i)\propto \ln x_i, \quad
  E_0(x_i)\propto \ln x_i,  \quad
  D'_0(x_i)\propto x_i, \quad 
  E'_0(x_i)\propto x_i. 
\end{equation}
This implies ``hard'' (quadratic) GIM suppression for processes 
governed by the
functions $S,B,C,D',E'$ provided the top quark contributions due 
to small CKM
factors can be neglected. In the case of $D(x_i)$ and $E(x_i)$ only ``soft''
(logarithmic) GIM suppression is present.
\par
\item
For large $x_t$ we have
\begin{equation}
S_0(x_t)\propto x_t, \quad 
  B_0(x_t)\propto {\rm const}, \quad
  C_0(x_t)\propto x_t
\end{equation}
\begin{equation}
D_0(x_t)\propto \ln x_t, \quad
  E_0(x_t)\propto {\rm const},  \quad
  D'_0(x_t)\propto {\rm const}, \quad 
  E'_0(x_t)\propto {\rm const}. 
\end{equation}
\end{itemize}
Thus for processes governed by top quark contributions, the GIM suppression is
not effective at the one loop level and in fact in the case of decays and
transitions receiving contributions from $S_0(x_t)$ and $C_0(x_t)$ some 
important
enhancement is possible. 
\par

The latter property emphasizes the special role of $K$ and $B$ decays with
regard to FCNC transitions. In these decays the appearance of the
top quark in the internal loop with $m_t>M_W\gg m_c,m_u$ removes the
GIM suppression, making $K$ and $B$ decays a particularly
useful place to test FCNC transitions and to study the physics of
the top quark. Of course the hierarchy of various FCNC transitions is also
determined by the hierarchy of the elements of the CKM matrix allowing
this way to perform sensitive tests of this sector of the Standard
Model. 

The FCNC decays of $D$--mesons are much stronger suppressed because only
$d$, $s$, and $b$ quarks with $m_d,m_s,m_b\ll M_W$ enter internal loops
and the GIM mechanism is much more effective. Also the known structure
of the CKM matrix is less favorable than in $K$ and $B$ decays. For these
reasons we will restrict our presentation to the latter.
In the extensions of the Standard Model, FCNC transitions are possible
at the tree level and the hierarchies discussed here may not apply.

The formalism developed so far is not complete because it does not
include QCD corrections. Moreover we did not address the
classification of the local operators $O_i$ and we have not shown how
to translate the calculations done in terms of quarks into predictions
for the decays of their bound states, the hadrons. These issues 
will be the topics of the following subsection.

\subsection{QCD, OPE and Renormalization Group}\label{SecRG}
\subsubsection{Preliminary Remarks}

An amplitude for a decay of a given meson $M=K,B,\ldots\;$ into a final
state $F$ is simply given by 
\begin{equation}
A(M\to F) = \langle F\mid{\cal H}_{\rm eff}\mid M\rangle\,,
\end{equation}
where ${\cal H}_{\rm eff}$ is the relevant Hamiltonian such as given
in (\ref{exp}). Since all Hamiltonians considered can be  written
as linear combinations of local four--fermion operators, the result
for the decay amplitude is generally given by
\begin{equation}\label{AMF}
A(M\to F) = \sum_i C_i\langle F\mid O_i\mid M\rangle .
\end{equation}
In the case of $B^0-\bar B^0$ mixing and $K^0-\bar K^0$ mixing,
 $\mid M\rangle$ and $\langle F\mid$ have to be
changed appropriately.

Before further discussing (\ref{AMF}) we have to elaborate on QCD
corrections to weak decays. Clearly these decays originate in weak
transitions mediated by $W^{\pm}$ and $Z^0$. However, the presence of
strong and electromagnetic interactions often has an important impact
on weak decays and consequently these interactions have a natural
place in the physics of quark mixing, CP violation and rare decays. We
have already seen the existence of FCNC transitions involving photonic
and gluonic penguin diagrams.
As far as electromagnetic interactions are  concerned,
it is sufficient to work to first order
in $\alpha$. However, the case of strong interactions is very
different and must be carefully investigated.

Due to the fact that $W^{\pm}$ and $Z^0$ are very massive, the basic
weak transitions take place at very short distance scales
${\cal O}(1/M_{W,Z}^2)$. The strong interactions being active at both short
and long distances change this picture in the case of hadron decays,
and generally weak decays of hadrons receive contributions from both short
and long distances.

Now due to asymptotic freedom present in QCD, its effective
coupling constant $\alpha_s(\mu)$ becomes small at $\mu=M_{W,Z}$. In
the two loop approximation this running coupling constant is given by
\begin{equation}\label{QCDC}
{{\alpha_s(\mu)}\over {4\pi}}= {{\bar g^2_s(\mu)}\over{16\pi^2}} =
{{1}\over{\beta_0 \ln(\mu^2/\Lambda^2_{\overline{MS}})}}
- {{\beta_1}\over{\beta^3_0}} {{\ln \ln (\mu^2/\Lambda^2_{\overline{MS}})}
\over
{\ln^2(\mu^2/\Lambda^2_{\overline{MS}})}} \,,
\end{equation} 
where $\beta_0=(33-2f)/3$ and $\beta_1=(306-38f)/3$ with $f$ being the
number of ``effective'' flavours. What ``effective'' means here will be
explained below.
Roughly speaking $f=6$ for $\mu\geq\mt$, $f=5$ for $\mb\leq\mu\leq\mt$,
$f=4$ for $\mc\leq\mu\leq\mb$ and $f=3$ for $\mu\leq\mc$.
$\Lambda_{\overline{MS}}$ 
is the QCD scale parameter \cite{BBDM} which generally depends on $f$.
Denoting by $\alpha^{(f)}_s$ the effective coupling constant for a
theory with $f$ effective flavours and by $\Lambda^{(f)}_{\overline{MS}}$ 
the corresponding QCD scale parameter, we have the following boundary
conditions which follow from the continuity of $\alpha_s$:
\begin{equation}
\alpha_s^{(6)}(\mt)=\alpha_s^{(5)}(\mt),
\qquad
\alpha_s^{(5)}(\mb)=\alpha_s^{(4)}(\mb),
\qquad
\alpha_s^{(4)}(\mc)=\alpha_s^{(3)}(\mc).
\end{equation}
These conditions allow to find values of $\Lambda^{(f)}_{\overline{MS}}$
for different $f$ once one particular $\Lambda^{(f)}_{\overline{MS}}$
is known. In table \ref{tab:alphas} we show different 
$\alpha^{(f)}_s(\mu)$ and
$\Lambda^{(f)}_{\overline{MS}}$ corresponding to
\begin{equation}
\alpha_s^{(5)}(\mz)=0.118 \pm 0.005,
\end{equation}
which is in the ball park of the present world average extracted from
different processes \cite{Schmelling}.
We observe that for $\mu\geq\mc $ the values of
$\alpha_s(\mu)$ are sufficiently small that
the effects of strong interactions can be
treated in perturbation theory. When one
moves to low energy scales, $\alpha_s$ increases and at
$\mu\approx {\cal O}(1~\gev)$ and high values of 
$\Lambda^{(3)}_{\overline{MS}}$  one finds
$\alpha_s^{(3)}(\mu)>0.5$. This signals breakdown of
perturbation theory for scales lower than $1~\gev$. Yet it is
gratifying that strong interaction contributions to weak decays coming
from scales higher than $1~\gev$ can be treated by perturbative
methods.

\begin{table}[thb]
\begin{center}
\begin{tabular}{|c|c|c|c|c|c|}\hline
 $\alpha_s^{(6)}(\mt)$& $0.1037$& $0.1054$& $0.1079$ & $0.1104$ &  $0.1120$ 
\\ \hline
 $\Lms^{(6)}[\mev]$& $66$& $76$& $92$ & $110$ &  $123$ 
\\ \hline\hline
 $\alpha_s^{(5)}(\mz)$& $0.113$& $0.115$& $0.118$ & $0.121$ &  $0.123$ 
\\ \hline
 $\Lms^{(5)}[\mev]$& $169$& $190$& $226$ & $267$ &  $296$ 
\\ \hline
 $\alpha_s^{(5)}(\mb)$& $0.204$& $0.211$& $0.222$ & $0.233$ &  $0.241$ 
\\ \hline\hline
 $\Lms^{(4)}[\mev]$& $251$& $278$& $325$ & $376$ &  $413$ 
\\ \hline
 $\alpha_s^{(4)}(\mc)$& $0.336$& $0.357$& $0.396$ & $0.443$ &  $0.482$ 
\\ \hline\hline
 $\Lms^{(3)}[\mev]$& $297$& $325$& $372$ & $421$ &  $457$ 
\\ \hline
 $\alpha_s^{(3)}(1\gev)$& $0.409$& $0.444$& $0.514$ & $0.605$ &  $0.690$ 
\\ \hline
 \end{tabular}
\end{center}
\caption[]{Values of $\alpha^{(f)}_s(\mu)$ and
$\Lambda^{(f)}_{\overline{MS}}$ corresponding to
$\alpha_s^{(5)}(\mz)=0.113$, 0.115, 0.118, 0.121, 0.123
with
$\mc=1.3~\gev$, $\mb=4.4~\gev$ and $\mt=170~\gev$
\label{tab:alphas}}
\end{table}

The impact of QCD effects on weak decays depends crucially on the
process considered, which is clearly seen when leptonic,
semi--leptonic and non--leptonic decays are compared with each other.

\begin{figure}[hbt]
\centerline{
\epsfysize=3.8in
\epsffile{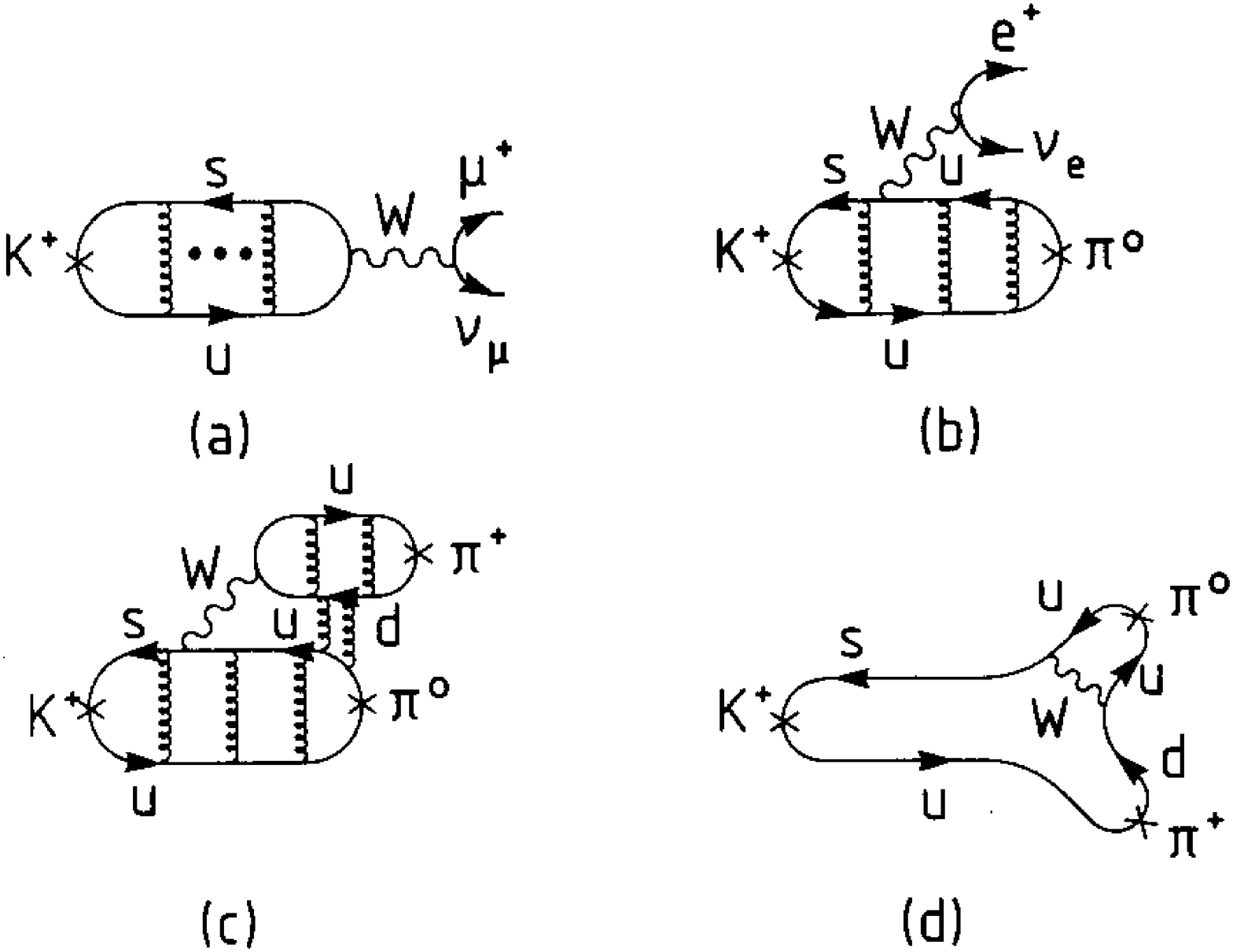}
}
\caption[]{
Examples of diagrams contributing to a) leptonic b)
semi--leptonic and c,d) non--leptonic decays. The curly lines are gluons.
\label{fig:10}}
\end{figure}

Consider for instance the {\it leptonic} decay $K^+\to\mu^+\nu_\mu$. One
has (fig.\ 10a)
\begin{equation}
A(K^+\to\mu^+\nu_\mu)= {{G_{\rm F}}\over{\sqrt2}} V_{us}^*\;(\bar\nu_\mu
\mu^-)_{V-A}\;\langle 0\mid(\bar su)_{V-A}\mid K^+\rangle.
\end{equation}
Since gluons do not connect the lepton and quark currents, this
factorized form of the amplitude (lepton current times the matrix
element of the quark current) remains valid in the presence of strong
interactions. In other words, cutting the $W$--propagator separates the
diagram into two simpler subdiagrams. The full effect of strong
interactions is then absorbed in the matrix element of the quark
current. Since there are no loops involving simultaneously $W^{\pm}$
and gluons, and the $K$ mass is low, the strong interaction effects
present in the current matrix element are purely long range and
must be treated by non--perturbative methods. Yet the leptonic decays
are the simplest ones because the effects of strong interactions can
be fully absorbed in the current matrix elements. The latter are
simple enough so that lattice calculations or QCD sum rules can give
plausible estimates for their values. Moreover they can be
determined experimentally. The knowledge of $\langle 0\mid(\bar
su)_{V-A}\mid K^+\rangle$ determines $F_K$ and the knowledge of analogous
matrix elements fixes the decay constants of other mesons.

{\it Semileptonic} decays are slightly more complicated. However, the
factorization of the amplitude into a lepton current and a matrix
element of the relevant quark current remains also true here as seen
in fig.~10b:
\begin{equation}
A(K^+\to\pi^o e^+\nu)= {{G_{\rm F}}\over{\sqrt2}} V_{us}^* (\bar\nu_e
e^-)_{V-A} \langle\pi^0\mid(\bar su)_{V-A}\mid K^+\rangle\,.
\end{equation}
Again the strong interactions are compactly collected in the matrix
element 
$\langle\pi^0\mid (\bar su)_{V-A} \mid K^+\rangle$ 
which can in
principle be extracted from experimental data or calculated by
non--per\-tur\-ba\-tive methods. Since this time the matrix element 
involves two meson states, its evaluation is more difficult and only
on the border of lattice capabilities. For these reasons several
models for these matrix elements have been invoked. 
Moreover, in the case of $K$ decays, chiral perturbation theory 
turns out to be useful. 
On the other hand, matrix elements involving $B$ mesons can be
efficiently studied in the Heavy Quark Effective Theory. 
Furthermore, in inclusive semi--leptonic decays of heavy quarks QCD 
corrections
resulting from real gluon emission can be calculated perturbatively.
These issues are discussed by Neubert in a separate chapter in this
book.

The {\it non--leptonic} decays such as $K\to\pi\pi$ or $B \to D K$ are more
complicated to analyze and to calculate because the factorization of
a given matrix element of a four--fermion operator into the product of
current matrix elements is no longer true. Indeed now the gluons can
connect the two quark currents (fig.\ 10c), and in addition the diagrams
of fig.\ 10d contribute. The breakdown of factorization in
non--leptonic decays is present both at short and long distances
simply because the effects of strong interactions are felt both at
large and small momenta. At large momenta, however, the QCD coupling
constant is small and the non--factorizable contributions can be
studied in perturbation theory. In order to accomplish this task, one
has to separate first short distance effects from long distance
effects. This is most elegantly done by means of the operator product
expansion approach (OPE) combined with the renormalization group. In
order to discuss these methods we have to say a few words about the
effective field theory picture which underlies our discussion presented
so far.

\subsubsection{Effective Field Theory Picture}
The basic framework for weak decays of hadrons containing $u$, $d$, $s$, $c$
and $b$ quarks is the effective field theory relevant for scales $ \mu \ll
M_W, M_Z , m_t$. This framework, as we have seen above, 
brings in local operators which govern ``effectively''
the transitions in question. From the point of view of the decaying
hadrons containing the lightest five quarks this is the only correct 
picture we know and also the most efficient one for studying the
presence of QCD. Furthermore it represents the generalization of the Fermi
theory as formulated by Sudarshan and Marshak \cite{SUMA} and 
Feynman and Gell-Mann \cite{GF} forty years ago.

Indeed the simplest effective Hamiltonian without QCD effects that
one would find 
from the first diagram of fig.~11 is (see (\ref{WE}))
\begin{equation}\label{Heff}
{\cal H}^{0}_{\rm eff} = \frac{G_{\rm F}}{\sqrt 2} V_{cb} V_{cs}^*
(\bar c b)_{V-A}(\bar s c)_{V-A}\,,
\end{equation}  
where $G_{\rm F}$ is the Fermi constant, $V_{ij}$ are the relevant CKM
factors and
\begin{equation}\label{OP}
(\bar c b )_{V-A}(\bar s c)_{V-A}\equiv
(\bar c \gamma_{\mu} (1-\gamma_5) b)
(\bar s \gamma_{\mu} (1-\gamma_5) c)=Q_2
\end{equation}
is a $(V-A)\cdot (V-A)$ current-current local operator  usually denoted
by $Q_2$. The situation in the Standard Model is, however, more complicated
because of the presence of additional interactions which effectively
generate new operators. These are in particular the gluon, photon and
$Z^0$-boson exchanges and internal top contributions as we have seen above.
 Some of the elementary
interactions of this type are shown this time for $B$ decays in fig.~11. 
Consequently the relevant effective Hamiltonian for $B$-meson decays involves 
generally several 
operators $Q_i$ with various colour and Dirac structures which are different
from $Q_2$. Moreover each operator is multiplied by a calculable
coefficient $C_i(\mu)$:
\begin{equation}\label{HOPE}
{\cal H}_{\rm eff} = \frac{G_{\rm F}}{\sqrt 2} V_{\rm CKM} \sum_i
   C_i (\mu) Q_i,   
\end{equation}
where the scale $\mu$ is discussed below and $V_{\rm CKM}$ denotes the 
relevant 
CKM factor. Analogous expressions apply to $K$ and $D$ decays with an
appropriate change of flavours.

At this stage it should be mentioned that the usual Feynman
diagram drawings of the type shown in fig.~11 containing full $W$-propagators,
$ Z^0-$propagators and top-quark propagators represent
really the happening at scales $ {\cal O}(\mw)$ whereas the true picture
of a decaying hadron is more correctly described by 
the local operators in question.
Thus, whereas at scales $ {\cal O}(\mw) $ we have to deal with 
the full six-quark theory
containing the photon, weak gauge bosons and gluons, 
at scales ${\cal O}(1\gev)$  the
relevant effective theory contains only three light quarks $u$, $d$ and $s$,
gluons and the photon. At intermediate energy scales $\mu={\cal O}(m_b)$
and $\mu={\cal O}(m_c)$
relevant for beauty and charm decays, effective five-quark
and effective four-quark theories have to be considered, respectively.
 
The usual procedure then is to start at a high energy scale 
${\cal O}(M_W) $
and consecutively integrate out the heavy degrees of freedom (heavy with
respect to the relevant scale $ \mu $) from explicitly appearing in the
theory. The word ``explicitly'' is very essential here. The heavy fields
did not disappear. Their effects are merely hidden in the effective
gauge coupling constants, running masses and most importantly in the
coefficients describing the ``effective'' strength of the operators at a
given scale $\mu $, the Wilson coefficient functions $C_i(\mu)$. 

\subsubsection{OPE and Renormalization Group}
The Operator Product Expansion (OPE) combined with the renormalization group
approach can be regarded as a mathematical formulation of the picture
outlined above. 
In this framework the amplitude for an {\it exclusive} decay $M\to F$ 
is written as
\begin{equation}\label{OPE}
 A(M \to F) = \frac{G_{\rm F}}{\sqrt 2} V_{\rm CKM} \sum_i
   C_i (\mu) \langle F \mid Q_i (\mu) \mid M \rangle,
\end{equation}
which generalizes (\ref{AMF}) to include QCD corrections. 
$ Q_i$ denote
the local operators generated by QCD and electroweak interactions.
$ C_i(\mu) $ stand for the Wilson
coefficient functions (c-numbers). 
The following comments should be made:
\begin{itemize}
\item
The scale $ \mu $ separates the physics contributions in the ``short
distance'' contributions (corresponding to scales higher than $\mu $)
contained in $ C_i(\mu) $ and the ``long distance'' contributions
(scales lower than $ \mu $) contained in 
$\langle F\mid Q_i(\mu)\mid M\rangle $.
 By evolving the scale from $ \mu ={\cal O}(M_W) $ down to 
lower values of $ \mu $ one transforms the physics information 
at scales higher
than $ \mu $ from the hadronic matrix elements into $ C_i(\mu) $. Since no
information is lost this way the full amplitude cannot depend on $ \mu $.
This is the essence of the renormalization group equations which govern the
evolution $ (\mu -dependence) $ of $ C_i(\mu) $. This $ \mu $-dependence 
must be cancelled by the one present in $\langle  Q_i (\mu)\rangle $.
 It should be
stressed, however, that this cancellation generally involves many
operators due to the operator mixing under renormalization.
\item
The set of basic operators entering
the OPE and ``driving'' a given weak decay can be specified at short
distances, i.e.\ without solving the difficult non--perturbative
problem. Similarly the Wilson coefficient functions of these operators
can be calculated by means of perturbative methods as long as $\mu$ is
not too small, say $\mu\ge 1~\gev$.
\item
In view of two vastly different scales entering the analysis
($M_W\gg\mu\approx{\cal O}(1-5\gev)$), 
the usual perturbative expansion has
to be improved, however. Indeed large logarithms $(\ln M_W/\mu)$
multiplying $\alpha_s$ have to be resummed to all orders in
$\alpha_s$ before a reliable estimate of the $C_i$ can be
obtained. This can be done very efficiently by means of the
renormalization group methods as will be discussed in a moment. 
The resulting ``renormalization group
improved'' perturbative expansion for the $C_i$'s in terms of the
effective QCD coupling of (\ref{QCDC}) does not involve large logarithms and
is more reliable.
\end{itemize}
Let us then say a few words about the $\mu$ dependence of the Wilson
coefficients which is governed by the renormalization group. Many more
details can be found in a recent review \cite{BBL}.

The general expression for $ C_i(\mu) $ is given by:
\begin{equation}\label{CV}
 \vec C(\mu) = \hat U(\mu,M_W) \vec C(M_W),
\end{equation}
where $ \vec C $ is a column vector built out of $ C_i $'s.
$\vec C(M_W)$ are the initial conditions which depend on the 
short distance physics at high energy scales.
In particular they depend on $m_t$ and are generally linear
combinations of the basic functions in (\ref{BF})-(\ref{BFF}).
$ \hat U(\mu,M_W) $, the evolution matrix,
is given as 
\begin{equation}\label{UM}
 \hat U(\mu,M_W) = T_g \exp \left[ 
   \int_{g(M_W)}^{g(\mu)}{dg' \frac{\hat\gamma^T(g')}{\beta(g')}}\right] 
\end{equation}
with $g=g_s$ denoting the QCD effective coupling constant. 
$T_g$ denotes the ordering in the coupling $g$ so that the couplings
increase from right to left (see (\ref{UMNLO})).
$ \beta(g) $
governs the evolution of $g$ and $ \hat\gamma $ is the anomalous dimension
matrix of the operators involved. The structure of this equation
makes it clear that the renormalization group approach goes
 beyond usual perturbation theory.
Indeed $ \hat  U(\mu,M_W) $ sums automatically large logarithms
$ \log (M_W/\mu) $ which appear for $ \mu<<M_W $. In the so-called leading
logarithmic approximation (LO) terms $ (g^2\log M_W/\mu)^n $ are summed.
The next-to-leading logarithmic correction (NLO) to this result involves
summation of terms $ (g^2)^n (\log M_W/\mu)^{n-1} $ and so on.
This hierarchic structure gives the renormalization group improved
perturbation theory.

As an example let us consider only QCD effects and the case of a single
operator so that (\ref{CV}) reduces to

\begin{equation}\label{CC}
  C(\mu) =  U(\mu,M_W)  C(M_W),   
\end{equation}
where $C(\mu)$ denotes the coefficient of the operator in question.
 Keeping the first two terms in the expansions of
 $\gamma(g)$ and $\beta(g)$ in powers of $g$:
\begin{equation}
\gamma (g) = \gamma^{(0)} \frac{\alpha_s}{4\pi} + \gamma^{(1)}
\frac{\alpha^2_s}{16\pi^2} 
\quad , \quad
 \beta (g) = - \beta_0 \frac{g^3}{16\pi^2} - \beta_1 
\frac{g^5}{(16\pi^2)^2}
\end{equation}
and inserting these expansions into (\ref{UM}) gives
\begin{equation}\label{UMNLO}
 U (\mu, \mw) = \Biggl\lbrack 1 + {{\alpha_s (\mu)}\over{4\pi}} J
\Biggl\rbrack \Biggl\lbrack {{\alpha_s (M_W)}\over{\alpha_s (\mu)}}
\Biggl\rbrack^P \Biggl\lbrack 1 - {{\alpha_s (M_W)}\over{4\pi}} J
\Biggl\rbrack,
\end{equation}
where
\begin{equation}
P = {{\gamma^{(0)}}\over{2\beta_0}},~~~~~~~~~ J = {{P}\over{\beta_0}}
\beta_1 - {{\gamma^{(1)}}\over{2\beta_0}}. 
\end{equation} 
General
formulae for $ \hat U (\mu, M_W) $ in the case of operator mixing and
valid also for electroweak effects can be found in \cite{BBL,BJLW}. 
The leading
logarithmic approximation corresponds to setting $ J = 0 $ in (\ref{UMNLO})
and dropping the second term in (\ref{QCDC}).

At this stage we should say a few words about the renormalization scheme
dependence. The initial conditions $C(\mw)$ depend at the NLO level on
the renormalization scheme for operators. Similarly NLO corrections in
$U(\mu, \mw)$, represented by $J$ in (\ref{UMNLO}), are scheme dependent
through the scheme dependence of the two-loop anomalous dimensions
$\gamma^{(1)}$. The scheme dependence in the last factor in (\ref{UMNLO})
is cancelled by the scheme dependence of $C(\mw)$ and the scheme dependence
of $C(\mu)$ is entirely given by the first factor in (\ref{UMNLO}). This
scheme dependence is cancelled by the one present in the matrix element
$\langle F|Q_i (\mu)| M \rangle$ so that the resulting physical amplitudes
are scheme independent.

In this review we will entirely work in the $\overline{MS}$ renormalization
scheme and the only scheme dependence will be signalled by two different
treatments of $\gamma_5$ in $D\not=4$ dimensions:
\begin{itemize}
\item
{\bf NDR-scheme}: anti-commuting $\gamma_5$
\item
{\bf HV-scheme}: non-anti-commuting $\gamma_5$ \cite{HV}
\end{itemize}
Details of these schemes are discussed in \cite{BW,BMBW}.

Clearly in order to calculate the full amplitude (\ref{OPE}) also
the hadronic matrix elements $\langle F|Q_i (\mu)| M \rangle$ have to
be evaluated. Since they involve long distance contributions one is
forced in this case to use non-perturbative methods such as 
lattice calculations, the
$1/N$ expansion, QCD sum rules or chiral perturbation theory. In the
case of semi-leptonic $B$ meson decays also the Heavy Quark Effective Theory
(HQET) turns out to be a useful tool. 
In HQET the matrix elements are
evaluated approximately in an expansion in $1/m_b$. 
Potential uncertainties in the calculation of the non-leading terms
in this expansion have been stressed recently \cite{MS}.
Needless 
to say, all these non-perturbative methods have some limitations.
 Consequently the dominant theoretical
uncertainties in the decay amplitudes reside in
the matrix elements of $Q_i$.

\begin{figure}[hbt]
\centerline{
\epsfysize=5in
\epsffile{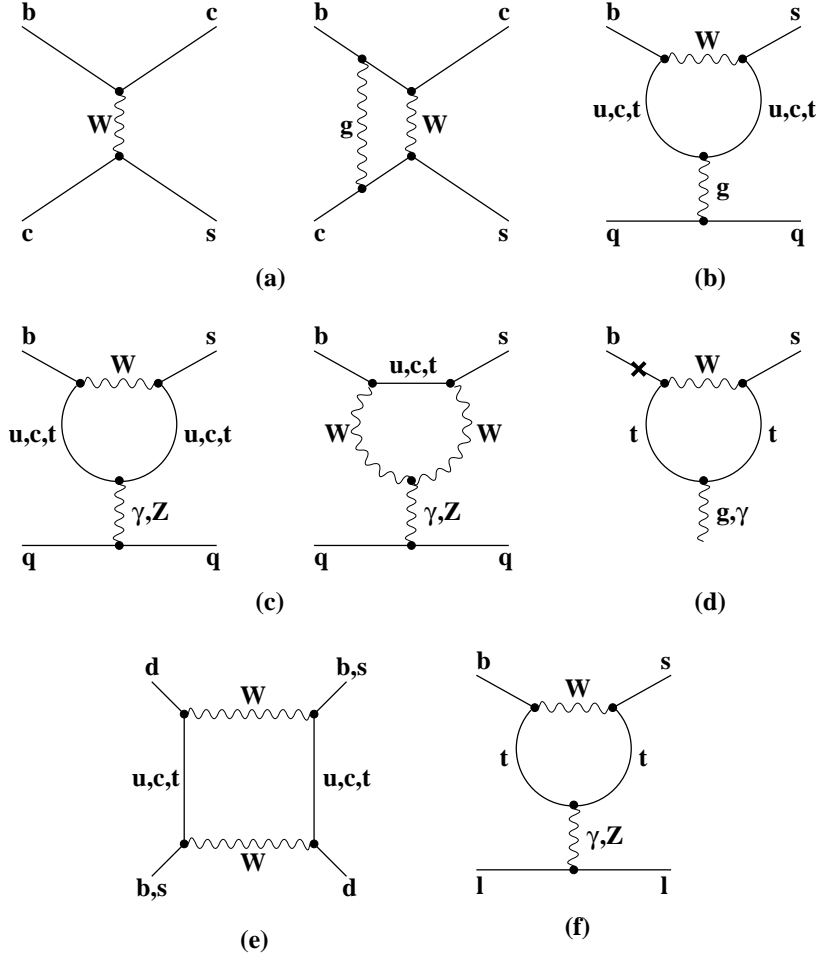}
}
\caption{Typical Penguin and Box Diagrams.}
\label{fig:fdia}
\end{figure}

\subsubsection{Classification of Operators}
Let us next systematically classify the operators which will appear
in the subsequent sections of this review and which play the
dominant role in the phenomenology of weak decays. Typical diagrams in
the full theory from which these operators originate are indicated
 and shown in fig.~11. The cross in fig.~11d indicates as in fig.~8
that magnetic penguins originate from the mass-term on the external
line in the usual QCD or QED penguin diagrams.
The six classes are given as follows ($\alpha$ and $\beta$ are colour
indices):

{\bf Current--Current (fig.~11a):}
\begin{equation}\label{O1} 
Q_1 = (\bar c_{\alpha} b_{\beta})_{V-A}\;(\bar s_{\beta} c_{\alpha})_{V-A}
~~~~~~Q_2 = (\bar c b)_{V-A}\;(\bar s c)_{V-A} 
\end{equation}

{\bf QCD--Penguins (fig.~11b):}
\begin{equation}\label{O2}
Q_3 = (\bar s b)_{V-A}\sum_{q=u,d,s,c,b}(\bar qq)_{V-A}~~~~~~   
 Q_4 = (\bar s_{\alpha} b_{\beta})_{V-A}\sum_{q=u,d,s,c,b}(\bar q_{\beta} 
       q_{\alpha})_{V-A} 
\end{equation}
\begin{equation}\label{O3}
 Q_5 = (\bar s b)_{V-A} \sum_{q=u,d,s,c,b}(\bar qq)_{V+A}~~~~~  
 Q_6 = (\bar s_{\alpha} b_{\beta})_{V-A}\sum_{q=u,d,s,c,b}
       (\bar q_{\beta} q_{\alpha})_{V+A} 
\end{equation}

{\bf Electroweak--Penguins (fig.~11c):}
\begin{equation}\label{O4} 
Q_7 = {3\over 2}\;(\bar s b)_{V-A}\sum_{q=u,d,s,c,b}e_q\;(\bar qq)_{V+A} 
~~~~~ Q_8 = {3\over2}\;(\bar s_{\alpha} b_{\beta})_{V-A}\sum_{q=u,d,s,c,b}e_q
        (\bar q_{\beta} q_{\alpha})_{V+A}
\end{equation}
\begin{equation}\label{O5} 
 Q_9 = {3\over 2}\;(\bar s b)_{V-A}\sum_{q=u,d,s,c,b}e_q(\bar q q)_{V-A}
~~~~~Q_{10} ={3\over 2}\;
(\bar s_{\alpha} b_{\beta})_{V-A}\sum_{q=u,d,s,c,b}e_q\;
       (\bar q_{\beta}q_{\alpha})_{V-A} 
\end{equation}

{\bf Magnetic--Penguins (fig.~11d):}
\begin{equation}\label{O6}
Q_{7\gamma}  =  \frac{e}{8\pi^2} m_b \bar{s}_\alpha \sigma^{\mu\nu}
          (1+\gamma_5) b_\alpha F_{\mu\nu}\qquad            
Q_{8G}     =  \frac{g}{8\pi^2} m_b \bar{s}_\alpha \sigma^{\mu\nu}
   (1+\gamma_5)T^a_{\alpha\beta} b_\beta G^a_{\mu\nu}  
\end{equation}

{\bf $\Delta S = 2 $ and $ \Delta B = 2 $ Operators (fig.\ 11e):}
\begin{equation}\label{O7}
Q(\Delta S = 2)  = (\bar s d)_{V-A} (\bar s d)_{V-A}~~~~~
 Q(\Delta B = 2)  = (\bar b d)_{V-A} (\bar b d)_{V-A} 
\end{equation}

{\bf Semi--Leptonic Operators (fig.~11f):}
\begin{equation}\label{9V}
Q_{9V}  = (\bar s b  )_{V-A} (\bar \mu\mu)_{V}~~~~~
Q_{10A}  = (\bar s b )_{V-A} (\bar \mu\mu)_{A}
\end{equation}

\begin{equation}\label{10V}
Q_{\nu\bar\nu}  = (\bar s b  )_{V-A} (\bar \nu\nu)_{V-A}~~~~~
Q_{\mu\bar\mu}  = (\bar s b )_{V-A} (\bar \mu\mu)_{V-A}
\end{equation}

 The above set of operators is characteristic for
any consideration of the interplay of QCD and electroweak effects, 
although, as we shall see in later chapters, on many occasions 
contributions of certain operators can be safely neglected. 
Moreover the set of operators $Q_1-Q_{10}$ given above has the
flavours relevant for $B$ decays. In $K$ decays the operators $Q_1$
and $Q_2$ should be replaced by
\begin{equation}\label{O1K} 
Q_1 = (\bar s_{\alpha} u_{\beta})_{V-A}\;(\bar u_{\beta} d_{\alpha})_{V-A}
~~~~~~Q_2 = (\bar s u)_{V-A}\;(\bar u d)_{V-A}.
\end{equation}
The relevant operators $Q_3-Q_{10}$ are then found by simply replacing
``$b$'' by ``$d$'' and summing only over $q=u,d,s$ in (\ref{O2})--(\ref{O5}).
Similarly ``$b$'' has to be replaced by ``$d$'' in (\ref{9V}) and (\ref{10V}).

\subsection{Manifestly Gauge Independent Formulation of FCNC Transitions}
Let us return to the basic functions in (\ref{BF})-(\ref{BFF}). 
The expressions
given there for the functions $B_0(x_t)$, $C_0(x_t)$ and $D_0(x_t)$
correspond to the 't Hooft--Feynman gauge ($\xi=1$). In an
arbitrary $R_\xi$ gauge, these functions 
are generalized as follows \cite{IL}:
\begin{equation}\label{T3}
 B_0(x_t,\xi,T_3)= \cases{B_0(x_t)+{1\over 8}\bar\varrho(x_t,\xi),&
~~~~$T_3=1/2$;\cr
B_0(x_t)+{1\over 2}\bar\varrho(x_t,\xi),&
~~~~$T_3=-1/2$;\cr}  
\end{equation}
\begin{equation}
C_0(x_t,\xi)=C_0(x_t)+{1\over 2}\bar\varrho(x_t,\xi) 
\end{equation}
\begin{equation}
 D_0(x_t,\xi)=D_0(x_t)- 2\bar\varrho(x_t,\xi), 
\end{equation}
where $\bar \varrho(x_t,\xi)$ summarizes the gauge dependence with
$\bar\varrho(x_t,1)=0$. Explicit formulae for $\bar\varrho(x_t,\xi)$ can be
found in \cite{IL,PBE0}. In (\ref{T3}), $T_3$ denotes the weak isospin of the
outgoing fermions in the flavour conserving part of the box vertex. We
observe that the $x_t$ dependence of the box vertices depends
generally on $T_3$ in the flavour conserving part and only for $\xi=1$
it reduces to the single function $B_0(x_t)$.

Now the initial conditions for the Wilson coefficient functions $C_j(M_W)$
are generally given
as linear combinations of the basic functions in (\ref{BF})-(\ref{BFF}).
Since physical amplitudes cannot depend on the chosen gauge 
the basic functions have to enter $C_j(M_W)$ in some
special combinations which are gauge independent. 
These special linear combinations turn out to be as follows \cite{PBE0}:
\begin{equation}
C_0(x_t,\xi)-4B_0(x_t,\xi,1/2)=C_0(x_t)-4B_0(x_t)=X_0(x_t)
\end{equation}
\begin{equation}
C_0(x_t,\xi)-B_0(x_t,\xi,-1/2)=C_0(x_t)-B_0(x_t)=Y_0(x_t)
\end{equation}
\begin{equation}
C_0(x_t,\xi)+{1\over 4}D_0(x_t,\xi)=C_0(x_t)+{1\over
                                  4}D_0(x_t)=Z_0(x_t).
\end{equation}
Explicitly then:
\begin{equation}\label{X0}
X_0(x_t)={{x_t}\over{8}}\;\left[{{x_t+2}\over{x_t-1}} 
+ {{3 x_t-6}\over{(x_t -1)^2}}\; \ln x_t\right] 
\end{equation}
\begin{equation}\label{Y0}
Y_0(x_t)={{x_t}\over8}\; \left[{{x_t -4}\over{x_t-1}} 
+ {{3 x_t}\over{(x_t -1)^2}} \ln x_t\right]
\end{equation}
\begin{equation}\label{Z0}
Z_0(x_t)=-{1\over9}\ln x_t + {{18x_t^4-163x_t^3 + 259x_t^2-108x_t}\over {144
(x_t-1)^3}} 
+{{32x_t^4-38x_t^3-15x_t^2+18x_t}\over{72(x_t-1)^4}}\ln x_t.
\end{equation}

It is also found that the functions $X_0, Y_0, Z_0$ multiply always
local operators of a particular structure (here $u$ represents
$u,c,t$; $e$ represents $e,\mu,\tau$ etc.):
\begin{center}
\begin{tabular}{ll}
$S_0(x_t)$: &$ (\bar s d)_{V-A} (\bar s d)_{V-A}$  \\
$X_0(x_t)$: &$ (\bar s d)_{V-A} (\bar u u)_{V-A},~~
         (\bar s d)_{V-A} (\bar \nu\nu)_{V-A}$ \\
$Y_0(x_t)$: &$ (\bar s d)_{V-A} (\bar d d)_{V-A},~~
         (\bar s d)_{V-A} (\bar e e)_{V-A}$ \\
$Z_0(x_t)$: &$ (\bar s d)_{V-A} (\bar u u)_V,~~
 (\bar s d)_{V-A} (\bar d d)_V,~~(\bar s d)_{V-A} (\bar e e)_V $\\   
$E_0(x_t)$: &$ (\bar s d)_{V-A} (\bar q q)_{V},~~
 (\bar s_{\alpha} d_{\beta})_{V-A} (\bar q_{\beta} q_{\alpha})_{V}$
 \\
$D'_0(x_t)$: &$ \bar s_{\alpha}\sigma^{\mu\nu}
(1+\gamma_5)b_{\alpha}F_{\mu\nu}$ \\
$E'_0(x_t)$: &$ \bar s_{\alpha}\sigma^{\mu\nu}T^a_{\alpha\beta}
 (1+\gamma_5)b_{\beta}G^a_{\mu\nu}$ 
\end{tabular}
\end{center}
Here $\alpha$ and $\beta$ are
colour indices, $F_{\mu\nu}$ is the electromagnetic field strength tensor and
$G^a_{\mu\nu}$ the gluonic field strength tensor.

We note that $ X_0(x_t) $ and $ Y_0(x_t) $ are
linear combinations of the $V-A$ components of $ Z^0$--penguin and
box--diagrams with final quarks or leptons having weak isospin $ T_3 $ 
equal to 1/2 and -- 1/2, respectively. $ Z_0(x_t) $ is a linear
combination of the vector component of the $ Z^0$--penguin and the
$\gamma$--penguin.

\subsection{Penguin--Box Expansion for FCNC Processes}
Having the set of gauge independet basic functions 
\begin{equation}\label{SXYZ}
S_0(x_t), \quad
X_0(x_t), \quad
Y_0(x_t), \quad
Z_0(x_t), \quad
E_0(x_t), \quad
D'_0(x_t), \quad
E'_0(x_t)
\end{equation}
at hand, let us return to the formal expression (\ref{OPE}) and rewrite it
in the form
\begin{equation}\label{USE}
A(M\to F)=\frac{G_{\rm F}}{\sqrt 2} V_{\rm CKM}
\sum_{i,k} \langle F\mid O_k(\mu)\mid M\rangle \;U_{kj}\;(\mu,M_W) 
          \; C_j(M_W),
\end{equation}
where $U_{kj}(\mu,M_W)$ is the renormalization group
transformation from $M_W$ down to $\mu$ given in (\ref{UM}). 

Now prior to the discussion of QCD effects we have formulated the FCNC
decays in terms of effective vertices.
This formulation demonstrates explicitly the universal character of
short distance interactions and
exhibits very clearly the dependence on
internal quark masses, in particular $m_t$, given by the process 
independent functions (\ref{SXYZ}). Yet as we have seen 
above, this universality and the transparent picture seems
to have been lost after the inclusion of QCD effects because these
effects are very different for different processes. Indeed when the
analysis is done in the framework of OPE, the basic functions of  
(\ref{SXYZ}) enter only the initial conditions of the renormalization
group analysis, i.e.\ the presence of effective vertices is only felt
in $C_j(M_W)$. The correspondence between $C_j(M_W)$ and
the effective vertices is, however, not simple because generally a given
diagram and the corresponding function contributes to several
coefficient functions of local operators and the $C_j(M_W)$ are
just linear combinations of them. Moreover, since the transformation
described by
$\hat U (\mu,M_W)$ is very complicated for non--leptonic decays, but
very simple for semi--leptonic decays, the resulting amplitudes have
no similarities. Indeed in the usual OPE analysis the amplitude 
(\ref{USE}) is
rewritten as in  (\ref{OPE}).
Thus, although the resulting coefficient functions evaluated at
$\mu={\cal O}(1\gev)$ remember the $m_t$ dependence acquired through the
effective vertices or basic functions, this dependence is hidden 
in a complicated
numerical evaluation of $U_{jk}$. In other words, the $m_t$
dependence of a given effective vertex is distributed among various
Wilson coefficient functions.

For phenomenological applications it is more elegant and more
convenient to have a formalism in which the final formulae for all
amplitudes are given explicitly in terms of the basic $m_t$-dependent
functions discussed above.

In \cite{PBE0} an approach was presented which
accomplishes this task. It gives the decay amplitudes as linear
combinations of the basic, universal, process independent but
$m_t$-dependent functions $F_r(x_t)$ of (\ref{SXYZ}) with
corresponding coefficients $P_r$ characteristic for the decay under
consideration. This approach termed ``Penguin Box Expansion'' (PBE) has
the following general form:
\begin{equation}
A({\rm decay}) = P_0({\rm decay}) + \sum_r P_r({\rm decay}) \, F_r(x_t),
\label{eq:generalPBE}
\end{equation}
where the sum runs over all possible functions contributing to a given
amplitude. In (\ref{eq:generalPBE}) we have separated a $m_t$-independent
term $P_0$ which summarizes contributions stemming from internal quarks
other than the top, in particular the charm quark.

Many examples of PBE appear in this review. Several decays or
transitions depend only on a single function out of the complete set
(\ref{SXYZ}). For completeness we give here the correspondence
between various processes and the basic functions:

\begin{center}
\begin{tabular}{lcl}
$B^0-\bar B^0$-mixing &\qquad\qquad& $S_0(x_t)$ \\
$K \to \pi \nu \bar\nu$, $B \to X_{d,s} \nu \bar\nu$ 
&\qquad\qquad& $X_0(x_t)$ \\
$K \to \mu \bar\mu$, $B \to l\bar l$ &\qquad\qquad& $Y_0(x_t)$ \\
$K_{\rm L} \to \pi^0 e^+ e^-$ &\qquad\qquad& $Y_0(x_t)$, $Z_0(x_t)$, 
$E_0(x_t)$ \\
$\varepsilon'$ &\qquad\qquad& $X_0(x_t)$, $Y_0(x_t)$, $Z_0(x_t)$,
$E_0(x_t)$ \\
$B \to X_s \gamma$ &\qquad\qquad& $D'_0(x_t)$, $E'_0(x_t)$ \\
$B \to X_s \mu^+ \mu^-$ &\qquad\qquad&
$Y_0(x_t)$, $Z_0(x_t)$, $E_0(x_t)$, $D'_0(x_t)$, $E'_0(x_t)$
\end{tabular}
\end{center}

In \cite{PBE0} an explicit transformation from OPE to
PBE has been made. This transformation and the relation between these
two expansions can be very clearly seen on the basis of (\ref{USE}). 
As we have seen, OPE puts the
last two factors in this formula together, mixing this way the physics
around $M_W$ with all physical contributions down to very low energy
scales. The PBE is realized on the other hand by putting the first two
factors together and rewriting $C_j(M_W)$ in terms of the basic
functions (\ref{SXYZ}). This results in the expansion of
(\ref{eq:generalPBE}). Further technical details and the methods for
the evaluation of the coefficients $P_r$ can be found in
\cite{PBE0}, where further virtues of PBE are discussed.

Finally, we give approximate formulae having power-like
dependence on $x_t$ for the basic, gauge independent functions of PBE:

\begin{equation}\label{PBE1}
 S_0(x_t)=0.784~x_t^{0.76},~~~~X_0(x_t)=0.660~x_t^{0.575},  
\end{equation}
\begin{equation}\label{PBE2}
 Y_0(x_t)=0.315~x_t^{0.78},\quad Z_0(x_t)=0.175~x_t^{0.93},  \quad
   E_0(x_t)=0.564~x_t^{-0.51}, 
\end{equation}
\begin{equation}
 D'_0(x_t)=0.244~x_t^{0.30},~~~~~~~~~~E'_0(x_t)=0.145~x_t^{0.19}.
\end{equation}
In the range $150\gev \le m_t \le 200\gev$ 
these approximations reproduce the
exact expressions to an accuracy better than 1\%. 
\subsection{Inclusive Decays}
So far we have discussed only {\it exclusive} decays. During 
recent years considerable progress has been made for {\it inclusive}
decays of heavy mesons. The starting point is again the effective
Hamiltonian in (\ref{HOPE}) which includes the short distance QCD
effects in $C_i(\mu)$. The actual decay described by the operators
$Q_i$ is then calculated in the spectator model corrected for
additional virtual and real gluon corrections.
Support for this approximation 
comes from heavy quark ($1/m_b $) expansions (HQE).
Indeed the spectator
model has been shown to correspond to the leading order approximation
in the $1/m_b$ expansion. 
The next corrections appear at the ${\cal O}(1/m_b^2)$
level. The latter terms have been studied by several authors
\cite{Chay,Bj,Bigi} with the result that they affect various
branching ratios by less than $10\%$ and often by only a few percent.
There is a vast literature on this subject and we can only refer here to
a few papers \cite{Bigi,Mannel} where further references can be found.
Of particular importance for this field was also the issue of the
renormalons which is nicely discussed in \cite{Beneke,Braun}.

\begin{table}[thb]
\begin{center}
\begin{tabular}{|l|l|}
\hline
\bf \phantom{XXXXXXXX} Decay & \bf Reference \\
\hline
\hline
\multicolumn{2}{|c|}{$\Delta F=1$ Decays} \\
\hline
current-current operators     & \cite{ALTA,BW} \\
QCD penguin operators         & \cite{BJLW1,BJLW,ROMA1,ROMA2} \\
electroweak penguin operators & \cite{BJLW2,BJLW,ROMA1,ROMA2} \\
magnetic penguin operators    & \cite{MisMu:94,Yao1,CZMM,GH97} \\
$Br(B)_{SL}$                  & \cite{ALTA,Buch:93,Bagan} \\
inclusive $\Delta S=1$ decays       & \cite{JP} \\
\hline
\multicolumn{2}{|c|}{Particle-Antiparticle Mixing} \\
\hline
$\eta_1$                   & \cite{HNa} \\
$\eta_2,~\eta_B$           & \cite{BJW} \\
$\eta_3$                   & \cite{HNb} \\
\hline
\multicolumn{2}{|c|}{Rare $K$- and $B$-Meson Decays} \\
\hline
$K^0_L \rightarrow \pi^0\nu\bar{\nu}$, $B \rightarrow l^+l^-$,
$B \rightarrow X_{\rm s}\nu\bar{\nu}$ & \cite{BB1,BB2} \\
$K^+   \rightarrow \pi^+\nu\bar{\nu}$, $K_{\rm L} \rightarrow \mu^+\mu^-$
                                      & \cite{BB3} \\
$K^+\to\pi^+\mu\bar\mu$               & \cite{BB5} \\
$K_{\rm L} \rightarrow \pi^0e^+e^-$         & \cite{BLMM} \\
$B\rightarrow X_s \mu^+\mu^-$           & \cite{Mis:94,BuMu:94} \\
$B\rightarrow X_s \gamma$      & \cite{AG2,Yao1,Pott,GREUB,CZMM,GH97} \\
\hline
\end{tabular}
\end{center}
\caption{References to NLO Calculations}
\label{TAB1}
\end{table}

\subsection{Weak Decays Beyond Leading Logarithms}
Until 1989 most of the calculations in the field of weak
decays were done in the leading logarithmic approximation.
An exception was the important work of Altarelli et al.~\cite{ALTA}
who  calculated NLO QCD corrections to the Wilson
coefficients of the current-current operators in 1981.
Today the effective Hamiltonians for weak decays are
available at the next-to-leading level for the most important
and interesting cases due to a series of publications devoted
to this enterprise written during the last six years.
The list of the existing calculations is given in table \ref{TAB1}. 
We will discuss some of the entries in this list below. 
A detailed review
of the existing NLO calculations is given in \cite{BBL}.

Let us recall why NLO calculations are important for the
phenomenology of weak decays:

\begin{itemize}
\item The NLO is first of all necessary to test the validity of
renormalization group improved perturbation theory.
\item Without going to NLO the QCD scale $\Lambda_{\overline{MS}}$
extracted from various high energy processes cannot be used 
meaningfully in weak decays.
\item 
Due to renormalization group invariance the physical
amplitudes do not depend on the scales $\mu$ present in $\alpha_s$
or in the running quark masses, in particular $m_t(\mu)$, 
$m_b(\mu)$ and $m_c(\mu)$. However,
in perturbation theory this property is broken through the truncation
of the perturbative series. Consequently one finds sizable scale
ambiguities in the leading order, which can be reduced considerably
by going to NLO.
\item In several cases the central issue of the top quark mass dependence
is strictly a NLO effect.
\end{itemize}

\section{Quark Mixing Matrix}
\setcounter{equation}{0}
\subsection{General Remarks}
Let us next discuss the stucture of the
quark-mixing-matrix $\hat V$ defined by (\ref{2.67}) in more
detail. In the case of $N$ generations, this matrix is given by a
unitary $N\times N$ matrix. The phase structure of the
quark-mixing-matrix is not unique since we have the freedom of
performing the following phase-transformations which are related to
phase-transformations of the corresponding quark fields:
\beq\label{e31}
V_{ij}\to \exp(i\xi_i)V_{ij}\exp(-i\tilde\xi_j).
\eeq
Note that there is no summation over the quark-flavour indices $i$ and
$j$ in this equation. Using the transformations (\ref{e31}), it
can be shown that the general $N$ generation quark-mixing-matrix is
described by $(N-1)^2$ parameters consisting of
\beq\label{e31a}
\frac{1}{2}N(N-1)
\eeq
Euler-type angles and
\beq\label{e31b}
\frac{1}{2}(N-1)(N-2)
\eeq
complex phases.

Consequently the quark-mixing-matrix is
real  in the two-generation case and takes the following standard 
form \cite{CAB,GIM1}:
\beq\label{e32}
\hat V_{\mbox{{\scriptsize C}}}=\left(\begin{array}{cc}
\cos\theta_{\mbox{{\scriptsize C}}} &\sin\theta_{\mbox{{\scriptsize
C}}}\\
-\sin\theta_{\mbox{{\scriptsize C}}}&\cos\theta_{\mbox{{\scriptsize
C}}} 
\end{array}\right),
\eeq
where $\sin\theta_{\mbox{{\scriptsize C}}}$ can be determined from 
semi-leptonic $K$-meson decays of the type $K\to\pi e^+\nu_e$ and is
given by $\sin\theta_{\mbox{{\scriptsize C}}}=0.22$.

On the other hand the $3\times 3$ 
quark-mixing-matrix of the three generation Standard Model --
the Cabibbo--Kobayashi--Maskawa--matrix (CKM matrix) \cite{KM} -- 
is parametrized by
three angles and a single complex phase.
This phase leading to an
imaginary part of the CKM matrix is a necessary ingredient to describe
CP violation within the framework of the Standard Model.

Many parametrizations of the CKM
matrix have been proposed in the literature.  We will use
two parametrizations in this review: the standard parametrization 
\cite{CHAU} recommended by the particle data group  \cite{PDG}  
and the Wolfenstein parametrization \cite{WO}.

\subsection{Standard Parametrization}
            \label{sec:sewm:stdparam}
Let us introduce the notation
$c_{ij}=\cos\theta_{ij}$ and $s_{ij}=\sin\theta_{ij}$ with $i$ and $j$
being generation labels ($i,j=1,2,3$). The standard parametrization is
then given as follows \cite{PDG}:
\begin{equation}\label{2.72}
V=
\left(\begin{array}{ccc}
c_{12}c_{13}&s_{12}c_{13}&s_{13}e^{-i\delta}\\ -s_{12}c_{23}
-c_{12}s_{23}s_{13}e^{i\delta}&c_{12}c_{23}-s_{12}s_{23}s_{13}e^{i\delta}&
s_{23}c_{13}\\ s_{12}s_{23}-c_{12}c_{23}s_{13}e^{i\delta}&-s_{23}c_{12}
-s_{12}c_{23}s_{13}e^{i\delta}&c_{23}c_{13}
\end{array}\right)\,,
\end{equation}
where $\delta$ is the phase necessary for CP violation. $c_{ij}$ and
$s_{ij}$ can all be chosen to be positive and $\delta$ may vary in the
range $0\le\delta\le 2\pi$. However, the measurements
of CP violation in $K$ decays force $\delta$ to be in the range
 $0<\delta<\pi$. 

The extensive phenomenology of the last years 
has shown that
$s_{13}$ and $s_{23}$ are small numbers: $\ord(10^{-3})$ and ${\cal
O}(10^{-2})$,
respectively. Consequently to an excellent accuracy $c_{13}=c_{23}=1$
and the four independent parameters are given as 
\begin{equation}\label{2.73}
s_{12}=| V_{us}|, \quad s_{13}=| V_{ub}|, \quad s_{23}=|
V_{cb}|, \quad \delta
\end{equation}
with the phase $\delta$ extracted from CP violating transitions or 
loop processes sensitive to $| V_{td}|$. The latter fact is based
on the observation that
 for $0\le\delta\le\pi$, as required by the analysis of CP violation
in the $K$ system,
there is a one--to--one correspondence between $\delta$ and $|V_{td}|$
given by
\begin{equation}\label{10}
| V_{td}|=\sqrt{a^2+b^2-2 a b \cos\delta},
\qquad
a=| V_{cd} V_{cb}|,
\qquad
b=| V_{ud} V_{ub}|\,.
\end{equation} 

What are the phenomenological advantages of (\ref{2.72}) \cite{HALE}? 
\begin{itemize}
\item
$| V_{ub} |$ is given by a single angle which is known to be very
small. Therefore $V_{ud},V_{us},V_{cb}$ and $V_{tb}$ are also given
each by a single parameter to an approximation better than four
significant figures. The relation between parameters and
experimentally measured quantities gets hence extremely simple.
\item
Each of the angles may then be characterized by a single physical
process, e.g.\ $\theta_{23}$ is directly measured by the $b\to c$
transition. 
\item
The CP violating phase is always multiplied by the very small
$s_{13}$. This shows clearly the suppression of CP violation.
\end{itemize}

For numerical evaluations the use of the standard parametrization
is strongly recommended. However once the four parameters in
(\ref{2.73}) have been determined it is often useful to make
a change of basic parameters in order to see the structure of
the result more transparently. This brings us to the Wolfenstein
parametrization \cite{WO} and its generalization given in 
\cite{BLO}.
\subsection{Wolfenstein Parameterization Beyond Leading Order}\label{Wolf-Par}
The original Wolfenstein parametrization \cite{WO}
is an approximate parametrization of the CKM matrix in which
each element is expanded as a power series in the small parameter
$\lambda=| V_{us}|=0.22$,
\begin{equation}\label{2.75} 
V=
\left(\begin{array}{ccc}
1-{\lambda^2\over 2}&\lambda&A\lambda^3(\varrho-i\eta)\\ -\lambda&
1-{\lambda^2\over 2}&A\lambda^2\\ A\lambda^3(1-\varrho-i\eta)&-A\lambda^2&
1\end{array}\right)
+\ord(\lambda^4)\,,
\end{equation}
and the set (\ref{2.73}) is replaced by
\begin{equation}\label{2.76}
\lambda, \qquad A, \qquad \varrho, \qquad \eta \, .
\end{equation}
Because of the
smallness of $\lambda$ and the fact that for each element 
the expansion parameter is actually
$\lambda^2$, it is sufficient to keep only the first few terms
in this expansion. 

The Wolfenstein parameterization
has several nice features. In particular it offers in conjunction with the
unitarity triangle a very transparent geometrical
representation of the structure of the CKM matrix and allows the derivation
of several analytic results to be discussed below. This turns out to be very
useful in the phenomenology of rare decays and of CP violation.

When using the Wolfenstein parametrization one should keep in mind that it
is an approximation and that in certain situations neglecting
${\cal O}(\lambda^4)$ terms may give wrong results. The question then
arises how to find ${\cal O}(\lambda^4)$ and higher order terms. The
point is that since (\ref{2.75}) is only an approximation the {\em exact}
definiton of the parameters in (\ref{2.76}) is not unique by terms of the 
neglected order
${\cal O}(\lambda^4)$. This is the reason why in different papers in the
literature different ${\cal O}(\lambda^4)$ terms can be found. They simply
correspond to different definitions of the parameters in (\ref{2.76}).
Obviously the physics does not depend on this choice.  Here
we will follow the definition given in \cite{BLO} which allows for simple
relations between the parameters (\ref{2.73}) and (\ref{2.76}).  This
will also restore the unitarity of the CKM matrix which in the
Wolfenstein parametrization as given in (\ref{2.75}) is not satisfied
exactly.

To this end
we go back to (\ref{2.72}) and following \cite{BLO} we impose 
the relations 
\begin{equation}\label{2.77} 
s_{12}=\lambda\,,
\qquad
s_{23}=A \lambda^2\,,
\qquad
s_{13} e^{-i\delta}=A \lambda^3 (\varrho-i \eta)
\end{equation}
to {\it  all orders} in $\lambda$. In view of the comments made above
this can certainly be done. It follows then that
\begin{equation}\label{2.84} 
\varrho=\frac{s_{13}}{s_{12}s_{23}}\cos\delta,
\qquad
\eta=\frac{s_{13}}{s_{12}s_{23}}\sin\delta.
\end{equation}
We observe that (\ref{2.77}) and (\ref{2.84}) represent simply
the change of variables from (\ref{2.73}) to (\ref{2.76}).
Making this change of variables in the standard parametrization 
(\ref{2.72}) we find the CKM matrix as a function of 
$(\lambda,A,\varrho,\eta)$ which satisfies unitarity exactly!
We also note that in view of $c_{13}=1-\ord(\lambda^6)$ the relations
between $s_{ij}$ and $| V_{ij}|$ in (\ref{2.73}) are 
satisfied to high accuracy. The relations in (\ref{2.84}) have
been used first in \cite{schmidtlerschubert:92}.
However, the improved treatment of the unitarity
triangle presented in \cite{BLO} and 
below goes beyond the analysis of these 
authors.

The procedure outlined above gives automatically the corrections to the
Wolfenstein parametrization in (\ref{2.75}).  Indeed expressing
(\ref{2.72}) in terms of Wolfenstein parameters by means of (\ref{2.77})
and then expanding in powers of $\lambda$ we recover the
matrix in (\ref{2.75}) and in addition find explicit corrections of
$\ord(\lambda^4)$ and higher order terms. $V_{ub}$ remains unchanged. The
corrections to $V_{us}$ and $V_{cb}$ appear only at $\ord(\lambda^7)$ and
$\ord(\lambda^8)$, respectively.  For many practical purposes the
corrections to the real parts can also be neglected.
The essential corrections to the imaginary parts are:
\begin{equation}\label{2.83g}
\Delta V_{cd}=-iA^2 \lambda^5\eta,
\qquad
\Delta V_{ts}=-iA\lambda^4\eta.
\end{equation}
The first of these corrections has to be included in the study of the CP
violating parameter $\varepsilon_K$. The second is important for direct
CP violation in certain $B$ decays.
On the other hand the imaginary part of $V_{cs}$, which in our expansion
in $\lambda$ appears only at $\ord(\lambda^6)$, can be fully neglected. 

In order to improve the accuracy of the unitarity triangle discussed
below one includes also the $\ord(\lambda^5)$ correction to $V_{td}$.
In summary then $V_{us}$, $V_{cb}$, $V_{ub}$, $V_{td}$ and $V_{ts}$
are given to an excellent approximation as follows:
\begin{equation}\label{CKM1}
V_{us}=\lambda, \qquad V_{cb}=A\lambda^2
\end{equation}
\begin{equation}\label{CKM2}
V_{ub}=A\lambda^3(\varrho-i\eta),
\qquad
V_{td}=A\lambda^3(1-\bar\varrho-i\bar\eta)
\end{equation}
\begin{equation}\label{2.83d}
 V_{ts}= -A\lambda^2+\frac{1}{2}A(1-2 \varrho)\lambda^4
-i\eta A \lambda^4 
\end{equation}
with
\begin{equation}\label{2.88d}
\bar\varrho=\varrho (1-\frac{\lambda^2}{2}),
\qquad
\bar\eta=\eta (1-\frac{\lambda^2}{2}).
\end{equation}
The advantage of this generalization of the Wolfenstein parametrization
over other generalizations found in the literature is the absence of
relevant corrections to $V_{us}$, $V_{cb}$ and $V_{ub}$ and an elegant
change in $V_{td}$ which allows a simple generalization of the unitarity
triangle as discussed in section 3.5.

It will turn out to be useful to have the following analytic expressions
for $\lambda_i=V_{id}V^*_{is}$ with $i=c,t$:
\begin{equation}\label{2.51}
 \IM\lambda_t= -\IM\lambda_c=\eta A^2\lambda^5=
\mid V_{ub}\mid \mid V_{cb} \mid \sin\delta 
\end{equation}
\begin{equation}\label{2.52}
 \RE\lambda_c=-\lambda (1-\frac{\lambda^2}{2})
\end{equation}
\begin{equation}\label{2.53}
 \RE\lambda_t= -(1-\frac{\lambda^2}{2}) A^2\lambda^5 (1-\bar\varrho) \,.
\end{equation}
Expressions (\ref{2.51}) and (\ref{2.52}) represent to an accuracy of
0.2\% the exact formulae obtained using (\ref{2.72}). The expression
(\ref{2.53}) deviates by at most 2\% from the exact formula in the
full range of parameters considered. 
In order to keep the analytic
expressions in the phenomenological applications in a transparent form
we have dropped a small $\ord(\lambda^7)$ term in deriving (\ref{2.53}).
After inserting the expressions (\ref{2.51})--(\ref{2.53}) in the exact
formulae for quantities of interest, a further expansion in $\lambda$
should not be made. 

\subsection{Unitarity Relations and Unitarity Triangles}

The unitarity of the CKM-matrix leads to
the following set of equations:
\bea
|V_{ud}|^2+|V_{cd}|^2+|V_{td}|^2 & = & 1\label{e41}\\
|V_{us}|^2+|V_{cs}|^2+|V_{ts}|^2 & = & 1\label{e42}\\
|V_{ub}|^2+|V_{cb}|^2+|V_{tb}|^2 & = & 1\label{e43}
\eea
\bea
|V_{ud}|^2+|V_{us}|^2+|V_{ub}|^2 & = & 1\label{e44}\\
|V_{cd}|^2+|V_{cs}|^2+|V_{cb}|^2 & = & 1\label{e45}\\
|V_{td}|^2+|V_{ts}|^2+|V_{tb}|^2 & = & 1\label{e46}
\eea
\bea
V_{ud}V_{us}^\ast+V_{cd}V_{cs}^\ast+V_{td}V_{ts}^\ast & = &
0\label{e47}\\
V_{ud}V_{ub}^\ast+V_{cd}V_{cb}^\ast+V_{td}V_{tb}^\ast & = &
0\label{e48}\\
V_{us}V_{ub}^\ast+V_{cs}V_{cb}^\ast+V_{ts}V_{tb}^\ast & = &
0\label{e49}
\eea
\bea
V_{ud}V_{cd}^\ast+V_{us}V_{cs}^\ast+V_{ub}V_{cb}^\ast & = &
0\label{e410}\\
V_{ud}V_{td}^\ast+V_{us}V_{ts}^\ast+V_{ub}V_{tb}^\ast & = &
0\label{e411}\\
V_{cd}V_{td}^\ast+V_{cs}V_{ts}^\ast+V_{cb}V_{tb}^\ast & = & 0.
\label{e412} 
\eea
Whereas (\ref{e41})-(\ref{e43}) and (\ref{e44})-(\ref{e46})
describe the normalization of the columns and rows of the CKM-matrix,
respectively, (\ref{e47})-(\ref{e49}) and
(\ref{e410})-(\ref{e412}) originate from the orthogonality of
different columns and rows, respectively. The orthogonality relations 
(\ref{e47})-(\ref{e412}) are of particular interest since they can be
represented as six ``unitarity'' triangles in the complex plane
\cite{js,akl}. Note that the set of equations (\ref{e41})-(\ref{e412}) is
invariant under the CKM phase-transformations specified in
(\ref{e31}). If one performs such transformations, the triangles
corresponding to (\ref{e47})-(\ref{e412}) are rotated in the
complex plane. Since the angles and the sides
(given by the moduli of the elements of the
mixing matrix)
 in these triangles remain 
unchanged
and do therefore not depend on the CKM-phase convention, these
quantities are physical observables.  

It can be shown that all six triangles
have the same area which is related to the measure of CP violation 
$J_{\rm CP}$
\cite{js}:
\begin{equation}
\mid J_{\rm CP} \mid = 2\cdot A_{\Delta},
\end{equation}
where $A_{\Delta}$ denotes the area of the unitarity triangles.
 
Let us briefly analyze the shape of the six unitarity triangles 
by using the original Wolfenstein parametrization. Then we find
that most of these triangles are very squashed ones, since the
Wolfenstein-structure both of eqs.~(\ref{e47})-(\ref{e49})
and (\ref{e410})-(\ref{e412}), respectively, is given as follows:
\bea
\order(\lambda)+\order(\lambda)+\order(\lambda^5)&=&0\label{e47a}\\
\order(\lambda^3)+\order(\lambda^3)+\order(\lambda^3)&=&0\label{e48a}\\
\order(\lambda^4)+\order(\lambda^2)+\order(\lambda^2)&=&0.\label{e49a}
\eea
Consequently, only in the unitarity triangles corresponding to
(\ref{e48}) and (\ref{e411}), all three sides are of comparable
magnitude $(\order(\lambda^3))$, while in those described by
(\ref{e47}), (\ref{e410}) and (\ref{e49}), (\ref{e412}) one side is
suppressed relative to the remaining ones by $\order(\lambda^4)$ and
$\order(\lambda^2)$, respectively. The triangles related to
(\ref{e48}) and (\ref{e411}) agree at the ${\cal O}(\lambda^3)$ level
and differ only through ${\cal O}(\lambda^5)$ corrections. Neglecting 
the latter subleading contributions they describe {\it the} 
unitarity triangle that appears usually in the literature. 

\subsection{Unitarity Triangle Beyond Leading Order}

Let us next concentrate on the most interesting unitarity triangle
described by
\begin{equation}\label{2.87h}
V_{ud}^{}V_{ub}^* + V_{cd}^{}V_{cb}^* + V_{td}^{}V_{tb}^* =0.
\end{equation}
Phenomenologically this triangle is very interesting as it involves
simultaneously the elements $V_{ub}$, $V_{cb}$ and $V_{td}$ which are
under extensive discussion at present.

In most analyses of the unitarity triangle present in the literature 
only terms ${\cal O}(\lambda^3)$ are kept in (\ref{2.87h}).
It is, however,
straightforward to include the next-to-leading $\ord(\lambda^5)$
terms \cite{BLO}. We note first that
\begin{equation}\label{2.88a}
V_{cd}^{}V_{cb}^*=-A\lambda^3+\ord(\lambda^7).
\end{equation}
Thus to an excellent accuracy $V_{cd}^{}V_{cb}^*$ is real with
$| V_{cd}^{}V_{cb}^*|=A\lambda^3$.
Keeping $\ord(\lambda^5)$ corrections and rescaling all terms in
(\ref{2.87h})
by $A \lambda^3$ 
we find
\begin{equation}\label{2.88b}
 \frac{1}{A\lambda^3}V_{ud}^{}V_{ub}^*
=\bar\varrho+i\bar\eta,
\qquad
\qquad
 \frac{1}{A\lambda^3}V_{td}^{}V_{tb}^*
=1-(\bar\varrho+i\bar\eta)
\end{equation}
with $\bar\varrho$ and $\bar\eta$ defined in (\ref{2.88d}). 
Thus we can represent (\ref{2.87h}) as the unitarity triangle 
in the complex $(\bar\varrho,\bar\eta)$
plane. This is  shown in fig.~12.
 The length of the side CB which lies
on the real axis equals unity when eq.~(\ref{2.87h}) is rescaled by
$V_{cd}^{}V_{cb}^*$. We observe that beyond the leading order
in $\lambda$ the point A {\it does not} correspond to  $(\varrho,\eta)$ but to
 $(\bar\varrho,\bar\eta)$.
Clearly within 3\% accuracy $\bar\varrho=\varrho$ and $\bar\eta=\eta$.
Yet in the distant future the accuracy of experimental results and
theoretical calculations may improve considerably so that the more
accurate formulation given in \cite{BLO} and here will be appropriate.

\begin{figure}[hbt]
\vspace{0.10in}
\centerline{
\epsfysize=2.1in
\epsffile{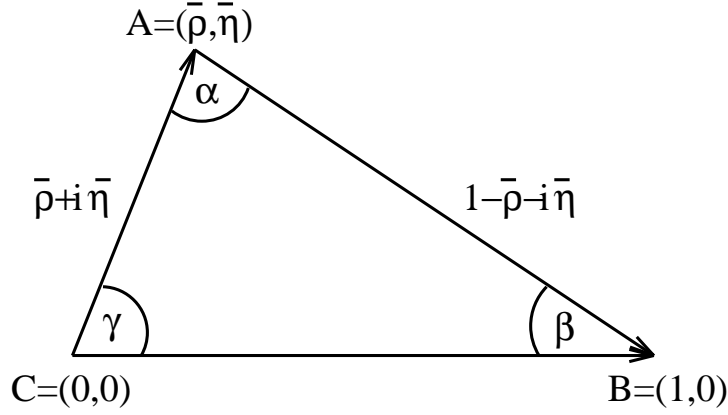}
}
\vspace{0.08in}
\caption{Unitarity Triangle.}\label{fig:utriangle}
\end{figure}

For numerical calculations the following procedure for the construction
of the unitarity triangle should be recommended:

\begin{itemize}
\item
Use the standard parametrization in phenomenological applications to
find $s_{12}$, $s_{13}$, $s_{23}$ and $\delta$.
\item
Translate to the set ($\lambda,~A,~\varrho,~\eta$) using 
(\ref{2.77}) and (\ref{2.84}).
\item
Calculate $\bar\varrho$ and $\bar\eta$ using (\ref{2.88d}).
\end{itemize}
It should be stressed that in calculations of quantities that are sensitive 
to $\RE\lambda_t$, like $\varepsilon_K$ or $Br(K^+\to\pi^+\nu\bar\nu)$ the
use of the original Wolfenstein parametrization may introduce additional
unnecessary errors in the predictions of order $5\%-7\%$.
 
Using simple trigonometry one can express $\sin(2\phi_i$), $\phi_i=
\alpha, \beta, \gamma$, in terms of $(\bar\varrho,\bar\eta)$ as follows:
\begin{equation}\label{2.89}
\sin(2\alpha)=\frac{2\bar\eta(\bar\eta^2+\bar\varrho^2-\bar\varrho)}
  {(\bar\varrho^2+\bar\eta^2)((1-\bar\varrho)^2
  +\bar\eta^2)}  
\end{equation}
\begin{equation}\label{2.90}
\sin(2\beta)=\frac{2\bar\eta(1-\bar\varrho)}{(1-\bar\varrho)^2 + \bar\eta^2}
\end{equation}
 \begin{equation}\label{2.91}
\sin(2\gamma)=\frac{2\bar\varrho\bar\eta}{\bar\varrho^2+\bar\eta^2}=
\frac{2\varrho\eta}{\varrho^2+\eta^2}.
\end{equation}
The lengths $CA$ and $BA$ in the
rescaled triangle of fig.~12 to be denoted by $R_b$ and $R_t$,
respectively, are given by
\begin{equation}\label{2.94}
R_b \equiv \frac{| V_{ud}^{}V^*_{ub}|}{| V_{cd}^{}V^*_{cb}|}
= \sqrt{\bar\varrho^2 +\bar\eta^2}
= (1-\frac{\lambda^2}{2})\frac{1}{\lambda}
\left| \frac{V_{ub}}{V_{cb}} \right|
\end{equation}
\begin{equation}\label{2.95}
R_t \equiv \frac{| V_{td}^{}V^*_{tb}|}{| V_{cd}^{}V^*_{cb}|} =
 \sqrt{(1-\bar\varrho)^2 +\bar\eta^2}
=\frac{1}{\lambda} \left| \frac{V_{td}}{V_{cb}} \right|.
\end{equation}
The expressions for $R_b$ and $R_t$ given here in terms of
$(\bar\varrho, \bar\eta)$ 
are excellent approximations. Clearly $R_b$ and $R_t$
can also be determined by measuring two of the angles $\phi_i$:
\begin{equation}\label{2.96}
R_b=\frac{\sin(\beta)}{\sin(\alpha)}=
\frac{\sin(\alpha+\gamma)}{\sin(\alpha)}=
\frac{\sin(\beta)}{\sin(\gamma+\beta)}
\end{equation}
\begin{equation}\label{2.97}
R_t=\frac{\sin(\gamma)}{\sin(\alpha)}=
\frac{\sin(\alpha+\beta)}{\sin(\alpha)}=
\frac{\sin(\gamma)}{\sin(\gamma+\beta)}.
\end{equation}

The angles $\beta$ and $\gamma$ of the unitarity triangle are related
directly to the complex phases of the CKM-elements $V_{td}$ and
$V_{ub}$, respectively, through
\beq\label{e417}
V_{td}=|V_{td}|e^{-i\beta},\quad V_{ub}=|V_{ub}|e^{-i\gamma}.
\eeq
The angle $\alpha$ can be obtained through the relation
\beq\label{e419}
\alpha+\beta+\gamma=180^\circ
\eeq
expressing the unitarity of the CKM-matrix.

The triangle depicted in fig.~12 together with $|V_{us}|$ and $\vcb$
gives a full description of the CKM matrix. 
Looking at the expressions for $R_b$ and $R_t$, we observe that within
the Standard Model the measurements of four CP
{\it conserving } decays sensitive to $\mid V_{us}\mid$, $\mid V_{ub}\mid$,   
$\mid V_{cb}\mid $ and $\mid V_{td}\mid$ can tell us whether CP violation
($\eta \not= 0$) is predicted in the Standard Model. 
This is a very remarkable property of
the Kobayashi-Maskawa picture of CP violation: quark mixing and CP violation
are closely related to each other. 

There is of course the very important question whether the KM picture
of CP violation is correct and more generally whether the Standard
Model offers a correct description of weak decays of hadrons. In order
to answer these important questions it is essential to calculate as
many branching ratios as possible, measure them experimentally and
check whether they all can be described by the same set of the parameters
$(\lambda,A,\varrho,\eta)$. In the language of the unitarity triangle
this means that the various curves in the $(\bar\varrho,\bar\eta)$ plane
extracted from different decays should cross each other at a single point
as shown in fig.~13.
Moreover the angles $(\alpha,\beta,\gamma)$ in the
resulting triangle should agree with those extracted one day from
CP-asymmetries in $B$-decays. More about this below. 
CP violation beyond the Standard Model is discussed in other
chapters in this book.

\begin{figure}[hbt]
\vspace{0.10in}
\centerline{
\epsfysize=6.3in
\rotate[r]{\epsffile{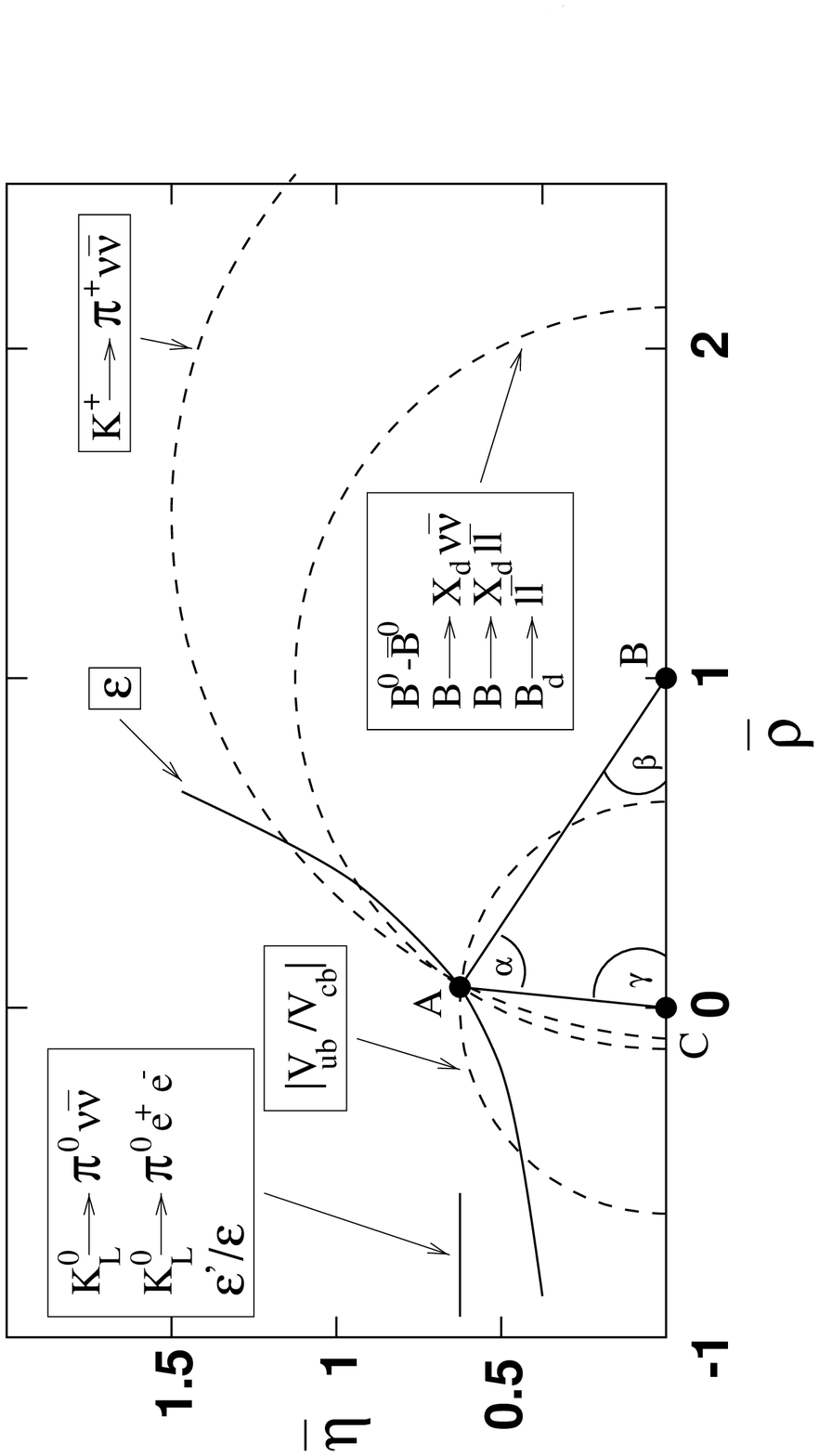}}
}
\vspace{0.08in}
\caption{The ideal Unitarity Triangle. For artistic reasons the value of
$\bar\eta$ has been chosen to be higher than the fitted central value
$\bar\eta\approx 0.4.$}\label{fig:2011}
\end{figure}

\subsection{CKM Matrix from Tree Level Decays and Unitarity}

In this review we are mainly dealing with the physics of heavy flavours.
Therefore we will comment only briefly on the determination of the
CKM elements describing the mixing between light quarks. The
interested reader should simply have a look at the 1996 report of
the Particle Data Group \cite{PDG}
where the subject is reviewed and further references can be found. 
The numbers quoted for the Cabibbo sector of the mixing matrix 
are taken from there. The remaining entries are sometimes 
different in view
of the most recent developments. We will also be very brief on
the determination of $|V_{ub}|$ and $\vcb$ from $B$ decays as this subject
is discussed by Neubert in another chapter of this book.
Concerning the top quark couplings $\vtd$, $\vts$ and $|V_{tb}|$
we will give here only the ranges following from tree level decays and
the unitarity of the CKM matrix.
The determination of $\vtd$, $\vts$ and of the parameters $\varrho$
and $\eta$ from FCNC processes will, however, be an important topic of 
the  subsequent sections. 

\subsubsection{Determination of $|V_{ud}|$}

$|V_{ud}|$ is mainly determined by comparing superallowed beta
decays, i.e.\ those with pure vector transitions, to muon decay. The
measurements are very accurate and therefore the
theoretical treatment requires a very careful consideration of radiative
corrections. The final result quoted in \cite{PDG} reads: 
\begin{equation}\label{vud}
|V_{ud}| = 0.9736 \pm 0.0010\,.
\end{equation}
A more accurate and slightly higher value
\begin{equation}\label{CHALK}
|V_{ud}|=0.9740\pm 0.0005
\end{equation}
has been obtained subsequently in a recent experiment on
$0^+\to 0^+$ superallowed beta decays at Chalk River Laboratory
\cite{Hagberg}.

\subsubsection{Determination of $|V_{us}|$}

There are mainly two ways to determine $|V_{us}|$: via
$K_{e3}$ decays and via semileptonic hyperon decays. We will first
deal with $K_{e3}$ decays, $K^+\to \pi^0 e^+ \nu_e$ and $K_{\rm L}^0 \to\pi^-
e^+ \nu_e$. Being pseudovector $\to$ pseudovector transitions,
these decays proceed via pure vector currents and therefore involve 
$SU(3)$ symmetry breaking in second order only. The
corrections have been calculated in chiral perturbation theory
\cite{LER1} yielding 
$|V_{us}| = 0.2196 \pm 0.0023$. 

Semileptonic hyperon decays are governed not only by vector but also
by axialvector currents. The latter break $SU(3)$ already in first
order which introduces considerably higher theoretical uncertainties
in the extraction of $|V_{us}|$  from experimental data than in
$K_{e3}$ decays. However, a careful calculation of $SU(3)$ symmetry
breaking effects \cite{DHK} allows to extract $|V_{us}|$ 
with reasonable accuracy from these decays. One finds \cite{PDG} 
$|V_{us}| = 0.222\pm 0.003$.

Combining these two determinations leads to the well known result
\begin{equation}\label{vus}
|V_{us}| = \lambda =  0.2205 \pm 0.0018\,.
\end{equation}
In view of the very
small error ($1\%$) we will set $\lambda=0.22$ in all 
numerical calculations.  

From (\ref{vud}), (\ref{vus}) and $|V_{ub}|$  given in  (\ref{vub2}) one
finds
\begin{equation}
|V_{ud}|^2 +|V_{us}|^2 +|V_{ub}|^2 =0.9965\pm0.0021\,,
\end{equation}
where the contribution of $|V_{ub}|^2$ is negligible.
Using (\ref{CHALK}) one finds \cite{Hagberg}
\begin{equation}
|V_{ud}|^2 +|V_{us}|^2 +|V_{ub}|^2 =0.9972\pm0.0013.
\end{equation}
Thus the departure from the unitarity relation (\ref{e44})
is by at least two standard deviations. The simplest
solution to this ``unitarity problem'' would be to double
the error in $|V_{ud}|$ or to increase its value. Since
the neutron decay data give, on the other hand, values
for the unitarity sum higher than unity \cite{Hagberg}, such a shift
is certainly possible. Clearly the current status
of the $|V_{ud}|$ determinations, in spite of small errors quoted
above, is unsatisfactory at present.
Further efforts should be made before one could conclude
that the failure to meet the unitarity constraint signals
some physics beyond the Standard Model.

\subsubsection{Determination of $|V_{cd}|$}

$|V_{cd}|$ is deduced from single charm production in deep
inelastic neutrino (antineutrino) -- nucleon scattering supplemented by
measurements of semileptonic branching
fractions of charmed mesons. The older value based mainly
on CDHS data ($|V_{cd}|=0.204\pm0.017$) has been shifted upwards
by the most recent Tevatron data \cite{DIS95} so that the final
value quoted in \cite{PDG} reads
\begin{equation}
|V_{cd}|=0.224\pm0.016.
\end{equation}

\subsubsection{Determination of $|V_{cs}|$}

Here data from deep--inelastic scattering cannot be used as
efficiently as in the case of $|V_{cd}|$ since one cannot
eliminate all unknown quantities but has to deal with the fairly
unknown strange--quark distribution. With conservative
assumptions only a very weak lower bound  $|V_{cs}|>0.59$ can
be obtained. Therefore one tries to determine
$|V_{cs}|$ from $D_{e3}$ decays, analogously to
$|V_{us}|$, by comparing the data with the theoretical decay width. 
This implies the use of model dependent
formfactors which introduce a considerable uncertainty in the
final result \cite{PDG}:
\begin{equation}
|V_{cs}| = 1.01 \pm 0.18.
\end{equation}
Especially in this case unitarity helps a lot to constrain the
allowed range as will be seen later.

\subsubsection{Determination of $\vcb$}

Clearly during the last two years there has been a considerable progress
done by experimentalists and theorists in the extraction of
$\vcb$ from exclusive and inclusive decays. In
particular we would like to mention important papers by
Shifman, Uraltsev and Vainshtein \cite{SUV},
Neubert \cite{Neubert} and 
Ball, Benecke and Braun \cite{Braun}
on the basis of which one is entitled to use the value
\begin{equation}\label{v23}
\vcb=0.040\pm0.003
\end{equation}
which should be compared with $\vcb=0.041\pm0.006$ used in
the {\it Top Quark Story} \cite{BH}.
The value in (\ref{v23}) 
is compatible with the value of Neubert ($\vcb=0.039\pm0.002$)
given in this book and the ones given in \cite{Rich,Gibbons,PDG}.
More details
can be found in the chapter by Neubert.

\subsubsection{Determination of $|V_{ub}|$}
In the case of $|V_{ub}|$
the situation is much worse but progress in the next few years is to be
expected in particular due to new information coming from
exclusive decays \cite{CLEOU,Gibbons}, the inclusive semileptonic
$b\to u$ rate \cite{SUV,Braun,URAL}, and the hadronic energy spectrum
in $ B\to X_u e \bar\nu_e$ \cite{GR96}. Combining the experimental and
theoretical uncertainties one has \cite{PDG}
\begin{equation}\label{v13}
\frac{|V_{ub}|}{\vcb}=0.08\pm0.02
\end{equation}
which should be compared with $|V_{ub}/V_{cb}|=0.13\pm0.04$ used in
the {\it Top Quark Story} \cite{BH}. Together with (\ref{v23}) this
implies
\begin{equation}\label{vub2}
|V_{ub}|=(3.2\pm0.8)\cdot 10^{-3}.
\end{equation}

\subsubsection{Determination of $|V_{td}|$, $|V_{ts}|$ and $|V_{tb}|$}

For completeness we would like to make already here a few remarks on 
the top quark couplings. 
A more extensive analysis of the couplings $|V_{td}|$ and 
$|V_{ts}|$ will be performed in subsequent sections.

Setting $\lambda=0.22$, scanning $\vcb$ and $|V_{ub}/V_{cb}|$ in
the ranges (\ref{v23}) and (\ref{v13}), respectively and $\cos\delta$ in
the range $-1\leq \cos\delta\leq 1$, we find the ranges
\begin{equation}\label{uni1}
4.5\cdot 10^{-3}\leq \vtd \leq 13.7\cdot 10^{-3}\,,
\qquad
0.0353\leq \vts \leq 0.0429
\end{equation}
and
\begin{equation}\label{uni2}
0.9991\leq |V_{tb}| \leq 0.9993.
\end{equation}
The last result should be compared with the direct measurement in top
quark decays at Tevatron yielding: $|V_{tb}|>0.58$ at 
$95\%$ C.L. \cite{Tipton}. 
From (\ref{uni1}) we observe
that the unitarity of the CKM matrix requires approximate equality of 
$\vts$ and $\vcb$:
\begin{equation}
0.954\leq \frac{|V_{ts}|}{\vcb} \leq 0.997
\end{equation}
which is evident if one compares (\ref{CKM1}) with (\ref{2.83d}). 
The determination of $\vtd$
will be considerably improved in the next section by using the
constraints from $B^0_d-\bar B^0_d$-mixing and CP violation in
the $K$-meson system.

\section{$\eps_K$, $B^0$-$\bar B^0$ Mixing and the Unitarity Triangle}
        \label{sec:epsBBUT}
\setcounter{equation}{0}
\subsection{Preliminaries}
Particle--antiparticle mixing has always been of fundamental
importance in testing the Standard Model and often has proven to be an
undefeatable challenge for suggested extensions of this model.
Particle--antiparticle mixing is responsible
for the small mass differences between the mass eigenstates of neutral
mesons. Being an FCNC process it involves heavy quarks in loops
and consequently it is a perfect 
testing ground for heavy flavour physics: from the calculation of the
$K_{\rm L}-K_{\rm S}$ mass difference, Gaillard and Lee \cite{GALE} 
were able to estimate the
value of the charm quark mass before charm discovery; $B_d^0-\bar B_d^0$
mixing \cite{ARGUS} gave the first indication of a large top quark mass. 
Particle--antiparticle mixing is also
closely related to the violation of the
CP symmetry which is experimentally known since 1964 \cite{CRONIN}. 

In this section we will deal almost exclusively with the parameter 
$\varepsilon_K$ 
describing so called {\it indirect} CP violation in the $K$ system and with 
the mass differences $\Delta M_{d,s}$  which
describe the size of $B_{d,s}^0-\bar B^0_{d,s}$ mixings. 
In the Standard Model all these phenomena
appear first at the one--loop level and as such they are
sensitive measures of the top quark couplings $V_{ti}(i=d,s,b)$ and 
of the top quark mass. 

We have seen in section 3 that tree level 
decays and the unitarity of the CKM
matrix give us already a good information about $V_{tb}$ and $V_{ts}$:
$V_{tb}\approx 1$ and $\mid V_{ts}\mid\;\approx\;\mid V_{cb}\mid$.
Similarly the value of the top quark mass measured by CDF and D0 (see below)
is known within $\pm 4\%$.
Consequently the
main new information to be gained from the quantities discussed here are 
the values of $|V_{td}|$ and of the phase $\delta$ in the CKM matrix.
This will allow us to construct the unitarity triangle.

Let us briefly recall the formalism of
particle--antiparticle mixing. We will here mainly discuss the $K$--system
as this mixing in the $B$--system is discussed in great detail in section 8.
Some formulae for $B_{d,s}^0-\bar B^0_{d,s}$ mixings, necessary for the
analysis of the unitarity triangle, are collected in subsection
\ref{subsec:BBformula}.

$K^0$ and $\bar K^0$ are flavour eigenstates which in the Standard Model
may mix via weak interactions through the box diagrams of fig.\ 11e. 
Constructing CP eigenstates and choosing the CKM
phase convention $(CP|K^0\rangle=|\bar K^0\rangle)$, we obtain 
\begin{equation}
K_1={1\over{\sqrt 2}}(K^0+\bar K^0),
  \qquad\qquad CP|K_1\rangle=|K_1\rangle
\end{equation}
\begin{equation}
K_2={1\over{\sqrt 2}}(K^0-\bar K^0),
  \qquad\qquad CP|K_2\rangle=-|K_2\rangle\,.
\end{equation}

Due to the complex phase in the CKM matrix $K_1$ and $K_2$ differ from
the physical mass eigenstates 
$K_{\rm S}$ and $K_{\rm L}$, respectively, by 
a small admixture of the
other CP eigenstate:
\begin{equation}
K_{\rm S}={{K_1+\bar\varepsilon K_2}
\over{\sqrt{1+\mid\bar\varepsilon\mid^2}}},
\qquad
K_{\rm L}={{K_2+\bar\varepsilon K_1}
\over{\sqrt{1+\mid\bar\varepsilon\mid^2}}}\,.
\end{equation}
The small parameter $\bar\varepsilon$ introduced here depends on the 
phase convention
chosen for $K^0$ and $\bar K^0$. Therefore it may not 
be taken as a physical measure of CP violation.

Since a two pion final state is CP even while a three pion final state is CP
odd, $K_{\rm S}$ and $K_{\rm L}$ preferably decay to $2\pi$ and $3\pi$, 
respectively
via the following CP conserving decay modes:
\begin{equation}
K_{\rm L}\to 3\pi {\rm ~~(via~K_2),}\qquad K_{\rm S}\to 2 
\pi {\rm ~~(via~K_1).}
\end{equation}
This difference is responsible for the large disparity in their
life-times.
However, since $K_{\rm L}$ and $K_{\rm S}$ are not CP eigenstates they 
may decay
with small branching fractions as follows:
\begin{equation}
K_{\rm L}\to 2\pi {\rm ~~(via~K_1),}\qquad K_{\rm S}\to 3 
\pi {\rm ~~(via~K_2).}
\end{equation}
Since these decays proceed not via explicit breaking of the CP symmetry in 
the decay itself but via the admixture of the CP state with opposite 
CP parity to the dominant
one, they are usually called ``indirect CP violating''. The measure for this
indirect CP violation is then defined as
\begin{equation}\label{ek}
\varepsilon_K
={{A(K_{\rm L}\rightarrow(\pi\pi)_{I=0}})\over{A(K_{\rm 
S}\rightarrow(\pi\pi)_{I=0})}},
\end{equation}
where $\varepsilon_K$ is, contrary to $\bar\varepsilon$ introduced above,
independent of the phase conventions and is measured to be \cite{PDG}:
\begin{equation}\label{eexp}
\varepsilon_K^{exp}
=(2.280\pm0.013)\cdot10^{-3}\;e^{i{\pi\over 4}}\,.
\end{equation}

\begin{figure}[hbt]
\centerline{
\epsfysize=1.5in
\epsffile{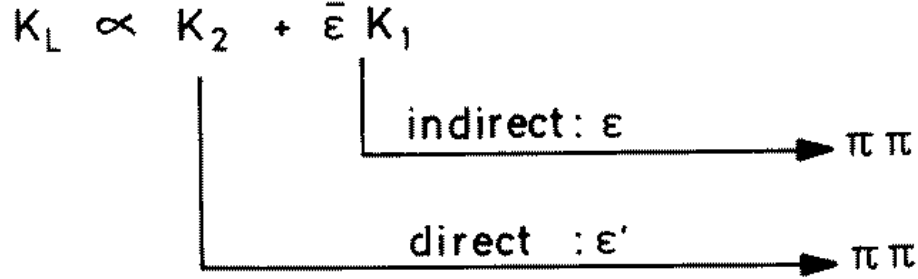}
}
\caption[]{
Indirect versus direct CP violation in $K_L \to \pi\pi$.
\label{fig:14}}
\end{figure}

While {\it indirect} CP violation reflects the fact that the mass
eigenstates are not CP eigenstates, so-called {\it direct} CP violation 
is realized via a 
direct transition of a CP odd to a CP even state or vice versa (see
fig.\ \ref{fig:14}). The parameter $\varepsilon'$  is defined as
\begin{equation}\label{eprime}
\varepsilon'={1\over {\sqrt 2}}\IM
\left({{A_2}\over{A_0}}\right) e^{i\Phi}\,,
\end{equation}
where
the isospin amplitudes $A_I$ in $K\to\pi\pi$
decays are introduced through
\begin{equation} 
A(K^+\rightarrow\pi^+\pi^0)=\sqrt{3\over 2} A_2 e^{i\delta_2}
\end{equation}
\begin{equation} 
A(K^0\rightarrow\pi^+\pi^-)=\sqrt{2\over 3} A_0 e^{i\delta_0}+ \sqrt{1\over
3} A_2 e^{i\delta_2}
\end{equation}
\begin{equation}
A(K^0\rightarrow\pi^0\pi^0)=\sqrt{2\over 3} A_0 e^{i\delta_0}-2\sqrt{1\over
3} A_2 e^{i\delta_2}\,.
\end{equation} 
Here the subscript $I=0,2$ denotes states with isospin $0,2$
equivalent to $\Delta I=1/2$ and $\Delta I = 3/2$ transitions,
respectively, and $\delta_{0,2}$ are the corresponding strong phases. 
Finally $\Phi=\pi/2+\delta_2-\delta_0 \approx \pi/4$.

Experimentally $\varepsilon_K\equiv \varepsilon$ and $\varepsilon'$
can be found by measuring the ratios
\begin{equation}
\eta_{00}={{A(K_{\rm L}\to\pi^0\pi^0)}\over{A(K_{\rm S}\to\pi^0\pi^0)}}
            \simeq \varepsilon-2\varepsilon',~~~~
  \eta_{+-}={{A(K_{\rm L}\to\pi^+\pi^-)}\over{A(K_{\rm S}\to\pi^+\pi^-)}}
            \simeq \varepsilon+ \varepsilon',
\end{equation}
\begin{equation}
\mid{{\eta_{00}}\over{\eta_{+-}}}\mid^2\simeq 1 -6\; 
\RE(\frac{\varepsilon'}{\varepsilon})\,.
\end{equation}

The strength of $K^0$--$\bar K^0$ mixing is described by the 
$K_{\rm L}-K_{\rm S}$
mass difference which is experimentally measured to be \cite{PDG}
\begin{equation}
\Delta M_K=M(K_{\rm L})-M(K_{\rm S}) = 
(3.491\pm 0.009) \cdot 10^{-15} \gev\,.
\end{equation}
In the Standard Model roughly $70\%$ of the measured $\Delta M_K$
is described by the real parts of the box diagrams in fig.\ 11e
\cite{HNa}. The rest is attributed to long distance contributions
which are difficult to estimate. For this reason in the theoretical
analysis of $\varepsilon_K$ it is customary to use the experimental value
of $\Delta M_K$. On the other hand the parameter $\varepsilon_K$ is
given (see below) by the imaginary part of the relevant off-diagonal
element $M_{12}$ in the neutral $K$-meson mass matrix. The latter
being related to CP violation and top quark physics should
be dominated by short distance contributions and well approximated
by the imaginary parts of the box diagrams in fig.\ 11e.

Similarly the strength of the $B^0_{d,s}-\bar B^0_{d,s}$ mixings
is described by the mass differences
\begin{equation}
\Delta M_{d,s}= M_H^{d,s}-M_L^{d,s}
\end{equation}
with ``H'' and ``L'' denoting heavy and light respectively. 
In contrast to $\Delta M_K$ , in this case the long distance contributions
are estimated to be very small and $\Delta M_{d,s}$ is very well
approximated by the relevant box diagrams as discussed below.

After these general remarks we are ready to enter the details.

\subsection{Basic Formula for $\eps_K$}
            \label{subsec:epsformula}
Indirect CP violation in $K_{\rm L} \to \pi\pi$ is described by 
the parameter $\eps_K$ defined in (\ref{ek}). 
The general formula for $\eps_K$ is given as follows:
\begin{equation}
\eps_K = \frac{\exp(i \pi/4)}{\sqrt{2} \Delta M_K} \,
\left( \IM M_{12} + 2 \xi \RE M_{12} \right),
\label{eq:epsdef}
\end{equation}
where
\begin{equation}
\xi = \frac{\IM A_0}{\RE A_0}
\label{eq:xi}
\end{equation}
with $A_0 \equiv A(K \to (\pi\pi)_{I=0})$ and $\Delta M_K$ denoting
the $K_{\rm L}$-$K_{\rm S}$ mass difference. The off-diagonal 
element $M_{12}$ in
the neutral $K$-meson mass matrix represents $K^0$-$\bar K^0$
mixing. It is given by
\begin{equation}
2 m_K M^*_{12} = \langle \bar K^0| \Heff(\Delta S=2) |K^0\rangle\,,
\label{eq:M12Kdef}
\end{equation}
where $\Heff(\Delta S=2)$ is the effective Hamiltonian for the 
$\Delta S=2$ transitions.

To lowest order these transitions are induced
through the box diagrams shown in fig.\ 11e.
Including QCD corrections, the
effective low energy Hamiltonian, to be derived from these diagrams,
can be written as follows $(\lambda_i = V_{is}^* V_{id}^{})$ \cite{BBL}:
\begin{eqnarray}\label{hds2}
{\cal H}^{\Delta S=2}_{\rm eff}&=&\frac{G^2_{\rm F}}{16\pi^2}M^2_W
 \left[\lambda^2_c\eta_1 S_0(x_c)+\lambda^2_t \eta_2 S_0(x_t)+
 2\lambda_c\lambda_t \eta_3 S_0(x_c, x_t)\right] \times
\nonumber\\
& & \times \left[\as^{(3)}(\mu)\right]^{-2/9}\left[
  1 + \frac{\as^{(3)}(\mu)}{4\pi} J_3\right]  Q(\Delta S=2) + h. c.
\end{eqnarray}
\\
This expression is valid for scales $\mu$
below the charm threshold $\mu_c=\ord(m_c)$. In this case
${\cal H}^{\Delta S=2}_{\rm eff}$ consists of a single four-quark operator
\begin{equation}\label{qsdsd}
Q(\Delta S=2)=(\bar sd)_{V-A}(\bar sd)_{V-A},
\end{equation}
which is multiplied by the corresponding coefficient function.
It is useful and customary to decompose this function into a
charm-, a top- and a mixed charm-top contribution as displayed
in (\ref{hds2}). This form is obtained upon eliminating $\lambda_u$
by means of the unitarity of the CKM matrix and setting $x_u=0$. The basic
electroweak loop contributions without QCD correction are then
expressed through the functions $S_0$ calculated in \cite{IL} and
given in (\ref{S0},\ref{BFF}).
\\
Short-distance QCD effects are described through the correction
factors $\eta_1$, $\eta_2$, $\eta_3$ and the explicitly
$\alpha_s$-dependent terms in (\ref{hds2}). 
The latter terms
are factored out to
exhibit the $\mu$-dependence of the coefficient function in the
$f=3$ regime which has to cancel the corresponding $\mu$-dependence
of the hadronic matrix element of $Q$ between meson states in
physical applications. A similar comment applies to the renormalization
 scheme
dependence present in $J_3$. In the NDR scheme $J_3=1.895$.
All these issues are discussed in detail in \cite{BBL}.
\\
Without QCD, i.e.\ in the limit $\alpha_s\to 0$, one has
$\eta_i [\alpha_s]^{-2/9}\to 1$. 
The NLO values of the QCD factors $\eta_1$ , $\eta_2$ and $\eta_3$ 
are given as follows \cite{HNa,BJW,HNb}:
\begin{equation}
\eta_1=1.38\pm 0.20,\qquad
\eta_2=0.57\pm 0.01,\qquad
  \eta_3=0.47\pm0.04.
\end{equation}

The quoted errors reflect the remaining theoretical uncertainties due to
$\Lambda_{\overline{MS}}$ and the quark
masses. The references to the leading order calculations can be found in
\cite{BBL}. The factor $\eta_1$ plays only a minor role in the analysis of
$\varepsilon_K$ but its enhanced value through NLO corrections \cite{HNa}
is essential for the $K_{\rm L}-K_{\rm S}$ mass difference.
\\
Defining the renormalization group invariant parameter
$B_K$ by
\begin{equation}
B_K = B_K(\mu) \left[ \alpha_s^{(3)}(\mu) \right]^{-2/9} \,
\left[ 1 + \frac{\alpha_s^{(3)}(\mu)}{4\pi} J_3 \right]
\label{eq:BKrenorm}
\end{equation}
\begin{equation}
\langle \bar K^0| (\bar s d)_{V-A} (\bar s d)_{V-A} |K^0\rangle
\equiv \frac{8}{3} B_K(\mu) F_K^2 m_K^2
\label{eq:KbarK}
\end{equation}
and using (\eqn{hds2}) one finds
\begin{equation}
M_{12} = \frac{G_{\rm F}^2}{12 \pi^2} F_K^2 B_K m_K \mw^2
\left[ {\lambda_c^*}^2 \eta_1 S_0(x_c) + {\lambda_t^*}^2 \eta_2 S_0(x_t) +
2 {\lambda_c^*} {\lambda_t^*} \eta_3 S_0(x_c, x_t) \right],
\label{eq:M12K}
\end{equation}
where $F_K$ is the $K$-meson decay constant and $m_K$
the $K$-meson mass. 

The last term in (\eqn{eq:epsdef}) constitutes at
most a 2\,\% correction to $\eps_K$ and consequently can be neglected
in view of other uncertainties, in particular those connected with
$B_K$.  Inserting (\eqn{eq:M12K}) into (\eqn{eq:epsdef}) one finds
\begin{equation}
\eps_K = C_{\eps} B_K \IM\lambda_t \left\{
\RE\lambda_c \left[ \eta_1 S_0(x_c) - \eta_3 S_0(x_c, x_t) \right] -
\RE\lambda_t \eta_2 S_0(x_t) \right\} \exp(i \pi/4)\,,
\label{eq:epsformula}
\end{equation}
where we have used the unitarity relation $\IM\lambda_c^* = {\rm
Im}\lambda_t$ and  have neglected $\RE\lambda_t/\RE\lambda_c
 = \ord(\lambda^4)$ in evaluating $\IM(\lambda_c^* \lambda_t^*)$.
The numerical constant $C_\eps$ is given by
\begin{equation}
C_\eps = \frac{G_{\rm F}^2 F_K^2 m_K \mw^2}{6 \sqrt{2} \pi^2 \Delta M_K}
       = 3.78 \cdot 10^4 \, .
\label{eq:Ceps}
\end{equation}
Using the standard parametrization of (\eqn{2.72}) to evaluate ${\rm
Im}\lambda_i$ and $\RE\lambda_i$, setting the values for $s_{12}$,
$s_{13}$, $s_{23}$ and $\mt$ in accordance with section 3
 and taking a value for $B_K$ (see below), one can
determine the phase $\delta$ by comparing (\eqn{eq:epsformula}) with the
experimental value for $\eps_K$.

Once $\delta$ has been determined in this manner one can find the
corresponding point $(\bar\varrho,\bar\eta)$ by using (\eqn{2.84}) and
(\eqn{2.88d}). Actually for a given set ($s_{12}$, $s_{13}$, $s_{23}$,
$\mt$, $B_K$) there are two solutions for $\delta$ and consequently two
solutions for $(\bar\varrho,\bar\eta)$. 
This will be evident from the analysis of the unitarity triangle
presented below.

Concerning the parameter $B_K$, the most recent analyses
using lattice methods summarized recently by 
Flynn \cite{Flynn} give
$B_K=0.90\pm 0.06$. The $1/N$ approach
 of \cite{BBG0}  gives  $B_K=0.70\pm 0.10$. A recent confirmation of this
result in a somewhat modified framework has been presented in 
\cite{Bijnens}.
Lower values for $B_K$ are obtained by using the QCD Hadronic Duality
approach \cite{Prades} ($B_K=0.39\pm 0.10$) or using the $SU(3)$ symmetry 
and  PCAC
($B_K=1/3$) \cite{Donoghue}. For $\vcb=0.040$ and $\vub=0.08$ such 
low values of
$B_K$ require $\mt>200\gev$ in order to explain the experimental
value of $\varepsilon_K$ \cite{AB,BLO,HNb}. The QCD sum rule results are in
the ball park of $B_K=0.60$ \cite{DENA}. In our numerical analysis presented 
below we will use 
\begin{equation}
B_K=0.75\pm 0.15 \,.
\end{equation}

\subsection{Basic Formula for $B^0$-$\bar B^0$ Mixing}
            \label{subsec:BBformula}
The strength of $B^0$-$\bar B^0$ mixing is described by
\begin{equation}
\Delta M_q= 2 |M_{12}^{(q)}|, \qquad q=d,s,
\label{eq:xdsdef}
\end{equation}
the mass difference between the mass
eigenstates in the $B_d^0-\bar B_d^0$ system and the $B_s^0-\bar B_s^0$
system, respectively. Equivalently one can use
\begin{equation}
x_q \equiv \frac{\Delta M_q}{\Gamma_{B_q}},
\end{equation}
where  $\Gamma_{B_q} = 1/\tau_{B_q}$ with
$\tau_{B_q}$ being the corresponding lifetimes.
In what follows we will work dominantly with $\Delta M_q$
as this avoids the experimental errors in lifetimes. Moreover 
this is the way the most recent experimental results for $B^0-\bar B^0$
mixing are quoted.

The off-diagonal
term $M_{12}$ in the neutral $B$-meson mass matrix is given by
\begin{equation}
2 m_{B_q} |M_{12}^{(q)}| = 
|\langle \bar B^0_q| \Heff(\Delta B=2) |B^0_q\rangle|,
\label{eq:M12Bdef}
\end{equation}
where $\Heff(\Delta B=2)$, relevant for scales scales $\mu_b=\ord(m_b)$,
is the effective Hamiltonian 
analogous to (\ref{hds2})  and given in the case of $B_d^0-\bar B_d^0$
mixing by
\cite{BJW}
\begin{equation}\label{hdb2}
{\cal H}^{\Delta B=2}_{\rm eff}=\frac{G^2_{\rm F}}{16\pi^2}M^2_W
 \left(V^\ast_{tb}V_{td}\right)^2 \eta_{B}
 S_0(x_t) \left[\alpha^{(5)}_s(\mu_b)\right]^{-6/23}\left[
  1 + \frac{\alpha^{(5)}_s(\mu_b)}{4\pi} J_5\right]  Q(\Delta B=2) + h. c.
\end{equation}
Here
\begin{equation}\label{qbdbd}
Q(\Delta B=2)=(\bar bd)_{V-A}(\bar bd)_{V-A}
\end{equation}
and
$\eta_B$ is the QCD factor analogous to $\eta_2$ and given
by  \cite{BJW}
\begin{equation}
\eta_B=0.55\pm0.01.
\end{equation}
 $J_5=1.627$ in the NDR scheme.
In the case of  $B_s^0-\bar B_s^0$ mixing one should simply replace
$d\to s$ in (\ref{hdb2}) and (\ref{qbdbd}) with all other quantities
unchanged.

Due to the particular hierarchy of the CKM matrix elements only the
top sector can contribute significantly to $B^0-\bar B^0$ mixing.
In contrast to the $K^0-\bar K^0$ case,
the charm sector and the mixed top-charm contributions are
entirely negligible here 
which considerably simplifies the analysis.

Defining the renormalization group invariant parameters $B_q$
by
\begin{equation}\label{Def-Bpar}
B_{B_q} = B_{B_q}(\mu) \left[ \as^{(5)}(\mu) \right]^{-6/23} \,
\left[ 1 + \frac{\as^{(5)}(\mu)}{4\pi} J_5 \right]
\label{eq:BBrenorm}
\end{equation}
\begin{equation}
\langle \bar B^0_q| (\bar b q)_{V-A} (\bar b q)_{V-A} |B^0_q\rangle
\equiv \frac{8}{3} B_{B_q}(\mu) F_{B_q}^2 m_{B_q}^2\,,
\label{eq:BbarB}
\end{equation}
where
$F_{B_q}$ is the $B_q$-meson decay constant
and using (\eqn{hdb2}) one finds
\begin{equation}
\Delta M_q = \frac{G_{\rm F}^2}{6 \pi^2} \eta_B m_{B_q} 
(B_{B_q} F_{B_q}^2 ) \mw^2 S_0(x_t) |V_{tq}|^2,
\label{eq:xds}
\end{equation}
which implies
\begin{equation}\label{DMD}
\Delta M_d=
0.50/{\rm ps}\cdot\left[ 
\frac{\sqrt{B_{B_d}}F_{B_d}}{200\mev}\right]^2
\left[\frac{\mtb(\mt)}{170~GeV}\right]^{1.52} 
\left[\frac{\vtd}{8.8\cdot10^{-3}} \right] 
\left[\frac{\eta_B}{0.55}\right]  
\end{equation}
and
\begin{equation}\label{DMS}
\Delta M_{s}=
15.1/{\rm ps}\cdot\left[ 
\frac{\sqrt{B_{B_s}}F_{B_s}}{240\mev}\right]^2
\left[\frac{\mtb(\mt)}{170~GeV}\right]^{1.52} 
\left[\frac{\vts}{0.040} \right] 
\left[\frac{\eta_B}{0.55}\right] \,.
\end{equation}

There is a vast literature on the lattice calculations of $F_{B_d}$ 
and $B_{B_d}$.
The most recent world averages  given by Flynn \cite{Flynn} are:
\begin{equation}
F_{B_d}=(175\pm 25)\mev\,, \qquad
B_{B_d}=1.31\pm 0.03\,.
\end{equation}
This result for $F_{B_d}$ is compatible with the results obtained 
with the help of 
QCD sum rules \cite{QCDSF}. An interesting upper bound 
$F_{B_d}<195\mev$ using QCD dispersion relations can be found in
\cite{BGL}. In our numerical analysis we will use
$F_{B_d}\sqrt{B_{B_d}}=(200\pm 40)\mev$. 
More details can be found in the chapter by Chris Sachrajda.
The experimental situation on
$\Delta M_d$ has been recently summarized by Gibbons \cite{Gibbons}
 and is given in table \ref{tab:inputparams}. 
For $\tau(B_d)=1.55~{\rm ps}$ one has  
$x_d= 0.72\pm 0.03$.

\subsection{Standard Analysis of the Unitarity Triangle}\label{UT-Det}

With all these formulae at hand we are now in a position to
discuss the standard analysis of the unitarity triangle.
It proceeds essentially in five
steps:

{\bf Step 1:}

{}From  $b\to c$ transition in inclusive and exclusive $B$ meson decays
one finds $\vcb$ and consequently the scale of the unitarity triangle:
\begin{equation}
\vcb\quad \Longrightarrow\quad\lambda \vcb= \lambda^3 A
\end{equation}

{\bf Step 2:}

{}From  $b\to u$ transition in inclusive and exclusive $B$ meson decays
one finds $\vub$ and consequently the side $CA=R_b$ of the UT:
\begin{equation}\label{rb}
\vub \quad\Longrightarrow \quad R_b=\sqrt{\bar\varrho^2+\bar\eta^2}=
4.44 \cdot \left| \frac{V_{ub}}{V_{cb}} \right|
\end{equation}

{\bf Step 3:}

{}From the observed indirect CP violation in $K \to \pi\pi$ described
experimentally by the parameter $\varepsilon_K$ (\ref{eexp}) 
and theoretically
by the formula (\ref{eq:epsformula}) one 
derives, using the approximations (\ref{2.51}-\ref{2.53}), the constraint
\begin{equation}\label{100}
\bar\eta \left[(1-\bar\varrho) A^2 \eta_2 S_0(x_t)
+ P_0(\varepsilon) \right] A^2 B_K = 0.226,
\end{equation}
where
\begin{equation}\label{102}
P_0(\varepsilon) = 
\left[ \eta_3 S_0(x_c,x_t) - \eta_1 x_c \right] \frac{1}{\lambda^4},
\qquad
x_t=\frac{\mt^2}{\mw^2}.
\end{equation}
 $P_0(\varepsilon)=0.31\pm0.02$ summarizes the contributions
of box diagrams with two charm quark exchanges and the mixed 
charm-top exchanges. $P_0(\varepsilon)$ depends very weakly on $m_t$ and
its range given above corresponds to $155\gev \le m_t \le 185\gev$.

Equation (\ref{100}) specifies 
a hyperbola in the $(\bar \varrho, \bar\eta)$
plane (see fig.~\ref{fig:2011}). 
This hyperbola intersects the circle found in step 2
in two points which correspond to the two solutions for
$\delta$ mentioned earlier.
The position of the hyperbola (\ref{100}) in the
$(\bar\varrho,\bar\eta)$ plane depends on $\mt$, $|V_{cb}|=A \lambda^2$
and $B_K$. With decreasing $\mt$, $|V_{cb}|$ and $B_K$ the
$\eps_K$-hyperbola moves away from the origin of the
$(\bar\varrho,\bar\eta)$ plane. When the hyperbola and the circle
(\ref{rb}) touch each other lower bounds consistent with $\eps_K^{\rm
exp}$ for $\mt$, $|V_{cb}|$, $|V_{ub}/V_{cb}|$ and $B_K$ can be found.
The lower bound on $\mt$ is discussed in \cite{AB}.
Corresponding results for $|V_{ub}/V_{cb}|$ and $B_K$ can be found in
\cite{HNb,BBL}.
Approximate analytic expressions for these bounds have been derived in
\cite{AB,BBL}.
One has
\begin{eqnarray}
(\mt)_{\rm min} &=& \mw \left[ \frac{1}{2 A^2} \left( \frac{1}{A^2 B_K
R_b} - 1.4 \right) \right]^{0.658}
\label{eq:mtmin} \\
\left| \frac{V_{ub}}{V_{cb}} \right|_{\rm min} &=&
\frac{\lambda}{1-\lambda^2/2} \,
\left[ A^2 B_K \left( 2 x_t^{0.76} A^2 + 1.4 \right) \right]^{-1}
\label{eq:Vubcbmin} \\
(B_K)_{\rm min} &=& \left[ A^2 R_b \left( 2 x_t^{0.76} A^2 + 1.4 \right)
                    \right]^{-1}.
\label{eq:BKmin}
\end{eqnarray}

We will return to the bound (\ref{eq:Vubcbmin}) below.

{\bf Step 4:}

{}From the observed $B^0_d-\bar B^0_d$ mixing described experimentally 
by the mass difference $\Delta M_d$ or by the
mixing parameter $x_d=\Delta M_d/\Gamma_B$
and theoretically by the formula (\ref{eq:xds}),
the side $BA=R_t$ of the unitarity triangle can be determined:
\begin{equation}\label{106}
 R_t= \frac{1}{\lambda}\frac{|V_{td}|}{\vcb} = 1.0 \cdot
\left[\frac{|V_{td}|}{8.8\cdot 10^{-3}} \right] 
\left[ \frac{0.040}{\vcb} \right]
\end{equation}
with
\begin{equation}\label{VT}
\vtd=
8.8\cdot 10^{-3}\left[ 
\frac{200\mev}{\sqrt{B_{B_d}}F_{B_d}}\right]
\left[\frac{170~GeV}{\mtb(\mt)} \right]^{0.76} 
\left[\frac{\Delta M_d}{0.50/{\rm ps}} \right ]^{0.5} 
\sqrt{\frac{0.55}{\eta_B}}.
\end{equation}

{\bf Step 5:}

{}The measurement of $B^0_s-\bar B^0_s$ mixing parametrized by $\Delta M_s$
together with $\Delta M_d$  allows to determine $R_t$ in a different
way. Using (\ref{eq:xds}) and setting $\Delta M^{{\rm max}}_d= 0.482/
\mbox{ps}$ and 
$|V_{ts}/V_{cb}|^{{\rm max}}=0.993$ (see table \ref{TAB2}) one finds 
a useful formula \cite{AJBW}:
\begin{equation}\label{107b}
(R_t)_{\rm max} = 1.0 \cdot \xi \sqrt{\frac{10.2/ps}{\Delta M_s}},
\qquad
\xi = 
\frac{F_{B_s} \sqrt{B_{B_s}}}{F_{B_d} \sqrt{B_{B_d}}},
\end{equation}
where $\xi=1$ in the  $SU(3)$--flavour limit.
Note that $\mt$ and $|V_{cb}|$ dependences have been eliminated this way
 and that $\xi$ should in principle 
contain much smaller theoretical
uncertainties than the hadronic matrix elements in $\Delta M_d$ and 
$\Delta M_s$ separately.  

The most recent values relevant for (\ref{107b}) are:
\begin{equation}\label{107c}
\Delta M_s > 9.2/ {\rm ps}\,,
\qquad
\xi=1.15\pm 0.05\,.
\end{equation}
The first number is the improved lower bound quoted in \cite{Gibbons}
based in particular on ALEPH and DELPHI results.
The second number comes from quenched lattice calculations summarized
by Flynn in \cite{Flynn}.
A similar result has been obtained using QCD sum rules \cite{NAR}.
On the other hand another recent quenched lattice calculation \cite{Soni}
not included in (\ref{107c}) finds
$\xi\approx 1.3 $. Moreover one expects that unquenching will increase
the value of $\xi$ in (\ref{107c}) by roughly 10\% so that values as
high as $\xi=1.25-1.30$ are certainly possible even from Flynn's point of
view. For such high values of $\xi$
the lower bound on $\Delta M_s$ in (\ref{107c}) implies $R_t\le 1.37$
which as we will see below is similar to the bound obtained on the basis of
the first four steps alone. On the other hand, for $\xi=1.15$ one finds
$R_t \le 1.21 $ which puts an additional constraint on the unitarity
triangle cutting lower values of $\bar\varrho$ and higher values 
of $|V_{td}|$. In
view of remaining large uncertainties in $\xi$ we will not use the
constraint from $\Delta M_s$ below.

\subsection{Messages for  UT Practitioners}
Before presenting the numerical results of a standard analysis 
we would like to make a few important messages
for  UT-practitioners.

\subsubsection{\bf Message 1}\label{Mess1}

The parameter $m_t$, the top quark mass, used in weak decays is not
equal to the one measured by CDF and D0 and used in the electroweak 
precision studies at LEP,
SLD or FNAL. In the latter investigations the so-called pole mass is used,
whereas in all the NLO calculations listed in table 1 $m_t$ refers
to the running current top quark mass normalized at $\mu=m_t$:
$\mtb(\mt)$. One has
\begin{equation}\label{POLE}
\mtb(\mt)=\mt^{{\rm Pole}}
\left[ 1-\frac{4}{3}\frac{\alpha_s(m_t)}{\pi}\right]
\end{equation}
so that for $\mt={\cal O}(170\gev)$, $\bar m_t(m_t)$ is typically
by $8\gev$ smaller than $m_t^{\rm Pole}$. 
This difference matters already
because the most recent pole mass value from CDF and D0 has a very 
small error,
 $(175\pm 6)\gev$ \cite{Tipton}, implying
$(167\pm 6)\gev$ for $\mtb(\mt)$.
 In this review we will often denote this mass by $\mt$. Note
that we do not include $\alpha_s^2$ corrections in (\ref{POLE}),
which have to be dropped if one works at the NLO level.

\subsubsection{\bf Message 2}

When using numerical values for $m_t$, $B_K$, $B_B$ and the QCD factors
$\eta_i$, care must be taken that they are used consistently. This
unfortunately is not always the case. As an example let us consider
the theoretical expression for $\Delta M_d$ which reads
\begin{equation}\label{x_d}
\Delta M_d= C_B F_{B_d}^2 B_{B_d}(\mu_b) 
\left[\alpha^{(5)}_s(\mu_b)\right]^{-6/23}\left[
  1 + \frac{\alpha^{(5)}_s(\mu_b)}{4\pi} J_5\right]
     \eta_B(\mu_t,\bar m_t(\mu_t)) S_0(\bar x_t(\mu_t))\vtd^2
\end{equation}
with $C_B$ being a numerical constant and all other quantities defined
before.
Two relevant scales are 
$\mu_b={\cal O}(m_b)$ and $\mu_t={\cal O}(m_t)$ which according to
the rules of the renormalization group game can be chosen for instance
in the ranges $2.5~\mbox{GeV}\le \mu_b \le 10~\mbox{GeV}$ and 
$100\gev \le \mu_t \le 300\gev$, respectively. Here $\mu_b$ is the scale at
which the relevant $\Delta B=2$ operator is normalized and $\mu_t$ is
the scale at which $\mt$ is defined. Clearly $\Delta M_d$ cannot depend on
$\mu_b$ and $\mu_t$. Combining the explicit $\alpha_s$ factors in (\ref{x_d})
with $B_{B_d}(\mu_b)$ as in (\ref{eq:BBrenorm}) and introducing 
the renormalization group
invariant $B_{B_d}$ removes $\mu_b$ from phenomenological expressions like
(\ref{VT}). On the other hand,
 the $\mu_t$ dependence cancels between the last
two terms as demonstrated explicitly in \cite{BJW}. 
To this end the NLO calculation
for $\eta_B$ is essential. Otherwise $\Delta M_d$ shows a sizable $\mu_t$
dependence. It turns out that for a choice $\mu_t=m_t$, $\eta_B$ and
similarly $\eta_2$ in (\ref{100}) are practically independent of $m_t$. This
is convenient and has been adopted in \cite{BJW} and in subsequent 
NLO calculations.
Then $\eta_B=0.55$ and $\eta_2=0.57$ independent of $m_t$.

In the past the explicit $\alpha_s$ factors in (\ref{x_d}) have been
combined with $\eta_B$ to give the corresponding $\mu_b$ dependent
QCD factor as high as 0.85. This change is compensated by 
$B_B(\mu_b)<B_B$. In view of the fact that most non-perturbative
results are given for $B_B$ and $B_K$, it is important that this older
definition is abandoned. 

Similar messages apply to $\eta_i$ in the case of $\varepsilon_K$.

\subsubsection{\bf Message 3}

It is sometimes stated in the literature that the QCD factors $\eta_B$
for $B^0_d-\bar B^0_d$  and $B^0_s-\bar B^0_s$ mixings are roughly
equal to each other. They are
equal. Indeed, $\eta_B$ resulting from short distance QCD calculations is
independent of whether $B^0_d-\bar B^0_d$  or $B^0_s-\bar B^0_s$ is considered.
Consequently the ratio $\Delta M_d/\Delta M_s$ is independent of
$m_t$ and short distance QCD corrections.
The only difference in these two mixings arises through different CKM
factors
and through different hadronic matrix elements of the relevant $\Delta B=2$
operators which corresponds to $m_{B_s}\not=m_{B_d}$, 
$F_{B_s}\not=F_{B_d}$ and 
$B_{B_s}\not=B_{B_d}$. 
The last two differences are explicitly summarized by $\xi$ in (\ref{107b}).

\subsection{Numerical Results}\label{sec:standard}
\subsubsection{Input Parameters}
In table \ref{tab:inputparams} we summarize the input parameters 
which have been used in the standard analysis presented below.  

\begin{table}[thb]
\begin{center}
\begin{tabular}{|c|c|c|}
\hline
{\bf Quantity} & {\bf Central} & {\bf Error} \\
\hline
$|V_{cb}|$ & 0.040 & $\pm 0.003$ \\
$|V_{ub}/V_{cb}|$ & 0.080 & $\pm 0.020$ \\
$B_K$ & 0.75 & $\pm 0.15$ \\
$\sqrt{B_d} F_{B_{d}}$ & $200\mev$ & $\pm 40\mev$ \\
$\sqrt{B_s} F_{B_{s}}$ & $240\mev$ & $\pm 40\mev$ \\
$\mt$ & $167\gev$ & $\pm 6\gev$ \\
$\Delta M_d$ & $0.464~\mbox{ps}^{-1}$ & $\pm 0.018~\mbox{ps}^{-1}$ \\
$\Lms^{(4)}$ & $325\mev$ & $\pm 80\mev$ \\
\hline
\end{tabular}
\caption[]{Collection of input parameters.\label{tab:inputparams}}
\end{center}
\end{table}

\subsubsection{\bf $\left| V_{ub}/V_{cb} \right|$,
$\left| V_{cb} \right|$ and $\varepsilon_K$}

The values for $\left| V_{ub}/V_{cb} \right|$ 
and $\left| V_{cb} \right|$ in table \ref{tab:inputparams}
are not correlated with
each other. On the other hand such a correlation is present in
the analysis of the CP violating parameter $\varepsilon_K$ which
is roughly proportional to the fourth power of $\left| V_{cb}\right|$
and linear in $\left|V_{ub}/V_{cb} \right|$. It follows
that not all values in table \ref{tab:inputparams} are simultaneously
consistent with the observed value of $\varepsilon_K$.
This 
has been emphasized last year by
Herrlich and Nierste \cite{HNb} and in \cite{BBL}. 
Explicitly one has using (\ref{eq:Vubcbmin}):

\begin{equation}
\left| \frac{V_{ub}}{V_{cb}} \right|_{\rm min}=
\frac{0.225}{B_K A^2(2 x_t^{0.76}A^2+1.4)}.
\end{equation}

\begin{figure}[hbt]
\vspace{0.10in}
\centerline{
\epsfysize=4.5in
\rotate[r]{\epsffile{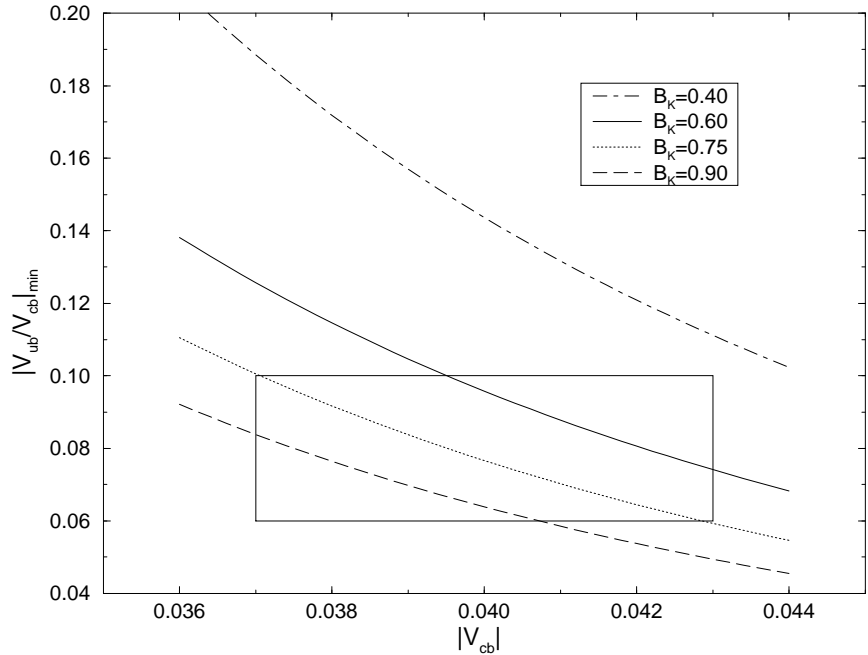}}
}
\vspace{0.08in}
\caption{Lower bound on $\vub$ from $\varepsilon_K$.}\label{fig:bound}
\end{figure}

This bound is shown as a function of $\vcb$ for different
values of $B_K$ and $\mt=173\gev$ in fig.\ \ref{fig:bound}. 
We observe that simultaneously
small values of $\left| V_{ub}/V_{cb} \right|$ and $\left| V_{cb} \right|$,
although still consistent with the ones given in 
table \ref{tab:inputparams}, are not allowed
by the size of indirect CP violation observed in $K \to \pi\pi$.

\subsubsection{Output of a Standard Analysis}
The output of the standard analysis depends to some extent on the
error analysis. This should always be remembered in view of the fact
that different authors use different procedures. In order to illustrate
this  we show in table \ref{TAB2}
 the results for various quantities of interest
using two types of error analyses:

\begin{itemize}
\item
Scanning: Both the experimentally measured numbers and the theoretical input
parameters are scanned independently within the errors given in
table~\ref{tab:inputparams}. 
\item
Gaussian: The experimentally measured numbers and the theoretical input 
parameters are used with Gaussian errors.
\end{itemize}
Clearly the ``scanning'' method is a bit conservative. On the other
hand using Gaussian distributions for theoretical input parameters
can certainly be questioned. 
Personally we think that
at present the conservative ``scanning'' method should be preferred.
In the future, however, when data and theory improve, it would be useful to  
find a less conservative estimate which most probably will give errors
somewhere inbetween these two error estimates. 
The analysis discussed here has been done by Matthias 
Jamin, Markus Lautenbacher and the first author. 
More details and more results can be found in 
\cite{BJL96b}.

\begin{table}[bth]
\begin{center}
\begin{tabular}{|c||c||c|}\hline
{\bf Quantity} & {\bf Scanning} & {\bf Gaussian} \\ \hline
$\mid V_{td}\mid/10^{-3}$ &$6.9 - 11.3$ &$ 8.6\pm 1.1$ \\ \hline
$\mid V_{ts}/V_{cb}\mid$ &$0.959 - 0.993$ &$0.976\pm 0.010$  \\ \hline
$\mid V_{td}/V_{ts}\mid$ &$0.16 - 0.31$ &$0.213\pm 0.034$  \\ \hline
$\sin(2\beta)$ &$0.36 - 0.80$ &$ 0.66\pm0.13 $ \\ \hline
$\sin(2\alpha)$ &$-0.76 - 1.0$ &$ 0.11\pm 0.55 $ \\ \hline
$\sin(\gamma)$ &$0.66 - 1.0 $ &$ 0.88\pm0.10 $ \\ \hline
$\IM \lambda_t/10^{-4}$ &$0.86 - 1.71 $ &$ 1.29\pm 0.22 $ \\ \hline
$\Delta M_s~ {\rm ps}$ &$ 8.0 - 25.4$ &$ 15.2 \pm 5.5 $ \\ \hline
\end{tabular}
\caption[]{Output of the Standard Analysis. 
 $\lambda_t=V^*_{ts} V_{td}$.\label{TAB2}}
\end{center}
\end{table}

Comparing the results for $\vtd$ with (\ref{uni1}) we observe that
the inclusion of the constraints from $\varepsilon$ and $\Delta M_d$
had a considerable impact on the allowed range for this CKM matrix
element. Similarly we observe that whereas $\sin 2\beta$ and $\sin\gamma$
are rather constrained, the uncertainty in $\sin 2\alpha$ is huge. Similarly
the uncertainties in $\IM\lambda_t$ and $\Delta M_s$ are large.

\begin{figure}[hbt]
\vspace{0.10in}
\centerline{
\epsfysize=5.5in
\rotate[r]{\epsffile{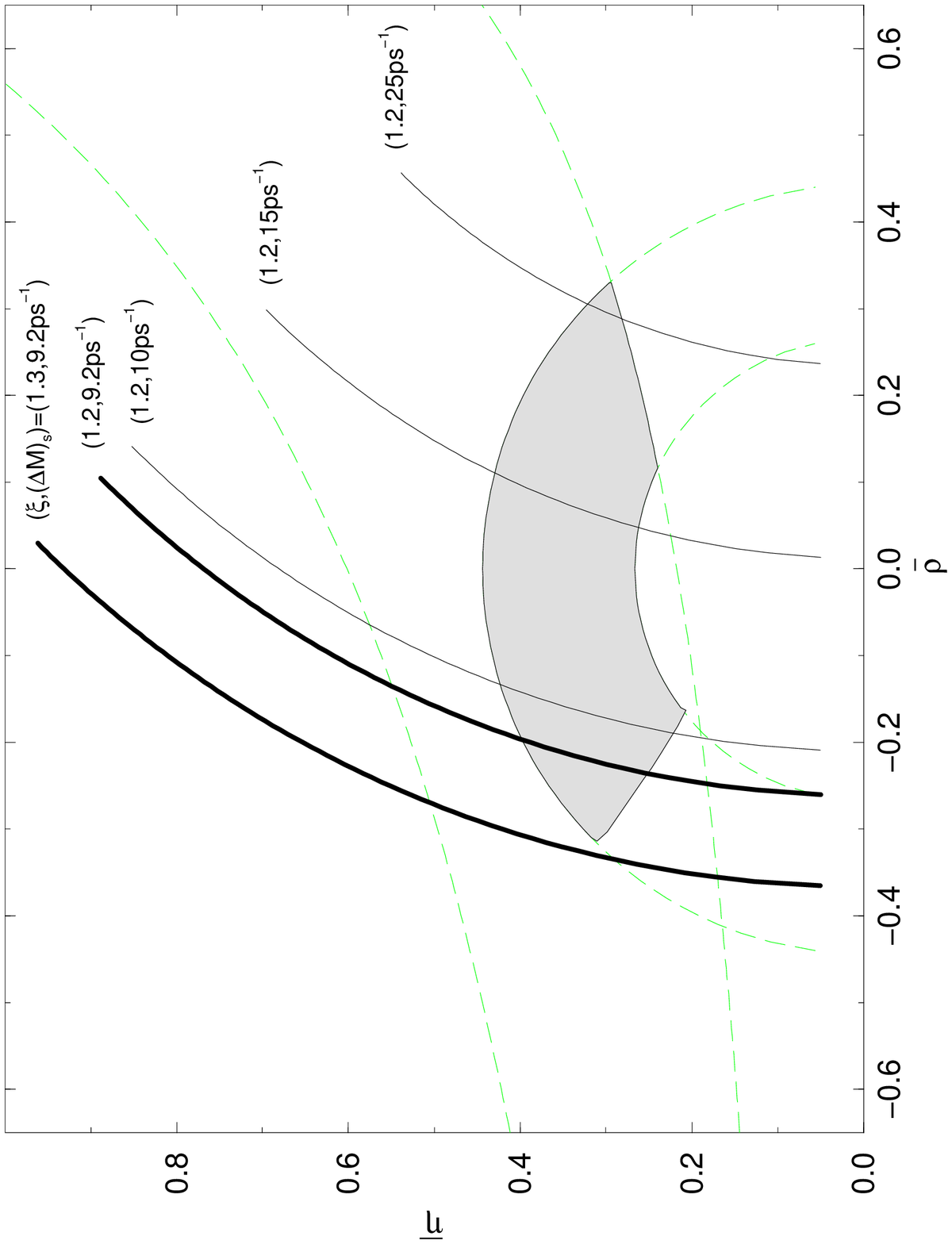}}
}
\vspace{0.08in}
\caption{Unitarity Triangle 1997.}
\label{fig:utdata}
\end{figure}

In fig.\ \ref{fig:utdata}  we show the range for the upper
corner A of the unitarity triangle. The solid thin lines correspond 
to $R_t^{\rm max}$ from 
(\ref{107b})
using $\xi=1.20$ and $\Delta M_s=10/{\rm ps},~15/{\rm ps}$ and 
$25/{\rm ps}$, respectively.
The allowed region has a typical ``banana'' shape which can be found
in many other analyses \cite{BLO,ciuchini:95,HNb,ALUT,PP,PW}. The size of
the banana and its position depends on the assumed input
parameters and on the error analysis which varies from paper
to paper. The results in fig.\ \ref{fig:utdata} correspond 
to a simple independent 
scanning of all parameters within one standard deviation.
Such an approach is more conservative than using
Gaussian distributions as done in some papers quoted above.
We show also the impact of the experimental bound 
$\Delta M_s>9.2/{\rm ps}$
with $\xi=1.20$ and the corresponding bound for $\xi=1.30$. In view
of the remaining uncertainty in $\xi$, in particular due to quenching
in lattice calculations,
 this bound has not been used in
obtaining the results in table \ref{TAB2}. 
It is evident, however, that $B^0_s-\bar
B^0_s$ mixing will have a considerable impact on the unitarity triangle
when the value of $\xi$ will be known better  and  data improves.
This is very desirable because as seen in fig.\ \ref{fig:utdata}
 our knowledge of
the unitarity triangle is still rather poor. 

\section{$\epe$ in the Standard Model}\label{EpsilonPrime}
\setcounter{equation}{0}
\subsection{Preliminaries}
Direct CP violation remains one of the important targets 
of contemporary particle physics. In this respect the search 
for direct CP violation in $K\to\pi\pi$ decays plays a special
role as already fifteen years have been devoted to this enterprise.
In this case,
a non-vanishing value of the ratio Re($\epe$) defined in (\ref{eprime}) 
would give the first
signal for direct CP violation ruling out superweak models.
The experimental situation of Re($\varepsilon'/\varepsilon$) is,
however, unclear
at present:
\begin{equation}\label{eprime1}
Re(\varepsilon'/\varepsilon) =\left\{ \begin{array}{ll}
(23 \pm 7)\cdot 10^{-4} & \cite{barr:93} \\
(7.4 \pm 5.9)\cdot 10^{-4} & \cite{gibbons:93}.\end{array} \right.
\end{equation}

While the result of the NA31 collaboration at CERN  \cite{barr:93}
clearly indicates direct CP violation, the value of E731 at Fermilab
\cite{gibbons:93} is compatible with superweak theories
\cite{wolfenstein:64} in which $\varepsilon'/\varepsilon = 0$.
 Hopefully, in about two years the experimental situation concerning
$\varepsilon'/\varepsilon$ will be clarified through the improved
measurements by the two collaborations at the $10^{-4}$ level and by
the KLOE experiment at  DA$\Phi$NE.

There is no question about that direct CP violation is present in
the Standard Model. Yet accidentally it could turn out that it will be
difficult to see it in $K \to \pi\pi$ decays.  Indeed in the Standard
Model $\varepsilon'/\varepsilon $ is governed by QCD penguins and
electroweak (EW) penguins. In spite of being suppressed by
$\alpha/\alpha_s$ relative to QCD penguin contributions, 
electroweak penguin contributions have to be included because of the
additional enhancement factor ${\rm Re}A_0/{\rm Re}A_2=22$ 
(see (\ref{eq:epsprim})-(\ref{eq:ReA0data})) relative
to QCD penguins. With increasing $\mt$ the EW penguins become
increasingly important \cite{flynn:89,buchallaetal:90} and, entering
$\varepsilon'/\varepsilon$ with the opposite sign to QCD penguins,
suppress this ratio for large $\mt$. For $\mt\approx 200\,\gev$ the ratio
can even be zero \cite{buchallaetal:90}.  Because of this strong
cancellation between two dominant contributions and due to uncertainties
related to hadronic matrix elements of the relevant local operators, a
precise prediction of $\varepsilon'/\varepsilon$ is not possible at
present. We will discuss this in detail below.

The first calculations of $\epe$ for $\mt \ll \mw$ and in the leading
order approximation can be found in \cite{GW79}. 
For $\mt \ll \mw$ only $QCD$
penguins play a substantial role. Over the eighties these calculations
were refined through the inclusion of isospin braking in the
quark masses \cite{donoghueetal:86,burasgerard:87,lusignoli:89},
the inclusion of QED penguin effects for $\mt \ll \mw$
\cite{BW84,donoghueetal:86,burasgerard:87}, 
and through improved estimates of hadronic matrix elements in
the framework of the $1/N$ approach \cite{bardeen:87}. 
This era of $\epe$ culminated
in the analyses in \cite{flynn:89,buchallaetal:90}, where QCD
penguins, electroweak penguins ($\gamma$ and $Z^0$ penguins)
and the relevant box diagrams were included for arbitrary
top quark masses. The strong cancellation between QCD penguins
and electroweak penguins for $m_t > 150~\gev$ found in these
papers was confirmed by other authors \cite{PW91}.

All these calculations were done in the leading logarithmic
approximation (e.g.\ one-loop anomalous dimensions of the relevant
operators) with the exception of the $\mt$-dependence which in 
the analyses \cite{flynn:89,buchallaetal:90,PW91} has been already
included at the NLO level. While such a procedure is not fully
consistent, it allowed for the first time to exhibit the strong
$\mt$-dependence of the electroweak penguin contributions,
which is not seen in a strict leading logarithmic approximation.

During the nineties considerable progrees has been made by
calculating complete NLO corrections to $\varepsilon'$
\cite{BJLW1,BJLW2,BJLW,ROMA1,ROMA2}. Together with the NLO
corrections to $\varepsilon_K$ and $B^0-\bar B^0$ mixing
discussed in the previous section, this allows
a complete NLO analysis of $\varepsilon'/\varepsilon$ including
constraints from the observed indirect CP violation ($\varepsilon_K$)
and  $B^0-\bar B^0$ mixing ($\Delta M_{d,s}$). The improved
determination of the $V_{ub}$ and $V_{cb}$ elements of the CKM matrix,
the improved estimates of hadronic matrix elements using the lattice 
approach and in particular the determination of the top quark mass
$\mt$ had of course also an important impact on
$\varepsilon'/\varepsilon$. 

After these general remarks we will now summarize the present status of
$\epe$ in explicit terms.
       
\subsection{Basic Formulae}
           \label{subsec:epeformulae}
The direct CP violation in $K \to \pi\pi$ is described by the parameter
$\varepsilon'$ defined in (\ref{eprime}).
It is is given in terms of the real and imaginary parts of 
the amplitudes $A_0 \equiv
A(K \to (\pi\pi)_{I=0})$ and $A_2 \equiv
A(K \to (\pi\pi)_{I=2})$ as follows:
\begin{equation}
\eps' = -\frac{\omega}{\sqrt{2}} \xi (1 - \Omega) \exp(i \Phi) \, ,
\label{eq:epsprim}
\end{equation}
where
\begin{equation}
\xi = \frac{\IM A_0}{\RE A_0} \, , \quad
\omega = \frac{\RE A_2}{\RE A_0} \, , \quad
\Omega = \frac{1}{\omega} \frac{\IM A_2}{\IM A_0}
\label{eq:xiomega}
\end{equation}
and $\Phi = \pi/2 + \delta_2 - \delta_0 \approx \pi/4$.
$\IM A_0$ is dominated by QCD penguins and is very weakly dependent
on $\mt$. $\IM A_2$ increases on the other hand strongly with $\mt$
and for large $\mt$ is dominated by electroweak penguins. It receives
also a sizable contribution from isospin braking $(m_u\not=m_d)$ which
conspires with electroweak penguins to cancel substantially the
QCD penguin contribution in $\IM A_0$. The factor $1/\omega\approx 22$
in $\Omega$ giving a large enhancement is to a large extend responsible
for this cancellation.

When using (\ref{eq:epsprim}) and (\ref{eq:xiomega}) in phenomenological
applications one usually takes $\RE A_0$ and $\omega$ from
experiment, i.e.
\begin{equation}
\RE A_0 = 3.33 \cdot 10^{-7}\gev,
\qquad
\RE A_2 = 1.50 \cdot 10^{-8}\gev,
\qquad
\omega = 0.045,
\label{eq:ReA0data}
\end{equation}
where the last relation reflects the so-called $\Delta I=1/2$ rule. The
main reason for this strategy is the unpleasant fact that until today
nobody succeded in fully explaining this rule which to a large extent is
believed to originate in the long-distance QCD contributions
\cite{DI12}. 
On the other hand the
imaginary parts of the amplitudes in (\ref{eq:xiomega}) being related to
CP violation and the top quark physics should be dominated by
short-distance contributions. Therefore $\IM A_0$ and $\IM A_2$ are
usually calculated using the effective Hamiltonian for $\Delta S=1$
transitions:

\begin{equation}
\Heff(\Delta S=1) = 
\frac{G_{\rm F}}{\sqrt{2}} V_{us}^* V_{ud}^{} \sum_{i=1}^{10}
\left( z_i(\mu) + \tau \; y_i(\mu) \right) Q_i(\mu) 
\label{eq:HeffdF1:1010}
\end{equation}
with $\tau=-V_{ts}^* V_{td}^{}/(V_{us}^* V_{ud}^{})$.

The operators $Q_i$ are the analogues of the ones given in 
(\ref{O1})-(\ref{O5}).
They are given explicitly  as follows:

{\bf Current--Current :}
\begin{equation}\label{OS1} 
Q_1 = (\bar s_{\alpha} u_{\beta})_{V-A}\;(\bar u_{\beta} d_{\alpha})_{V-A}
~~~~~~Q_2 = (\bar s u)_{V-A}\;(\bar u d)_{V-A} 
\end{equation}

{\bf QCD--Penguins :}
\begin{equation}\label{OS2}
Q_3 = (\bar s d)_{V-A}\sum_{q=u,d,s}(\bar qq)_{V-A}~~~~~~   
 Q_4 = (\bar s_{\alpha} d_{\beta})_{V-A}\sum_{q=u,d,s}(\bar q_{\beta} 
       q_{\alpha})_{V-A} 
\end{equation}
\begin{equation}\label{OS3}
 Q_5 = (\bar s d)_{V-A} \sum_{q=u,d,s}(\bar qq)_{V+A}~~~~~  
 Q_6 = (\bar s_{\alpha} d_{\beta})_{V-A}\sum_{q=u,d,s}
       (\bar q_{\beta} q_{\alpha})_{V+A} 
\end{equation}

{\bf Electroweak--Penguins :}
\begin{equation}\label{OS4} 
Q_7 = {3\over 2}\;(\bar s d)_{V-A}\sum_{q=u,d,s}e_q\;(\bar qq)_{V+A} 
~~~~~ Q_8 = {3\over2}\;(\bar s_{\alpha} d_{\beta})_{V-A}\sum_{q=u,d,s}e_q
        (\bar q_{\beta} q_{\alpha})_{V+A}
\end{equation}
\begin{equation}\label{OS5} 
 Q_9 = {3\over 2}\;(\bar s d)_{V-A}\sum_{q=u,d,s}e_q(\bar q q)_{V-A}
~~~~~Q_{10} ={3\over 2}\;
(\bar s_{\alpha} d_{\beta})_{V-A}\sum_{q=u,d,s}e_q\;
       (\bar q_{\beta}q_{\alpha})_{V-A} \,.
\end{equation}
Here, $e_q$ denotes the electrical quark charges reflecting the
electroweak origin of $Q_7,\ldots,Q_{10}$. 

The Wilson coefficient functions $z_i(\mu)$ and $ y_i(\mu)$
were calculated including
the complete next-to-leading order (NLO) corrections in
\cite{BJLW1,BJLW2,BJLW,ROMA1,ROMA2}. The details
of these calculations can be found there and in the review
\cite{BBL}. Only the coefficients $ y_i(\mu)$ enter the evaluation
of $\epe$. Examples of their numerical values are given in table 
\ref{tab:wc10smu13}.
Extensive tables for $ y_i(\mu)$ can be found in \cite{BBL}.

\begin{table}[htb]
\begin{center}
\begin{tabular}{|c|c|c|c||c|c|c||c|c|c|}
\hline
& \multicolumn{3}{c||}{$\Lms^{(4)}=245\mev$} &
  \multicolumn{3}{c||}{$\Lms^{(4)}=325\mev$} &
  \multicolumn{3}{c| }{$\Lms^{(4)}=405\mev$} \\
\hline
Scheme & LO & NDR & HV & LO & 
NDR & HV & LO & NDR & HV \\
\hline
$z_1$ & -0.550 & -0.364 & -0.438 & -0.625 & 
-0.415 & -0.507 & -0.702 & -0.469 & -0.585 \\
$z_2$ & 1.294 & 1.184 & 1.230 & 1.345 & 
1.216 & 1.276 & 1.399 & 1.251 & 1.331 \\
\hline
$y_3$ & 0.029 & 0.024 & 0.027 & 0.034 & 
0.029 & 0.033 & 0.039 & 0.034 & 0.039 \\
$y_4$ & -0.054 & -0.050 & -0.052 & -0.061 & 
-0.057 & -0.060 & -0.068 & -0.065 & -0.068 \\
$y_5$ & 0.014 & 0.007 & 0.014 & 0.015 & 
0.005 & 0.016 & 0.016 & 0.002 & 0.018 \\
$y_6$ & -0.081 & -0.073 & -0.067 & -0.096 & 
-0.089 & -0.081 & -0.113 & -0.109 & -0.097 \\
\hline
$y_7/\aem$ & 0.032 & -0.031 & -0.030 & 0.039 & 
-0.030 & -0.028 & 0.045 & -0.029 & -0.026 \\
$y_8/\aem$ & 0.100 & 0.111 & 0.120 & 0.121 & 
0.136 & 0.145 & 0.145 & 0.166 & 0.176 \\
$y_9/\aem$ & -1.445 & -1.437 & -1.437 & -1.490 & 
-1.479 & -1.479 & -1.539 & -1.528 & -1.528 \\
$y_{10}/\aem$ & 0.588 & 0.477 & 0.482 & 0.668 & 
0.547 & 0.553 & 0.749 & 0.624 & 0.632 \\
\hline
\end{tabular}
\end{center}
\caption[]{$\Delta S=1 $ Wilson coefficients at $\mu=\mc=1.3\gev$ for
$\mt=170\gev$ and $f=3$ effective flavours.
$|z_3|,\ldots,|z_{10}|$ are numerically irrelevant relative to
$|z_{1,2}|$. $y_1 = y_2 \equiv 0$.
\label{tab:wc10smu13}}
\end{table}

Using the Hamiltonian in (\ref{eq:HeffdF1:1010}) and the experimental
values for $\varepsilon$, $\RE A_0$ and $\omega$ the ratio $\epe$ can be
written as follows:
\begin{equation}
\frac{\varepsilon'}{\varepsilon} = 
\IM \lambda_t\cdot \left[ P^{(1/2)} - P^{(3/2)} \right],
\label{eq:epe}
\end{equation}
where
\begin{eqnarray}
P^{(1/2)} & = & r \sum y_i \langle Q_i\rangle_0
(1-\Omega_{\eta+\eta'})
\label{eq:P12} \\
P^{(3/2)} & = &\frac{r}{\omega}
\sum y_i \langle Q_i\rangle_2~~~~~~
\label{eq:P32}
\end{eqnarray}
with
\begin{equation}
r = \frac{G_{\rm F} \omega}{2 |\eps| \RE A_0} \, 
\qquad
\langle Q_i\rangle_I \equiv \langle (\pi\pi)_I | Q_i | K \rangle .
\label{eq:repe}
\end{equation}
Note that the overall phases in $\varepsilon'$ and $\varepsilon$ cancel
in the ratio to an excellent approximation.
The sum in (\ref{eq:P12}) and (\ref{eq:P32}) runs over all contributing
operators. $P^{(3/2)}$ is fully dominated by electroweak penguin
contributions. $P^{(1/2)}$ on the other hand is governed by QCD penguin
contributions which are suppressed by isospin breaking in the quark
masses ($m_u \not= m_d$). The latter effect is described by

\begin{equation}
\Omega_{\eta+\eta'} = \frac{1}{\omega} \frac{(\IM A_2)_{\rm
I.B.}}{\IM A_0}\,.
\label{eq:Omegaeta}
\end{equation}
For $\Omega_{\eta+\eta'}$ we will take
\begin{equation}
\Omega_{\eta+\eta'} = 0.25 \pm 0.05\,,
\label{eq:Omegaetadata}
\end{equation}
which is in the ball park of the values obtained in the $1/N_c$ approach
\cite{burasgerard:87} and in chiral perturbation theory
\cite{donoghueetal:86,lusignoli:89}. $\Omega_{\eta+\eta'}$ is
independent of $\mt$.

The main source of uncertainty in the calculation of
$\epe$ are the hadronic matrix elements $\langle Q_i \rangle_I$.
They depend generally
on the renormalization scale $\mu$ and on the scheme used to
renormalize the operators $Q_i$. These two dependences are canceled by
those present in the Wilson coefficients $y_i(\mu)$ so that the
resulting physical $\epe$ does not (in principle) depend on $\mu$ and on the
renormalization scheme of the operators.  Unfortunately the accuracy of
the present non-perturbative methods used to evalutate $\langle Q_i
\rangle_I$, like lattice methods or the $1/N_c$ expansion, is not
sufficient to obtain the required $\mu$ and scheme dependences of
$\langle Q_i \rangle_I$. A review of the existing methods and their
comparison can be found in \cite{BJLW}, \cite{ciuchini:95}.

In view of this situation it has been suggested in \cite{BJLW} to
determine as many matrix elements $\langle Q_i \rangle_I$ as possible
from the leading CP conserving $K \to \pi\pi$ decays, for which the
experimental data are summarized in (\ref{eq:ReA0data}). To this end it
turned out to be very convenient to determine $\langle Q_i \rangle_I$
at a scale $\mu = \mc$.  Using the renormalization group evolution one
can then find $\langle Q_i \rangle_I$ at any other scale $\mu \not=
\mc$. The details of this procedure can be found in
\cite{BJLW}. We will briefly summarize the most important results of this
work below.

\subsection{Hadronic Matrix Elements}

It is customary to express the matrix elements
$\langle Q_i \rangle_I$ in terms of non-perturbative parameters
$B_i^{(1/2)}$ and $B_i^{(3/2)}$ as follows:
\begin{equation}
\langle Q_i \rangle_0 \equiv B_i^{(1/2)} \, \langle Q_i
\rangle_0^{\rm (vac)}\,,
\qquad
\langle Q_i\rangle_2 \equiv B_i^{(3/2)} \, \langle Q_i
\rangle_2^{\rm (vac)} \,.
\label{eq:1}
\end{equation}
The label ``vac'' stands for the vacuum
insertion estimate of the hadronic matrix elements in question. 
It suffices to give here only a few examples \cite{BJLW}:
\begin{eqnarray}
\langle Q_1 \rangle_0 &=& -\,\frac{1}{9} X B_1^{(1/2)} \, ,
\label{eq:Q10} \\
\langle Q_2 \rangle_0 &=&  \frac{5}{9} X B_2^{(1/2)} \, ,
\label{eq:Q20} \\
\langle Q_6 \rangle_0 &=&  -\,4 \sqrt{\frac{3}{2}} 
\left[ \frac{m_{\rm K}^2}{\ms(\mu) + \md(\mu)}\right]^2
\frac{F_\pi}{\kappa} \,B_6^{(1/2)} \, ,
\label{eq:Q60}\\ 
\langle Q_1 \rangle_2 &=& 
\langle Q_2 \rangle_2 = \frac{4 \sqrt{2}}{9} X B_1^{(3/2)} \, ,
\label{eq:Q122} \\
\langle Q_i \rangle_2 &=&  0 \, , \qquad i=3,\ldots,6 \, ,
\label{eq:Q362} \\
\langle Q_8 \rangle_2 &=& 
  -\left[ \frac{\kappa}{2 \sqrt{2}} \langle \overline{Q_6} \rangle_0
          + \frac{\sqrt{2}}{6} X
   \right] B_8^{(3/2)} \, ,
\label{eq:Q82} \\
\langle Q_9 \rangle_2 &=& 
   \langle Q_{10} \rangle_2 = \frac{3}{2} \langle Q_1 \rangle_2 \, ,
\label{eq:Q9102}
\end{eqnarray}
where
\begin{equation}
\kappa = 
         \frac{F_\pi}{F_{\rm K} - F_\pi} \, ,
\qquad
X = \sqrt{\frac{3}{2}} F_\pi \left( m_{\rm K}^2 - m_\pi^2 \right) \, ,
\label{eq:XQi}
\end{equation}
and
\begin{equation}
\langle \overline{Q_6} \rangle_0 =
   \frac{\langle Q_6 \rangle_0}{B_6^{(1/2)}} \, .
\label{eq:Q60bar}
\end{equation}
In the vacuum insertion method $B_i=1$ independent of $\mu$. In QCD,
however, the hadronic parameters $B_i$ generally depend on the
renormalization scale $\mu$ and the renormalization scheme considered.

In view of the smallness of $\tau=\ord(10^{-4})$ entering
(\ref{eq:HeffdF1:1010}), the
real amplitudes in (\ref{eq:ReA0data}) are governed by the coefficients
$z_i(\mu)$. The method of extracting some of the matrix elements
from the data as proposed in \cite{BJLW} relies then on the
fact that due to the GIM mechanism the coefficients $z_i(\mu)$ of
the penguin operators (i=3....10) vanish for $\mu=m_c$ in the HV
scheme and are negligible in the NDR scheme. This allows to
to find
\begin{equation}
\langle Q_1(m_c) \rangle_2 = \langle Q_2(m_c) \rangle_2 =
\frac{10^6\gev^2}{1.77} \frac{\RE A_2}{z_+(m_c)} =
\frac{8.47 \cdot 10^{-3}\gev^3}{z_+(m_c)}
\label{eq:Q122data}
\end{equation}
with $z_+=z_1+z_2$ and
\begin{equation}
\langle Q_1(\mc) \rangle_0 = \frac{10^6\gev^2}{1.77} \frac{\RE
A_0}{z_1(\mc)} - \frac{z_2(\mc)}{z_1(\mc)} \langle Q_2(\mc) \rangle_0\,.
\label{eq:Q10mc}
\end{equation}

Comparing (\ref{eq:Q122data})  with (\ref{eq:Q122}) one finds immediately
\begin{equation}
B_1^{(3/2)}(m_c) = \frac{0.363}{z_+(m_c)}\,,
\label{eq:B321}
\end{equation}
which using table \ref{tab:wc10smu13} gives for $\mc=1.3\gev$ and 
$\Lms^{(4)}=325\mev$
\begin{equation}
B_{1,NDR}^{(3/2)}(\mc) =  0.453\,,
\qquad
B_{1,HV}^{(3/2)}(\mc) =  0.472 \, .
\label{eq:B321mc}
\end{equation}
The extracted values for $B_1^{(3/2)}$ are by more than a factor of two
smaller than the vacuum insertion estimate.
They are compatible with the $1/N_c$ value $B_1^{(3/2)}(1\gev) \approx
0.55$ \cite{bardeen:87} and are somewhat smaller than the lattice result
$B_1^{(3/2)}(2\gev) \approx 0.6$ \cite{ciuchini:95}.
As analyzed in \cite{BJLW},
$B_1^{(3/2)}(\mu)$ decreases slowly with increasing $\mu$.
As seen in (\ref{eq:Q9102}), this analysis gives also
$\langle Q_9(\mc) \rangle_2$ and $\langle Q_{10}(\mc) \rangle_2$.

In order to extract $B_1^{(1/2)}(\mc)$ and $B_2^{(1/2)}(\mc)$ from
(\ref{eq:Q10mc})
one can make the very plausible assumption $\langle Q_-(\mc) \rangle_0 \ge
\langle Q_+(\mc) \rangle_0 \ge 0$, where $Q_\pm=(Q_2\pm Q_1)/2$ which is
valid in known non-perturbative approaches. 
This gives  for $\Lms^{(4)}=325\mev$
\begin{equation}
B_{2,NDR}^{(1/2)}(\mc) =  6.6 \pm 1.0,
\qquad
B_{2,HV}^{(1/2)}(\mc) =  6.2 \pm 1.0 \, .
\label{eq:B122mc}
\end{equation}
The extraction of $B_1^{(1/2)}(\mc)$ and of analogous parameters
$B_{3,4}^{(1/2)}(\mc)$ are presented in detail in \cite{BJLW}.
$B_1^{(1/2)}(\mc)$ depends very sensitively on $B_2^{(1/2)}(\mc)$ and
its central value is as high as 15. $B_4^{(1/2)}(\mc)$ is 
typically by (10--15)\,\% lower than $B_2^{(1/2)}(\mc)$. In any case
this analysis shows very large deviations from the results of the
vacuum insertion method.

The matrix elements of the $(V-A) \otimes (V+A)$ operators $Q_5$--$Q_8$
cannot be constrained by CP conserving data and one has to rely on
existing non-perturbative methods to calculate them. 
This is rather unfortunate because the QCD penguin operator $Q_6$
and the electroweak penguin operator $Q_8$, having large Wilson
coefficients and large hadronic matrix elements, play the dominant
role in $\epe$. 

The values of $B_i$ factors describing the matrix elements of
$Q_5-Q_8$ operators are equal to unity in the 
vacuum insertion method.
The same result is found in the large $N$ limit
\cite{bardeen:87,burasgerard:87}. Also lattice calculations give
similar results: $B^{1/2}_{5,6}=1.0 \pm 0.2$ \cite{kilcup:91,sharpe:91} 
and $B^{3/2}_{7,8}= 1.0 \pm 0.2$ 
\cite{kilcup:91}-\cite{francoetal:89},
$B^{3/2}_8= 0.81(1)$ \cite{gupta:96}. These are the values used in
\cite{BJLW,ciuchini:95,BBL,BJL96a}.
In the chiral quark model one finds \cite{bertolinietal:95}: 
$B^{1/2}_6=1.0\pm0.4$,
$B^{3/2}_8=2.2\pm1.5$ and generally $B^{3/2}_8>B^{1/2}_6$. 
On the other hand the Dortmund group
\cite{heinrichetal:92, paschos:96} advocates  
$B^{1/2}_6>B^{3/2}_8$. From \cite{paschos:96} $B^{1/2}_6=1.3$ and 
$B^{3/2}_8=0.7$ can be extracted.
Concerning $B_{7,8}^{(1/2)}$ one can simply set $B_{7,8}^{(1/2)}=1$ as
the matrix elementes $\langle Q_{7,8} \rangle_0$ play only a minor role
in the $\epe$ analysis.

As demonstrated in \cite{BJLW}, the parametrs $B_{5,6}^{(1/2)}$ and
$B_{7,8}^{(3/2)}$ depend only very weakly on the renormalization scale
$\mu$ when $\mu > 1\gev$ is considered. The $\mu$ dependence of the
matrix elements $\langle Q_{5,6} \rangle_0$
and $\langle Q_{7,8} \rangle_2$ is then given to excellent acurracy
by the $\mu$ dependence of
$\ms(\mu)$. For $\langle Q_6 \rangle_0$ and $\langle Q_8 \rangle_2$
this property has been first found in the $1/N_c$ approach
\cite{burasgerard:87}: in the large-$N_c$ limit the anomalous
dimensions of $Q_6$ and $Q_8$ are simply twice the anomalous dimension
of the mass operator leading to $\sim 1/\ms^2(\mu)$ for the
corresponding matrix elements. In the numerical
renormalization study in \cite{BJLW} the
factors $B_{5,6}^{(1/2)}$ and $B_{7,8}^{(3/2)}$  have been set to unity 
at $\mu=\mc$.
Subsequently the evolution of the matrix elements in the range $1\gev
\le \mu \le 4\gev$ has been calculated showing that for the NDR scheme
$B_{5,6}^{(1/2)}$ and $B_{7,8}^{(3/2)}$ were $\mu$ independent within
an accuracy of (2--3)\,\%. The $\mu$ dependence in the HV scheme has
been found to be stronger but still below 10\,\%.

In summary the treatment of $\langle Q_i \rangle_{0,2}$, $i=5,\ldots 8$
in \cite{BJLW,BBL,BJL96a} is to set
\begin{equation}
B_{7,8}^{(1/2)}(\mc) = 1,
\qquad
B_5^{(1/2)}(\mc) = B_6^{(1/2)}(\mc),
\qquad
B_7^{(3/2)}(\mc) = B_8^{(3/2)}(\mc)
\label{eq:B1278mc}
\end{equation}
and to treat $B_6^{(1/2)}(\mc)$ and $B_8^{(3/2)}(\mc)$ as free
parameters in the range
\begin{equation}
B_6^{(1/2)}(\mc)=1.0 \pm 0.2,
\qquad
B_8^{(3/2)}(\mc)=1.0\pm 0.2
\label{eq:B78mc}
\end{equation}
suggested by lattice calculations.
Then the main uncertainty in the values of $\langle Q_i \rangle_{0,2}$,
$i=5,\ldots 8$ results from the value of the strange quark mass
$\ms(\mc)$.

It seems therefore appropriate to summarize now the present status of 
the value of the strange quark mass.
The most recent results of QCD sum
rule (QCDSR) calculations \cite{jaminmuenz:95,chetyrkinetal:95,narison:95}
obtained at $\mu=1$ GeV correspond to $\ms(\mc)=(170\pm20)\,\mev$ with
$\mc=1.3\,\gev$. The lattice calculation of \cite{alltonetal:94} finds
$\ms(2\,\gev)=(128\pm18)\,\mev$ which corresponds to 
$\ms(\mc)=(150\pm20)\,\mev$,
in rather good agreement with the QCDSR result. In summer 1996 a new lattice
result has been presented by Gupta and Bhattacharya \cite{gupta:96}.
In the quenched approximation they
find $\ms(2\,\gev)=(90\pm20)\,\mev$ corresponding to $\ms(\mc)=
(105\pm20)\,\mev$.
For $n_f=2$ the value is found to be even lower: $\ms(2\,\gev)=
(70\pm15)\,\mev$
corresponding to $\ms(\mc)=(82\pm17)\,\mev$. Similar results are found
by the lattice group at FNAL \cite{FNAL:96}.

The situation with the strange quark mass is therefore unclear at present
and it is useful to present the results for its low and high values.
Such an analysis has been done recently in \cite{BJL96a} in which the
values
\begin{equation}
\ms(\mc)=(150\pm20)\,\mev
\quad
{\rm and}
\quad
\ms(\mc)=(100\pm20)\,\mev
\end{equation}
have been used. The results presented below are the results of this
paper. For convenience we also provide in 
table \ref{tab:ms} the dictionary between the values of $\ms$ normalized at
different scales \cite{BJL96a}. To this end the standard renormalization
group formula at the two-loop level with $\Lms^{(4)}=325~\mev$ has been used.

\begin{table}[thb]
\begin{center}
\begin{tabular}{|c|c|c|c|c|c|}\hline
  $\ms(\mc)$& $ ~75$& $100$& $125$ & $150$ &  $175$ \\ \hline
 $\ms(2~\gev)$& $ ~64$& $~86$& $107$ & $129$ &  $150$ \\ \hline
 $\ms(1~\gev)$& $ ~86$& $115$& $144$ & $173$ &  $202$ \\ \hline
 \end{tabular}
\end{center}
\caption[]{The dictionary between the values of $\ms$ in units of
$\mev$ normalized at different scales with $\mc=1.3~\gev$.
\label{tab:ms}}
\end{table}

Finally one should remark that the decomposition of the relevant hadronic
matrix elements of penguin operators into a product of $B_i$ factors times
$1/m_s^2$, although useful in the $1/N_c$ approach, is in principle 
unnecessary in a brute
force method like the lattice approach. It is to be expected that the
future lattice calculations will directly give the relevant hadronic 
matrix elements and the issue of $\ms$ in connection with $\epe$ will
effectively disappear.

\subsection{An Analytic Formula for $\epe$}
           \label{subsec:epeanalytic}
As shown in \cite{buraslauten:93}, it is possible to cast the formal
expression for $\epe$ in (\ref{eq:epe})
into an analytic formula which exhibits the $\mt$ dependence
together with the dependence on $\ms$, $\Lms^{(4)}$,
 $B_6^{(1/2)}$ and $B_8^{(3/2)}$.
Such an analytic formula should be useful for those phenomenologists
and experimentalists who are not interested in getting involved with
the technicalities discussed above.

In order to find an analytic expression for $\epe$, which exactly
reproduces the numerical results based on the formal OPE method,
one uses the PBE presented in section 2.7. The updated  
analytic formula for $\epe$ of \cite{buraslauten:93} 
presented recently in \cite{BJL96a}
is given as follows:

\begin{equation}
\frac{\varepsilon'}{\varepsilon} = {\rm Im}\lambda_t \cdot F(x_t) \, ,
\label{eq:3}
\end{equation}
where
\begin{equation}
F(x_t) =
P_0 + P_X \, X_0(x_t) + P_Y \, Y_0(x_t) + P_Z \, Z_0(x_t) 
+ P_E \, E_0(x_t) 
\label{eq:3b}
\end{equation}
and
\begin{equation}
{\rm Im}\lambda_t = {\rm Im} V_{ts}^*V_{td} = |V_{\rm ub}| \, 
|V_{\rm cb}| \, \sin \delta = \eta \, \lambda^5 \, A^2
\label{eq:4}
\end{equation}
in the standard parameterization of the CKM matrix
(\ref{2.72}) and in the Wolfenstein parameterization
(\ref{2.75}), respectively. 

The $\mt$-dependent functions in (\ref{eq:3b}) are given in 
(\ref{E0}), (\ref{X0}), (\ref{Y0}) and (\ref{Z0}).
The coefficients $P_i$ are given in terms of $B_6^{(1/2)} \equiv
B_6^{(1/2)}(\mc)$, $B_8^{(3/2)} \equiv B_8^{(3/2)}(\mc)$ and $\ms(\mc)$
as follows:
\begin{equation}
P_i = r_i^{(0)} + \left[ \frac{158\mev}{\ms(\mc)+\md(\mc)} \right]^2
\left(r_i^{(6)} B_6^{(1/2)} + r_i^{(8)} B_8^{(3/2)} \right) \, .
\label{eq:pbePi}
\end{equation}
The $P_i$ are renormalization scale and scheme independent. They depend,
however, on $\Lms^{(4)}$. In table~\ref{tab:pbendr} we give the numerical
values of $r_i^{(0)}$, $r_i^{(6)}$ and $r_i^{(8)}$ for different values
of $\Lms^{(4)}$ at $\mu=\mc$ in the NDR renormalization scheme. 
The
coefficients $r_i^{(0)}$, $r_i^{(6)}$ and $r_i^{(8)}$ depend only very
weakly on
$\ms(\mc)$ as the dominant $\ms$ dependence has been factored out. The
numbers given in table~\ref{tab:pbendr} correspond to $\ms(\mc)=150\,\mev$.
However, even for $\ms(\mc)\approx100\mev$, the analytic expressions given
here reproduce the numerical calculations of $\epe$ given below 
to better than $4\%$.
For different scales $\mu$ the numerical values in the tables change
without modifying the values of the $P_i$'s as it should be. To this
end also $B_6^{(1/2)}$ and $B_8^{(3/2)}$ have to be modified as they
depend albeit weakly on $\mu$.

Concerning the scheme dependence only the $r_0$ coefficients
are scheme dependent at the NLO level. Their values in the HV
scheme are given in the last row of table~\ref{tab:pbendr}.
The coefficients $r_i$, 
$i=X, Y, Z, E$ are on the other hand scheme independent at NLO. 
This is related to the fact that the $\mt$
dependence in $\epe$ enters first at the NLO level and consequently all
coefficients $r_i$ in front of the $\mt$ dependent functions must be
scheme independent. 
Consequently, when changing the renormalization scheme, one is only
obliged to change appropriately $B_6^{(1/2)}$ and $B_8^{(3/2)}$ in the
formula for $P_0$ in order to obtain a scheme independence of $\epe$.
In calculating $P_i$ where $i \not= 0$, $B_6^{(1/2)}$ and $B_8^{(3/2)}$
can in fact remain unchanged, because their variation in this part
corresponds to higher order contributions to $\epe$ which would have to
be taken into account in the next order of perturbation theory.

For similar reasons the NLO analysis of $\epe$ is still insensitive to
the precise definition of $\mt$. In view of the fact that the NLO
calculations needed to extract $\IM \lambda_t$ (see previous section) 
have been done with $\mt=\overline{m}_t(\mt)$ we will also use  this 
definition in calculating $F(x_t)$.

\begin{table}[thb]
\begin{center}
\begin{tabular}{|c||c|c|c||c|c|c||c|c|c|}
\hline
& \multicolumn{3}{c||}{$\Lms^{(4)}=245\mev$} &
  \multicolumn{3}{c||}{$\Lms^{(4)}=325\mev$} &
  \multicolumn{3}{c| }{$\Lms^{(4)}=405\mev$} \\
\hline
$i$ & $r_i^{(0)}$ & $r_i^{(6)}$ & $r_i^{(8)}$ &
      $r_i^{(0)}$ & $r_i^{(6)}$ & $r_i^{(8)}$ &
      $r_i^{(0)}$ & $r_i^{(6)}$ & $r_i^{(8)}$ \\
\hline
0 &
   --2.674 &   6.537 &   1.111 &
   --2.747 &   8.043 &   0.933 &
   --2.814 &   9.929 &   0.710 \\
$X$ &
    0.541 &   0.011 &       0 &
    0.517 &   0.015 &       0 &
    0.498 &   0.019 &       0 \\
$Y$ &
    0.408 &   0.049 &       0 &
    0.383 &   0.058 &       0 &
    0.361 &   0.068 &       0 \\
$Z$ &
    0.178 &  --0.009 &  --6.468 &
    0.244 &  --0.011 &  --7.402 &
    0.320 &  --0.013 &  --8.525 \\
$E$ &
    0.197 &  --0.790 &   0.278 &
    0.176 &  --0.917 &   0.335 &
    0.154 &  --1.063 &   0.402 \\
\hline
0 &
   --2.658 &   5.818 &   0.839 &
   --2.729 &   6.998 &   0.639 &
   --2.795 &   8.415 &   0.398 \\
\hline
\end{tabular}
\end{center}
\caption[]{PBE coefficients for $\epe$ for various $\Lms^{(4)}$ in 
the NDR scheme.
The last row gives the $r_0$ coefficients in the HV scheme.
\label{tab:pbendr}}
\end{table}

The inspection of table~\ref{tab:pbendr} shows
that the terms involving $r_0^{(6)}$ and $r_Z^{(8)}$ dominate the ratio
$\epe$. The function $Z_0(x_t)$ representing a gauge invariant
combination of $Z^0$- and $\gamma$-penguins grows rapidly with $\mt$
and due to $r_Z^{(8)} < 0$ these contributions suppress $\epe$ strongly
for large $\mt$ \cite{flynn:89,buchallaetal:90} as stressed at the
beginning of this section. 

\subsection{Numerical Results for $\epe$}
In order to complete the analysis of $\epe$ one needs the value of ${\rm
Im}\lambda_t$. Since this value has been already determined in 
section \ref{sec:standard}, we are ready to present the results for $\epe$. 
Here we follow \cite{BJL96a} were the the same input parameters
as in this review have been used.

For $m_s( \mc)=150\pm20\mev$ one finds \cite{BJL96a}
\begin{equation}
-1.2 \cdot 10^{-4} \le \epe \le 16.0 \cdot 10^{-4}
\label{eq:eperangenew}
\end{equation}
and
\begin{equation}
\epe= ( 3.6\pm 3.4) \cdot 10^{-4}
\label{eq:eperangefinal}
\end{equation}
for the ``scanning'' method and the ``gaussian'' method discussed
in section \ref{sec:standard}, respectively.

The result in (\ref{eq:eperangefinal}) agrees rather well with
the 1995 analysis of the Rome group \cite{ciuchini:95} which gave
 $\varepsilon'/\varepsilon=(3.1\pm 2.5)\cdot 10^{-4}$ and with
the recent update of this group \cite{ciuchini:96}:
$(4.6\pm 3.0)\cdot 10^{-4}$.
On the other hand the range in (\ref{eq:eperangenew}) shows that for
particular choices of the input parameters, values for $\epe$ as high as
$16\cdot 10^{-4}$ cannot be excluded at present. Such high values are
found if simultaneously  $\vub=0.10$, $B_6^{(1/2)}=1.2$, 
$B_8^{(3/2)}=0.8$, $B_K=0.6$,
$\ms(\mc)=130$ MeV, $\Lms^{(4)}=405\mev$ and low values of $\mt$ still
consistent with $\varepsilon_K$ and the observed $B_d^0-\bar B_d^0$ 
mixing
are chosen. It is, however, evident from  the comparision of
(\ref{eq:eperangenew}) and (\ref{eq:eperangefinal})  that such 
high values of $\epe$ and generally values above $10^{-3}$ 
are very improbable for $\ms(\mc)={\cal O}(150\mev)$.

\begin{table}[thb]
\begin{center}
\begin{tabular}{|c|c|c|c|c||c|}\hline
  $|V_{ub}/V_{cb}|$& $\Lms[MeV]$& $B^{(1/2)}_6$& $B^{(3/2)}_8$ & 
$\ms(\mc)[\mev]$ &
 $\epe[10^{-4}]$ \\ \hline
       &       &      &     & $~75$ & $16.8$ \\
       &       &      &     & $100$ & $~9.1$ \\
$0.08$ & $325$ & $1.0$&$1.0$& $125$ & $~5.3$ \\
       &       &      &     & $150$ & $~3.2$ \\
       &       &      &     & $175$ & $~1.8$ \\ \hline\hline
       &       &      &     & $~75$ & $27.8$ \\
       &       &      &     & $100$ & $15.6$ \\
$0.08$ & $325$ & $1.2$&$0.8$& $125$ & $~9.6$ \\
       &       &      &     & $150$ & $~6.2$ \\
       &       &      &     & $175$ & $~4.1$ \\ \hline\hline
       &       &      &     & $~75$ & $39.8$ \\
       &       &      &     & $100$ & $22.5$ \\
$0.10$ & $405$ & $1.2$&$0.8$& $125$ & $14.0$ \\
       &       &      &     & $150$ & $~9.2$ \\
       &       &      &     & $175$ & $~6.2$ \\ \hline
\end{tabular}
\end{center}
\caption[]{ Values of $\epe$ in units of $10^{-4}$ 
for specific values of various input parameters at $\mt=167\,\gev$, 
$V_{cb}=0.040$ and $B_K=0.75$.
\label{tab:31731}}
\end{table}

The authors of \cite{bertolinietal:95} calculating the $B_i$ factors
in the chiral quark model find using the scanning method
 a rather large range $-50 \cdot 10^{-4}\le \epe \le 14 \cdot 10^{-4}$.
 In particular they find in contrast to
\cite{BJLW,ciuchini:95,BBL,BJL96a} 
that negative values for $\epe$ as large as $-5\cdot
10^{-3}$ are possible. 
The Dortmund group
\cite{heinrichetal:92} advocating on the other hand $B_6>B_8$ finds
$\epe=(9.9\pm 4.1)\cdot 10^{-4}$ for $\ms(\mc)=150~\mev$ \cite{paschos:96}.

\begin{table}[thb]
\begin{center}
\begin{tabular}{|c|c|c|c||c|}\hline
  {\bf Reference}& $B^{(1/2)}_6$& $B^{(3/2)}_8$ & $\ms(\mc)[\mev]$ &
 $\epe[10^{-4}]$ \\ \hline
\cite{BJL96a}& $1.0\pm 0.2$ &$1.0\pm0.2$ & $150\pm20$ & $-1.2\to 16.0$ (S) \\
\cite{BJL96a}& $1.0\pm 0.2$ &$1.0\pm0.2$ & $150\pm20$ & $3.6\pm 3.4$ (G) \\
\cite{BJL96a}& $1.0\pm 0.2$ &$1.0\pm0.2$ & $100\pm20$ & $0.0\to 43.0$ (S) \\
\cite{BJL96a}& $1.0\pm 0.2$ &$1.0\pm0.2$ & $100\pm20$ & $10.4\pm 8.3$ (G) \\
\hline
\cite{ciuchini:96}& $1.0\pm 0.2$ &$1.0\pm0.2$ & $150\pm20$ & 
$4.6\pm 3.0$ (G) \\
\hline
\cite{bertolinietal:95}& $1.0\pm 0.4$ &$2.2\pm1.5$ & $-$ & 
$-50\to 14$ (S) \\
\hline
\cite{heinrichetal:92,paschos:96} 
& $\sim 1.3$ &$\sim 0.7$ & $150$ & $9.9\pm 4.1$ \\
\hline
\end{tabular}
\end{center}
\caption[]{ Results for $\epe$ in units of $10^{-4}$ obtained
by various groups. The labels (S) and (G) in the last column
stand for ``Scanning'' and ``Gaussian'' respectively, as discussed
in the text. 
\label{tab:31738}}
\end{table}

The situation concerning $\epe$ in the Standard Model may, however, change
if the value for $m_s$ is as low as found in \cite{gupta:96,FNAL:96}. 
Using $\ms(\mc)=(100\pm20)\mev$ one finds \cite{BJL96a}
\begin{equation}
0 \le \epe \le 43.0 \cdot 10^{-4}
\label{eq:eperangenewa}
\end{equation}
and
\begin{equation}
\epe= ( 10.4\pm 8.3) \cdot 10^{-4}
\label{eq:eperangefinala}
\end{equation}
for the ``scanning'' method and the ``gaussian'' method, respectively.
We observe that the ``gaussian'' result agrees well with the E731
value and,
as stressed in \cite{BJL96a}, the decrease of $\ms$
 with $\ms(\mc)\geq 100$ MeV alone is insufficient to bring 
the Standard Model in agreement with
the NA31 result. However, for $B_6>B_8$, sufficiently large values of
$|V_{ub}/V_{cb}|$ and $\Lms$, and small values of $\ms$, the values
of $\epe$ in the Standard Model can be as large as $(2-4)\cdot 10^{-3}$
and consistent with the NA31 result.
In order to see this explicitly we present in table \ref{tab:31731} the
values of $\epe$ for five choices of $\ms(\mc)$ and for selective
sets of other input parameters keeping $V_{cb}=0.040$, $\mt=167\,\gev$
and $B_K=0.75$ fixed \cite{BJL96a}.

\subsection{Summary}
The fate of $\epe$ in the Standard Model after the
improved measurement of $\mt$, depends sensitively on the values of
$|V_{ub}/V_{cb}|$, $\Lms$ and in particular on $B_6$, $B_8$ and $\ms$.
For $\ms(\mc)={\cal O}(150\,\mev)$, $\epe$ is generally below $10^{-3}$
in agreement with E731 with central values in the ball park of
a few $10^{-4}$. However, if
the low values of $\ms(\mc)={\cal O}(100\,\mev)$ found in 
\cite{gupta:96,FNAL:96}
are confirmed by other groups in the future, a conspiration of
other parameters may give values as large as $(2-4)\cdot 10^{-3}$ in the
ball park of the NA31 result. The predictions for $\epe$ obtained by
various groups are summarized in table \ref{tab:31738}. 

Let us hope that the future experimental and theoretical results will
be sufficiently accurate to be able to see whether $\epe\not=0$ and
whether the Standard Model agrees with the data. In any case the
coming years should be very exciting. 

\section{The Decays $K_{\rm L}\to\pi^0 e^+e^-$, 
$B\to X_s \gamma$ and $B \to X_s l^+l^-$}
\label{sec:KLpee}
\setcounter{equation}{0}
\subsection{General Remarks}
In this section we will discuss three well known decays:
$K_{\rm L}\to\pi^0 e^+e^-$, 
$B\to X_s \gamma$ and $B \to X_s l^+l^-$.
The reason for collecting these decays in one section is
related to the fact that their effective Hamiltonians
constitute three different generalizations of the
effective $\Delta S=1$ ($\Delta B=1$) Hamiltonian considered in the
previous section in the absence of electroweak penguin
operators $Q_7 .... Q_{10}$. In principle electroweak penguin operators
can contribute here. However, their effect is tiny and can be safely
neglected.

Thus  the effective Hamiltonian for $K_{\rm L}\to\pi^0 e^+e^- $ 
given in (\ref{eq:HeffKpe}) includes in addition to the
operators $Q_1 .... Q_6$ the semi-leptonic operators $Q_{7V}$ 
and $Q_{7A}$ defined in (\ref{q7v7a}).
Next, the effective
Hamiltonian for $B\to X_s\gamma$ given in
(\ref{Heff_at_mu}) includes in addition to the
operators $Q_1 .... Q_6$
the magnetic penguin operators $Q_{7\gamma}$ and $Q_{8G}$ defined
in (\ref{O6B}).
Finally the effective Hamiltonian
for $B\to X_s \mu^+ \mu^- $ given in (\ref{Heff2_at_mu}) 
can be considered as the generalization of the effective Hamiltonian
for $B\to X_s\gamma$ to include the semi-leptonic operators $Q_{9V}$ and
$Q_{10A}$ defined in (\ref{Q9V}).  

\subsection{$K_{\rm L}\to\pi^0 e^+e^-$}
\subsubsection{The Effective Hamiltonian}
The effective Hamiltonian for $K\to\pi^0 e^+e^-$ at scales 
$ \mu < m_c$ is given as follows:
\begin{equation}\label{eq:HeffKpe}
{\cal H}_{\rm eff}(K\to\pi^0 e^+e^-) = 
\frac{G_{\rm F}}{\sqrt{2}} V_{us}^* V_{ud}
 \left[\sum_{i=1}^{6,7V} \left[ z_i(\mu)+\tau y_i(\mu)\right] Q_i 
+\tau y_{7A}(M_W)Q_{7A}\right]\,,
\end{equation}
where $Q_1,...Q_6$ are the operators present also in the discussion
of $\epe$ and the new operators $Q_{7V}$ and $Q_{7A}$ are given by 
\begin{equation}\label{q7v7a} 
Q_{7V}=(\bar sd)_{V-A}(\bar ee)_V\,,  \qquad
Q_{7A}=(\bar sd)_{V-A}(\bar ee)_A \,.
\end{equation}
 
Whereas in $K \to \pi \pi$ decays the CP violating
contribution is a tiny part of the full amplitude and direct CP
violation is expected to be at least by three orders of magnitude
smaller than indirect CP violation, the corresponding hierarchies
are very different for the rare decay $K_{\rm L}\to\pi^o e^+e^-$ .
At lowest order in
electroweak interactions (single photon, single Z-boson or double
W-boson exchange), this decay takes place only if the CP symmetry is
violated \cite{GIL1}.
The CP conserving contribution to the amplitude
comes from a two photon exchange which,
although higher order in $\alpha$, could in principle be sizable.

The three contributions: CP conserving, {\it indirectly} CP violating 
and {\it directly} CP violating are all expected to be
${\cal O}(10^{-12})$. Unfortunately out of these three contributions
only the directly CP violating one can be calculated reliably. Let
us discuss these three contributions one by one.

\subsubsection{CP Conserving Contribution}
The estimate of the CP conserving part is very difficult as it can
only be done outside the perturbative framework. The most recent estmates
give:
\begin{equation}
Br(K_{\rm L} \to \pi^0 e^+ e^-)_{{\rm cons}}\approx\left\{ \begin{array}{ll}
(0.3-1.8)\cdot 10^{-12} & \hbox{\cite{cohenetal:93}} \\
     4.0 \cdot 10^{-12} & \hbox{\cite{heiligerseghal:93}} \\
(5 \pm 5)\cdot 10^{-12} & \hbox{\cite{donoghuegabbiani:95}.}
\end{array} \right.
\label{eq:BKLtheo}
\end{equation}
The details can be found in the original papers and in a recent
review by Pich \cite{Pich96}.
The measurement of the
branching ratio
\begin{equation}
Br(K_{\rm L} \to \pi^0 \gamma\gamma) =
\left\{ \begin{array}{ll}
(1.7\pm 0.3) \cdot 10^{-6} & \hbox{\cite{barretal:92}} \\
(2.0\pm 1.0) \cdot 10^{-6} & \hbox{\cite{papadimitriou:91}}
\end{array} \right.
\label{eq:BKLexp}
\end{equation}
and of the shape of the $\gamma\gamma$ mass spectrum play an important
role in these estimates.
The authors of \cite{cohenetal:93} compute first the amplitude for
$K_{\rm L} \to \pi^0 \gamma\gamma$ in chiral perturbation theory and
estimate the CP conserving two-photon contribution $K_{\rm L}\to \pi^0e^+e^-$
by taking the {\it absorptive} part due to two-photon discontinuity as
an educated guess of the actual size of the complete amplitude. As
chiral perturbation theory, after the inclusion of ${\cal O}(p^6)$
unitarity corrections and resonance contributions 
\cite{cohenetal:93,CAM,KAHO},
is capable of describing both the shape of the $\gamma\gamma$ mass spectrum
and the branching ratio in (\ref{eq:BKLexp}), one could expect that the
estimate in \cite{cohenetal:93} is reasonable. The large uncertainty
in this ``absorptive'' estimate $(0.3-1.8)\cdot 10^{-12}$ could then
be reduced in the future by a more careful analysis of 
$K_{\rm L} \to \pi^0 \gamma\gamma$ taking the experimental acceptance
into account. On the other hand, as stressed in \cite{heiligerseghal:93,
donoghuegabbiani:95}, the poorly understood {\it dispersive} pieces
could considerably modify the estimate of \cite{cohenetal:93}
giving a much larger CP-conserving part. Consequently a better 
understanding of the dispersive part is very desirable.

It should be noted that there is no interference in the rate between
the CP conserving and CP violating contributions discussed below.

\subsubsection{The Indirectly CP Violating Contribution}

The indirectly CP violating amplitude is given by the
$K_{\rm S} \to \pi^0 e^+ e^-$ amplitude times the CP parameter $\eps_K$.
The amplitude
$A(K_{\rm S}\to\pi^0e^+e^-)$ can be written as 

\begin{equation}\label{akspee}
A(K_{\rm S}\to\pi^0e^+e^-)=\langle\pi^0e^+e^-|{\cal H}_{\rm eff}| 
K_{\rm S}\rangle
\end{equation} 
with ${\cal H}_{\rm eff}$ given in (\ref{eq:HeffKpe}).
The coefficients of $Q_{7V}$ and $Q_{7A}$ are
$\ord(\alpha)$. Their hadronic matrix elements   
$\langle\pi^0e^+e^-|Q_{7V,A}| K_{\rm S}\rangle$ are $\ord(1)$. 
In the case of $Q_i$ ($i=1,\ldots,
6$) the situation is reversed: the Wilson coefficients are $\ord(1)$,
but the matrix elements $\langle\pi^0e^+e^-| Q_i| K_{\rm S}\rangle$ are
$\ord(\alpha)$. Consequently at $\ord(\alpha)$ all operators contribute
to $A(K_{\rm S}\to\pi^0e^+e^-)$. However because $K_{\rm S}\to\pi^0e^+e^-$ 
is CP
conserving, the coefficients $y_i$ multiplied by $\tau=\ord(\lambda^4)$
can be fully neglected and the operator $Q_{7A}$ drops out in this
approximation.  Now whereas $\langle\pi^0e^+e^-| Q_{7V}| K_{\rm S}\rangle$
can be trivially calculated, this is not the case for
$\langle\pi^0e^+e^-| Q_i| K_{\rm S}\rangle$ with $i=1,\ldots, 6$ which can
only be evaluated using non-perturbative methods. Moreover it is clear
from the short-distance analysis of \cite{BLMM} that the
inclusion of $Q_i$ in the estimate of $A(K_{\rm S}\to\pi^0e^+e^-)$ cannot be
avoided. Indeed, whereas $\langle\pi^0e^+e^-| Q_{7V}| K_{\rm S}\rangle$ is
independent of $\mu$ and the renormalization scheme, the coefficient
$z_{7V}$ shows very strong renormalization scheme and 
$\mu$-dependences.  They can only
be canceled by the contributions from the four-quark operators $Q_i$.
All this demonstrates that the estimate of indirect CP violation in
$K_{\rm L}\to\pi^0e^+e^-$ cannot be done very reliably at present.

Using chiral
perturbation theory it is, however, possible to get an estimate by
relating $K_{\rm S} \to \pi^0 e^+ e^-$ to the $K^+ \to \pi^+ e^+ e^-$
transition \cite{eckeretal:88}. To this end one
can write 

\begin{equation}
Br(K_{\rm L} \to \pi^0 e^+ e^-)_{\rm indir}=Br(K^+ \to \pi^+e^+e^-)
\frac{\tau(K_{\rm L})}{\tau(K^+)} |\eps_K|^2 r^2\,,
\label{eq:BKLindir1}
\end{equation}
where
\begin{equation}
r^2=\frac{\Gamma(K_{\rm S} \to \pi^0 e^+ e^-)}{\Gamma(K^+ 
\to \pi^+ e^+ e^-)}\,.
\label{eq:r2}
\end{equation}
With $Br(K^+ \to \pi^+e^+e^-)=(2.74\pm 0.23)\cdot 10^{-7}$
\cite{alliegro:92} and the  chiral perturbation theory
estimate $|r| \le 0.5 $ \cite{eckeretal:88,brunoprades:93},
one has
\begin{equation}
Br(K_{\rm L} \to \pi^0 e^+ e^-)_{\rm indir}=(5.9\pm 0.5)\cdot 10^{-12} r^2
 \le 1.6\cdot 10^{-12},
\label{eq:BKLindir2}
\end{equation}
i.e.\ a rather small contribution. 
Yet, as emphasized in \cite{donoghuegabbiani:95} and also in
\cite{heiligerseghal:93}, the knowledge of $r$ is very uncertain at
present. In particular the estimate in (\eqn{eq:BKLindir2}) is based on
a relation between two non-perturbative parameters, which is rather
ad hoc and certainly not a consequence of chiral symmetry. As shown
in \cite{donoghuegabbiani:95}, a small deviation from this relation
increases $r$ to values above unity so that $Br(K_{\rm L} \to \pi^0
e^+e^-)_{\rm indir}$ could be as high as $ 5\cdot 10^{-12} $.
In summary, the following ranges can be found in the literature
 \cite{eckeretal:88,brunoprades:93,heiligerseghal:93,donoghuegabbiani:95}:
\begin{equation}
Br(K_{\rm L} \to \pi^0 e^+ e^-)_{\rm indir}=(1-5)\cdot 10^{-12} \,.
\label{eq:BKLindir3}
\end{equation}

A much better assessment of the importance of the
indirect CP violation in $K_{\rm L}\to\pi^0e^+e^-$ will become possible after
a measurement of $Br(K_{\rm S}\to\pi^0e^+e^-)$.  Bounding the latter
branching ratio below $ 1 \cdot 10^{-9}$ or $ 1 \cdot 10^{-10}$ would
bound the indirect CP contribution below $ 3 \cdot 10^{-12}$ and
 $ 3 \cdot 10^{-13}$, respectively. The present bounds
$ 1.1 \cdot 10^{-6}$ (NA31) and  $ 3.9 \cdot 10^{-7}$ (E621) are still 
too weak. On the other hand, KLOE at DA${\Phi}$NE
could make an important contribution here.  

\subsubsection{Directly CP Violating Contribution}
Fortunately the directly
CP violating contribution can be fully calculated as a function of
$m_t$, CKM parameters and the QCD coupling constant $\alpha_s$. There are
practically no theoretical uncertainties related to hadronic matrix
elements because $\langle\pi^0|(\bar sd)_{V-A}| K_{\rm L}\rangle$ can be
extracted using isospin symmetry from the well measured decay
$K^+\to\pi^0e^+\nu$. 

The directly CP violating contribution is governed by the coefficients
$y_i$ which vanish for $i=1,2$. Consequently only the penguin operators 
$Q_3,\ldots, Q_6$,
$Q_{7V}$ and $Q_{7A}$ have to be considered. Now 
$y_i=\ord(\as)$ for $i=3,\ldots, 6$ and the contribution of
QCD penguins to $Br(K_{\rm L}\to\pi^0e^+e^-)_{\rm dir}$ is really
$\ord(\alpha\as)$ to be compared with the $\ord(\alpha)$
contributions of $Q_{7V}$ and $Q_{7A}$. 
Furthermore the following relation for the
quark-level matrix elements
\begin{equation}\label{qillq7v2}
\sum_{i=3}^6 y_i(\mu)\langle d e^+e^-|Q_i| s \rangle\ll
y_{7V}(\mu)\langle d e^+e^-|Q_{7V}| s \rangle
\end{equation}
can be easily verified perturbatively. Consequently the contribution
of QCD penguin operators can be safely neglected.
This is compatible with the scheme
and $\mu$-independence of the resulting branching ratio.  Indeed
$y_{7A}$ does not depend on $\mu$ and the renormalization scheme at all
and the corresponding dependences in $y_{7V}$ are at the level of $\pm
1\%$ \cite{BLMM}.

Using PBE of section 2.7 and introducing 
\begin{equation}
y_i = \frac{\alpha}{2 \pi} \tilde{y_i} 
\label{eq:ytilde}
\end{equation}
one finds \cite{BLMM}
\begin{equation}\label{9a}
Br(K_{\rm L} \to \pi^0 e^+ e^-)_{\rm dir} = 6.3\cdot 10^{-6}(\IM\lambda_t)^2
(\tilde y_{7A}^2 + \tilde y_{7V}^2)\,,
\end{equation}
where
$\IM \lambda_t = \IM (V_{td} V^*_{ts})$,

\begin{equation}
\label{kappae}
6.3 \cdot 10^{-6}=\frac{1}{V_{us}^2}\frac{\tau(K_{\rm L})}{\tau(K^+)}
\left( \frac{\alpha}{2 \pi} \right)^2 Br(K^+\to\pi^0e^+\nu)
\equiv\kappa_e
\end{equation}
and
\begin{equation}\label{y7vpbe}
\tilde{y}_{7V} =
P_0 + \frac{Y_0(x_t)}{\sin^2\Theta_{\rm W}} - 4 Z_0(x_t)+ P_E E_0(x_t)
\end{equation}
\begin{equation}\label{y7apbe}
\tilde{y}_{7A}=-\frac{1}{\sin^2\Theta_{\rm W}} Y_0(x_t)
\end{equation}
with $Y_0$, $Z_0$ and $E_0$ given in (\ref{Y0}), (\ref{Z0}) and
(\ref{E0}), respectively. 
$P_E$ is $\ord(10^{-2})$ and consequently the last
term in (\ref{y7vpbe}) can be neglected. 
The next-to-leading QCD corrections to the coefficients above enter
only $P_0$. They
have been calculated in \cite{BLMM} 
reducing certain ambiguities present in 
leading order analyses \cite{GIL2,FR89} and enhancing the leading 
order value typically from $P_0(LO)=1.9$ to $P_0(NLO)=3.0$.
Partially this enhancement is  due
to the fact that for $\Lambda_{LO}=\Lambda_{\overline{MS}}$ the QCD
coupling constant in the leading order is $20-30\%$ larger than its
next-to-leading order value. 
In any case the inclusion of NLO QCD
effects and a meaningful use of $\Lambda_{\overline{MS}}$ show that the
next-to-leading order effects weaken the QCD suppression of $y_{7V}$.
$P_0$ is given for different
values of $\mu$ and $\Lambda_{\overline{MS}}$ in table
\ref{tab:P0klpee} \cite{BBL}.  
There we also show the leading order results and
the case without QCD corrections.
As seen in table \ref{tab:P0klpee}, the suppression of $P_0$ by QCD
corrections amounts to about $15\%$ in the complete next-to-leading
order calculation.

\begin{table}[htb]
\begin{center}
\begin{tabular}{|c|c||c|c|c|}
\hline
\multicolumn{2}{|c||}{} & \multicolumn{3}{c|}{$P_0$} \\
\hline
$\Lms^{(4)} [\mev]$ & $\mu [\gev]$ & LO & NDR & HV \\
\hline
\  & 0.8 & 2.012 & 3.138 & 3.088 \\
245 & 1.0 & 1.987 & 3.111 & 3.060 \\
\  & 1.2 & 1.965 & 3.089 & 3.037 \\
\hline
\  & 0.8 & 1.863 & 3.080 & 3.024 \\
325 & 1.0 & 1.834 & 3.053 & 2.996 \\
\  & 1.2 & 1.811 & 3.028 & 2.970 \\
\hline
\  & 0.8 & 1.723 & 3.009 & 2.949 \\
405 & 1.0 & 1.692 & 2.991 & 2.927 \\
\  & 1.2 & 1.666 & 2.965 & 2.900 \\
\hline
\end{tabular}
\end{center}
\caption{PBE coefficient $P_0$ of $y_{7V}$ for various values of
$\Lms^{(4)}$ and $\mu$. In the absence of QCD $P_0=8/9\;\ln(M_W/m_c)
= 3.664$ holds universally.
\label{tab:P0klpee}}
\end{table}

\begin{figure}[hbt]
\vspace{0.10in}
\centerline{
\epsfysize=5in
\rotate[r]{\epsffile{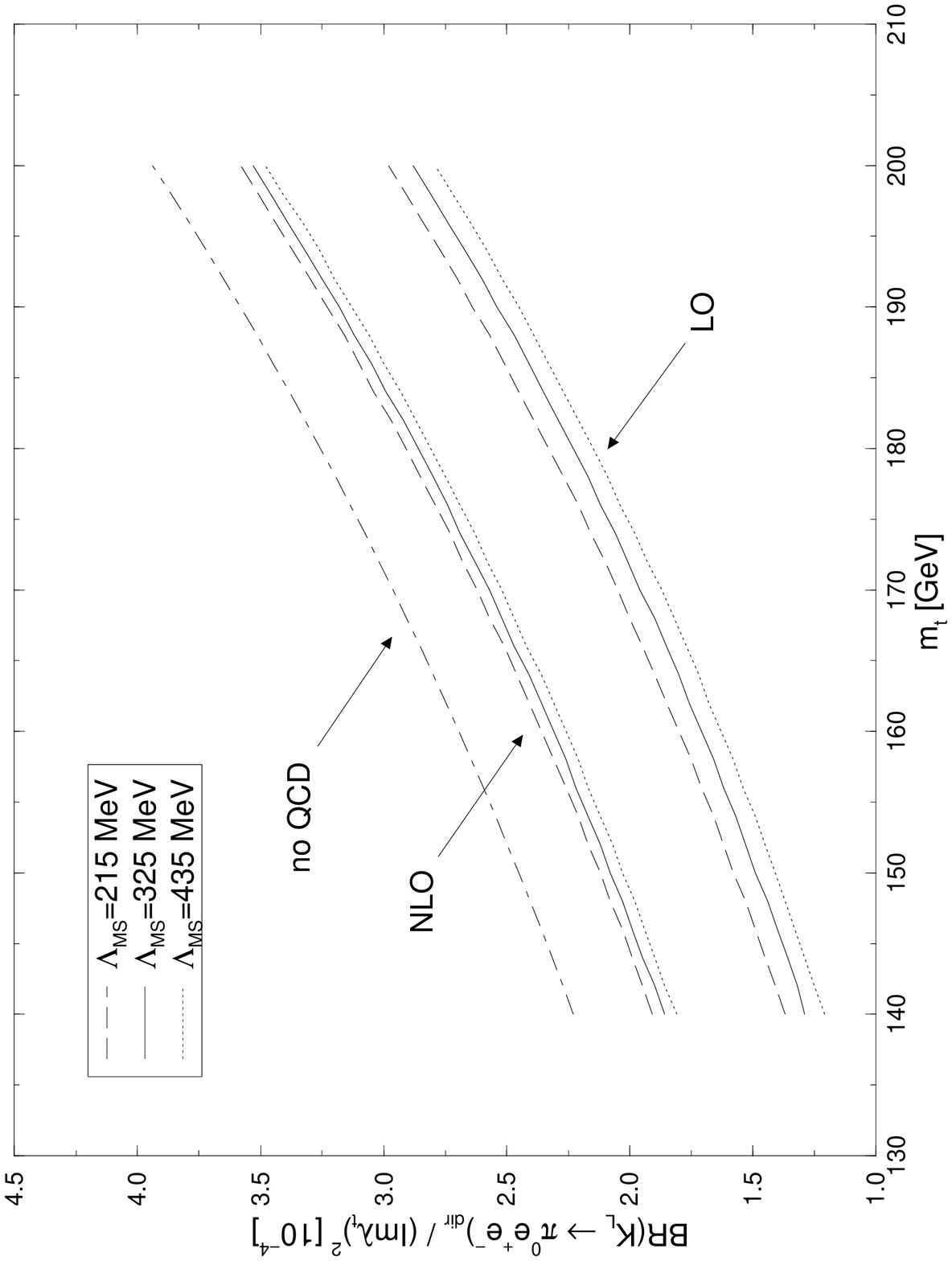}}
}
\vspace{0.10in}
\caption{$Br(K_{\rm L} \to \pi^0 e^+e^-)_{\rm dir}/(\IM\lambda_t)^2$
as a function of $\mt$ for various values of $\Lms^{(4)}$ at scale
$\mu =1.0\gev $.}\label{fig:mtbrimltAll}
\end{figure}

The
dominant $m_t$-dependence of $Br(K_{\rm L}\to\pi^0e^+e^-)_{\rm dir}$ 
originates from the coefficient of the operator $Q_{7A}$ although
only  for $m_t > 175\gev$ one finds
$y_{7A}> y_{7V}$.
In fig.\ \ref{fig:mtbrimltAll} the ratio
$Br(K_{\rm L}\to\pi^0e^+e^-)_{\rm dir}/({\rm Im}\lambda_t)^2$
is shown as a function of $m_t$ \cite{BLMM,BBL}. 
The enhancement of the directly
CP violating contribution through NLO corrections relatively to the
LO estimate is clearly visible on this plot. However,
due to large uncertainties present in ${\rm Im}\lambda_t$, this
enhancement cannot yet be fully appreciated phenomenologically.

The very weak dependence on $\Lambda_{\overline{MS}}$ should be
contrasted with the very strong dependence found in the case of
$\varepsilon^\prime/\varepsilon$. Therefore, provided the other two
contributions to $K_{\rm L}\to\pi^0e^+e^-$ can be shown to be small or can
be reliably calculated one day, the measurement of
$Br(K_{\rm L}\to\pi^0e^+e^-)$ should offer a good determination of
${\rm Im}\lambda_t$ or $\eta$.

Using the numerical results presented in fig.\ \ref{fig:mtbrimltAll}
one finds to a very good approximation:

\begin{equation}\label{KT}
Br(K_{\rm L} \to \pi^0 e^+ e^-)_{\rm dir}=
4.9\cdot 10^{-12}\left [ 
\frac{\eta}{0.39}\right ]^2
\left [\frac{|V_{cb}|}{0.040} \right ]^4 
\left [\frac{(\mtb(\mt)}{170\gev} \right ]^2 \,.
\end{equation}

Next we would like to comment on the possible uncertainties due to the
definition of $\mt$. In the formulae above we have set $\mt=\mtb(\mt)$
in accordance with the NLO analysis of CKM parameters in section 4.
However, at the level of accuracy at which we work one
cannot fully address this question yet. In order to be able to do it,
one needs to know the perturbative QCD corrections to $Y_0(x_t)$ and
$Z_0(x_t)$ and for consistency an additional order in the
renormalization group improved calculation of $P_0$.  Since the
$m_t$-dependence of $y_{7V}$ is rather moderate, the main concern in
this issue is the coefficient $y_{7A}$ whose $m_t$-dependence is fully
given by $Y(x_t)$. Fortunately the QCD corrected function $Y(x_t)$ is
known from the analysis of $K_{\rm L}\to\mu^+\mu^-$ and can be directly used
here. As we will discuss in section 7, for $\mt=\mtb(\mt)$
the QCD corrections to $Y_0(x_t)$ are around 2\%. On this basis we
believe that if $\mt=\mtb(\mt)$ is chosen, the additional  QCD
corrections to $Br(K_{\rm L}\to\pi^0e^+e^-)_{\rm dir}$ should be small.

Using the input parameters of table \ref{tab:inputparams} 
and performing two
types of error analysis one finds \cite{BJL96b}
\begin{equation}
Br(K_{\rm L}\to\pi^0 e^+ e^-)_{\rm dir}=\left\{ \begin{array}{ll}
(4.5 \pm 2.6)\cdot 10^{-12} & {\rm Scanning} \\
(4.2 \pm 1.4) \cdot 10^{-12} & {\rm Gaussian,} \end{array} \right.
\end{equation}
where the error comes dominantly from the uncertainties in the CKM
parameters. 
These results are compatible with those found in \cite{BLMM,
donoghuegabbiani:95, kohlerpaschos:95} with differences
originating in various choices of CKM parameters and the fact that
\cite{BJL96b} uses the latest values of $\mt$.

Comparing with the estimates of the CP conserving contribution 
in (\ref{eq:BKLtheo})
and of the indirectly CP violating contribution in (\ref{eq:BKLindir3}),
we observe that
the directly CP violating contribution is comparable
to the other two contributions. It is, however, possible that the direct
CP violation dominates in this decay (see (\ref{eq:BKLindir2}))
which is of course very exciting.
In order to see whether this is indeed the case  improved estimates of the
other two contributions are necessary.

Finally it should also be stressed that in reality the CP indirect amplitude
may interfere with the vector part of the CP direct amplitude.  The
full CP violating amplitude can then be written following
\cite{GIL2} as follows:
\begin{equation}
Br(K_{\rm L} \to \pi^0 e^+ e^-)_{\rm CP}=| 2.43\cdot 10^{-6} r e^{i\pi/4}-
i\sqrt{\kappa_e} Im\lambda_t\tilde y_{7V}|^2+
\kappa_e (Im\lambda_t)^2\tilde y_{7A}^2\,.
\label{eq:BKLCP}
\end{equation}
Numerical analyses of (\ref{eq:BKLCP}) in \cite{donoghuegabbiani:95,BLMM}
show that for $0 \le r \le 1$ the dependence of $Br(K_{\rm L} \to
\pi^0 e^+e^-)_{\rm CP}$ on $r$ is moderate. It is rather strong
otherwise and already for $r < -0.6$ values as high as $10^{-11}$ are
found.

\subsubsection{Summary and Outlook}
The results presented above indicate that within the Standard Model
$Br(K_{\rm L} \to \pi^0 e^+ e^-)$ could be as high as $1\cdot 10^{-11}$.
Moreover the direct CP violating contribution is found to be important
and could even be dominant. Unfortunately the large uncertainties in
the remaining two contributions will probably not allow an easy
identification of the direct CP violation by measuring the branching
ratio only. The future measurements of $Br(K_{\rm S} \to \pi^0 e^+e^-)$ and
improvements in the estimate of the CP conserving part may of course
change this unsatisfactory situation. Alternatively the measurements
of the electron energy asymmetry \cite{heiligerseghal:93}, 
\cite{donoghuegabbiani:95} and the study of the time evolution of $K^0 \to
\pi^0 e^+e^-$ \cite{littenberg:89b}, \cite{donoghuegabbiani:95},
\cite{kohlerpaschos:95} could allow for a refined study of CP violation
in this decay.

The present experimental bounds
\begin{equation}
Br(K_{\rm L}\to\pi^0 e^+ e^-) \leq\left\{ \begin{array}{ll}
4.3 \cdot 10^{-9} & \cite{harris} \\
5.5 \cdot 10^{-9} & \cite{ohl} \end{array} \right.
\end{equation}
are still by three orders of magnitude away from the theoretical
expectations in the Standard Model. Yet the prospects of getting the
required sensitivity of order $10^{-11}$--$10^{-12}$ by 1999 are
encouraging \cite{CPRARE}. More details on this interesting decay can
be found in the original papers and in the review by Pich \cite{Pich96}.

\subsection{$B\to X_s\gamma$} 
\subsubsection{General Remarks}
The rare decay $B\to X_s\gamma$ plays an important role in 
present day phenomenology. 
The effective Hamiltonian for $B\to X_s\gamma$ at scales $\mu={\cal O}(m_b)$
is given by
\begin{equation} \label{Heff_at_mu}
{\cal H}_{\rm eff}(b\to s\gamma) = - \frac{G_{\rm F}}{\sqrt{2}} V_{ts}^* V_{tb}
\left[ \sum_{i=1}^6 C_i(\mu) Q_i + C_{7\gamma}(\mu) Q_{7\gamma}
+C_{8G}(\mu) Q_{8G} \right]\,,
\end{equation}
where in view of $\mid V_{us}^*V_{ub} / V_{ts}^* V_{tb}\mid < 0.02$
we have neglected the term proportional to $V_{us}^* V_{ub}$.
Here $Q_1....Q_6$ are the usual four-fermion operators whose
explicit form is given in (\ref{O1}-\ref{O3}). The remaining two operators,
characteristic for this decay, are the {\bf Magnetic--Penguins}
\begin{equation}\label{O6B}
Q_{7\gamma}  =  \frac{e}{8\pi^2} m_b \bar{s}_\alpha \sigma^{\mu\nu}
          (1+\gamma_5) b_\alpha F_{\mu\nu},\qquad            
Q_{8G}     =  \frac{g}{8\pi^2} m_b \bar{s}_\alpha \sigma^{\mu\nu}
   (1+\gamma_5)T^a_{\alpha\beta} b_\beta G^a_{\mu\nu}  
\end{equation}
originating in the diagrams of fig.\ 11d.
In order to derive the contribution of $Q_{7\gamma}$ to the
Hamiltonian in (\ref{Heff_at_mu}), in the absence of QCD corrections,
one multiplies the vertex in (\ref{MGP})
by ``i'' and makes the replacement 
\begin{equation}
2i\sigma_{\mu\nu}q^\nu\to-\sigma^{\mu\nu}F_{\mu\nu}.
\end{equation}
Analogous
procedure gives the contribution of $Q_{8G}$.

It is the magnetic $\gamma$-penguin which plays the crucial role in
this decay. However, the role of the dominant current-current
operator $Q_2$ should not be underestimated.
Indeed the perturbative QCD effects involving in particular the
mixing between $Q_2$  and $Q_{7\gamma}$ are very important in this decay.
They are known
\cite{Bert,Desh} to enhance $C_{7\gamma}(\mu)$ for $\mu={\cal O}(m_b)$
substantially, so that the resulting branching ratio
$Br(B\to X_s\gamma)$ turns out to be by a factor 
of 2--3 higher than it would be without QCD effects.
Since the first analyses
in \cite{Bert,Desh} a lot of progress has been made in calculating
these important  QCD effects beginning with the work in \cite{Grin,Odon}. 
We will briefly summarize this progress.

A peculiar feature of the renormalization group analysis 
in $B\to X_s\gamma$ is that the mixing under infinite renormalization 
between
the set $(Q_1...Q_6)$ and the operators $(Q_{7\gamma},Q_{8G})$ vanishes
at the one-loop level. Consequently in order to calculate 
the coefficients
$C_{7\gamma}(\mu)$ and $C_{8G}(\mu)$ in the leading logarithmic
approximation, two-loop calculations of ${\cal{O}}(e g^2_s)$ 
and ${\cal{O}}(g^3_s)$
are necessary. The corresponding NLO analysis requires the evaluation
of the mixing in question at the three-loop level. 
This peculiar feature caused
that the first fully correct calculation of the leading  anomalous
dimension matrix relevant for this decay
has been obtained only in 1993 \cite{CFMRS:93,CFRS:94}.
It has been
confirmed subsequently in \cite{CCRV:94a,CCRV:94b,Mis:94}.

In 1996 the NLO corrections have been completed.
It was a joint effort of many groups. The two-loop
mixing involving the operators
$Q_1.....Q_6$ and the two-loop mixing
in the sector $(Q_{7\gamma},Q_{8G})$ has been calculated in 
\cite{ALTA,BW,BJLW1,BJLW,ROMA1,ROMA2} 
and \cite{MisMu:94}, respectively. The $O(\alpha_s)$
corrections to $C_{7\gamma}(M_W)$ and $C_{8G}(M_W)$ have been first
calculated in \cite{Yao1} and recently confirmed in \cite{GH97}. 
One-loop matrix elements 
$\langle s\gamma {\rm gluon}|Q_i| b\rangle$ have been calculated in 
\cite{AG2,Pott}. The very difficult two-loop corrections to 
$\langle s\gamma |Q_i| b\rangle$ have been presented in \cite{GREUB}.
Finally after a heroic effort  the three loop mixing between
the set $(Q_1...Q_6)$ and the operators $(Q_{7\gamma},Q_{8G})$
 has been completed at the end of 1996 \cite{CZMM}.
As a byproduct the authors of \cite{CZMM} confirmed the existing
two-loop anomalous dimension matrix in the $Q_1...Q_6$ sector.

In order to appreciate the importance of NLO calculations for this
decay it is instructive to discuss first the leading logarithmic
approximation.

\subsubsection{The Decay $B\to X_s\gamma$ in the Leading Log Approximation}
         \label{sec:Heff:Bsgamma:lo}

In calculating $Br(B\to X_s\gamma)$ it is customary to use the
spectator model in which the inclusive decay $B\to X_s\gamma$
is approximated by the partonic decay $b\to s\gamma$. That is
one uses the following approximate equality: 

\begin{equation}\label{ratios}
\frac{\Gamma(B \to X_s \gamma)}
     {\Gamma(B \to X_c e \bar{\nu}_e)}
 \simeq                                                     
\frac{\Gamma(b \to s \gamma)}
     {\Gamma(b \to c e \bar{\nu}_e)} \equiv R_{{\rm quark}},
\end{equation}
where the quantities on the r.h.s are calculated in the spectator model
corrected for short-distance QCD effects. The normalization to the
semileptonic rate is usually introduced in order to cancel the
uncertainties due to the CKM matrix
elements and factors of $\mb^5$ in the r.h.s. of (\ref{ratios}).
Additional support for the approximation given above comes from the
heavy quark expansions.  Indeed the spectator model has been shown to
correspond to the leading order approximation of an expansion in
$1/\mb$.  The first corrections appear at the ${\cal O}(1/\mb^2)$
level and will be discussed at the end of this section.

The leading
logarithmic calculations 
\cite{Grin,CFRS:94,CCRV:94a,Mis:94,AG1,BMMP:94} 
can be summarized in a compact form
as follows:
\begin{equation}\label{main}
R_{{\rm quark}} = 
 \frac{|V_{ts}^* V_{tb}^{}|^2}{|V_{cb}|^2} 
\frac{6 \alpha}{\pi f(z)} |C^{(0){\rm eff}}_{7\gamma}(\mu)|^2\,,
\end{equation}
where
\begin{equation}\label{g}
f(z) = 1 - 8z^2 + 8z^6 - z^8 - 24z^4 \ln z           
\quad\mbox{with}\quad
z =
\frac{\mc}{\mb}
\end{equation}
is the phase space factor in the semileptonic $b$-decay.

The crucial quantity in (\ref{main}) is the effective coefficient 
$C^{(0){\rm eff}}_{7\gamma}(\mu)$ given explicitly as follows: 
\begin{equation}
\label{C7eff}
C_{7\gamma}^{(0){\rm eff}}(\mu)  =  
\eta^\frac{16}{23} C_{7\gamma}^{(0)}(\mw) + \frac{8}{3}
   \left(\eta^\frac{14}{23} - \eta^\frac{16}{23}\right) C_{8G}^{(0)}(\mw) +
    C_2^{(0)}(\mw)\sum_{i=1}^8 h_i \eta^{a_i}\,,
\end{equation}
where

\begin{equation}\label{c2}
C^{(0)}_2(\mw) = 1                               
\end{equation}
\begin{equation}\label{c7}
C^{(0)}_{7\gamma} (\mw) = \frac{3 x_t^3-2 x_t^2}{4(x_t-1)^4}\ln x_t + 
   \frac{-8 x_t^3 - 5 x_t^2 + 7 x_t}{24(x_t-1)^3}
   \equiv -\frac{1}{2} D'_0(x_t)
\end{equation}
and
\begin{equation}
\eta  =  \frac{\as(\mw)}{\as(\mu)}.
\end{equation}
For completeness we give also
\begin{equation}
\label{C7Geff}
C_{8G}^{(0){\rm eff}}(\mu)  =  
\eta^\frac{14}{23} C_{8G}^{(0)}(\mw) 
   + C_2^{(0)}(\mw) \sum_{i=1}^8 \bar h_i \eta^{a_i},
\end{equation}
which is relevant for $b \to s~ {\rm gluon}$ transition. Here
\begin{equation}\label{c8}
C^{(0)}_{8G}(\mw) = \frac{-3 x_t^2}{4(x_t-1)^4}\ln x_t +
   \frac{-x_t^3 + 5 x_t^2 + 2 x_t}{8(x_t-1)^3}                               
   \equiv -\frac{1}{2} E'_0(x_t).
\end{equation}
The functions $D'_0(x_t)$ and
$E'_0(x_t)$ appeared already in section 2.
The numbers $a_i$, $h_i$ and $\bar h_i$ are
given in table \ref{tab:akh}.

\begin{table}[htb]
\begin{center}
\begin{tabular}{|r|r|r|r|r|r|r|r|r|}
\hline
$i$ & 1 & 2 & 3 & 4 & 5 & 6 & 7 & 8 \\
\hline
$a_i $&$ \frac{14}{23} $&$ \frac{16}{23} $&$ \frac{6}{23} $&$
-\frac{12}{23} $&$
0.4086 $&$ -0.4230 $&$ -0.8994 $&$ 0.1456 $\\
$h_i $&$ 2.2996 $&$ - 1.0880 $&$ - \frac{3}{7} $&$ -
\frac{1}{14} $&$ -0.6494 $&$ -0.0380 $&$ -0.0185 $&$ -0.0057 $\\
$\bar h_i $&$ 0.8623 $&$ 0 $&$ 0 $&$ 0
 $&$ -0.9135 $&$ 0.0873 $&$ -0.0571 $&$ 0.0209 $\\
\hline
\end{tabular}
\end{center}
\caption[]{Magic Numbers.
\label{tab:akh}}
\end{table}

Using the leading $\mu$-dependence of $\as$:
\begin{equation} 
\as(\mu) = \frac{\as(\mz)}{1 - \beta_0 \as(\mz)/2\pi \, \ln(\mz/\mu)} 
\label{eq:asmumz}
\end{equation} 
one finds the results in table \ref{tab:c78effnum}.

\begin{table}[htb]
\begin{center}
\begin{tabular}{|c||c|c||c|c||c|c|}
\hline
& \multicolumn{2}{c||}{$\as^{(5)}(\mz) = 0.113$} &
  \multicolumn{2}{c||}{$\as^{(5)}(\mz) = 0.118$} &
  \multicolumn{2}{c| }{$\as^{(5)}(\mz) = 0.123$} \\
\hline
$\mu [\gev]$ & 
$C^{(0){\rm eff}}_{7\gamma}$ & $C^{(0){\rm eff}}_{8G}$ &
$C^{(0){\rm eff}}_{7\gamma}$ & $C^{(0){\rm eff}}_{8G}$ &
$C^{(0){\rm eff}}_{7\gamma}$ & $C^{(0){\rm eff}}_{8G}$ \\
\hline
 2.5 & --0.328 & --0.155 & --0.336 & --0.158 & --0.344 & --0.161 \\
 5.0 & --0.295 & --0.142 & --0.300 & --0.144 & --0.306 & --0.146 \\
 7.5 & --0.277 & --0.134 & --0.282 & --0.136 & --0.286 & --0.138 \\
10.0 & --0.265 & --0.130 & --0.269 & --0.131 & --0.273 & --0.133 \\
\hline
\end{tabular}
\end{center}
\caption[]{Wilson coefficients $C^{(0){\rm eff}}_{7\gamma}$ and 
$C^{(0){\rm eff}}_{8G}$
for $\mt = 170 \gev$ and various values of $\as^{(5)}(\mz)$ and $\mu$.
\label{tab:c78effnum}}
\end{table}

Two features of these results should be emphasised:
\begin{itemize}
\item
The strong enhancement of the
coefficient $C^{(0){\rm eff}}_{7\gamma}$ by short distance QCD effects which 
we illustrate by the relative numerical importance of the three terms in
expression (\ref{C7eff}).
For instance, for $\mt = 170\gev$, $\mu = 5\gev$ and $\as^{(5)}(\mz)
=0.118$ one obtains
\begin{eqnarray}
C^{(0){\rm eff}}_{7\gamma}(\mu) &=&
0.695 \; C^{(0)}_{7\gamma}(\mw) +
0.085 \; C^{(0)}_{8G}(\mw) - 0.158 \; C^{(0)}_2(\mw)
\nn\\
 &=& 0.695 \; (-0.193) + 0.085 \; (-0.096) - 0.158 = -0.300 \, .
\label{eq:C7geffnum}
\end{eqnarray}

In the absence of QCD we would have $C^{(0){\rm eff}}_{7\gamma}(\mu) =
C^{(0)}_{7\gamma}(\mw)$ (in that case one has $\eta = 1$). Therefore, the
dominant term in the above expression (the one proportional to
$C^{(0)}_2(\mw)$) is the additive QCD correction that causes the
enormous QCD enhancement of the \Bsg rate \cite{Bert,Desh}.
It originates solely from the two-loop diagrams. On the other hand, the
multiplicative QCD correction (the factor 0.695 above) tends to
suppress the rate, but fails in the competition with the additive
contributions.

In the case of $C^{(0){\rm eff}}_{8G}$ a similar enhancement is observed
\begin{eqnarray}
C^{(0){\rm eff}}_{8G}(\mu) &=&
0.727 \; C^{(0)}_{8G}(\mw) - 0.074 \; C^{(0)}_2(\mw)
\nn \\
 &=& 0.727 \; (-0.096) - 0.074 = -0.144 \, .
\label{eq:C8Geffnum}
\end{eqnarray}
\item
A strong $\mu$-dependence of both coefficients   
as first stressed by Ali and Greub \cite{AG1} and confirmed
in \cite{BMMP:94}. 
One can see that when $\mu$ is varied by a factor of 2 in both
directions around $\mb \simeq 5\gev$, the ratio (\ref{main}) changes
by around $\pm 25\%$, i.e.\ the ratios $R_{{\rm quark}}$ 
obtained for $\mu=2.5\gev$
and $\mu=10\gev$ differ by a factor of 1.6 \cite{AG1}. 
Since \Bsg is dominated by QCD effects, it is not 
surprising 
that this scale-uncertainty in the leading order 
is particularly large.
\end{itemize}

A critical analysis of theoretical and
experimental
uncertainties present in the prediction for Br(\Bsg) based on the
formula (\ref{main}) has been made in \cite{BMMP:94} giving
\begin{equation}
Br(B \to X_s\gamma)_{{\rm TH}} = (2.8 \pm 0.8) \times 10^{-4}
\label{theo}
\end{equation}
where the error is dominated by the uncertainty in 
the choice of the renormalization scale
$m_b/2<\mu<2 m_b$ discussed above. To this end
$Br(B \to X_c e \bar{\nu}_e) = (10.43 \pm 0.24)\%$
has been used. Similar result has been found in \cite{AG1}.

In 1994 the first measurement of the inclusive rate was
presented by CLEO \cite{CLEO2}:
\begin{equation}\label{EXP}
Br(B \to X_s\gamma) = (2.32 \pm 0.57 \pm 0.35) \times 10^{-4}\,,
\label{incl}
\end{equation}
where the first error is statistical and the second is systematic.
We will report on the status of the exclusive measurements such as
$B \to K^* \gamma$ later on.

The result in (\ref{incl}) agrees with (\ref{theo}) although 
the large theoretical and experimental errors do not allow
for a definitive conclusion and to see
whether some contributions beyond the Standard Model, such as present in the
Two-Higgs-Doublet Model (2HDM) 
or in the Minimal Supersymmetric Standard
Model (MSSM), are required. In any case the agreement of the
theory with data is consistent with the large QCD enhancement
of \Bsg. Without this enhancement the theoretical prediction
would be at least by a factor of 2 below the data. 
  
Since the theoretical result in (\ref{theo}) is dominated by
the scale ambiguities present in the leading order approximation,
it was clear already in 1993 that  a  complete
NLO analysis is very desirable. 
Such a complete next-to-leading
calculation of \Bsg was described in \cite{BMMP:94} in general terms. 
As demonstrated formally there, the cancellation of the dominant 
$\mu$-dependence in the leading
order can then be  achieved. While this formal NLO analysis was
very encouraging with respect to the reduction of the $\mu$-dependence,
it could obviously not provide the actual size of Br(\Bsg) after
the inclusion of NLO corrections. Fortunately three years later such
a complete NLO analysis exists and the impact of NLO corrections on
Br(\Bsg) can be analysed in explicit terms.

\subsubsection{\Bsg Beyond Leading Logarithms}
         \label{sec:Heff:Bsgamma:nlo}
The formula (\ref{main}) modifies after the inclusion of NLO
corrections as follows \cite{CZMM}:
\be \label{ration}
R_{{\rm quark}} = 
\frac{|V_{ts}^* V_{tb}|^2}{|V_{cb}|^2} 
\frac{6 \alpha}{\pi f(z)} F \left( |D|^2 + A \right)\,,
\ee
where
\be \label{factor}
F = \f{1}{\kappa(z)} \left( \f{m_b(\mu=m_b)}{m_{b,{\rm pole}}} \right)^2 = 
    \f{1}{\kappa(z)} \left( 1 - \f{8}{3} \f{\as(m_b)}{\pi} \right),
\ee
 
\be \label{Dvirt}
D = C_{7\gamma}^{(0){\rm eff}}(\mu_b) + \frac{\as(\mu_b)}{4 \pi} \left\{ 
C_{7\gamma}^{(1){\rm eff}}(\mu_b) + \sum_{i=1}^8 C_i^{(0){\rm eff}}(\mu_b) 
\left[ r_i + \gamma_{i7}^{(0){\rm eff}} \ln \frac{m_b}{\mu_b} 
\right] \right\}
\ee
and $A$ is discussed below. Here $\mu_b={\cal O}(m_b)$.

We will now explain the origin of various new contributions:
\begin{itemize}
\item
First $\kappa(z)$
is the QCD correction to the semileptonic decay
\cite{CM78}. To a good approximation it is given by \cite{KIMM}
\be \label{kap}
\kappa(z) = 1 - \frac{2 \as (m_b)}{3 \pi}
\left[(\pi^2-\frac{31}{4})(1-z)^2+\frac{3}{2}\right] \,.
\ee
An exact analytic formula for $\kappa(z)$ can be found in \cite{N89}.
\item

The second factor in (\ref{factor}) originates as follows.
The \Bsg rate is proportional to $m_{b,{\rm pole}}^3$ present in the
two body phase space and to $m_b(\mu=m_b)^2$ present in 
$<s\gamma|Q_{7\gamma}|B>^2$. On the other hand the semileptonic
rate is is proportional to $m_{b,{\rm pole}}^5$ present in the
three body phase space. Thus the $m_b^5$ factors present in
both rates differ by a ${\cal O}(\as)$ correction which has
been consistently omitted in the leading logarithmic approximation.
\item
For similar reason the variable $z$ entering $f(z)$ and $\kappa(z)$
can be more precisely specified at the NLO level to be 
\cite{GREUB,CZMM}:
\be \label{g(z)}
 z = \frac{m_{c,{\rm pole}}}{m_{b,{\rm pole}}}=0.29\pm0.02 
\ee
which is obtained from $m_{b,{\rm pole}} = 4.8 \pm 0.15$~GeV and
$m_{b,{\rm pole}}-m_{c,{\rm pole}}=3.40$~GeV. 
This gives
\be \label{kf}
\kappa(z)=0.879\pm0.002\approx 0.88\,,
\qquad
f(z)=0.54\pm 0.04\,.
\end{equation}
\item
The amplitude $D$ in (\ref{Dvirt}) includes two types of new
contributions. The first $\as$-correction originates in
the NLO correction to the Wilson coefficients of $Q_{7\gamma}$:
\be \label{C.expanded}
C^{\rm eff}_{7\gamma}(\mu) = C^{(0){\rm eff}}_{7\gamma}(\mu) + 
\frac{\as(\mu)}{4 \pi} C^{(1){\rm eff}}_{7\gamma}(\mu)\,. 
\ee
It is this correction which requires the calculation of the
three-loop anomalous dimensions \cite{CZMM}. An explicit formula for
$C^{(1){\rm eff}}_{7\gamma}(\mu)$ can be found in \cite{CZMM}. It has
a similar structure to (\ref{C7eff}) with $C_{7\gamma}^{(0)}(\mw)$
and $C_{8G}^{(0)}(\mw)$ replaced by the corresponding NLO corrections
calculated in \cite{Yao1,GH97}. 
Also the $C_2(\mw)$-part in (\ref{C7eff}) 
has to be modified.

The two remaining corrections in (\ref{Dvirt}) come from one-loop
matrix elements $<s\gamma|Q_{7\gamma}|B>$ and 
$<s\gamma|Q_{8G}|B>$ and from two-loop matrix elements
$<s\gamma|Q_i|B>$ of the remaining operators. These two-loop
matrix elements have been calculated in \cite{GREUB}. The coefficients
of the logarithm are the relevant elements in the leading
anomalous dimension matrix. The explicit logarithmic 
$\mu_b$ dependence in the last term in $D$ cancels the
leading $\mu_b$ dependence present in the first term in (\ref{Dvirt}) 
as already pointed out in \cite{BMMP:94}.

Now $C^{(1){\rm eff}}_{7\gamma}(\mu)$ is renormalization scheme dependent.
This scheme dependence is cancelled by the one present in the
constant terms $r_i$. Actually ref. \cite{GREUB} does not provide
the matrix elements of the QCD-penguin operators and consequently
$r_i~(i=3-6)$ are unknown. However, the Wilson coefficients
of QCD-penguin operators are very small and this omission is immaterial.

\item
The term $A$ in (\ref{ration}) originates from the bremsstrahlung
corrections and the necessary virtual corrections needed for the
cancellation of the infrared divergences. These have been
calculated in \cite{AG2,Pott} and are also considered in 
\cite{CZMM,GREUB} in the
context of the full analysis.  Since the virtual corrections
are also present in the terms $r_i$ in $D$, care must be taken
in order to avoid double counting. This is discussed in detail
in \cite{CZMM} where an explicit formula for $A$ can be found.
Actually $A$ depends on  an explicit lower cut on the
photon energy 
\be
E_{\gamma} > 
( 1 - \delta ) E_{\gamma}^{{\rm max}} \equiv ( 1 - \delta ) \frac{m_b}{2}.
\ee
Moreover $A$ is divergent in the limit $\delta \to 1$.
In order to cancel this divergence one would have to consider
the sum of \Bsg and $b{\to}X_s$ decay rates.
However, the divergence at $\delta{\to}1$ is very slow. 
In order to
allow an easy comparison with previous experimental and theoretical
publications the authors in \cite{CZMM} choose $\delta = 0.99$.
Further details on the $\delta$-dependence can be found in this paper.

\item
Finally the values of $\as(\mu)$ in all the
above formulae are calculated with the use of the NLO expression 
for the strong coupling constant:
\be \label{alphaNLL}
\as(\mu) = \frac{\as(M_Z)}{v(\mu)} \left[1 - \f{\beta_1}{\beta_0} 
           \frac{\as(M_Z)}{4 \pi}    \f{\ln v(\mu)}{v(\mu)} \right],
\ee
where 
\be \label{v(mu)}
v(\mu) = 1 - \beta_0 \frac{\as(M_Z)}{2 \pi} 
\ln \left( \frac{M_Z}{\mu} \right),
\ee
$\beta_0 = \frac{23}{3}$ and $\beta_1 = \frac{116}{3}$.
\end{itemize}

The main uncertainty in the leading order formulae related
to the $\mu$-dependence has been thus considerably reduced
at the NLO level. This reduction comes dominantly through
the inclusion of the two-loop matrix element of the
operator $Q_2$ calculated in \cite{GREUB}. However, as we stated
above, the  calculation of the three loop anomalous
dimensions present in $C^{(1){\rm eff}}_{7\gamma}(\mu)$ is
necessary for the complete analysis and in particular
for the cancellation of the renormalization scheme dependence.
As analysed in \cite{GREUB,CZMM} the $\mu$ dependence in the
final result for Br(\Bsg) is reduced from $\pm 25\%$ in
LO down to $\pm 6\%$. Other sources of uncertainties are
given in table \ref{tab:RQ} taken from \cite{CZMM}. We observe that the 
dominant sources of remaining uncertainties are $m_c/m_b$ and $\mu_b$. 

\begin{table}[htb]
\begin{center}
\begin{tabular}{|c|c|c|c|c|c|c|}
\hline
$\as(M_Z)$ & $m_t$ & $\mu_b$ & $m_{c,{\rm pole}}/m_{b,{\rm pole}}$ & 
$m_{b,{\rm pole}}$ & $\alpha$ & CKM angles \\
\hline
2.5\%   &    1.7\%   & 6.2\% &   5.2\%        &   0.5\%    &    1.9\% 
 & 2.1\%\\
\hline
\end{tabular} 
\end{center}
\caption[]{Uncertainties in $R_{{\rm quark}}$ due to various sources.
\label{tab:RQ}}
\end{table}

Finally one has to pass from the calculated
$b$-quark decay rates to the $B$-meson decay rates. Relying on the
Heavy Quark Expansion (HQE) calculations one finds \cite{CZMM}
\be \label{BR} 
Br(B{\to}X_s \gamma) = Br(B{\to}X_c e \bar{\nu_e})
\cdot R_{{\rm quark}} 
\left( 1 - \frac{\delta^{NP}_{sl}}{m_b^2}
         + \frac{\delta^{NP}_{rad}}{m_b^2} \right),
\ee
where $\delta^{{\rm NP}}_{{\rm sl}}$ and $\delta^{{\rm NP}}_{{\rm rad}} $
parametrize nonperturbative corrections to the semileptonic and
radiative $B$-meson decay rates, respectively. 

According to \cite{BBB95}, $\delta^{{\rm NP}}_{{\rm sl}} 
= -(1.05 \pm 0.10)\;{\rm GeV}^2$. Next, following \cite{FLS96}, one can 
express $\delta^{{\rm NP}}_{{\rm rad}}$
in terms of the HQET parameters $\lambda_1$ and $\lambda_2$:
\be
\delta^{{\rm NP}}_{{\rm rad}}  = \f{1}{2} \lambda_1 - \f{9}{2} \lambda_2.
\ee
The value of $\lambda_2$ is known from $B^*$--$B$ mass splitting
\be
\lambda_2 = \f{1}{4} ( m_{B^*}^2 - m_B^2 ) \simeq 0.12\;{\rm GeV}^2.
\ee
The value of $\lambda_1$ is controversial. In \cite{CZMM},
$\lambda_1 = -(0.6 \pm 0.1)\;{\rm GeV}^2$ taken from
\cite{BBB95} has been used.

The two nonperturbative corrections in (\ref{BR}) are
both around $4\%$ in magnitude and tend to cancel each other. In
effect, they sum up to only $(1 \pm 0.5) \%$. As stressed in \cite{CZMM},
such a small number has
to be taken with caution. If values of $\lambda_1$ around $-0.1\;{\rm
GeV}^2$ were accepted \cite{FLS96}, the total nonperturbative
correction in (\ref{BR}) would change by one or two percent.
Moreover, one has to remember that the four-quark operators 
$Q_1...... Q_6$ have not been included in the calculation of
$\delta^{{\rm NP}}_{{\rm rad}}$. Contributions from these operators could
potentially give one- or two-percent effects. Nevertheless, it seems
reasonable to conclude that the total nonperturbative $1/m_b^2$ 
correction to
(\ref{BR}) is well below 10\%, i.e.\ it is smaller than the
inaccuracy of the perturbative calculation of $R_{{\rm quark}}$.

Using $Br(B{\to}X_c e \bar{\nu_e}) = ( 10.4 \pm 0.4)\%$~\cite{PDG}, 
the authors of \cite{CZMM} find
the following prediction for $Br(B{\to}X_s \gamma)$ :
\be\label{theon}
Br(B{\to}X_s \gamma) = (3.28 \pm 0.33) \times 10^{-4}.
\ee
Similar result has been obtained in \cite{GREUB}. Comparing with the leading
order estimate (\ref{theo}) we observe that the central value
has been shifted upwards by roughly $15\%$ and the theoretical
uncertainty has been reduced by more than a factor of two.
We also note that
the central value of the NLO prediction is outside the $1\sigma$
experimental error bar in (\ref{EXP}).
However, the experimental and theoretical error bars
practically touch each other. Therefore we conclude that the present
$B{\to}X_s\gamma$ measurement remains in agreement with the Standard
Model.  
\subsubsection{Long Distance Contributions}
The long distance contributions to $B\to X_s\gamma$ are not easy
to calculate and the present estimates are based on phenomenological
models. These long distance contributions are expected to arise dominantly 
from transitions $B \to \sum_i V_i+X_s \to \gamma X_s$ where
$V_i=J/\psi,\psi^\prime,...$. In view of space limitations we will not
 discuss these contributions which are generally expected to be below
$10\%$ \cite{LDGAMMA}. A better estimate of these effects is
desirable. 

A related issue are the ${\cal O}(1/m^2_c)$ corrections pointed out
recently by Voloshin \cite{VOL96} and also discussed by other
authors \cite{LRW97}. These non-perturbative corrections
originate in the photon coupling to a virtual $c\bar c$ loop and
their general structure is given by 
$$
(\Lambda_{\rm QCD}^2/\mc^2)(\Lambda_{\rm QCD}\mb/\mc^2)^n
$$
with $(n=0,1..)$. The term $n=0$ can be estimated reliably 
and amounts to a $3\%$ reduction of the rate. Since
$\Lambda_{\rm QCD}\mb/\mc^2\approx 0.6 $, the terms with $n>0$
are not necessarily much smaller. Although the presence of
unknown matrix elements in these contributions does not
allow a definite estimate of their actual size, the analyses in
\cite{LRW97} indicate, however, that these higer order contributions
are substantially smaller than the $n=0$ term. 
Less optimistic view can be found in \cite{VOL96}.

\subsubsection{$B\to K^*\gamma$ and other Exclusive Modes}
  
In 1993
CLEO reported \cite{CLEO} 
$Br(B \to K^* \gamma) = (4.5 \pm 1.5 \pm 0.9) \times 10^{-5}.$
This was in fact the first observation of the $b \to s \gamma$
transition and equivalently of magnetic $\gamma$-penguins.
The corresponding 1996 value has a substantially smaller error
\cite{CLEO96}:
\begin{equation}
Br(B \to K^* \gamma) = (4.2 \pm 0.8 \pm 0.6) \times 10^{-5}
\label{excl}
\end{equation}
implying an improved measurement of $R_{K^*}$:

\begin{equation}
R_{K^*}=\frac{\Gamma (B \to K^* \gamma)}{\Gamma(B \to X_s \gamma)}
 = 0.181 \pm 0.068
\label{K*}
\end{equation}

This result puts some constraints on various formfactor models
listed in \cite{CLEO96}. There one can also find 90\% C.L. bounds: 
$Br (B^0\to\rho^0\gamma)< 3.9\cdot 10^{-5}$,
$Br (B^0\to\omega\gamma)< 1.3\cdot 10^{-5}$,
$Br (B^-\to\rho^-\gamma)< 1.1\cdot 10^{-5}$.
Combined with (\ref{excl}) these
bounds imply 
\begin{equation}
\frac{\vtd}{\vts}\le 0.45-0.56 \,,
\end{equation}
where the uncertainty in
the bound reflects the model dependence. Clearly this bound is
still higher by a factor of two than the value obtained in the
standard analysis of section 4. 
In 1996 also DELPHI \cite{DEL96a} provided two bounds:
$Br (B^0_d\to K^*\gamma)< 2.1\cdot 10^{-4}$
and
$Br (B^0_s\to\phi\gamma)< 7.0\cdot 10^{-4}$.
The second bound is the only existing bound on this decay.

\subsubsection{Summary and Outlook}
The rare decay $B\to X_s\gamma$ plays at present together with
$B^0_{d,s}-\bar B^0_{d,s}$ mixing the central role
in loop induced transitions in the $B$-system. On the theoretical
side considerable progress has been made last year by calculating
NLO corrections, thereby reducing the large $\mu_b$ uncertainties
present in the leading order. This way the error in the
{\it short distance} prediction for $Br(B\to X_s\gamma)$ as
given in (\ref{theon}) has been decreased down to roughly
$\pm 10\%$ compared with $\pm (25-30)\%$ in the leading order.
The precise size of long distance contributions due to
the intermediate $J/\psi$ state is difficult to obtain but
generally these contributions are found to be below $10\%$.
The $1/m_b^2$ and $1/m_c^2$ corrections amount to a few percent.

On the experimental side considerable progress has been made
by CLEO in the case of $Br (B^0_d\to K^*\gamma)$. It is very
desirable to obtain now an improved measurement of 
$Br(B\to X_s\gamma)$.
As advertised at the HEP conference in Warsaw by 
Browder, Kagan and Kagan, CLEO II
should be able to discover $b \to s~ {\rm gluon}$ transition soon.
In view of this and the fact that here the theoretical uncertainties
are large, theorists should sharpen their tools. 

More
on \Bsg, in particular on the photon spectrum and the determination
of $\vtd/\vts$ from $B\to X_{s,d}\gamma$, can be found in
\cite{ALUT,ALIB,Photon}. The impact of new physics is discussed
in another chapter of this book \cite{MPR}.
CP violation in $B \to K^* \gamma$ and $B \to \varrho \gamma$ 
is discussed in \cite{GSW95}.
 
\subsection{$B\to X_{\lowercase{s}} \lowercase{l}^+\lowercase{l}^-$}
         \label{sec:Heff:BXsee:nlo}
\subsubsection{General Remarks} 
         \label{sec:Heff:BXsee:nlo:rem}
The rare decays $B \to X_s l^+ l^-$ have been the subject of 
many theoretical studies
in the framework of the Standard Model and its extensions such as the
two Higgs doublet models and models involving supersymmetry
\cite{HWS:87}-\cite{CMW:96}.  In
particular the strong dependence of $B \to X_s l^+ l^-$ on $\mt$ has 
been stressed in \cite{HWS:87}. It is clear that once 
these decays have been observed, they
will offer a useful test of the Standard Model and its extensions.
We will concentrate here on the predictions within the Standard
Model and in particular on $B \to X_s \mu^+ \mu^-$.
The impact of new physics is discussed
in another chapter in this book \cite{MPR}.

The central element in any analysis of  $B \to X_s \mu^+ \mu^-$
is the effective Hamiltonian  at scales $\mu=O(m_b)$
 given by
\begin{equation} \label{Heff2_at_mu}
{\cal H}_{\rm eff}(b\to s \mu^+\mu^-) =
{\cal H}_{\rm eff}(b\to s\gamma)  - \frac{G_{\rm F}}{\sqrt{2}} V_{ts}^* V_{tb}
\left[ C_{9V}(\mu) Q_{9V}+
C_{10A}(M_W) Q_{10A}    \right]\,,
\end{equation}
where we have again neglected the term proportional to $V_{us}^*V_{ub}$
and ${\cal H}_{\rm eff}(b\to s\gamma)$ is given in (\ref{Heff_at_mu}).
In addition to the operators relevant for $B\to X_s\gamma$,
there are two new operators:
\begin{equation}\label{Q9V}
Q_{9V}    = (\bar{s} b)_{V-A}  (\bar{\mu}\mu)_V\,,         
\qquad
Q_{10A}  =  (\bar{s} b)_{V-A}  (\bar{\mu}\mu)_A\,,
\end{equation}
where $V$ and $A$ refer to $\gamma_{\mu}$ and $ \gamma_{\mu}\gamma_5$,
respectively. They are generated through the electroweak
penguin diagrams of fig.\ 11f and the related box diagrams needed mainly
to keep gauge invariance. 

The actual calculation of $Br(B \to X_s \mu^+ \mu^-)$ involves the
evaluation of the Wilson coefficients of the relevant local operators and
the calculation of the corresponding matrix elements of these
operators relevant for the decay in question. As in the case of 
$B \to X_s \gamma$, the latter part of the analysis can be
done in the spectator model which, as indicated by the heavy quark
expansion, should offer a good approximation to QCD for $B$-decays. One can
also include the non-perturbative ${\cal O}(1/\mb^2)$ corrections to
the spectator model which we will briefly discuss at the end of this
section.
A realistic phenomenological analysis should also
include the long-distance contributions which are mainly due to the
$J/\psi$ and $\psi'$ resonances \cite{LMS:89}-\cite{KS96b}.
We will first concentrate our presentation on the short 
distance contributions. The impact of the long distance contributions
will be briefly discussed subsequently.

\subsubsection{Wilson Coefficients $C_{9V}(\mu)$ and $C_{10A}(\mu)$}
         \label{sec:Heff:BXsee:wc}
The coefficient $C_{10A}(\mu)$ is given by
\begin{equation} \label{C10}
C_{10A}(\mw) =  \frac{\aem}{2\pi} \Ctilde_{10}(\mw), \qquad
\Ctilde_{10}(\mw) = - \frac{Y_0(x_t)}{\sin^2\Theta_{\rm W}}
\end{equation}
with $Y_0(x)$ given in (\eqn{Y0}). Since $Q_{10A}$ does not renormalize
under QCD, its coefficient does not depend on $\mu\approx {\cal
O}(\mb)$. The only renormalization scale dependence in (\ref{C10})
enters through the definition of the top quark mass. We will return to
this issue below.

The main issue of QCD corrections to $B\to X_s \mu^+\mu^-$
centers around the coefficient $C_{9V}(\mu)$.
The special feature of $C_{9V}(\mu)$ compared to the coefficients
of the remaining operators contributing to $B\to X_s \mu^+\mu^-$ is the
large logarithm represented by $1/\as$ in $P_0$ in the formula 
(\ref{P0NDR}) given below.
 Consequently the renormalization group improved
perturbation theory for $C_{9V}$ has the structure $ {\cal O}(1/\as) +
{\cal O}(1) + {\cal O}(\as)+ \ldots$ whereas the corresponding series
for the remaining coefficients is $ {\cal O}(1) + {\cal O}(\as)+
\ldots$. Therefore in order to find the next-to-leading ${\cal O}(1)$
term in the branching ratio for $B\to X_s \mu^+\mu^-$,
 the full two-loop renormalization group analysis 
 has to be performed in order to find $C_{9V}$, but the
coefficients of the remaining operators should be taken in the leading
logarithmic approximation. In particular at the NLO level one should
only include the leading term  $C^{(0){\rm eff}}_{7\gamma}(\mu)$ in
(\ref{C.expanded}). The recently calculated scheme dependent correction
$C^{(1){\rm eff}}_{7\gamma}(\mu)$ is a part of next-to-NLO correction
to $B\to X_s \mu\bar\mu$ 
and should be omitted in a consistent NLO calculation. 

The QCD corrections to $C_{9V}(\mu)$ have been calculated 
 over the last years with increasing precision by several
groups \cite{GSW:89,GDSN:89,CRV:91,Mis:94} culminating in two complete
next-to-leading QCD calculations 
\cite{Mis:94,BuMu:94} which agree with each other.
 Defining
$\tilde C_{9}$ by
\begin{equation} \label{C9}
C_{9V}(\mu) = \frac{\aem}{2\pi} \Ctilde_9(\mu) 
\end{equation}
one finds \cite{BuMu:94} in the NDR scheme
\begin{equation}\label{C9tilde}
\Ctilde_9^{\rm NDR}(\mu)  =  
P_0^{\rm NDR} + \frac{Y_0(x_t)}{\sin^2\Theta_{\rm W}} -4 Z_0(x_t) +
P_E E_0(x_t)
\end{equation}
with
\begin{eqnarray}
\label{P0NDR}
P_0^{\rm NDR} & = & \frac{\pi}{\as(\mw)} (-0.1875+ \sum_{i=1}^8 p_i
\eta^{a_i+1}) \nn \\ 
          &   & + 1.2468 +  \sum_{i=1}^8 \eta^{a_i} \lbrack
r^{\rm NDR}_i+s_i \eta \rbrack \\ 
\label{PE}
P_E & = & 0.1405 +\sum_{i=1}^8 q_i\eta^{a_i+1}  \, .
\end{eqnarray}
The formula (\ref{C9tilde}) has an identical structure to (\ref{y7vpbe}) 
relevant for
$K_{\rm L} \to \pi^0 e^+ e^-$ with different numerical values for 
$P_0$ and $P_E$ due to different scales $\mu$ involved. 
However, because of the one step evolution from
$\mu=\mw$ down to $\mu=\mb$ without the charm threshold in between, it was
possible to find an analytic formula for $P_0$ here which
was not possible in the case of $K_{\rm L} \to \pi^0 e^+ e^-$.

$Y_0(x)$, $Z_0(x)$ and
$E_0(x)$ are defined in (\ref{Y0}), (\ref{Z0}) and (\ref{E0}), 
respectively.
The powers $a_i$ are the same as in table
\ref{tab:akh}.  The coefficients $p_i$, $r^{\rm NDR}_i$, $s_i$, and $q_i$
can be found in table \ref{tab:prsq}.  $P_E$ is ${\cal O}(10^{-2})$ and
consequently the last term in (\ref{C9tilde}) can be neglected. We keep
it however in our numerical analysis. 

\begin{table}[htb]
\begin{center}
\begin{tabular}{|r|r|r|r|r|r|r|r|r|}
\hline
$i$ & 1 & 2 & 3 & 4 & 5 & 6 & 7 & 8 \\
\hline
$p_i $&$ 0, $&$ 0, $&$ -\frac{80}{203}, $&$  \frac{8}{33}, $&$
0.0433 $&$  0.1384 $&$ 0.1648 $&$ - 0.0073 $\\
$r^{\rm NDR}_{i} $&$ 0 $&$ 0 $&$ 0.8966 $&$ - 0.1960 $&$
- 0.2011 $&$ 0.1328 $&$ - 0.0292 $&$ - 0.1858 $\\
$s_i $&$ 0 $&$ 0 $&$ - 0.2009 $&$  -0.3579 $&$
0.0490 $&$ - 0.3616 $&$ -0.3554 $&$ 0.0072 $\\
$q_i $&$ 0 $&$ 0 $&$ 0 $&$  0 $&$
0.0318 $&$ 0.0918 $&$ - 0.2700 $&$ 0.0059 $\\
\hline
$r^{\rm HV}_{i} $&$ 0 $&$ 0 $&$ -0.1193 $&$ 0.1003 $&$
- 0.0473 $&$ 0.2323 $&$ - 0.0133 $&$ - 0.1799 $\\
\hline
\end{tabular}
\end{center}
\caption[]{Additional Magic Numbers.
\label{tab:prsq}}
\end{table}

In the HV scheme only the coefficients $r_i$ are changed. 
They are given on the last line of table \ref{tab:prsq}.
Equivalently we can write
\begin{equation} \label{P0HV}
P_0^{k} = P_0^{\rm NDR} + \xi_{k} \frac{4}{9} \left( 3 C_1^{(0)} +
C_2^{(0)} - C_3^{(0)} -3 C_4^{(0)} \right)
\end{equation}
with
\begin{equation} \label{xi}
\xi_k = \left\{
\begin{array}{rl}
0  &\quad k=\mbox{\rm NDR} \\
-1 &\quad k=\mbox{HV}
\end{array} \, .
\right.
\end{equation}
and $C_i^{(0)}$ denoting the LO coefficients. Their numerical values
are given in table \ref{tab:XXX}.

In table \ref{tab:p0C9} we show the constant $P_0$ in (\ref{P0NDR}) for
different $\mu$ and $\Lms$ in the leading order corresponding to the
first term in (\ref{P0NDR}) and for the NDR and HV schemes as given by
(\ref{P0NDR}) and (\ref{P0HV}), respectively. In table \ref{tab:BXsee:C9} we
show the corresponding values for $\Ctilde_9(\mu)$. To this end
we set $\mt= 170 \gev$. These results are essentially the same as in 
\cite{BuMu:94,BBL} except for an update in $\Lms$.

\begin{table}[htb]
\begin{center}
\begin{tabular}{|c||c|c|c||c|c|c||c|c|c|}
\hline
& \multicolumn{3}{c||}{$\Lms^{(5)} = 160 \mev$} &
  \multicolumn{3}{c||}{$\Lms^{(5)} = 225 \mev$} &
  \multicolumn{3}{c| }{$\Lms^{(5)} = 290 \mev$} \\
\hline
$\mu [\gev]$ & LO & NDR & HV & LO & NDR & HV & LO & NDR & HV \\
\hline
2.5 & 2.022 & 2.907 & 2.787 & 1.933 & 2.846 & 2.759 & 1.857 & 2.791 &
2.734 \\
5.0 & 1.835 & 2.616 & 2.402 & 1.788 & 2.591 & 2.395 & 1.748 & 2.568 &
2.390 \\
7.5 & 1.663 & 2.386 & 2.127 & 1.632 & 2.373 & 2.127 & 1.605 & 2.361 &
2.128 \\
10.0 & 1.517 & 2.201 & 1.913 & 1.494 & 2.194 & 1.917 & 1.475 & 2.185 &
1.920\\
\hline
\end{tabular}
\end{center}
\caption[]{The coefficient $P_0$ of $\widetilde C_9$ for various values
of $\Lms^{(5)}$ and $\mu$.
\label{tab:p0C9}}
\end{table}

\begin{table}[htb]
\begin{center}
\begin{tabular}{|c||c|c|c||c|c|c||c|c|c|}
\hline
& \multicolumn{3}{c||}{$\Lms^{(5)} = 160 \mev$} &
  \multicolumn{3}{c||}{$\Lms^{(5)} = 225 \mev$} &
  \multicolumn{3}{c| }{$\Lms^{(5)} = 290 \mev$} \\
\hline
$\mu [\gev]$ & LO & NDR & HV & LO & NDR & HV & LO & NDR & HV \\
\hline
2.5 & 2.022 & 4.472 & 4.352 & 1.933 & 4.410 & 4.323 & 1.857 & 4.355 &
4.298 \\
5.0 & 1.835 & 4.182 & 3.968 & 1.788 & 4.156 & 3.961 & 1.748 & 4.134 &
3.955 \\
7.5 & 1.663 & 3.954 & 3.694 & 1.632 & 3.940 & 3.694 & 1.605 & 3.928 &
3.695 \\
10.0 & 1.517 & 3.769 & 3.481 & 1.494 & 3.761 & 3.485 & 1.475 & 3.754 &
3.487 \\
\hline
\end{tabular}
\end{center}
\caption[]{Wilson coefficient $\widetilde C_9$ for $\mt = 170 \gev$ and
various values of $\Lms^{(5)}$ and $\mu$.
\label{tab:BXsee:C9}}
\end{table}

\noindent
Let us briefly discuss these numerical results. We observe:
\begin{itemize}
\item
The NLO corrections to $P_0$ enhance this constant relatively to the
LO result by roughly 45\% and 35\% in the NDR and HV schemes,
respectively. This enhancement is analogous to the one found in the
case of $K_{\rm L} \to \pi^0 e^+ e^-$.
\item
It is tempting to compare $P_0$ in table \ref{tab:p0C9} with that found in
the absence of QCD corrections. In the limit $\as \to 0$ we
find $P_0^{\rm NDR} = 8/9 \, \ln(\mw/\mu) + 4/9$ and $P_0^{\rm HV} = 8/9\,
\ln(\mw/\mu)$ which for $\mu = 5 \gev$ give $P_0^{\rm NDR} = 2.91$ and
$P_0^{\rm HV} = 2.46$. Comparing these values with table~\ref{tab:p0C9} we
conclude that the QCD suppression of $P_0$ present in the leading order
approximation is considerably weakened in the NDR treatment of
$\gamma_5$ after the inclusion of NLO corrections. It is essentially
removed for $\mu > 5 \gev$ in the HV scheme.
\item
The NLO corrections to $\Ctilde_9$, which include also the
$\mt$-dependent contributions, are large as seen in table
\ref{tab:BXsee:C9}. The results in HV and NDR schemes are by more than
a factor of two larger than the leading order result $\Ctilde_9 =
P_0^{\rm LO}$ which consistently should not include $\mt$-contributions.
This demonstrates very clearly the necessity of NLO calculations which
allow a consistent inclusion of the important $\mt$-contributions. 
\item
  
The $\mu$ dependence of $\Ctilde_9$ is sizable: 
$\sim 15\%$ in the range of $\mu$
considered. On the other hand its $\Lms$ dependence  is rather weak.  
Also
the $\mt$ dependence of $\Ctilde_9$ is weak. Varying
$\mt$ between $150 \gev$ and $190 \gev$ changes $\Ctilde_9$ by at most
10\%. This weak $\mt$ dependence of $\Ctilde_9$ originates in the
partial cancelation of $\mt$ dependences between $Y_0(x_t)$ and
$Z_0(x_t)$ in (\ref{C9tilde}) as already seen in the case of 
$K_{\rm L} \to \pi^0 e^+ e^-$.
Finally, the difference between
$\Ctilde_9^{\rm NDR}$ and $\Ctilde_9^{\rm HV}$ is small and amounts to 
roughly 5\%.
\item
The dominant $\mt$-dependence in this decay originates, similarly to
$K_{\rm L} \to \pi^0 e^+ e^-$, in the $\mt$
dependence of $\tilde C_{10}(\mw)$. In fact,
$\tilde C_{10}(\mw)=2\pi y_{7A}/\alpha$ with $y_{7A}$
present in $K_{\rm L} \to \pi^0 e^+ e^-$. 
\end{itemize}

\subsubsection{The Differential Decay Rate}
         \label{sec:Heff:BXsee:nlo:rate}
We are now ready to present the results for the differential decay
rate based on the effective Hamiltonian in (\ref{Heff2_at_mu}) and
the spectator model for the matrix elements of $Q_i$.
Introducing
\begin{equation} \label{invleptmass}
\hat s = \frac{(p_{\mu^+} + p_{\mu^-})^2}{\mb^2}, \qquad z =
\frac{\mc}{\mb}
\end{equation}
and calculating the one-loop matrix elements of $Q_i$ using the
spectator model in the NDR scheme one finds \cite{Mis:94,BuMu:94}
\begin{eqnarray} \label{rateee}
& &
R(\hat s) \equiv \frac{{d}/{d\hat s} \, 
\Gamma (b \to s \mu^+\mu^-)}{\Gamma
(b \to c e\bar\nu)} = \frac{\alpha^2}{4\pi^2}
\left|\frac{V_{ts}}{V_{cb}}\right|^2 \frac{(1-\hat s)^2}{f(z)\kappa(z)}
\times \\ 
& &
\biggl[(1+2\hat s)\left(|\Ctilde_9^{\rm eff}|^2 + |\Ctilde_{10}|^2\right) + 
4 \left( 1 + \frac{2}{\hat s}\right) |C_{7\gamma}^{(0){\rm eff}}|^2 + 12
C_{7\gamma}^{(0){\rm eff}} \ \RE\,\Ctilde_9^{\rm eff}  \biggr]\,,
\nn
\end{eqnarray}
where
\begin{eqnarray} \label{C9eff}
\Ctilde_9^{\rm eff} & = & \Ctilde_9^{\rm NDR} \tilde\eta(\hat s) + h(z, \hat
s)\left( 3 C_1^{(0)} + C_2^{(0)} + 3 C_3^{(0)} + C_4^{(0)} + 3
C_5^{(0)} + C_6^{(0)} \right) - \nn \\
& & \frac{1}{2} h(1, \hat s) \left( 4 C_3^{(0)} + 4 C_4^{(0)} + 3
C_5^{(0)} + C_6^{(0)} \right) - \\
& & \frac{1}{2} h(0, \hat s) \left( C_3^{(0)} + 3 C_4^{(0)} \right) +
\frac{2}{9} \left( 3 C_3^{(0)} + C_4^{(0)} + 3 C_5^{(0)} + C_6^{(0)}
\right) \, .
\nn
\end{eqnarray}
The general expression (\ref{rateee}) with $\kappa(z)=1$ has been first
presented by \cite{GSW:89} who in their approximate leading
order renormalization group analysis kept only the operators $Q_1, Q_2$
and $Q_{7\gamma},Q_{9V}, Q_{10A}$. The generalization of (\ref{rateee}),
which includes small
$\ms^2/\mb^2$ corrections, can be found in \cite{AGM:94}.

The various entries in (\ref{rateee}) are given as follows:
\begin{eqnarray} \label{phasespace}
h(z, \hat s) & = & -\frac{8}{9}\ln\frac{\mb}{\mu} - \frac{8}{9}\ln z +
\frac{8}{27} + \frac{4}{9} x - \\
& & \frac{2}{9} (2+x) |1-x|^{1/2} \left\{
\begin{array}{ll}
\left( \ln\left| \frac{\sqrt{1-x} + 1}{\sqrt{1-x} - 1}\right| - i\pi \right),
 & \mbox{for } x \equiv 4 z^2/\hat s < 1 \nn \\
2 \arctan \frac{1}{\sqrt{x-1}}, & \mbox{for } x \equiv 4 z^2/\hat s > 1,
\end{array}
\right. \\
h(0, \hat s) & = &
\frac{8}{27} -\frac{8}{9} \ln\frac{\mb}{\mu} - \frac{4}{9} \ln
\hat s + \frac{4}{9} i\pi. \\ 
\tilde\eta(\hat s) & = & 1 + \frac{\as(\mu)}{\pi}\, \omega(\hat s)
\end{eqnarray}
with
\begin{eqnarray} \label{omega}
\omega(\hat s) & = & - \frac{2}{9} \pi^2 - \frac{4}{3}\mbox{Li}_2(s) -
\frac{2}{3}
\ln s \ln(1-s) - \frac{5+4s}{3(1+2s)}\ln(1-s) - \nn \\
& &  \frac{2 s (1+s) (1-2s)}{3(1-s)^2 (1+2s)} \ln s + \frac{5+9s-6s^2}{6
(1-s) (1+2s)}.
\end{eqnarray}
Next $f(z)$ is the phase-space factor for $b \to c e \bar\nu$ given
in (\ref{g}) and
$\kappa(z)$ is the corresponding single
gluon QCD correction given already in (\ref{kap}).
Numerical values of $f(z)$ and $\kappa(z)$ are given in (\ref{kf}).
$\tilde\eta$  represents single gluon
corrections to the matrix element of $Q_{9V}$ with $\ms = 0$ 
\cite{JK:89,Mis:94}. For consistency reasons this correction should only
multiply the leading logarithmic term in $\tilde{C}_9^{\rm NDR}$.
The values of $C_i^{(0)}$ are given in table \ref{tab:XXX}.

In the HV scheme the one-loop matrix elements are different and one
finds an additional explicit contribution to (\ref{C9eff}) given by
\cite{BuMu:94}
\begin{equation} \label{MEHV}
- \xi^{\rm HV} \frac{4}{9} \left( 3 C_1^{(0)} + C_2^{(0)} - C_3^{(0)} - 3
C_4^{(0)} \right).
\end{equation}
However, $\Ctilde_9^{\rm NDR}$ has to be replaced by $\Ctilde_9^{\rm HV}$ 
given in
(\ref{C9tilde}) and (\ref{P0HV}) and consequently $\Ctilde_9^{\rm eff}$ is the
same in both schemes.

The first term in the function $h(z, \hat s)$ in (\ref{phasespace})
represents the leading $\mu$-dependence in the matrix elements. It is
canceled by the $\mu$-dependence present in the leading logarithm in
$\tilde C_{9}$.

\subsubsection{Numerical Analysis}
         \label{sec:Heff:BXsee:nlo:num}
\begin{figure}[hbt]
\vspace{0.10in}
\centerline{
\epsfysize=7in
\rotate[r]{\epsffile{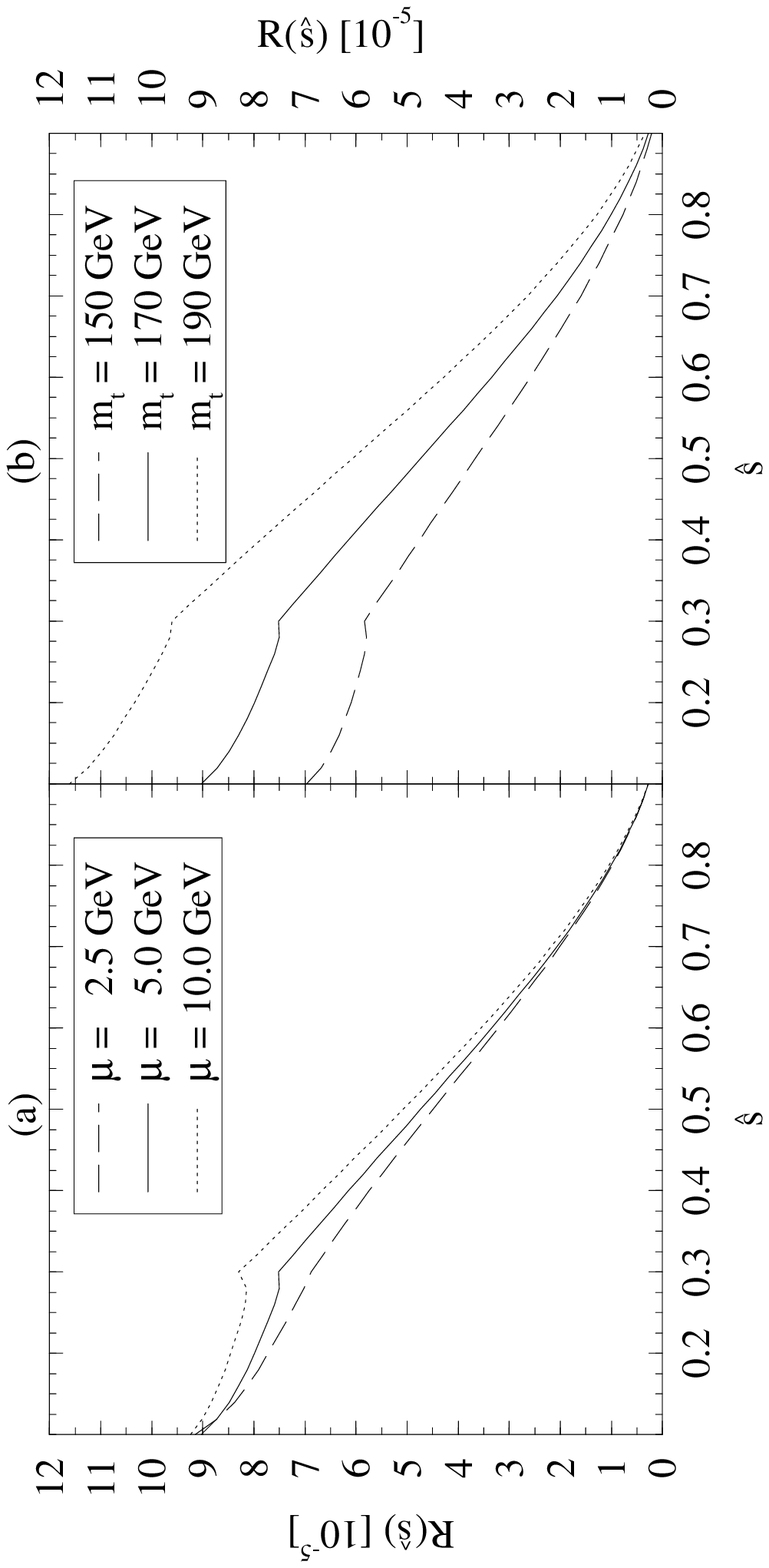}}
}
\vspace{0.08in}
\caption{(a) $R(\hat{s})$ for $\mt=170\gev$, $\Lms^{(5)}=225\mev$ 
and differents values of $\mu$. 
(b) $R(\hat{s})$ for $\mu=5\gev$, $\Lms^{(5)}=225\mev$ and various
values of $\mt$.}\label{fig:bsee:rs}
\end{figure}

\begin{figure}[hbt]
\vspace{0.10in}
\centerline{
\epsfysize=4in
\rotate[r]{\epsffile{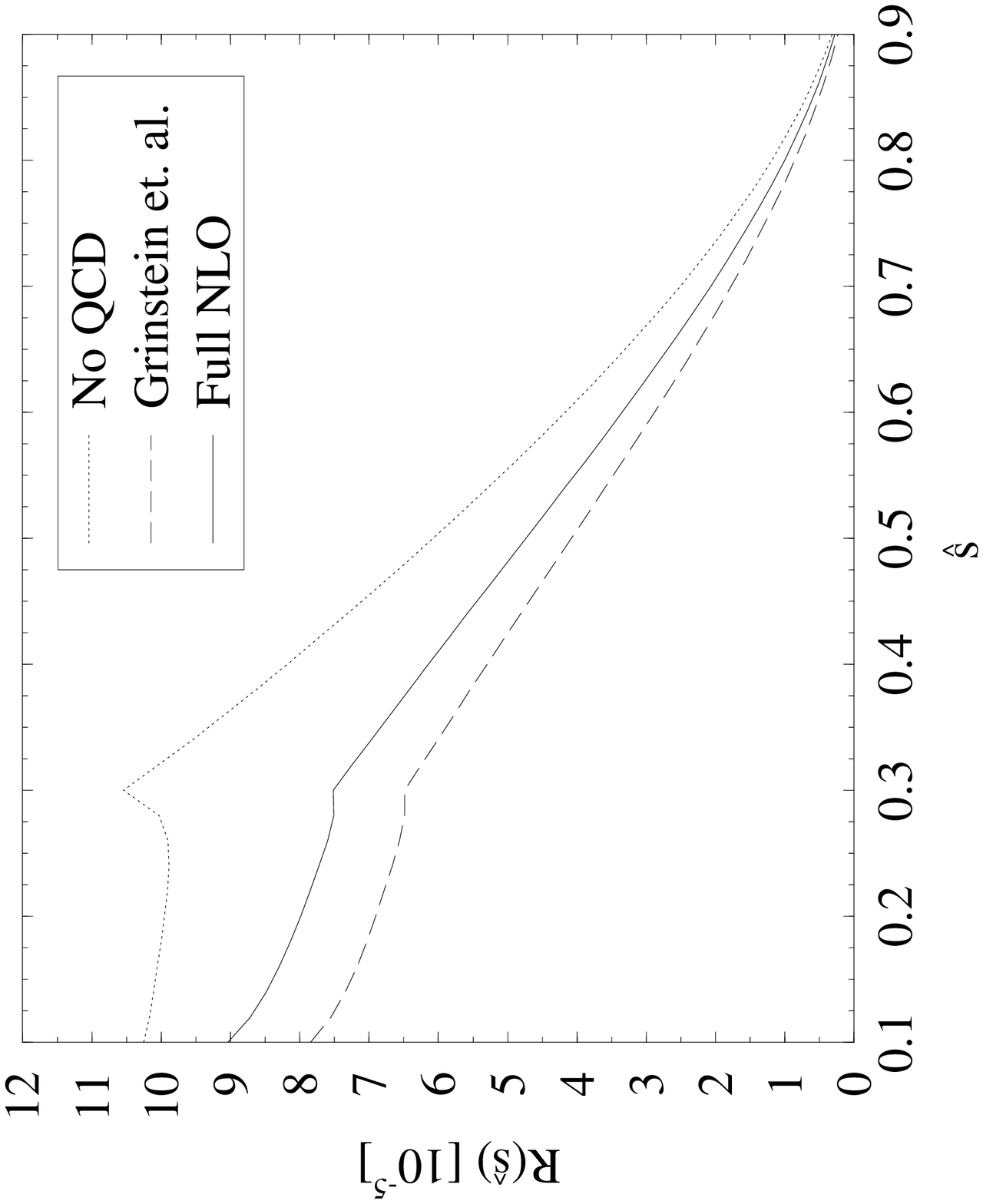}}
}
\vspace{0.08in}
\caption{$R(\hat{s})$ for $\mt=170\gev$, $\Lms^{(5)}=225\mev$ and 
$\mu=5\gev$.}\label{fig:bsee:rs2}
\end{figure}

In figs.\ \ref{fig:bsee:rs} and \ref{fig:bsee:rs2} 
we show the results of
a detailed numerical analysis of the formulae given above 
\cite{BuMu:94}. To this end we set for simplicity 
$|V_{ts}/V_{cb}|  = 1$ which in view of the results of section 4 is a good
approximation.  
In fig.\ \ref{fig:bsee:rs}\,(a) we show $R(\hat s)$ for $\mt = 170 \gev$,
$\Lms = 225 \mev$ and different values of $\mu$. In fig.\
\ref{fig:bsee:rs}\,(b) we set $\mu = 5 \gev$ and vary $\mt$ from $150 \gev$
to $190 \gev$. Finally, in fig.\ \ref{fig:bsee:rs2} we show $R(\hat s)$ 
for $\mu = 5
\gev$, $\mt = 170 \gev$ and $\Lms = 225 \mev$ compared to the case of
no QCD corrections and to the results \cite{GSW:89} would obtain
for our set of parameters using their approximate leading order
formulae. 
We observe:
\begin{itemize}
\item
The remaining $\mu$ dependence is rather weak and
amounts to at most $\pm 6\%$ in the full range of parameters
considered. This considerable
reduction in the $\mu$-dependence of the resulting branching ratio
through the inclusion of NLO corrections should be considered as
an important result. Indeed
in LO an uncertainty as large as $\pm 20\%$ can be found.
\item
The $\mt$ dependence of $R(\hat s)$ is sizeable. Varying
$\mt$ between $150\gev$ and $190\gev$ changes $R(\hat s)$ by typically
60--65\% which in this range of $\mt$ corresponds to $R(\hat s) \sim
\mt^2$. 
For 
$Br(B \to X_c e \bar{\nu}_e)=10.4\%$ the resulting
non-resonant part of the branching ratio 
can be then well approximated by
\begin{equation}
Br(B\to X_s \mu^+\mu^-)_{\rm NR}=
6.2\cdot 10^{-6} 
\left [\frac{|V_{ts}|}{|V_{cb}|} \right ]^2 
\left [\frac{(\mtb(\mt)}{170\gev} \right ]^2 \,.
\label{MUMU}
\end{equation}
It is easy to verify that this strong $\mt$ dependence
originates in the coefficient $\Ctilde_{10}$ given in (\ref{C10}).
We do not show the $\Lms$ dependence as it is very weak. Typically,
changing $\Lms^{(5)}$ from $160\mev$ to $290\mev$ decreases $R(\hat s)$ by
about 5\%. 
\item
Based on fig.\ \ref{fig:bsee:rs2} we conclude that
the NLO branching ratio turns out to
be enhanced by $10\%$ over its LO value.
\end{itemize}

\begin{figure}[hbt]
\vspace{0.10in}
\centerline{
\epsfysize=4in
\rotate[r]{\epsffile{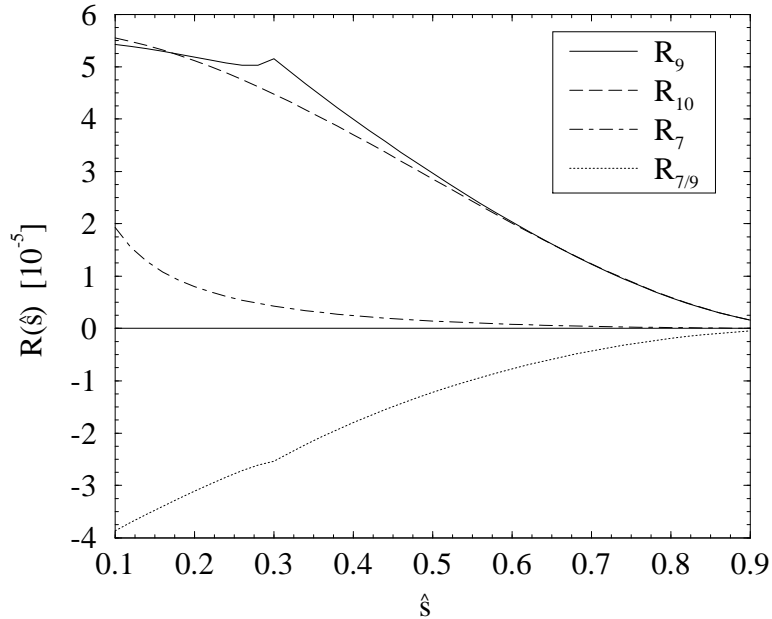}}
}
\vspace{0.08in}
\caption{Comparison of the four different contributions to $R(\hat{s})$ 
according to eq.\ \ref{rateee}.}\label{fig:bsee:rscmp}
\end{figure}

As seen in (\ref{rateee}),
$R(\hat s)$ is governed by three coefficients, $\Ctilde_9^{\rm eff}$,
$\Ctilde_{10}$ and $C_{7\gamma}^{(0){\rm eff}}$.  The importance of various
contributions is illustrated in fig.\ \ref{fig:bsee:rscmp} \cite{BuMu:94},
where
$\Lms^{(5)}=225\gev$, $\mt=170\gev$ and $\mu = 5 \gev$ has been chosen.
We show there $R(\hat s)$ keeping only
$\Ctilde_9^{\rm eff}$, $\Ctilde_{10}$, $C_{7\gamma}^{(0){\rm eff}}$ and the
$C_{7\gamma}^{(0){\rm eff}}$--$\Ctilde_9^{\rm eff}$ interference term,
respectively.  Denoting these contributions by $R_9$, $R_{10}$, $R_7$
and $R_{7/9}$ we observe that the term $R_7$ plays only a minor role in
$R(\hat s)$. On the other hand the presence of $C_{7\gamma}^{(0){\rm eff}}$
cannot be ignored because the interference term $R_{7/9}$ is
significant. In fact the presence of this large interference term could
be used to measure experimentally the relative sign of
$C_{7\gamma}^{(0){\rm eff}}$ and $\mbox{Re}\,\Ctilde_9^{\rm eff}$
\cite{GSW:89,JW:90,AMM:91,GIW:94,AGM:94}, 
which with our conventions is negative
in the Standard Model. However, the most important contributions are
$R_9$ and $R_{10}$ in the full range of $\hat s$ considered. For $\mt
\approx 170 \gev$ these two contributions are roughly of the same size.
Due to a strong $\mt$ dependence of $R_{10}$, this contribution
dominates for higher values of $\mt$ and is less important than $R_9$
for $\mt < 170 \gev$. This behaviour is again similar to the one
found in the case of $K_{\rm L} \to \pi^0 e^+ e^-$.

Finally varying the input parameters according to table \ref{tab:inputparams}
 one finds
for the non-resonant part of the branching ratio \cite{BAER}
\begin{equation}\label{MUR}
Br(B\to X_s \mu^+\mu^-)_{\rm NR}=
(5.7\pm 0.9)\cdot 10^{-6}
\end{equation}
with a similar result in \cite{ALUT96}.
This should be compared with the most recent preliminary
upper bound from 
D0 \cite{DARIA}:
\begin{equation}\label{NREXP}
Br(B\to X_s \mu^+\mu^-)_{\rm NR} <
3.6 \cdot 10^{-5}
\end{equation}
which improves the 1991 bound of UA1 \cite{UA1} by roughly a 
factor of two. It is exciting that the experimental bound is
only by a factor of five above the Standard Model expectations.
D0 should be able to measure this branching ratio during the Run II
at Tevatron.

One also finds \cite{ALUT96}:
\begin{equation}
Br(B\to X_s e^+ e^-)_{\rm NR}=
(8.4\pm 2.3)\cdot 10^{-6} 
\end{equation}
\begin{equation}
Br(B\to X_s \tau^+ \tau^-)_{\rm NR}=
(2.6\pm 0.5)\cdot 10^{-6} 
\end{equation}
with  similar results in \cite{BAER}.
\subsubsection{Long Distance Contributions}
The most recent discussions of the long-distance contributions 
to $B \to X_s l^+l^-$ can be found in \cite{Ahmady}-\cite{KS96b}
and in \cite{ALUT96}.
These contributions are due to the
$J/\psi$ and $\psi'$ resonances as well as $c\bar c$ continuum.
They affect only the coefficient $\tilde C^{\rm eff}_9$. Clearly
these contributions complicate the theoretical analysis. One
possibility as done by experimentalists is to remove the
resonance contributions from their final result in (\ref{NREXP}).
Another possibility is to leave out in the integral over the
lepton pair mass the regions dominated by $J/\psi$ and $\psi^\prime$:
say $2.9\gev-3.3\gev$ and $3.6\gev-3.8\gev$, respectively.
Finally one can include the resonances using phenomenological
models. Details on all this can be found in the papers quoted
above.

Another issue are the $1/m_b^2$ corrections. They have been
first calculated in \cite{FALK} with the result that they
increase the entire dilepton mass spectrum by typically $10\%$.
However, in this estimate the older values of the HQET parameters
$\lambda_1$ and $\lambda_2$ have been used. In particular
a positive value of $\lambda_1$ has been used. The use
of the more recent negative values of $\lambda_1$ makes
the corrections considerably smaller \cite{ALUT96,PBAER}. Moreover the
authors in \cite{ALUT96} do not confirm the formulae in \cite{FALK}
and the net effect of $1/m^2_b$-corrections found in
\cite{ALUT96} is a suppression of $Br(B \to X_sl^+l^-)$ 
by about $1.5\%$.

\subsubsection{Other Distributions and Exclusive Decays}
Clearly the calculation of $Br(B\to X_s \mu^+\mu^-)$ and of 
the invariant dilepton mass spectrum is only a small
part of the activities present in the literature in connection
with this decay. The forward-backward charge asymmetry and 
various lepton polarization asymmetries (in particular in
$B\to X_s\tau^+\tau^-$) should enable a detailed study 
of the dynamics of the Standard Model and the search for new 
physics beyond it  \cite{ASYMM,ALUT,ALIB,GLN96}. 
While the CP violation in $B \to X_s l^+l^-$ is strongly 
suppressed in the Standard Model \cite{ADIP}, CP asymmetries
of order $5\%$ are expected in $B \to X_d e^+e^-$ \cite{KSa}.
However, due to the expected branching ratio ${\cal O}(10^{-7})$ such an
analysis is a formidable task.

The Standard Model predictions for exclusive channels 
$B_d\to K^* e^+e^-$ and $B_d\to K^* \mu^+\mu^-$ amount to
$(2.3\pm0.9)\cdot 10^{-6}$ and $(1.5\pm 0.6)\cdot 10^{-6}$,
respectively \cite{ALUT}. This should be compared with the $90\%$ C.L.
upper bounds
$1.6\cdot 10^{-5}$ and $2.5\cdot 10^{-5}$ by CLEO \cite{CLEOC} 
and CDF \cite{CDFMU}, respectively.
Sensitivity of $3\cdot 10^{-7}$ should be reached for
$B_d\to K^* \mu^+\mu^-$ by CDF during Run II. The exclusive channels,
although not as clean as the inclusive ones, should also offer  
some insight in the dynamics involved \cite{BURM}.

\subsubsection{Summary}
The decays $B\to X_{s,d} l^+l^-$ remain to be an important
arena for tests of the Standard Model and of its
extensions. The discovery of the top quark, the calculations
of NLO short distance QCD corrections and the estimate of
$1/m_b^2$ corrections improved considerably the accuracy of
the expected branching ratio and of various asymmetries and
distributions.

Better estimates of long distance contributions, or equivalently
efficient methods of their removal from the experimental data,
are certainly desirable. It is exciting that the most recent
D0 upper bound on $B \to X_s \mu^+\mu^-$ is only a factor of
five away from the Standard Model expectations and that D0
as well as BABAR, BELLE, CLEO, CDF and HERA-B should be able
to measure the branching ratio and the related distributions and
asymmetries at the beginning of the next decade.

\section{ Rare $K$- and $B$-Decays}
         \label{sec:HeffRareKB}
\setcounter{equation}{0}
\subsection{General Remarks}
            \label{sec:HeffRareKB:overview}
We will now move to discuss
the semileptonic rare FCNC
transitions $\kpn$,  $K_{\rm L}\to\pi^0\nu\bar\nu$, $B\to X_{s,
d}\nu\bar\nu$, $B_{s,d}\to l^+l^-$ and $\klm$.
These decay modes are very
similar in their structure which differs considerably from the one
encountered in
the decays $K \to \pi\pi$, $K \to \pi e^+
e^-$, $B \to X_s \gamma$ and $B \to X_s \mu^+ \mu^-$ discussed in previous
sections. In particular: 

\begin{itemize}
\item
Within the Standard Model all the decays listed above are loop-induced
semileptonic FCNC processes determined by $Z^0$-penguin and box diagrams
(fig.~\ref{fig:fdia}\,(d) and (e)).\\
Thus, a distinguishing feature of the present class of decays
is the absence of a photon penguin contribution. For the decay modes
with neutrinos in the final state this is obvious, since the photon
does not couple to neutrinos. For the mesons decaying into a charged
lepton pair the photon penguin amplitude vanishes due to vector current
conservation. Consequently the decays in question are governed by the
functions $X_0(x_t)$ and $Y_0(x_t)$ (see (\ref{X0}) and (\ref{Y0}))
which as seen in (\ref{PBE1}) and (\ref{PBE2}) exhibit strong 
$\mt$ dependences.

\item
A particular and very important advantage of these decays
is their clean theoretical character.
This is related to the fact that
the low energy hadronic
matrix elements required are just the matrix elements of quark currents
between hadron states, which can be extracted from the leading
(non-rare) semileptonic decays. Other long-distance contributions
(with the exception of the decay $\klm$)
are negligibly small. As a consequence of these features,
the scale ambiguities, inherent to perturbative QCD, essentially
constitute (except for $\klm$) the only theoretical uncertainties 
present in the analysis of these decays.
These theoretical uncertainties have been considerably reduced
through the inclusion of
the next-to-leading QCD corrections 
 \cite{BB1,BB2,BB3} as we will demonstrate below. 
The decay $\klm$ receives important contributions from
the two-photon intermediate state, which are difficult to calculate
reliably. However, the short-distance part $(\klm)_{\rm SD}$ alone can
be calculated reliably.
\item
The investigation of these low energy rare decay processes in
conjunction with their theoretical cleanliness, allows to probe,
albeit indirectly, high energy scales of the theory and in particular
to measure the top quark couplings $V_{ts}$ and $V_{td}$.
Moreover $\klpn$  offers
a clean determination of the Wolfenstein parameter $\eta$ and 
as we will stress in section 11 offers the cleanest measurement
of $\IM\lambda_t= \IM V^*_{ts} V_{td}$ which governs all  
CP violating  $K$-decays. 
However, the very fact
that these processes are based on higher order electroweak effects
implies
that their branching ratios are expected to be very small and not easy to
access experimentally.
\end{itemize}

\begin{table}[htb]
\begin{center}
\begin{tabular}{|r|c|c|c|c|}
\hline
&$\kpn$&$K_{\rm L}\to\pi^0\nu\bar\nu$&$B\to X_s\nu\bar\nu$&
$B\to X_d\nu\bar\nu$\\
&$(\klm)_{\rm SD}$&&$B_s\to l^+l^-$&$B_d\to l^+l^-$\\  \hline
$\lambda_c$&$\sim\lambda$&(${\rm Im}\lambda_c\sim\lambda^5$)&
$\sim\lambda^2$&$\sim\lambda^3$\\  \hline
$\lambda_t$&$\sim\lambda^5$&(${\rm Im}\lambda_t\sim\lambda^5$)&
$\sim\lambda^2$&$\sim\lambda^3$ \\
\hline
\end{tabular}
\end{center}
\caption[]{
Order of magnitude of CKM parameters relevant for the various decays,
expressed in powers of the Wolfenstein parameter $\lambda=0.22$. In the
case of $K_{\rm L}\to\pi^0\nu\bar\nu$, which is CP-violating, only the
imaginary parts of $\lambda_{c, t}$ contribute.
\label{tab:lambdaexp}}
\end{table}

The effective Hamiltonians governing the decays
$\kpn$, $(\klm)_{\rm SD}$, $K_{\rm L}\to\pi^0\nu\bar\nu$,
$B\to X_{s, d}\nu\bar\nu$, $B\to l^+l^-$
resulting from the $Z^0$-penguin and box-type contributions can all be
written in the following general form:
\begin{equation}\label{hnr} 
{\cal H}_{\rm eff}={G_{\rm F} \over{\sqrt 2}}{\alpha\over 2\pi 
\sin^2\Theta_{\rm W}}
 \left( \lambda_c F(x_c) + \lambda_t F(x_t)\right)
 (\bar nn^\prime)_{V-A}(\bar rr)_{V-A}\,,  \end{equation}
where $n$, $n^\prime$ denote down-type quarks
($n, n^\prime=d, s, b$ but $n\not= n^\prime$) and $r$ leptons,
$r=l, \nu_l$ ($l=e, \mu, \tau$). The $\lambda_i$ are products of CKM elements,
in the general case $\lambda_i=V^*_{in}V_{in^\prime}^{}$. Furthermore
$x_i=m^2_i/M^2_W$.
The functions $F(x_i)$ describe the dependence on the internal
up-type quark masses $m_i$ (and on lepton masses if necessary)
and are understood to include QCD corrections.
They are increasing functions of the quark masses, a property that is
particularly important for the top contribution.
Since $F(x_c)/F(x_t)\approx
\ord(10^{-3})\ll 1$ the top contributions are by far dominant unless there
is a partial compensation through the CKM factors $\lambda_i$. 
 As seen in
table~\ref{tab:lambdaexp} such a partial compensation takes place in
$\kpn$ and $(\klm)_{\rm SD}$ and consequently in these decays internal
charm contributions, albeit smaller than the top contributions,
have to be kept. On the other hand in the remaining decays the
charm contributions can be safely neglected. Since the charm contributions
involve QCD corrections with $\as(m_c)$, the scale uncertainties in 
$\kpn$ and $(\klm)_{\rm SD}$ are found to be larger 
than in the remaining decays in which the QCD effects enter only
through $\as(m_t) < \as(m_c)$.

\subsection{The Decay \kpnn}
            \label{sec:HeffRareKB:kpnn}
\subsubsection{The effective Hamiltonian}
The effective Hamiltonian for $\kpn$  can
be written as
\begin{equation}\label{hkpn} 
{\cal H}_{\rm eff}={G_{\rm F} \over{\sqrt 2}}{\alpha\over 2\pi 
\sin^2\Theta_{\rm W}}
 \sum_{l=e,\mu,\tau}\left( V^{\ast}_{cs}V_{cd} X^l_{\rm NL}+
V^{\ast}_{ts}V_{td} X(x_t)\right)
 (\bar sd)_{V-A}(\bar\nu_l\nu_l)_{V-A} \, .
\end{equation}
The index $l$=$e$, $\mu$, $\tau$ denotes the lepton flavour.
The dependence on the charged lepton mass resulting from the box-graph
is negligible for the top contribution. In the charm sector this is the
case only for the electron and the muon but not for the $\tau$-lepton.

The function $X(x)$ relevant for the top part is given by

\begin{equation}\label{xx} 
X(x_t)=X_0(x_t)+\aspi X_1(x_t) 
\end{equation}
with the leading contribution $X_0(x)$ given in (\ref{X0})
and the QCD correction \cite{BB2}
\begin{eqnarray}\label{xx1}
X_1(x)=&-&{23x+5x^2-4x^3\over 3(1-x)^2}+{x-11x^2+x^3+x^4\over (1-x)^3}\ln x
\nonumber\\
&+&{8x+4x^2+x^3-x^4\over 2(1-x)^3}\ln^2 x-{4x-x^3\over (1-x)^2}L_2(1-x)
\nonumber\\
&+&8x{\partial X_0(x)\over\partial x}\ln x_\mu\,,
\end{eqnarray}
where $x_\mu=\mu^2/M^2_W$ with $\mu=\mu_t=\ord(m_t)$ and
\begin{equation}\label{l2} 
L_2(1-x)=\int^x_1 dt {\ln t\over 1-t}   \,.
\end{equation}
The $\mu$-dependence of the last term in (\ref{xx1}) cancels to the
considered order the $\mu$-dependence of the leading term $X_0(x(\mu))$.
The leftover $\mu$-dependence in $X(x_t)$ is tiny and will be given
in connection with the discussion of the branching ratio below.

The function $X$ in (\ref{xx})
can also be written as
\begin{equation}\label{xeta}
X(x)=\eta_X\cdot X_0(x), \qquad\quad \eta_X=0.985,
\end{equation}
where $\eta_X$ summarizes the NLO corrections represented by the second
term in (\ref{xx}).
With $\mt\equiv \mtb(\mt)$ the QCD factor $\eta_X$
is practically independent of $m_t$ and $\Lambda_{\overline{MS}}$.

The expression corresponding to $X(x_t)$ in the charm sector is the function
$X^l_{\rm NL}$. It results from the NLO calculation \cite{BB3} and is given
explicitly in \cite{BB3,BBL}.
The inclusion of NLO corrections reduced considerably the large
$\mu_c$ dependence
(with $\mu_c={\cal O}(m_c)$) present in the leading order expressions
for the charm contribution
 \cite{novikovetal:77,ellishagelin:83,dibetal:91,PBE0}.
Varying $\mu_c$ in the range $1\gev\le\mu_c\le 3\gev$ changes $X_{\rm NL}$
by roughly $24\%$ after the inclusion of NLO corrections to be compared
with $56\%$ in the leading order. Further details can be found in
\cite{BB3,BBL}. The impact of the $\mu_c$ uncertainties on the
resulting branching ratio $Br(\kpn)$ is discussed below.

The
numerical values for $X_{\rm NL}$ for $\mu = \mc$ and several values of
$\Lms^{(4)}$ and $\mc(\mc)$ are given in table \ref{tab:xnlnum}. 
The net effect of QCD corrections is to suppress the charm contribution
by roughly $30\%$.

\begin{table}[htb]
\begin{center}
\begin{tabular}{|c|c|c|c|c|c|c|}
\hline
& \multicolumn{3}{c|}{$X^e_{\rm NL}/10^{-4}$} &
  \multicolumn{3}{c|}{$X^\tau_{\rm NL}/10^{-4}$} \\
\hline
$\Lms^{(4)}\ [\mev]\;\backslash\;\mc\ [\gev]$ &
1.25 & 1.30 & 1.35 & 1.25 & 1.30 & 1.35 \\
\hline
245 & 10.32  & 11.17  & 12.04 & 6.94 & 7.63 & 8.36 \\
285 & 10.02  & 10.86  & 11.73 & 6.64 & 7.32 & 8.04 \\
325 &  9.71  & 10.55  & 11.41 & 6.32 & 7.01 & 7.72 \\
365 &  9.38  & 10.22  & 11.08 & 6.00 & 6.68 & 7.39 \\
405 &  9.03  &  9.87  & 10.72 & 5.65 & 6.33 & 7.04 \\
\hline
\end{tabular}
\end{center}
\caption[]{The functions $X^e_{\rm NL}$ and $X^\tau_{\rm NL}$
for various $\Lms^{(4)}$ and $\mc$.
\label{tab:xnlnum}}
\end{table}

\subsubsection{Basic Phenomenology}

We are now ready to present the expression for the branching fraction
$Br(\kpn)$ and to collect various formulae relevant for phenomenological
applications.
Since the relevant hadronic
matrix element of the weak current $(\bar sd)_{V-A}$ can be measured
in the leading decay $K^+\to\pi^0e^+\nu$, the resulting theoretical
expression for  the branching fraction $Br(K^+\to\pi^+\nu\bar\nu)$ can
be related to the experimentally well known quantity
$Br(K^+\to\pi^0e^+\nu)$ using isospin symmetry. 
Using the effective Hamiltonian (\ref{hkpn}) and summing over the three
neutrino flavours one finds
\begin{equation}\label{bkpn}
Br(\kpn)=\kappa_+\cdot\left[\left({\imlt\over\lambda^5}X(x_t)\right)^2+
\left({\relc\over\lambda}P_0(X)+{\relt\over\lambda^5}X(x_t)\right)^2
\right]
\end{equation}

\begin{equation}\label{kapp}
\kappa_+=r_{K^+}{3\alpha^2 Br(K^+\to\pi^0e^+\nu)\over 2\pi^2
\sin^4\Theta_{\rm W}}
 \lambda^8=4.11\cdot 10^{-11}\,,
\end{equation}
where we have used
\begin{equation}\label{alsinbr}
\alpha=\frac{1}{129},\qquad \sin^2\Theta_{\rm W}=0.23, \qquad
Br(K^+\to\pi^0e^+\nu)=4.82\cdot 10^{-2}\,.
\end{equation}
Here $\lambda_i=V^\ast_{is}V_{id}$ with $\lambda_c$ being
real to a very high accuracy. $r_{K^+}=0.901$ summarizes isospin
breaking corrections in relating $\kpn$ to $K^+\to\pi^0e^+\nu$.
These isospin breaking corrections are due to quark mass effects and 
electroweak radiative corrections and have been calculated in
\cite{MP}. Next
\begin{equation}\label{p0k}
P_0(X)=\frac{1}{\lambda^4}\left[\frac{2}{3} X^e_{\rm NL}+\frac{1}{3}
 X^\tau_{\rm NL}\right]
\end{equation}
with the numerical values for $X_{\rm NL}^l$ given in table \ref{tab:xnlnum}.
The corresponding values for $P_0(X)$ as a function of
$\Lambda_{\overline{MS}}$ and $m_c\equiv m_c(m_c)$ are collected in
table \ref{tab:P0Kplus}. We remark that a negligibly small term
$\sim (X^e_{\rm NL}-X^{\tau}_{\rm NL})^2 $ has been discarded in
(\ref{bkpn}).

\begin{table}[htb]
\begin{center}
\begin{tabular}{|c|c|c|c|}
\hline
&\multicolumn{3}{c|}{$P_0(X)$}\\
\hline
$\Lms^{(4)}$ $\backslash$ $m_c$ & $1.25\gev$ & $1.30\gev$ & $1.35\gev$  \\
\hline
$245\mev$ & 0.393 & 0.426 & 0.462 \\
$285\mev$ & 0.380 & 0.413 & 0.448 \\
$325\mev$ & 0.366 & 0.400 & 0.435 \\
$365\mev$ & 0.352 & 0.386 & 0.420 \\
$405\mev$ & 0.337 & 0.371 & 0.405 \\ 
\hline
\end{tabular}
\end{center}
\caption[]{The function $P_0(X)$ for various $\Lms^{(4)}$ and $m_c$.
\label{tab:P0Kplus}}
\end{table}

Using the improved Wolfenstein parametrization and the approximate
formulae (\ref{2.51}) -- (\ref{2.53}) we can next put 
(\ref{bkpn}) into a more transparent form \cite{BLO}:
\begin{equation}\label{108}
Br(K^{+} \to \pi^{+} \nu \bar\nu) = 4.11 \cdot 10^{-11} A^4 X^2(x_t)
\frac{1}{\sigma} \left[ (\sigma \bar\eta)^2 +
\left(\varrho_0 - \bar\varrho \right)^2 \right]\,,
\end{equation}
where
\begin{equation}\label{109}
\sigma = \left( \frac{1}{1- \frac{\lambda^2}{2}} \right)^2\,.
\end{equation}

The measured value of $Br(K^{+} \to \pi^{+} \nu \bar\nu)$ then
determines  an ellipse in the $(\bar\varrho,\bar\eta)$ plane  centered at
$(\varrho_0,0)$ with 
\begin{equation}\label{110}
\varrho_0 = 1 + \frac{P_0(X)}{A^2 X(x_t)}
\end{equation}
and having the squared axes
\begin{equation}\label{110a}
\bar\varrho_1^2 = r^2_0, \qquad \bar\eta_1^2 = \left( \frac{r_0}{\sigma}
\right)^2
\end{equation}
where
\begin{equation}\label{111}
r^2_0 = \frac{1}{A^4 X^2(x_t)} \left[
\frac{\sigma \cdot Br(K^{+} \to \pi^{+} \nu \bar\nu)}
{4.11 \cdot 10^{-11}} \right]\,.
\end{equation}
Note that $r_0$ depends only on the top contribution.
The departure of $\varrho_0$ from unity measures the relative importance
of the internal charm contributions.

The ellipse defined by $r_0$, $\varrho_0$ and $\sigma$ given above
intersects with the circle (\ref{2.94}).  This allows to determine
$\bar\varrho$ and $\bar\eta$  with 
\begin{equation}\label{113}
\bar\varrho = \frac{1}{1-\sigma^2} \left( \varrho_0 - \sqrt{\sigma^2
\varrho_0^2 +(1-\sigma^2)(r_0^2-\sigma^2 R_b^2)} \right), \qquad
\bar\eta = \sqrt{R_b^2 -\bar\varrho^2}
\end{equation}
and consequently
\begin{equation}\label{113aa}
R_t^2 = 1+R_b^2 - 2 \bar\varrho,
\end{equation}
where $\bar\eta$ is assumed to be positive.

In the leading order of the Wolfenstein parametrization
\begin{equation}\label{113ab}
\sigma \to 1, \qquad \bar\eta \to \eta, \qquad \bar\varrho \to \varrho
\end{equation}
and $Br(K^+ \to \pi^+ \nu \bar\nu)$ determines a circle in the
$(\varrho,\eta)$ plane centered at $(\varrho_0,0)$ and having the radius
$r_0$ of (\ref{111}) with $\sigma =1$. Formulae (\ref{113}) and
(\ref{113aa}) then simplify to \cite{BB3}
\begin{equation}\label{113a}
R_t^2 = 1 + R_b^2 + \frac{r_0^2 - R_b^2}{\varrho_0} - \varrho_0, \qquad
\varrho = \frac{1}{2} \left( \varrho_0 + \frac{R_b^2 - r_0^2}{\varrho_0}
\right).
\end{equation}
Given $\bar\varrho$ and $\bar\eta$ one can determine $V_{td}$:
\begin{equation}\label{vtdrhoeta}
V_{td}=A \lambda^3(1-\bar\varrho-i\bar\eta),\qquad
|V_{td}|=A \lambda^3 R_t.
\end{equation}
At this point a few remarks are in
order:
\begin{itemize}
\item
The long-distance contributions to $\kpn$ have been studied in
\cite{RS} and found to be
very small: a few percent of the charm contribution to the amplitude at
most, which is savely negligible.

\item
The determination of $|V_{td}|$ and of the unitarity triangle requires
the knowledge of $V_{cb}$ (or $A$) and of $|V_{ub}/V_{cb}|$. Both
values are subject to theoretical uncertainties present in the existing
analyses of tree level decays. Whereas the dependence on
$|V_{ub}/V_{cb}|$ is rather weak, the very strong dependence of
$Br(\kpn)$ on $A$ or $V_{cb}$ makes a precise prediction for this
branching ratio difficult at present. We will return to this below.
\item
The dependence of $Br(\kpn)$ on $\mt$ is also strong. However $\mt$
is known already  within $\pm 4\%$ and
consequently the related uncertainty in 
$Br(\kpn)$ is substantialy smaller than the corresponding uncertainty 
due to $V_{cb}$.
\item
Once $\varrho$ and $\eta$ are known precisely from CP asymmetries in
$B$ decays, some of the uncertainties present in (\ref{108}) related
to $|V_{ub}/V_{cb}|$ (but not to $V_{cb}$) will be removed.
\item
A very clean determination of $\sin 2\beta$ without essentially
any dependence on $m_t$ and $V_{cb}$ can be made by combining
$Br(\kpn)$ with $Br(\klpn)$ discussed below. 

\end{itemize}

\subsubsection{Numerical Analysis of \kpnn}
\label{sec:Kpnn:NumericalKp}
Let us begin the numerical analysis by  investigating the uncertainties 
in the prediction for $Br(\kpn)$ and in the determination of  $|V_{td}|$
related to the choice of the renormalization scales $\mu_t$
and $\mu_c$ in the top part and the charm part, respectively. To this end we
will fix the remaining parameters as follows:
\begin{equation}\label{mcmtnum}
\mc\equiv \mcb(\mc)=1.3\gev, \qquad \mt\equiv \mtb(\mt)=170\gev
\end{equation}
\begin{equation}\label{vcbubnum}
V_{cb}=0.040, \qquad |V_{ub}/V_{cb}|=0.08\,.
\end{equation}
In the case of $Br(\kpn)$ we need the values of both $\bar\varrho$
and $\bar\eta$. Therefore in this case we will work with
\begin{equation}\label{rhetnum}
\bar\varrho=0, \qquad\quad  \bar\eta=0.36
\end{equation}
rather than with $|V_{ub}/V_{cb}|$. Finally we will set
$\Lambda_{\overline{MS}}^{(4)}=0.325\gev$ and
$\Lambda_{\overline{MS}}^{(5)}=0.225\gev$ for the charm part and top
part, respectively.
We then vary the scales $\mu_c$ and $\mu_t$ entering $m_c(\mu_c)$
and $m_t(\mu_t)$, respectively, in the ranges
\begin{equation}\label{muctnum}
1\gev\leq\mu_c\leq 3\gev, \qquad 100\gev\leq\mu_t\leq 300\gev\,.
\end{equation}

The results of such an analysis are as follows \cite{BBL}:
The uncertainty in $Br(\kpn)$
\begin{equation}\label{varbkpnLO}
0.68\cdot 10^{-10}\leq Br(\kpn)\leq 1.08\cdot 10^{-10}
\end{equation}
present in the leading order is reduced to
\begin{equation}\label{varbkpnNLO}
0.79\cdot 10^{-10}\leq Br(\kpn)\leq 0.92\cdot 10^{-10}
\end{equation}
after including NLO corrections. 
The difference in the numerics compared to \cite{BBL} results
from $r_{K^+}=1$ used there.
Similarly one finds
\begin{equation}\label{varvtdLO}
8.24\cdot 10^{-3}\leq |V_{td}|\leq 10.97\cdot 10^{-3} \qquad {\rm LO}
\end{equation}
\begin{equation}\label{varvtdNLO}
9.23\cdot 10^{-3}\leq |V_{td}|\leq 10.10\cdot 10^{-3}  \qquad {\rm NLO}\,,
\end{equation}
where $Br(\kpn)=0.9\cdot 10^{-10}$ has been set. We observe that including
the full next-to-leading corrections reduces the uncertainty in the
determination of $|V_{td}|$ from $\pm 14\%$ (LO) to $\pm 4.6\%$ (NLO)
in the present example. The main bulk of this theoretical error stems
from the charm sector. Indeed, keeping $\mu_c=m_c$ fixed and varying
only $\mu_t$, the uncertainties in the determination of $|V_{td}|$
would shrink to $\pm 4.7\%$ (LO) and $\pm 0.6\%$ (NLO).
Similar comments apply to $Br(\kpn)$ where, as seen in
(\ref{varbkpnLO}) and (\ref{varbkpnNLO}), the theoretical uncertainty
due to $\mu_{c,t}$ is reduced from $\pm 22\%$ (LO) to $\pm 7\%$ (NLO).

Finally using the input parameters of table \ref{tab:inputparams}
 and performing two
types of error analysis one finds \cite{BJL96b}
\begin{equation}\label{kpnr}
Br(\kpn)=\left\{ \begin{array}{ll}
(9.1 \pm 3.2)\cdot 10^{-11} & {\rm Scanning} \\
(8.0 \pm 1.5) \cdot 10^{-11} & {\rm Gaussian}\,, \end{array} \right.
\end{equation}
where the error comes dominantly from the uncertainties in the CKM
parameters. 

\subsubsection{Summary and Outlook}
The accuracy of the Standard Model prediction for $Br(\kpn)$ has
improved considerably during the last five years. Indeed in the
{\it Top Quark Story} \cite{BH} a range $(5-80)\cdot 10^{-11}$
can still be found. This progress can be traced back to the
improved values of $\mt$ and $\vcb$ and to the inclusion of NLO
QCD corrections which considerably reduced the scale uncertainties
in the charm sector. It is expected \cite{BBL} that further progress
in the determination of CKM parameters via the standard analysis of
section \ref{sec:standard} could reduce 
the errors in (\ref{kpnr}) by at least a
factor of two during the next five years.

The present experimental bound on $Br(K^+\to \pi^+\nu\bar\nu)$
is \cite{Adler95}:
\begin{equation}\label{EXPkpn}
Br(\kpn)<2.4 \cdot 10^{-9}\,.
\end{equation}
This is about a factor of 25 above the Standard Model expectations
(\ref{kpnr}).
A new bound $ 2 \cdot 10^{-10}$ for 
this decay  is expected from E787 at AGS in Brookhaven in 1997.
In view of the clean character of this decay a measurement of its
branching ratio at this level would signal the presence of physics
beyond the Standard Model. The Standard Model sensitivity is
expected to be reached at AGS around the year 2000 \cite{AGS2}.
Recently also an experiment has been proposed to measure $\kpn$ 
at the Fermilab Main Injector \cite{Cooper}.

\subsection{The Decay $K_{\rm L}\to\pi^0\nu\bar\nu$}
            \label{sec:HeffRareKB:klpinn1}
\subsubsection{The effective Hamiltonian}
The effective
Hamiltonian for $K_{\rm L}\to\pi^0\nu\bar\nu$
is given as follows:
\begin{equation}\label{hxnu}
{\cal H}_{\rm eff} = {G_{\rm F}\over \sqrt 2} {\alpha \over
2\pi \sin^2 \Theta_{\rm W}} V^\ast_{ts} V_{td}
X (x_t) (\bar sd)_{V-A} (\bar\nu\nu)_{V-A} + h.c.\,,   
\end{equation}
where the function $X(x_t)$, present already in $\kpn$,
includes NLO corrections and is given in (\ref{xx}). 

Since $\klpn$  proceeds in the Standard Model almost
entirely through CP violation \cite{littenberg:89}, it
is completely dominated by short-distance loop diagrams with top quark
exchanges. The charm contribution, as we discussed above, can be fully
neglected and the theoretical uncertainties present in $\kpn$ due to
$m_c$, $\mu_c$ and $\Lambda_{\overline{MS}}$ are absent here. 
Consequently the rare decay $\klpn$ is even cleaner than $\kpn$
and is very well suited for the determination of 
the Wolfenstein parameter $\eta$ and $\imlt$.

Before going into the details it is appropriate to clarify one point
\cite{NIR96,BUCH96}. It is usually stated in the literature that the
decay $\klpn$ is dominated by {\it direct} CP violation. Now
the standard definition of the direct CP violation (see section 8
and e.g.\ (\ref{ED84})) requires the presence of strong phases which are
completely neglegible in $\klpn$. Consequently the violation of
CP symmetry in $\klpn$ arises through the interference between
$K^0-\bar K^0$ mixing and the decay amplitude. This type of CP
violation is often called {\it mixing-induced} CP violation.
However, as already pointed out by Littenberg \cite{littenberg:89},
the contribution of CP violation to $\klpn$ via $K^0-\bar K^0$ mixing 
alone is tiny. It gives $Br(\klpn) \approx 5\cdot 10^{-15}$.
Consequently, in this sence,  CP violation in $\klpn$ with
$Br(\klpn) = {\cal O}(10^{-11})$ is a manifestation of CP violation
in the decay and as such deserves the name of {\it direct} CP violation.
In other words the difference in the magnitude of CP violation in
$K_{\rm L}\to\pi\pi~(\varepsilon_K)$ and $\klpn$ is a signal of direct
CP violation and measuring $\klpn$ at the expected level would
rule out superweak scenarios. More details on this
issue can be found in \cite{NIR96,BUCH96,BB96}.

\subsubsection{Master Formulae for $Br(\klpn)$}
\label{sec:Kpnn:MasterKL}
Using the effective Hamiltonian (\ref{hxnu}) and summing over three
neutrino flavours one finds
\begin{equation}\label{bklpn}
Br(K_{\rm L}\to\pi^0\nu\bar\nu)=\kappa_{\rm L}\cdot
\left({\imlt\over\lambda^5}X(x_t)\right)^2
\end{equation}
\begin{equation}\label{kapl}
\kappa_{\rm L}=\frac{r_{K_{\rm L}}}{r_{K^+}}
 {\tau(K_{\rm L})\over\tau(K^+)}\kappa_+ =1.80\cdot 10^{-10}
\end{equation}
with $\kappa_+$ given in (\ref{kapp}) and
$r_{K_{\rm L}}=0.944$ summarizing isospin
breaking corrections in relating $\klpn$ to $K^+\to\pi^0e^+\nu$
\cite{MP}.

Using the Wolfenstein
parametrization we can rewrite (\ref{bklpn}) as
\begin{equation}\label{bklpnwol1}
Br(\klpn)=1.80\cdot 10^{-10} \eta^2 A^4 X^2(x_t)
\end{equation}
or
\begin{equation}\label{bklpnwol2}
Br(\klpn)=3.29\cdot 10^{-5} \eta^2 |V_{cb}|^4 X^2(x_t)
\end{equation}
or using 
\begin{equation}\label{xxappr}
X(x_t)=0.65\cdot x_t^{0.575}
\end{equation}

\begin{equation}
Br(K_{\rm L}\to\pi^0\nu\bar\nu)=
3.0\cdot 10^{-11}
\left [ \frac{\eta}{0.39}\right ]^2
\left [\frac{\mtb(\mt)}{170~GeV} \right ]^{2.3} 
\left [\frac{\mid V_{cb}\mid}{0.040} \right ]^4 \,.
\label{bklpn1}
\end{equation}

A few remarks are in order:
\begin{itemize}
\item
The determination of $\eta$ using $Br(\klpn)$ requires the knowledge
of $V_{cb}$ and $m_t$. The very strong dependence on $V_{cb}$ or $A$
makes a precise prediction for this branching ratio difficult at
present.
\item
It was pointed out in \cite{AJB94} that the strong
dependence of $Br(\klpn)$ on $V_{cb}$, together with the clean nature of
this decay, can be used to determine this element without any hadronic
uncertainties. To this end $\eta$ and $m_t$ have to be known with
sufficient precision in addition to $Br(\klpn)$. 
Inverting (\ref{bklpn1})
one finds
\begin{equation}\label{vcbklpn}
|V_{cb}|=40.0\cdot 10^{-3} \sqrt{\frac{0.39}{\eta}}
\left[\frac{170\gev}{\mtb(\mt)}\right]^{0.575}
\left[\frac{Br(\klpn)}{3\cdot 10^{-11}}\right]^{1/4}\,.
\end{equation}
We note that the weak dependence of $V_{cb}$ on $Br(\klpn)$ allows
to achieve a high precision for this CKM element even when $Br(\klpn)$
is known with only relatively moderate accuracy, e.g.\ 10--15\%.
A numerical analysis of (\ref{vcbklpn}) can be found in
\cite{AJB94,BB96} and will be presented in section 11.
\end{itemize}

\subsubsection{Numerical Analysis of \klpnn}
\label{sec:Kpnn:NumericalKL}
The $\mu_t$-uncertainties present in the function $X(x_t)$ have 
already been
discussed in connection with $\kpn$. After the inclusion of NLO
corrections they are so small that they can be neglected for all
practical purposes. 
At the level of $Br(\klpn)$ the ambiguity in the choice of $\mu_t$ is
reduced from $\pm 10\%$ (LO) down to $\pm 1\%$ (NLO), which
considerably increases the predictive power of the theory. Varying
$\mu_t$ according to (\ref{muctnum}) and using the input parameters
as in the case of $\kpn$ we find that the uncertainty
in $Br(\klpn)$
\begin{equation}\label{varbklpnLO}
2.53\cdot 10^{-11}\leq Br(\klpn)\leq 3.08\cdot 10^{-11}
\end{equation}
present in the leading order is reduced to
\begin{equation}\label{varbklpnNLO}
2.64\cdot 10^{-11}\leq Br(\klpn)\leq 2.72\cdot 10^{-11}
\end{equation}
after including NLO corrections. This means that the theoretical
uncertainty in the determination of $\eta$ amounts to only $\pm 0.7\%$
which is safely negligible.

Using the input parameters of table \ref{tab:inputparams}
one finds \cite{BJL96b}
\begin{equation}\label{klpnr4}
Br(\klpn)=\left\{ \begin{array}{ll}
(2.8 \pm 1.7)\cdot 10^{-11} & {\rm Scanning} \\
(2.6 \pm 0.9) \cdot 10^{-11} & {\rm Gaussian} \end{array} \right.
\end{equation}
where the error comes dominantly from the uncertainties in the CKM
parameters. 

\subsubsection{Summary and Outlook}
The accuracy of the Standard Model prediction for $Br(\klpn)$ has
improved considerably during the last five years. Indeed in the
{\it Top Quark Story} \cite{BH} values as high as $15\cdot 10^{-11}$
can be found. This progress can be traced back mainly to the
improved values of $\mt$ and $\vcb$ and to some extent to 
the inclusion of NLO QCD corrections.
It is expected \cite{BBL} that further progress
in the determination of CKM parameters via the standard analysis of
section 4 could reduce the errors in (\ref{klpnr4}) by at least a
factor of two during the next five years.

The present upper bound on $Br(K_{\rm L}\to \pi^0\nu\bar\nu)$ from
FNAL experiment E731 \cite{WEAVER} is 
\begin{equation}\label{KLD}
Br(\klpn)<5.8 \cdot 10^{-5}\,.
\end{equation}
This is about six orders of magnitude above the Standard Model expectation
(\ref{klpnr4}).

How large could $Br(\klpn)$ really be? As shown recently in \cite{NIR96}
one can easily derive by means of isospin symmetry the following 
{\it model independent} bound:
\begin{equation}
Br(\klpn) < 4.4 \cdot Br(\kpn)
\end{equation}
which through (\ref{EXPkpn})  gives
\begin{equation}\label{B108}
Br(\klpn) < 1.1 \cdot 10^{-8}
\end{equation}
This bound is much stronger than the direct experimental bound in
(\ref{KLD}).
With the upper bound $Br(\kpn)<2\cdot 10^{-10}$ expected this year
from BNL, the bound in (\ref{B108}) can be improved to $ 9 \cdot 10^{-10}$.

Now FNAL-E799 expects to reach
the accuracy ${\cal O}(10^{-8})$ and
a very interesting new experiment
at Brookhaven (BNL E926) \cite{AGS2} 
expects to reach the single event sensitivity $2\cdot 10^{-12}$
allowing a $10\%$ measurement of the expected branching ratio. 
There are furthermore plans
to measure this gold-plated  decay with comparable sensitivity
at Fermilab \cite{FNALKL} and KEK \cite{KEKKL}.

\begin{figure}[hbt]
\vspace{0.10in}
\centerline{
\epsfysize=2.7in
\epsffile{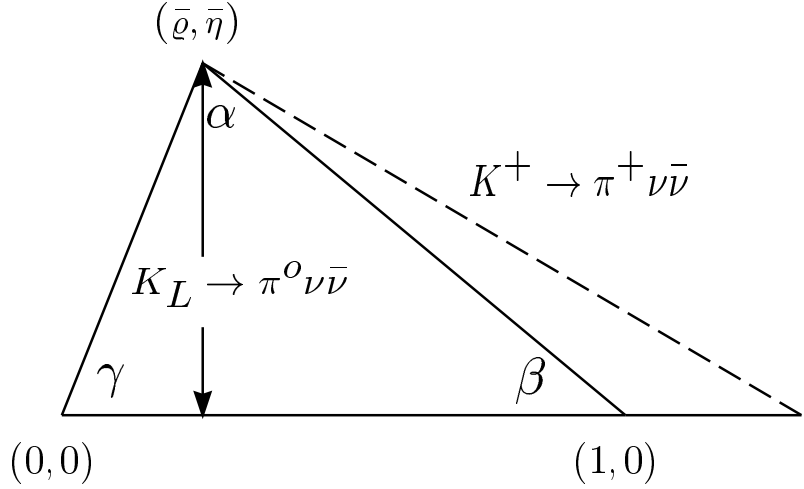}
}
\vspace{0.08in}
\caption{Unitarity triangle from $K\to\pi\nu\bar\nu$.}\label{fig:KPKL}
\end{figure}

\subsection{Unitarity Triangle and $\sin 2\beta$ from $K\to\pi\nu\bar\nu$}
\label{sec:Kpnn:Triangle}
The measurement of $Br(\kpn)$ and $Br(\klpn)$ can determine the
unitarity triangle completely, (see fig. \ref{fig:KPKL}), 
provided $\mt$ and $V_{cb}$ are known \cite{BH}.
Using these two branching ratios simultaneously allows to eliminate
$|V_{ub}/V_{cb}|$ from the analysis which removes a considerable
uncertainty. Indeed it is evident from (\ref{bkpn}) and
(\ref{bklpn}) that, given $Br(\kpn)$ and $Br(\klpn)$, one can extract
both $\imlt$ and $\relt$. One finds \cite{BB4,BBL}
\begin{equation}\label{imre}
\imlt=\lambda^5{\sqrt{B_2}\over X(x_t)}\qquad
\relt=-\lambda^5{{\relc\over\lambda}P_0(X)+\sqrt{B_1-B_2}\over X(x_t)}\,,
\end{equation}
where we have defined the ``reduced'' branching ratios
\begin{equation}\label{b1b2}
B_1={Br(\kpn)\over 4.11\cdot 10^{-11}}\qquad
B_2={Br(\klpn)\over 1.80\cdot 10^{-10}}\,.
\end{equation}
Using next the expressions for $\imlt$, $\relt$ and $\relc$ given
in (\ref{2.51}) -- (\ref{2.53}) we find
\begin{equation}\label{rhetb}
\bar\varrho=1+{P_0(X)-\sqrt{\sigma(B_1-B_2)}\over A^2 X(x_t)}\,,\qquad
\bar\eta={\sqrt{B_2}\over\sqrt{\sigma} A^2 X(x_t)}
\end{equation}
with $\sigma$ defined in (\ref{109}). An exact treatment of the CKM
matrix shows that the formulae (\ref{rhetb}) are rather precise
\cite{BB4}. The error in $\bar\eta$ is below 0.1\% and
$\bar\varrho$ may deviate from the exact expression by at most
$\Delta\bar\varrho=0.02$ with essentially negligible error for
$0\leq\bar\varrho\leq 0.25$.

Using (\ref{rhetb}) one finds subsequently \cite{BB4}
\begin{equation}\label{sin}
r_s=r_s(B_1, B_2)\equiv{1-\bar\varrho\over\bar\eta}=\cot\beta\,, \qquad
\sin 2\beta=\frac{2 r_s}{1+r^2_s}
\end{equation}
with
\begin{equation}\label{cbb}
r_s(B_1, B_2)=\sqrt{\sigma}{\sqrt{\sigma(B_1-B_2)}-P_0(X)\over\sqrt{B_2}}\,.
\end{equation}
Thus within the approximation of (\ref{rhetb}) $\sin 2\beta$ is
independent of $V_{cb}$ (or $A$) and $m_t$. An exact treatment of
the CKM matrix confirms this finding to a high accuracy. The
dependence on $V_{cb}$ and $m_t$ enters only at order
$\ord(\lambda^2)$ and as a numerical analysis shows this
dependence can be fully neglected.

It should be stressed that $\sin 2\beta$ determined this way depends
only on two measurable branching ratios and on the function
$P_0(X)$ which is completely calculable in perturbation theory.
Consequently this determination is free from any hadronic
uncertainties and its accuracy can be estimated with a high degree
of confidence.

An extensive numerical analysis of the formulae above has been presented
in \cite{BB4,BB96}. We summarize the results of the latter paper.
Assuming that the branching ratios are known to within $\pm 10\%$
\begin{equation}\label{bkpkl}
Br(\kpn)=(1.0\pm 0.1)\cdot 10^{-10}\,,\qquad
Br(\klpn)=(3.0\pm 0.30)\cdot 10^{-11}
\end{equation}
and choosing 
\begin{equation}\label{mtcv}
m_t=(170\pm 3)\gev,\quad P_0(X)=0.40\pm0.06,\quad
|V_{cb}|=0.040\pm 0.002
\end{equation}
one finds the results given in the second column of table 
\ref{tabkb1}.
In the third coulumn the results for the choice
$|V_{cb}|=0.040\pm 0.001$ are shown.
It should be remarked that the quoted errors for the input parameter
are quite reasonable if one keeps in mind
that it will take  five years to achieve the accuracy
assumed in (\ref{bkpkl}). The error in $P_0(X)$ in (\ref{mtcv}) 
results from the errors (see table \ref{tab:P0Kplus} and (\ref{muctnum})) 
in $\Lms^{(4)}$, $m_c$ and $\mu_c$ added quadratically.
Doubling the error in $m_c$ would give $P_0(X)=0.40\pm 0.09$ and
an increase of the errors in $|V_{td}|/10^{-3}$, $\bar\varrho$ and
$\sin 2\beta$ by at most $\pm 0.2$, $\pm 0.02$ and $\pm 0.01$
respectively, without any changes in $\bar\eta$ and 
${\rm Im}\lambda_t$.

\begin{table}
\begin{center}
\begin{tabular}{|c||c|c|}\hline
&$|V_{cb}|=0.040\pm 0.002$&$|V_{cb}|=0.040\pm 0.001$.\\ 
\hline
\hline
$|V_{td}|/10^{-3}$&$10.3\pm 1.1$&$10.3\pm 0.9$\\ 
\hline
$|V_{ub}/V_{cb}|$&$0.089\pm 0.017$
&$0.089\pm 0.011$ \\
\hline 
$\bar\varrho$&$-0.10\pm 0.16$ &$-0.10\pm 0.12$\\
\hline
$\bar\eta$&$0.38\pm 0.04$&$0.38\pm 0.03$\\
\hline
$\sin 2\beta$&$0.62\pm 0.05$&$0.62\pm 0.05$\\
\hline
${\rm Im}\lambda_t/10^{-4}$&$1.37\pm 0.07$
&$1.37\pm 0.07$ \\
\hline
\end{tabular}
\end{center}
\caption[]{Illustrative example of the determination of CKM
parameters from $K\to\pi\nu\bar\nu$ for two choices of
$V_{cb}$ and other parameters given in the text.
\label{tabkb1}}
\end{table}

We observe that respectable determinations of all considered 
quantities except for 
$\bar\varrho$ can be obtained.
Of particular interest are the accurate determinations of
$\sin 2\beta$ and of ${\rm Im}\lambda_t$.
The latter quantity as seen in (\ref{imre}) 
can be obtained from
$K_{\rm L}\to\pi^0\nu\bar\nu$ alone and does not require knowledge
of $V_{cb}$.

As pointed out in \cite{BB96} and discussed in section 11,
$K_{\rm L}\to\pi^0\nu\bar\nu$ appears to be the best decay to 
measure ${\rm Im}\lambda_t$; even better than the CP asymmetries
in $B$ decays discussed in the following sections.
The importance of measuring accurately  ${\rm Im}\lambda_t$ is evident.
It plays a central role in the phenomenology of CP violation
in $K$ decays and is furthermore equivalent to the 
Jarlskog parameter $J_{\rm CP}$ \cite{CJ}, 
the invariant measure of CP violation in the Standard Model, 
$J_{\rm CP}=\lambda(1-\lambda^2/2){\rm Im}\lambda_t$.

The accuracy to which $\sin 2\beta$ can be obtained from
$K\to\pi\nu\bar\nu$ is, in the  example discussed above, 
comparable to the one expected
in determining $\sin 2\beta$ from CP asymmetries in $B$ decays prior to
LHC experiments.  In this case $\sin 2\beta$ is determined best by
measuring CP violation in $B_d\to J/\psi K_{\rm S}$ 
as we will discuss in detail in the following sections.
Using the formula (\ref{Gold-int}) for the corresponding time-integrated 
CP asymmetry one finds an
interesting connection between rare $K$ decays and $B$ physics \cite{BB4}
\begin{equation}\label{kbcon}
{2 r_s(B_1,B_2)\over 1+r^2_s(B_1,B_2)}=
-a_{\mbox{{\scriptsize CP}}}(B_d\to J/\psi K_{\mbox{{\scriptsize S}}})
{1+x^2_d\over x_d}
\end{equation}
which must be satisfied in the Standard Model. We stress that except
for $P_0(X)$ given in table \ref{tab:P0Kplus} all quantities in
(\ref{kbcon}) can be directly measured in experiment and that this
relationship is essentially independent of $m_t$ and $V_{cb}$.
Due to very small theoretical uncertainties in (\ref{kbcon}), this
relation is particularly suited for tests of CP violation in the
Standard Model and offers a powerful tool to probe the physics
beyond it.
Further comparision between the potential of $K \to \pi \nu\bar\nu$ and
CP asymmetries in $B$ decays will be given in section 11.

\subsection{The Decays $B\to X_{s,d}\nu\bar\nu$}
            \label{sec:HeffRareKB:klpinn2}

\subsubsection{Effective Hamiltonian}
The decays $B\to X_{s,d}\nu\bar\nu$ are the theoretically
cleanest decays in the field of rare $B$-decays.
They are dominated by the same $Z^0$-penguin and box diagrams
involving top quark exchanges which we encountered already
in the case of $\kpn$ and $\klpn$ except for the appropriate
change of the external quark flavours. Since the change of external
quark flavours has no impact on the $m_t$ dependence,
the latter is fully described by the function $X(x_t)$ in
(\ref{xx}) which includes
the NLO corrections \cite{BB2}. The charm contribution as
discussed at the beginning of this section is fully neglegible
here and the resulting effective Hamiltonian is very similar to
the one for $\klpn$ given in (\ref{hxnu}). 
For the decay $B\to X_s\nu\bar\nu$ it reads
\begin{equation}\label{bxnu}
{\cal H}_{\rm eff} = {G_{\rm F}\over \sqrt 2} {\alpha \over
2\pi \sin^2 \Theta_{\rm W}} V^\ast_{tb} V_{ts}
X (x_t) (\bar bs)_{V-A} (\bar\nu\nu)_{V-A} + h.c.   
\end{equation}
with $s$ replaced by $d$ in the
case of $B\to X_d\nu\bar\nu$.
 
The theoretical uncertainties related to the renormalization
scale dependence are as in $\klpn$ and 
can be essentially neglected.
On the other hand $B\to X_{s,d}\nu\bar\nu$ are CP conserving and
consequently the relevant branching ratios are sensitive to 
$\vtd$ and $\vts$ as opposed to $Br(\klpn)$ in which
$\IM (V^\ast_{ts} V_{td})$ enters. As we will stress below the
measurement of both
$B\to X_{s}\nu\bar\nu$ and $B\to X_{d}\nu\bar\nu$ offers the
cleanest determination of the ratio $\vtd/\vts$.

\subsubsection{The Branching Ratios}
The calculation of the branching fractions for $B\to X_{s,d}\nu\bar\nu$ 
can be done similarly to $B\to X_s \gamma$ and $B \to X_s \mu^+\mu^-$
in the spectator model corrected for short distance QCD effects.
Normalizing as in these latter decays 
to $Br(B\to X_c e\bar\nu)$ and summing over three neutrino 
flavours one finds

\begin{equation}\label{bbxnn}
\frac{Br(B\to X_s\nu\bar\nu)}{Br(B\to X_c e\bar\nu)}=
\frac{3 \alpha^2}{4\pi^2\sin^4\Theta_{\rm W}}
\frac{|V_{ts}|^2}{|V_{cb}|^2}\frac{X^2(x_t)}{f(z)}
\frac{\bar\eta}{\kappa(z)}\,.
\end{equation}
Here $f(z)$ is the phase-space factor for $B\to X_c
e\bar\nu$ defined already in (\eqn{g}) and $\kappa(z)$ is the
corresponding QCD correction given in (\eqn{kap}). The
factor $\bar\eta$ represents the QCD correction to the matrix element
of the $b\to s\nu\bar\nu$ transition due to virtual and bremsstrahlung
contributions and is given by the well known expression
\begin{equation}\label{etabar}
\bar\eta=\kappa(0)=
1+\frac{2\alpha_s(m_b)}{3\pi}\left(\frac{25}{4}-\pi^2\right)
\approx 0.83\,.
\end{equation}
In the case of $B\to X_d\nu\bar\nu$ one has to replace $V_{ts}$ by
$V_{td}$ which results in a decrease of the branching ratio by
roughly an order of magnitude.

It should be noted that $Br(B \to X_s \nu\bar\nu)$ as given in
(\eqn{bbxnn}) is in view of $|V_{ts}/V_{cb}|^2 \approx 0.95 \pm 0.03$
essentially independent of the CKM parameters and the main uncertainty
resides in the value of $\mt$ which is already rather precisely
known. Setting $Br(B\to X_ce\bar\nu)=10.4\%$, $f(z)=0.54$,
$\kappa(z)=0.88$ and using the values in (\ref{alsinbr})
 we have
\begin{equation}
Br(B \to X_s \nu\bar\nu) = 3.7 \cdot 10^{-5} \,
\frac{|V_{ts}|^2}{|V_{cb}|^2} \,
\left[ \frac{\mtb(\mt)}{170\gev} \right]^{2.30} \, .
\label{eq:bxsnnnum}
\end{equation}

Taking next, in accordance with (\ref{kf}), $\kappa(z)=0.88$,
$f(z)=0.54\pm 0.04$ and 
$Br(B\to X_ce\bar\nu)=(10.4\pm 0.4)\%$
and using the input parameters of table \ref{tab:inputparams}
one finds \cite{BJL96b}
\begin{equation}\label{klpnr3}
Br(B \to X_s \nu\bar\nu)=\left\{ \begin{array}{ll}
(3.4 \pm 0.7)\cdot 10^{-5} & {\rm Scanning} \\
(3.2 \pm 0.4) \cdot 10^{-5} & {\rm Gaussian}\,. \end{array} \right.
\end{equation}
These values are by $10\%$ lower than the ones given in \cite{BBL}
where $f(z)=0.49$ has been used.

What about the data? 
One of the high-lights of FCNC-1996 was the upper bound:
\begin{equation}\label{124}
Br(B\to X_s \nu\bar\nu) < 7.7\cdot 10^{-4} 
\quad
(90\%\,\,\mbox{C.L.})
\end{equation}
obtained for the first time by ALEPH \cite{Aleph96}.
This is only a factor of 20 above the Standard Model expectation.
Even if the actual measurement of this decay is extremly difficult,
all efforts should be made to measure it. One should also 
make attempts to measure $Br(B\to X_d \nu\bar\nu)$. Indeed 

\begin{equation}\label{bratio}
\frac{Br(B\to X_d\nu\bar\nu)}{Br(B\to X_s\nu\bar\nu)}=
\frac{|V_{td}|^2}{|V_{ts}|^2}
\end{equation} 
offers the
cleanest direct determination of $\vtd/\vts$ as all uncertainties related
to $\mt$, $f(z)$ and $Br(B\to X_ce\bar\nu)$ cancel out.

Meanwhile the bound in 
(\ref{124}) puts some constraints on exotic physics beyond the
Standard Model \cite{Ligetti}.
Finally, we would like to mention the new $90 \%$ C.L. bounds for 
the exclusive channels: 
$Br(B_d\to K^*\nu\bar\nu)<1\cdot 10^{-3}$
and $Br(B_s\to \phi\nu\bar\nu)<5.4\cdot 10^{-3}$ from DELPHI
\cite{Delphi} which should be compared with ${\cal O}(10^{-5})$ in the
Standard Model. As usual the exclusive channels are subject to
hadronic uncertainties.

\subsection{The Decays $B_{s,d}\to l^+l^-$}
\subsubsection{The Effective Hamiltonian}
The decays $B_{s,d}\to l^+l^-$ are after $B\to X_{s,d}\nu\bar\nu$ 
the theoretically cleanest decays in the field of rare $B$-decays.
They are dominated by the $Z^0$-penguin and box diagrams
involving top quark exchanges which we encountered already
in the case of $B\to X_{s,d}\nu\bar\nu$   except that due to
charged leptons in the final state the charge flow in the
internal lepton line present in the box diagram is reversed.
This results in a different $\mt$ dependence summarized
by the function  $Y(x_t)$, the NLO generalization \cite{BB2}
of the function $Y_0(x_t)$ given in (\ref{Y0}).
The charm contributions as
discussed at the beginning of this section are fully negligible
here and the resulting effective Hamiltonian is given 
for $B_s\to l^+l^-$ as follows:

\begin{equation}\label{hyll}
{\cal H}_{\rm eff} = -{G_{\rm F}\over \sqrt 2} {\alpha \over
2\pi \sin^2 \Theta_{\rm W}} V^\ast_{tb} V_{ts}
Y (x_t) (\bar bs)_{V-A} (\bar ll)_{V-A} + h.c.   \end{equation}
with $s$ replaced by $d$ in the
case of $B_d\to l^+l^-$.

The function $Y(x)$ is given by
\begin{equation}\label{yy}
Y(x_t) = Y_0(x_t) + \aspi Y_1(x_t)\,,
\end{equation}
where $Y_0(x_t)$ can be found in (\ref{Y0})
and \cite{BB2}
\begin{eqnarray}\label{yy1}
Y_1(x) = &&{4x + 16 x^2 + 4x^3 \over 3(1-x)^2} -
           {4x - 10x^2-x^3-x^4\over (1-x)^3} \ln x\nonumber\\
         &+&{2x - 14x^2 + x^3 - x^4\over 2(1-x)^3} \ln^2 x
           + {2x + x^3\over (1-x)^2} L_2(1-x)\nonumber\\
         &+&8x {\partial Y_0(x) \over \partial x} \ln x_\mu\,.
\end{eqnarray}
The $\mu$-dependence of the last term in (\ref{yy1}) cancels to the
considered order the one of the leading term $Y_0(x(\mu))$.
The leftover $\mu$-dependence in $Y(x_t)$ is tiny and amounts to
an uncertainty of $\pm 1\%$ at the level of the branching ratio.

The function $Y(x)$ of (\ref{yy}) can also be written as
\begin{equation}\label{yeta}
Y(x)=\eta_Y\cdot Y_0(x)\,, \qquad\quad \eta_Y=1.026\pm 0.006\,,
\end{equation}
where $\eta_Y$ summarizes the NLO corrections.
With $\mt\equiv \mtb(\mt)$ this QCD factor
depends only very weakly on $m_t$. The range in (\ref{yeta})
corresponds to $150\gev\leq m_t\leq 190\gev$. The dependence on
$\Lambda_{\overline{MS}}$ can be neglected. 

\subsubsection{The Branching Ratios}
The branching ratio for $B_s\to l^+l^-$ is given by \cite{BB2}
\begin{equation}\label{bbll}
Br(B_s\to l^+l^-)=\tau(B_s)\frac{G^2_{\rm F}}{\pi}
\left(\frac{\alpha}{4\pi\sin^2\Theta_{\rm W}}\right)^2 F^2_{B_s}m^2_l m_{B_s}
\sqrt{1-4\frac{m^2_l}{m^2_{B_s}}} |V^\ast_{tb}V_{ts}|^2 Y^2(x_t)
\end{equation}
where $B_s$ denotes the flavour eigenstate $(\bar bs)$ and $F_{B_s}$ is
the corresponding decay constant. Using
(\ref{alsinbr}), (\ref{yeta}) and (\ref{PBE2}) we find in the
case of $B_s\to\mu^+\mu^-$
\begin{equation}\label{bbmmnum}
Br(B_s\to\mu^+\mu^-)=3.5\cdot 10^{-9}\left[\frac{\tau(B_s)}{1.6
\mbox{ps}}\right]
\left[\frac{F_{B_s}}{210\mev}\right]^2 
\left[\frac{|V_{ts}|}{0.040}\right]^2 
\left[\frac{\mtb(\mt)}{170\gev}\right]^{3.12}
\end{equation}

The main uncertainty in this branching ratio results from
the uncertainty in $F_{B_s}$.
Using the input parameters of table \ref{tab:inputparams}
together with $\tau(B_s)=1.6$ ps and $F_{B_s}=(210\pm 30)\mev$ 
one finds \cite{BJL96b}
\begin{equation}\label{klpnr1}
Br(B_s\to\mu^+\mu^-)=\left\{ \begin{array}{ll}
(3.6 \pm 1.9)\cdot 10^{-9} & {\rm Scanning} \\
(3.4 \pm 1.2) \cdot 10^{-9} & {\rm Gaussian.} \end{array} \right.
\end{equation}

For $B_d\to\mu^+\mu^-$ a similar formula holds with obvious
replacements of labels $(s\to d)$. Provided the decay constants
$F_{B_s}$ and $F_{B_d}$ will have been calculated reliably by
non-perturbative methods or measured in leading leptonic decays one
day, the rare processes $B_{s}\to\mu^+\mu^-$ and $B_{d}\to\mu^+\mu^-$
should offer clean determinations of $|V_{ts}|$ and $|V_{td}|$. 
In particular the ratio
\begin{equation}
\frac{Br(B_d\to\mu^+\mu^-)}{Br(B_s\to\mu^+\mu^-)}
=\frac{\tau(B_d)}{\tau(B_s)}
\frac{m_{B_d}}{m_{B_s}}
\frac{F^2_{B_d}}{F^2_{B_s}}
\frac{|V_{td}|^2}{|V_{ts}|^2}
\end{equation}
having smaller theoretical uncertainties than the separate
branching ratios should offer a useful measurement of
$\vtd/\vts$. Since $Br(B_d\to\mu^+\mu^-)= {\cal O}(10^{-10})$
this is, however, a very difficult task. For $B_s \to \tau^+\tau^-$
and $B_s\to e^+e^-$ one expects branching ratios ${\cal O}(10^{-6})$
and ${\cal O}(10^{-13})$, respectively, with the corresponding branching 
ratios for $B_d$-decays by one order of magnitude smaller.

We should also remark that in conjunction with a future measurement of 
$x_s$, the branching
ratio $Br(B_s\to \mu\bar\mu)$ could help to determine 
the non-perturbative parameter $B_{B_s}$ and consequently allow
a test of existing non-perturbative methods \cite{B95}:
\begin{equation}
 B_{B_s}=
\left[\frac{x_s}{22.1}\right]
\left[\frac{\mtb(\mt)}{170~\mbox{GeV}} \right]^{1.6} 
\left[\frac{4.2\cdot 10^{-9}}{Br(B_s\to \mu\bar\mu)} \right] \,.
\end{equation}

\subsubsection{Outlook}
What about the data?

The bounds on $B_{s,d}\to l\bar l$ are still
many orders of magnitude away from Standard Model expectations.
One has:

\begin{equation}\label{MUBOUND}
Br(B_s\to\mu^+\mu^-)\le \left\{ \begin{array}{ll}
8.4\cdot 10^{-6}~({\rm CDF}) & \cite{CDFMU} \\
 8.0\cdot 10^{-6}~({\rm D0}) & 
\cite{DARIA} \end{array} \right.
\end{equation}
and $Br(B_d\to\mu\bar\mu)< 1.6\cdot 10^{-6}$ (CDF),
where
the D0 result in (\ref{MUBOUND}) is really an upper bound 
on $(B_s+B_d)\to \mu\bar\mu$. CDF should reach in Run II the
sensitivity of $1\cdot 10^{-8}$ and $4\cdot 10^{-8}$ for
$B_d\to \mu\bar\mu$ and $B_s\to \mu\bar\mu$, respectively \cite{Lewis}.
It is hoped that these decays will be observed at
LHC-B. The experimental status of $B\to\tau^+\tau^-$ and its
usefulness in tests of the physics beyond the Standard Model
is discussed in \cite{GLN96}.

\subsection{$K_{\rm L}\to \mu\bar\mu$}
\subsubsection{General Remarks}
The rare decay $K\to\mu\bar\mu$ is CP conserving and in addition to its
short-distance part, given by Z-penguins and box diagrams, receives 
important contributions from the
two-photon intermediate state which are difficult to calculate
reliably \cite{gengng:90}-\cite{eeg}.
This latter fact is rather unfortunate because the
short-distance part is, similarly to $K\to\pi\nu\bar\nu$, free of hadronic
uncertainties and if extracted from the existing data would give a useful
determination of the Wolfenstein parameter $\varrho$.
 As we will discuss below, the separation
of the short-distance piece from the long-distance piece in the measured
rate is very difficult, however.
We will first discuss the short distance part.

\subsubsection{Effective Hamiltonian}
The analysis of the short distance part proceeds in essentially the same
manner as for $K\to \pi\nu\bar\nu$. The only difference enters through 
the lepton line in the box contribution which brings in the function
$Y(x_t)$ discussed in connection with $B_{s,d}\to l\bar l$. 
The decay $K_{\rm L}\to \mu\bar\mu$ receives also a non-negligible internal
charm contributions and consequently
the effective Hamiltonian including NLO corrections can be written as
follows \cite{BB3}:
\begin{equation}\label{hklm}
{\cal H}_{\rm eff}=-{G_{\rm F} \over{\sqrt 2}}{\alpha\over 2\pi 
\sin^2\Theta_{\rm W}}
 \left( V^{\ast}_{cs}V_{cd} Y_{\rm NL}+
V^{\ast}_{ts}V_{td} Y(x_t)\right)
 (\bar sd)_{V-A}(\bar\mu\mu)_{V-A} + h.c. 
\end{equation}
The function $Y(x)$ is given in
(\ref{yy}).
The function $Y_{\rm NL}$ representing the charm contribution,
an analogue of $X_{\rm NL}$ in the case of $\kpn$, includes
the next-to-leading QCD corrections 
calculated in \cite{BB3}. This calculation
reduced the theoretical uncertainty
due to the choice of the renormalization scales present in the
leading order expression for the branching ratio 
from $\pm 24\%$ to $\pm 10\%$. The
remaining scale uncertainty is larger than in 
$K^+ \rightarrow \pi^+ \nu \bar{\nu}$ because of a particular
feature of the perturbative expansion in the charm contribution
to this decay \cite{BB3}.
The numerical
values for $Y_{\rm NL}$ for $\mu=\mc$ and several values of $\Lms^{(4)}$
and $\mc\equiv\mcb(\mc)$ are given in table \ref{tab:ynlnum}.
Further details on the theoretical structure of $Y_{\rm NL}$ can be found
in \cite{BB3,BBL}

\begin{table}[htb]
\begin{center}
\begin{tabular}{|c|c|c|c|}\hline
&\multicolumn{3}{c|}{$Y_{\rm NL}/10^{-4}$}\\
\hline
$\Lms^{(4)} [\mev]\,/\,\mc\ [\gev]$ & 1.25 & 1.30 & 1.35 \\
\hline
245 & 3.15 & 3.36 & 3.59 \\
325 & 3.27 & 3.50 & 3.73 \\
405 & 3.37 & 3.61 & 3.85 \\
\hline
\end{tabular}
\end{center}
\caption[]{The function $Y_{\rm NL}$ for various $\Lms^{(4)}$ and $\mc$.
\label{tab:ynlnum}}
\end{table}

\subsubsection{$Br(K_{\rm L} \to \mu^+ \mu^-)_{\rm SD}$}
\label{sec:KLmm:MasterKL}
Using the effective Hamiltonian (\ref{hklm}) and relating
$\langle 0|(\bar sd)_{V-A}|K_{\rm L}\rangle$ to $Br(K^+\to\mu^+\nu)$
one finds \cite{BB3,BBL}
\begin{equation}\label{bklm}
Br(\klm)_{\rm SD}=\kappa_\mu\left[\frac{\relc}{\lambda}P_0(Y)+
\frac{\relt}{\lambda^5} Y(x_t)\right]^2
\end{equation}
\begin{equation}\label{kapm}
\kappa_\mu=\frac{\alpha^2 Br(K^+\to\mu^+\nu)}{\pi^2\sin^4\Theta_{\rm W}}
\frac{\tau(K_{\rm L})}{\tau(K^+)}\lambda^8=1.68\cdot 10^{-9}\,,
\end{equation}
where we have used
\begin{equation}\label{klmpar}
\alpha=\frac{1}{129}\,,\qquad \sin^2\Theta_{\rm W}=0.23\,,\qquad
Br(K^+\to\mu^+\nu)=0.635\,.
\end{equation}
The values of
\begin{equation}\label{p0kldef}
P_0(Y)=\frac{Y_{\rm NL}}{\lambda^4}
\end{equation}
as a function of $\Lambda^{(4)}_{\overline{MS}}$
and $m_c\equiv m_c(m_c)$ are collected in table \ref{tab:P0KL}.

\begin{table}[htb]
\begin{center}
\begin{tabular}{|c|c|c|c|}
\hline
&\multicolumn{3}{c|}{$P_0(Y) $}\\
\hline
$\Lms^{(4)}$ / $m_c$  & $1.25\gev$ & $1.30\gev$ & $1.35\gev$\\
\hline
$245\mev$ & 0.134 & 0.144 & 0.153 \\
$325\mev$ & 0.140 & 0.149 & 0.159 \\
$435\mev$ & 0.144 & 0.154 & 0.164 \\
\hline
\end{tabular}
\end{center}
\caption[]{The function $P_0(Y)$ for various $\Lms^{(4)}$ and $m_c$.
\label{tab:P0KL}}
\end{table}

Using the improved Wolfenstein parametrization and the approximate
formulae (\ref{2.51}) -- (\ref{2.53}) we can next write
\begin{equation}\label{bklmnum}
Br(\klm)_{\rm SD}=1.68\cdot 10^{-9} A^4 Y^2(x_t)\frac{1}{\sigma}
\left(\bar\varrho_0-\bar\varrho\right)^2
\end{equation}
with
\begin{equation}\label{rhosig}
\bar\varrho_0=1+\frac{P_0(Y)}{A^2 Y(x_t)}\,,\qquad
\sigma=\left(\frac{1}{1-\frac{\lambda^2}{2}}\right)^2\,.
\end{equation}
The ``experimental'' value of $Br(\klm)_{\rm SD}$ determines the value of
$\bar\varrho$ given by
\begin{equation}\label{rhor0}
\bar\varrho=\bar\varrho_0-\bar r_0\,,
\qquad\qquad
\bar r_0^2=\frac{1}{A^4 Y^2(x_t)}\left[
\frac{\sigma Br(\klm)_{\rm SD}}{1.68\cdot 10^{-9}}\right]\,.
\end{equation}
Similarly to $r_0$ in the case of $\kpn$, the value of $\bar r_0$
is fully determined by the top contribution which has only a very
weak renormalization scale ambiguity after the inclusion of
$\ord(\as)$ corrections. The main scale ambiguity resides in
$\bar\varrho_0$ whose departure from unity measures the relative
importance of the charm contribution.

Using (\ref{bklmnum}) one can find the following approximate expression
valid for $150\gev\le \mt\le 190\gev$:
\begin{equation}\label{bkmumu}
Br(K_{\rm L}\to\mu\bar\mu)_{\rm SD}=
0.9\cdot 10^{-9}\left (1.2 - \bar\varrho \right )^2
\left [\frac{\mtb(\mt)}{170~GeV} \right ]^{3.1} 
\left [\frac{\mid V_{cb}\mid}{0.040} \right ]^4 \,.
\end{equation}
In the absence of the charm contribution, ``1.2'' in the first parenthesis
would be replaced by ``1.0".

The main uncertainty in the short distance part results from
the uncertainty in $|V_{cb}|$.
Using the input parameters of table \ref{tab:inputparams}
one finds \cite{BJL96b}
\begin{equation}\label{BSD}
Br(K_{\rm L}\to\mu\bar\mu)_{\rm SD}=\left\{ \begin{array}{ll}
(1.23 \pm 0.57)\cdot 10^{-9} & {\rm Scanning} \\
(1.02 \pm 0.25) \cdot 10^{-9} & {\rm Gaussian}\,. \end{array} \right.
\end{equation}
\subsubsection{The Full Branching Ratio}

Now the full branching ratio can be written generally as follows:
\begin{equation}\label{BSDa}
Br(K_{\rm L}\to\mu\bar\mu)= |\RE A|^2+|\IM A|^2\,,\qquad
\RE A = A_{\rm SD}+A_{\rm LD}
\end{equation}
with $\RE A$ and $\IM A$ denoting the dispersive and absorptive contributions,
respectively. The absorptive contribution can be calculated using the
data for $K_{\rm L}\to \gamma\gamma$ and is known under the name of the 
unitarity bound \cite{Unitary}. 
One finds $(6.81\pm 0.32)\cdot 10^{-9}$ which is very close to the
experimental measurements

\begin{equation}
Br(K_{\rm L}\to \bar\mu\mu) =\left\{ \begin{array}{ll}
(6.86\pm0.37)\cdot 10^{-9}~({\rm BNL} 791) & \cite{PRINZ} \\
(7.9\pm 0.6 \pm 0.3)\cdot 10^{-9}~({\rm KEK} 137) & 
\cite{Akagi} \end{array} \right.
\end{equation}
which give the world average \cite{PDG}:
\begin{equation}\label{princ}
Br(K_{\rm L}\to \bar\mu\mu) = (7.2\pm 0.5)\cdot 10^{-9}\,.
\end{equation}
The accuracy of this result is impressive $(\pm 7\%)$. It will be
reduced to $(\pm 1\%)$ at BNL in the next years.

The BNL791 group using their data and the unitarity bound extracts
$|\RE A|^2\le 0.6\cdot 10^{-9}$ at $90\%$ C.L. This is a bit lower than the
short distance prediction in (\ref{BSD}). Unfortunately in order to use
this result for the determination of $\varrho$ the long distance dispersive 
part $A_{\rm LD}$ resulting from the intermediate off-shell two photon states 
should be known.
The present estimates of $A_{\rm LD}$ are too uncertain to obtain a useful
information on $\varrho$. It is believed that the measurement of
$Br(K_{\rm L}\to e\bar e \mu\bar\mu)$ should help in estimating this part.
The present result $(2.9+6.7-2.4)\cdot 10^{-9}$ from E799 
should therefore be improved.

More details on this decay can be found in 
\cite{PRINZ,BB3,CPRARE,Singer,eeg}.
More promising from theoretical point of view is the parity-violating
asymmetry in $K^+\to \pi^+\mu^+\mu^-$ \cite{GENG,BB5,BLO}.          
Finally the longitudinal
polarization in this decay is rather sensitive to contributions
beyond the Standard Model \cite{EKPICH}.

\section{CP Violation in the $B$ System}\label{CP-SEC}
\setcounter{equation}{0}
\subsection{General Remarks}
At present the observed CP-violating effects arising in the neutral
$K$-meson system can be described successfully by the Standard Model
of electroweak interactions. However, since only a single 
CP-violating observable, i.e.\ $\varepsilon$, has to be fitted, many 
different ``non-standard'' model descriptions of CP violation are imaginable. 
While a measurement of a non-vanishing $\varepsilon'/\varepsilon$ will
exclude superweak scenarios, the large hadronic uncertainties in this
ratio will not allow a stringent test of the Standard Model. More promising
in this respect is the rare decay $K_{\rm L}\to\pi^0\nu\bar\nu$. Yet it is 
clear that the $K$-system by itself cannot provide the full picture of
CP-violating phenomena and it is essential to study CP violation outside
this system. In this respect the $B$-meson system appears to be most
promising. Indeed, as we will work out in detail in the following sections, 
the $B$-meson system represents
a very fertile ground for testing the Standard Model description of CP 
violation.
Concerning such tests, the central target is again the unitarity triangle.

For the following discussion it is useful to have a parametrization of
the CKM matrix that makes the dependence on the angles of the
unitarity triangle explicit. It can be obtained from the original Wolfenstein 
parametrization (\ref{2.75}) by using (\ref{e417}) and is given by
\begin{equation}\label{wolf2}
\hat V_{\mbox{{\scriptsize CKM}}} =\left(\begin{array}{ccc}
1-\frac{1}{2}\lambda^2 & \lambda & A\lambda^3 R_b\, e^{-i\gamma} \\
-\lambda & 1-\frac{1}{2}\lambda^2 & A\lambda^2\\
A\lambda^3R_t\,e^{-i\beta} & -A\lambda^2 & 1
\end{array}\right)+\,{\cal O}(\lambda^4)
\end{equation}
with $R_b$ and $R_t$ defined in (\ref{2.94}) and (\ref{2.95}), respectively.
The $3^{\mbox{{\scriptsize rd}}}$ angle $\alpha$ of the unitarity triangle 
can be obtained
straightforwardly through the relation
\begin{equation}
\alpha+\beta+\gamma=180^\circ.
\end{equation}

As discussed in subsection~\ref{UT-Det}, at present the 
unitarity triangle can only be 
constrained indirectly through experimental
data from CP-violating effects in the neutral $K$-meson system,
$B^0_d-\overline{B^0_d}$ mixing, and from certain tree decays measuring
$|V_{cb}|$ and $|V_{ub}|/|V_{cb}|$. It should, however, be possible
to determine the three angles $\alpha$, $\beta$ and $\gamma$ of the 
unitarity triangle
independently in a {\it direct} way at future $B$ physics
facilities by measuring CP-violating effects in $B$ decays. Obviously one 
of the most exciting questions related to these measurements is whether 
the results for $\alpha$, $\beta$, $\gamma$ will be compatible with 
each other and with the results obtained from the $K$-system. Any
incompatibilities would signal ``New Physics''
beyond the Standard Model \cite{NIR96,newphys}.

\begin{figure}
\centerline{
\epsfxsize=12.5truecm
\epsffile{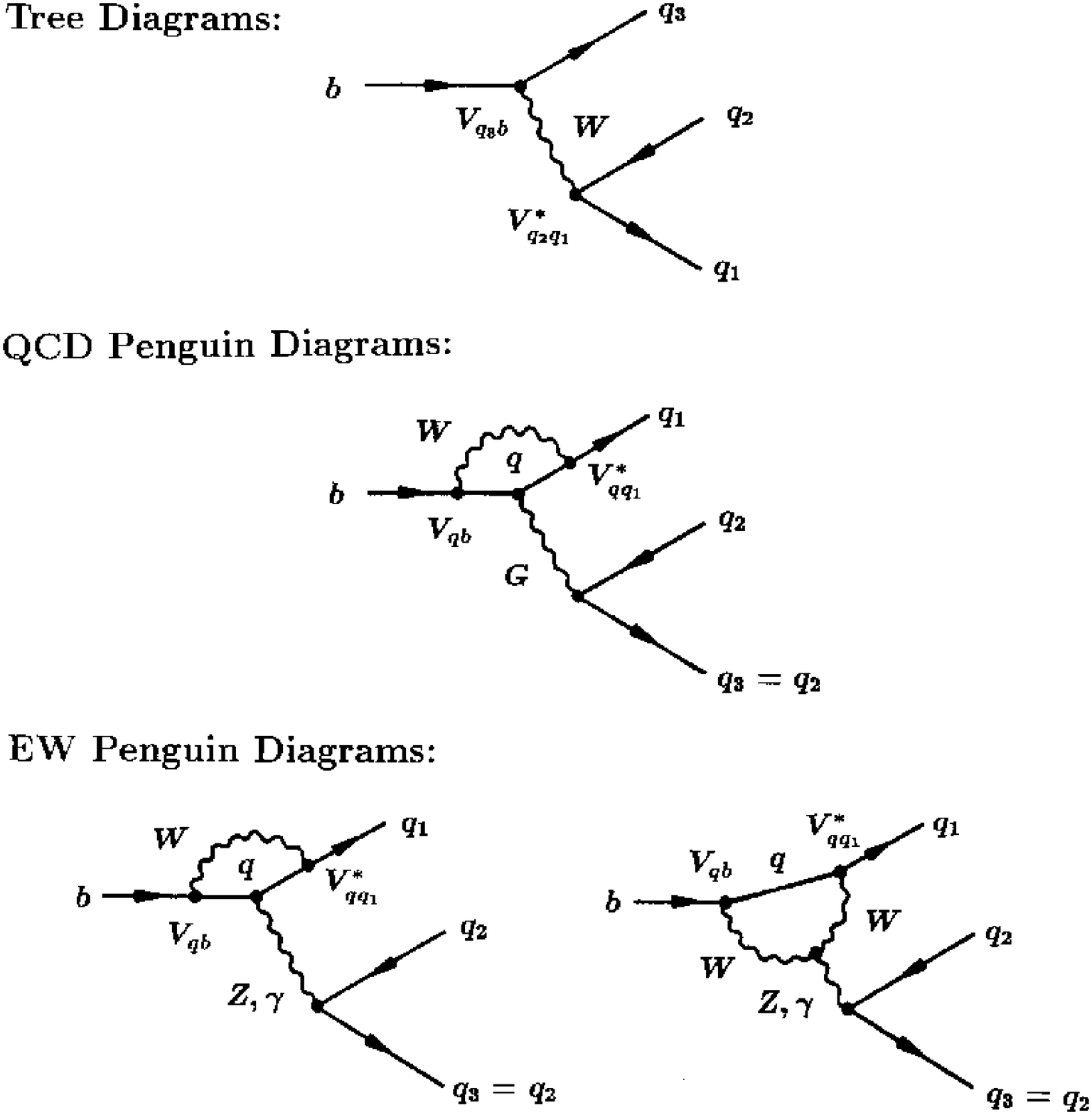}
}
\caption{Lowest order contributions to non-leptonic $b$-quark
decays $(q\in\{u,c,t\})$.}\label{feyndiags}
\end{figure}

In view of such measurements starting at the end of this millennium 
it is mandatory to search for decays
that should allow interesting insights both into the mechanism of CP
violation and into the structure of electroweak interactions in general.
Since non-leptonic $B$-meson decays play the central role in respect to
CP violation and extracting angles of the unitarity triangle, 
let us have a closer
look at these transitions in the following subsection.

\subsection{Classification of Non-leptonic $B$ Decays and Low Energy
Effective Hamiltonians}\label{Class-Ham}
Non-leptonic $B$ decays are caused by $b$-quark transitions of
the type $b\to q_1\,\overline{q}_2\,q_3$ with $q_1\in\{d,s\}$ and
$q_2,q_3\in\{u,d,c,s\}$ and can be divided into three classes:
\begin{itemize}
\item[i)]$q_2=q_3\in\{u,c\}$: both tree and penguin diagrams contribute.
\item[ii)]$q_2=q_3\in\{d,s\}$: only penguin diagrams contribute.
\item[iii)]$q_2\not=q_3\in\{u,c\}$: only tree diagrams contribute.
\end{itemize}
The corresponding lowest order Feynman diagrams are shown in
fig.~\ref{feyndiags}. As we have seen in section 2, there are two types 
of penguin topologies: {\it gluonic} (QCD) and {\it electroweak} (EW) 
penguins originating from strong and electroweak interactions, respectively. 
Such penguin diagrams play not only an important role in $K$-meson decays 
as we have seen in the previous sections but also in non-leptonic $B$ 
decays.  

\begin{table}[t]
\begin{displaymath}
\begin{array}{|c|c|c|c|c|}
\hline
\mbox{Quark-Decay}&\mbox{Exclusive Decay}&\mbox{Discussed in}&
\mbox{Probe of}&\mbox{Cleanliness}\\
\hline
\hline
b\to d\bar uu &
\begin{array}{c}
B_d\to\pi^+\pi^-\\
B_s\to\rho^0K_{\mbox{{\scriptsize S}}}
\end{array} &
\begin{array}{l}
\ref{Bdpipi-Alpha}\\
\ref{Bsrk-Gamma}
\end{array} &
\begin{array}{c}
\alpha\\
\gamma
\end{array} &
\begin{array}{c}
++\\
-
\end{array} \\
\hline
b\to d\bar cc &
\begin{array}{c}
B_d\to D^+D^-\\
B_s\to J/\psi\,K_{\mbox{{\scriptsize S}}}
\end{array} &
\begin{array}{l}
\ref{Bdpipi-Alpha}\\
\ref{gold}
\end{array} &
\begin{array}{c}
\beta\\
\lambda^2\eta
\end{array} &
\begin{array}{c}
++\\
+-
\end{array}\\
\hline
b\to s\bar uu &
\begin{array}{c}
B_{u,d}\to \pi K\\
B_s\to K^+K^-,K^{\ast+}K^{\ast-} 
\end{array} &
\begin{array}{l}
\ref{SIMalp-gam},\, \ref{SU3REL}\\
\ref{BsiLoDG},\, \ref{kkbar}
\end{array} &
\begin{array}{c}
\alpha,\, \gamma\\
\gamma
\end{array} &
\begin{array}{c}
+-\\
++
\end{array}\\
\hline
b\to s\bar cc &
\begin{array}{c}
B_d\to J/\psi\,K_{\mbox{{\scriptsize S}}}\\
B_s\to J/\psi\,\phi,D_s^{\ast+}D_s^{\ast-}
\end{array} &
\begin{array}{l}
\ref{Beta-Ext}\\
\ref{gold}
\end{array}&
\begin{array}{c}
\beta\\
\lambda^2\eta
\end{array} &
\begin{array}{c}
+++\\
++
\end{array}\\
\hline
\end{array}
\end{displaymath}
\caption{Examples for non-leptonic $B$ decays belonging to decay
class i) receiving both tree and penguin contributions.}\label{Class-i}
\end{table}

\begin{table}[t]
\begin{displaymath}
\begin{array}{|c|c|c|c|c|}
\hline
\mbox{Quark-Decay}&\mbox{Exclusive Decay}&\mbox{Discussed in}&
\mbox{Probe of}&\mbox{Cleanliness}\\
\hline
\hline
b\to d\bar ss &
\begin{array}{c}
B_d\to K^0\overline{K^0}
\end{array} &
\begin{array}{l}
\ref{Bdpipi-Alpha}
\end{array} &
\begin{array}{c}
\mbox{QCD Pen's}
\end{array} &
\begin{array}{c}
\mbox{}
\end{array} \\
\hline
b\to s\bar ss &
\begin{array}{c}
B_d\to\phi\,K_{\mbox{{\scriptsize S}}}
\end{array} &
\begin{array}{l}
\ref{Zoo}
\end{array} &
\begin{array}{c}
\beta
\end{array} &
\begin{array}{c}
++
\end{array} \\
\hline
b\to s\bar dd &
\begin{array}{c}
B_s\to K^0\overline{K^0},\,K^{\ast0}\overline{K^{\ast0}}
\end{array} &
\begin{array}{l}
\ref{BsiLoDG},\,\ref{kkbar}
\end{array} &
\begin{array}{c}
\mbox{QCD Pen's}
\end{array} &
\begin{array}{c}
\mbox{}
\end{array} \\
\hline
\end{array}
\end{displaymath}
\caption{Examples for non-leptonic $B$ decays belonging to decay
class ii) receiving only penguin contributions.}\label{Class-ii}
\end{table}

Concerning CP violation, decay classes i) and ii) are very promising.
These modes, which are usually referred to as $|\Delta B|=1$, $\Delta
C=\Delta U=0$ transitions, will hence play the major role in the 
present section. We have collected examples of exclusive decays belonging
to these categories in tables~\ref{Class-i} and \ref{Class-ii}. There
we have listed where these modes are discussed in the present section and 
which weak phases they probe. We have also given a classification of their 
theoretical cleanliness in respect of extracting these quantities. 
To analyze such transitions we shall use appropriate
low energy effective Hamiltonians calculated in renormalization group 
improved perturbation theory. In the case of $|\Delta B|=1$, $\Delta 
C=\Delta U=0$ transitions we have
\begin{equation}
{\cal H}_{\mbox{{\scriptsize eff}}}={\cal H}_{\mbox{{\scriptsize eff}}}
(\Delta B=-1)+{\cal H}_{\mbox{{\scriptsize eff}}}(\Delta B=-1)^\dagger
\end{equation}
with
\begin{equation}\label{LEham}
{\cal H}_{\mbox{{\scriptsize eff}}}(\Delta B=-1)=\frac{G_{\mbox{{\scriptsize
F}}}}{\sqrt{2}}\left[\sum\limits_{j=u,c}V_{jq}^\ast V_{jb}\left\{\sum
\limits_{k=1}^2Q_k^{jq}\,C_k(\mu)+\sum\limits_{k=3}^{10}Q_k^{q}\,C_k(\mu)
\right\}\right],
\end{equation}
where $\mu={\cal O}(m_b)$. In writing this effective Hamiltonian we
have generalized the notation of subsection \ref{SecRG} in order to
exhibit different cases. We have 
introduced two quark flavour labels $j$ and $q$ to parametrize 
$b\to j\bar j q$ quark level transitions, i.e.\ $q\in\{d,s\}$ distinguishes
between $b\to d$ and $b\to s$ transitions, respectively. These labels will
turn out to be useful for the following discussion. Consequently
\begin{itemize}
\item current-current operators:
\begin{equation}\label{cc-def}
\begin{array}{rcl}
Q_{1}^{jq}&=&(\bar{q}_{\alpha}j_{\beta})_{\mbox{{\scriptsize V--A}}}
(\bar{j}_{\beta}b_{\alpha})_{\mbox{{\scriptsize V--A}}}\\
Q_{2}^{jq}&=&(\bar{q}_\alpha j_\alpha)_{\mbox{{\scriptsize
V--A}}}(\bar{j}_\beta b_\beta)_{\mbox{{\scriptsize V--A}}}.
\end{array}
\end{equation}
\item QCD penguin operators:
\begin{equation}\label{qcd-def}
\begin{array}{rcl}
Q_{3}^q&=&(\bar{q}_\alpha b_\alpha)_{\mbox{{\scriptsize V--A}}}
\sum\limits_{q'=u,d,s,c,b}
(\bar{q}'_\beta q'_\beta)_{\mbox{{\scriptsize V--A}}}\\
Q_{4}^q&=&(\bar{q}_{\alpha}b_{\beta})_{\mbox{{\scriptsize V--A}}}
\sum\limits_{q'=u,d,s,c,b}(\bar{q}'_{\beta}q'_{\alpha})_{\mbox{{\scriptsize 
V--A}}}\\
Q_{5}^q&=&(\bar{q}_\alpha b_\alpha)_{\mbox{{\scriptsize V--A}}}
\sum\limits_{q'=u,d,s,c,b}
(\bar{q}'_\beta q'_\beta)_{\mbox{{\scriptsize V+A}}}\\
Q_{6}^q&=&(\bar{q}_{\alpha}b_{\beta})_{\mbox{{\scriptsize V--A}}}
\sum\limits_{q'=u,d,s,c,b}(\bar{q}'_{\beta}q'_{\alpha})_{\mbox{{\scriptsize 
V+A}}}.
\end{array}
\end{equation}
\item EW penguin operators:
\begin{equation}\label{ew-def}
\begin{array}{rcl}
Q_{7}^q&=&\frac{3}{2}(\bar{q}_\alpha b_\alpha)_{\mbox{{\scriptsize V--A}}}
\sum\limits_{q'=u,d,s,c,b}e_{q'}(\bar{q}'_\beta q'_\beta)_{\mbox{{\scriptsize 
V+A}}}\\
Q_{8}^q&=&\frac{3}{2}(\bar{q}_{\alpha}b_{\beta})_{\mbox{{\scriptsize V--A}}}
\sum\limits_{q'=u,d,s,c,b}e_{q'}(\bar{q}_{\beta}'
q'_{\alpha})_{\mbox{{\scriptsize V+A}}}\\
Q_{9}^q&=&\frac{3}{2}(\bar{q}_\alpha b_\alpha)_{\mbox{{\scriptsize V--A}}}
\sum\limits_{q'=u,d,s,c,b}e_{q'}(\bar{q}'_\beta q'_\beta)_{\mbox{{\scriptsize 
V--A}}}\\
Q_{10}^q&=&\frac{3}{2}(\bar{q}_{\alpha}b_{\beta})_{\mbox{{\scriptsize V--A}}}
\sum\limits_{q'=u,d,s,c,b}e_{q'}
(\bar{q}'_{\beta}q'_{\alpha})_{\mbox{{\scriptsize V--A}}}.
\end{array}
\end{equation}
\end{itemize}
Let us stress that one has to be very careful using NLO Wilson coefficient 
functions. The point is that renormalization scheme dependences arising 
in $C_k(\mu)$ beyond LO require the inclusion of certain matrix elements 
calculated at $\mu={\cal O}(m_b)$ in order to cancel this 
dependence \cite{rf1,rfewp1,kps}. Numerical values of the Wilson 
coefficient functions are given in table \ref{tab:XXX}. We note the
large value of the coefficient $C_9$. A remarkable feature of this
coefficient is its very weak renormalization scheme dependence. We will
see below that the operator $Q_9$ plays an important role in certain
non-leptonic $B$ decays \cite{rfewp1,rfewp2,rfewp3}.

\begin{table}[htb]
\begin{center}
\begin{tabular}{|c|c|c|c||c|c|c||c|c|c|}
\hline
& \multicolumn{3}{c||}{$\Lms^{(5)}=160\mev$} &
  \multicolumn{3}{c||}{$\Lms^{(5)}=225\mev$} &
  \multicolumn{3}{c| }{$\Lms^{(5)}=290\mev$} \\
\hline
Scheme & LO & NDR & HV & LO & 
NDR & HV & LO & NDR & HV \\
\hline
$C_1$ & -0.283 & -0.171 & -0.209 & -0.308 & 
-0.185 & -0.228 & -0.331 & -0.198 & -0.245 \\
$C_2$ & 1.131 & 1.075 & 1.095 & 1.144 & 
1.082 & 1.105 & 1.156 & 1.089 & 1.114 \\
\hline
$C_3$ & 0.013 & 0.013 & 0.012 & 0.014 & 
0.014 & 0.013 & 0.016 & 0.016 & 0.014 \\
$C_4$ & -0.028 & -0.033 & -0.027 & -0.030 & 
-0.035 & -0.029 & -0.032 & -0.038 & -0.032 \\
$C_5$ & 0.008 & 0.008 & 0.008 & 0.009 & 
0.009 & 0.009 & 0.009 & 0.009 & 0.010 \\
$C_6$ & -0.035 & -0.037 & -0.030 & -0.038 & 
-0.041 & -0.033 & -0.041 & -0.045 & -0.036 \\
\hline
$C_7/\aem$ & 0.043 & -0.003 & 0.006 & 0.045 & 
-0.002 & 0.005 & 0.047 & -0.002 & 0.005 \\
$C_8/\aem$ & 0.043 & 0.049 & 0.055 & 0.048 & 
0.054 & 0.060 & 0.053 & 0.059 & 0.065 \\
$C_9/\aem$ & -1.268 & -1.283 & -1.273 & -1.280 & 
-1.292 & -1.283 & -1.290 & -1.300 & -1.293 \\
$C_{10}/\aem$ & 0.302 & 0.243 & 0.245 & 0.328 & 
0.263 & 0.266 & 0.352 & 0.281 & 0.284 \\
\hline
\end{tabular}
\end{center}
\caption{$\Delta B=1$ Wilson coefficients at 
$\mu=\overline{m}_{\rm b}(\mb)=
4.40\gev$ for $\mt=170\gev$.
\label{tab:XXX}}
\end{table}

Decays belonging to class iii) allow in some cases clean extractions of 
the angle $\gamma$ of the unitarity triangle without any hadronic 
uncertainties and are 
therefore also very important. We have given examples of such modes 
in table~\ref{Class-iii}. In the case of these transitions only 
current-current operators 
contribute. The structure of the corresponding low energy effective 
Hamiltonians is completely analogous to (\ref{LEham}). We have simply to 
replace both the CKM factors $V_{jq}^\ast V_{jb}$ and the flavour contents 
of the current-current operators straightforwardly, and have
to omit the sum over penguin operators. We shall come back to the
resulting Hamiltonians in our discussion of $B_s$ decays originating from 
$\bar b\to\bar uc\bar s$ ($b\to c\bar u s$) quark-level transitions that is 
presented in \ref{nonCP}.

Whereas CP-violating asymmetries in charged $B$ decays suffer in general
from large hadronic uncertainties and are hence mainly interesting in
respect of ruling out superweak models \cite{wolfenstein:64} of CP violation,
the neutral $B_q$-meson systems $(q\in\{d,s\})$ provide excellent 
laboratories to perform stringent tests of the Standard Model 
description of CP 
violation \cite{cp-revs}. This feature is mainly due to ``mixing-induced'' 
CP violation which is absent in the charged $B$ system and arises from 
interference between decay- and $B^0_q-\overline{B^0_q}$ mixing-processes. 
We have discussed $B^0_q-\overline{B^0_q}$ mixing briefly in subsection
\ref{subsec:BBformula}. In order to derive the formulae for the
CP-violating asymmetries, we have to extend this discussion considerably.

\begin{table}
\begin{displaymath}
\begin{array}{|c|c|c|c|c|}
\hline
\mbox{Quark-Decay}&\mbox{Exclusive Decay}&\mbox{Discussed in}&
\mbox{Probe of}&\mbox{Cleanliness}\\
\hline
\hline
b\to s\left\{\begin{array}{c}\bar uc\\
\bar cu
\end{array}\right\}&
\begin{array}{c}
B_s\to D_s K,\,D\phi\\
B_{u,d}\to DK
\end{array} &
\begin{array}{l}
\ref{nonCP}\\
\ref{BDKtri}
\end{array} &
\begin{array}{c}
\gamma\\
\gamma
\end{array} &
\begin{array}{c}
+++\\
+++
\end{array}\\
\hline
\end{array}
\end{displaymath}
\caption{Examples for non-leptonic $B$ decays belonging to decay
class iii) receiving only tree contributions. The corresponding exclusive 
$b\to d$ modes are not promising in respect of CP violation and the
extraction of CKM phases since interference between the $b\to d\bar uc$ and
$b\to d\bar cu$ amplitudes is highly CKM-suppressed.}\label{Class-iii}
\end{table}

\subsection{More about $B^0_q-\overline{B^0_q}$ Mixing}\label{B0B0bar-Mix}
\begin{figure}[t]
\centerline{
\epsfysize=2.7truecm
\epsffile{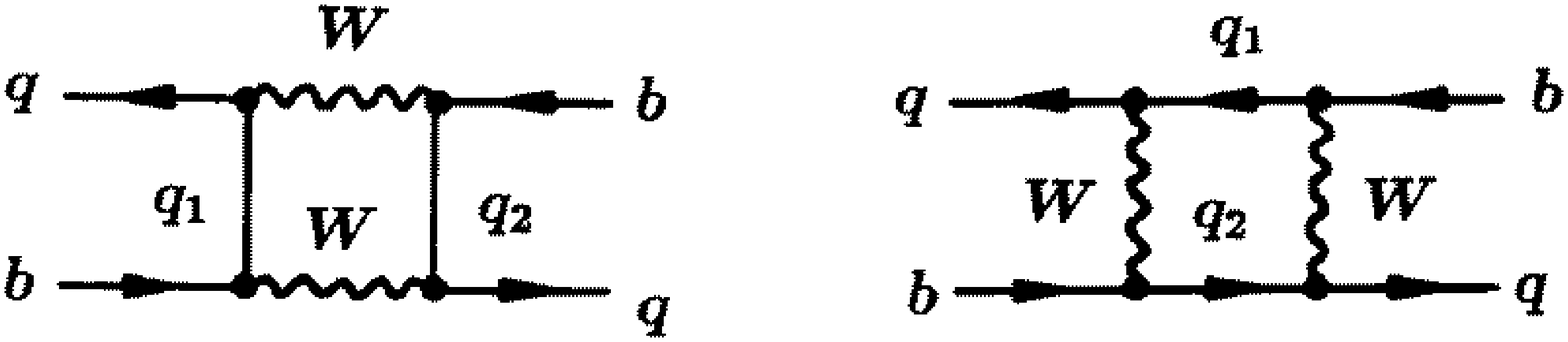}}
\caption{Box diagrams contributing to $B^0_q-\overline{B^0_q}$
mixing $(q_1,q_2\in\{u,c,t\})$.}\label{b-b.bar-mix}
\end{figure}

Within the Standard Model, $B^0_q-\overline{B^0_q}$ mixing is induced at 
lowest order
through the box diagrams shown in fig.~\ref{b-b.bar-mix}. Applying a
matrix notation, the Wigner-Weisskopf formalism \cite{wigwei} yields an
effective Schr{\"o}dinger equation of the form
\begin{equation}\label{e71}
i\,\frac{\partial}{\partial t}\left(\begin{array}{c} a(t)\\ b(t)
\end{array}
\right)=
\left[\left(\begin{array}{cc}
M_{0}^{(q)} & M_{12}^{(q)}\\ M_{12}^{(q)\ast} & M_{0}^{(q)}
\end{array}\right)-
\frac{i}{2}\left(\begin{array}{cc}
\Gamma_{0}^{(q)} & \Gamma_{12}^{(q)}\\
\Gamma_{12}^{(q)\ast} & \Gamma_{0}^{(q)}
\end{array}\right)\right]
\cdot\left(\begin{array}{c}
a(t)\\ b(t)
\end{array}
\right)
\end{equation}
describing the time evolution of the state vector
\begin{equation}\label{e72}
\left\vert \psi_q(t)\right\rangle=a(t)\left\vert B^{0}_q\right\rangle+
b(t)\left\vert\overline{B^{0}_q}\right\rangle.
\end{equation}
The special form of the mass and decay matrices in (\ref{e71})
follows from invariance under CPT transformations. It is an easy
exercise to evaluate the eigenstates $\left\vert B_{\pm}^{(q)}\right\rangle$
with eigenvalues $\lambda_{\pm}^{(q)}$ of that Hamilton operator. They are
given by
\begin{equation}\label{e73}
\left\vert B_{\pm}^{(q)} \right\rangle  =
\frac{1}{\sqrt{1+\vert \alpha_q\vert^{2}}}
\left(\left\vert B^{0}_q\right\rangle\pm\alpha_q\left\vert
\overline{B^{0}_q}\right\rangle\right)
\end{equation}
\begin{equation}\label{e74}
\lambda_{\pm}^{(q)}  =
\left(M_{0}^{(q)}-\frac{i}{2}\Gamma_{0}^{(q)}
\right)\pm
\left(M_{12}^{(q)}-\frac{i}{2}\Gamma_{12}^{(q)}\right)\alpha_q,
\end{equation}
where
\begin{equation}\label{e75}
\alpha_q  = \sqrt{\frac{4\vert M_{12}^{(q)}\vert^{2}
e^{-i2\delta\Theta_{M/\Gamma}^{(q)}}+\vert\Gamma_{12}^{(q)}\vert^{2}}
{4\vert M_{12}^{(q)}\vert^{2}+\vert\Gamma_{12}^{(q)}\vert^{2}- 4\vert
M_{12}^{(q)}\vert\vert\Gamma_{12}^{(q)}\vert\sin\delta\Theta_{M
/\Gamma}^{(q)}}}e^{-i\left(\Theta_{\Gamma_{12}}^{(q)}+n'
\pi\right)}.
\end{equation}
Here the notations $M_{12}^{(q)}\equiv e^{i\Theta_{M_{12}}^{(q)}}\vert
M_{12}^{(q)}\vert$, $\Gamma_{12}^{(q)}\equiv
e^{i\Theta_{\Gamma_{12}}^{(q)}}\vert\Gamma_{12}^{(q)}
\vert$ and
$\delta\Theta_{M/\Gamma}^{(q)}\equiv\Theta_{M_{12}}^{(q)}-
\Theta_{\Gamma_{12}}^{(q)}$ have been introduced and $n'\in {\rm Z}$
parametrizes the sign of the square root appearing in that expression.
Calculating the dispersive and absorptive parts of the box diagrams
depicted in fig.~\ref{b-b.bar-mix} one obtains \cite{buslst}
\begin{equation}\label{e711}
M_{12}^{(q)}=
\frac{G_{\mbox{{\scriptsize F}}}^{2}M_{W}^{2}\eta_B m_{B_q}
B_{B_q}F_{B_q}^{2}}{12\pi^{2}}v_{t}^{(q)2}S_0(x_{t})\,e^{i(\pi-
\phi_{\mbox{{\scriptsize CP}}}(B_q))}
\end{equation}
and
\begin{eqnarray}
\lefteqn{\Gamma_{12}^{(q)}=\frac{G_{\mbox{{\scriptsize F}}}^{2}m_{b}^{2}
m_{B_q}B_{B_q}F_{B_q}^{2}}{8\pi}
\left[v_{t}^{(q)2}+\frac{8}{3}v_{c}^{(q)}v_{t}^{(q)}
\left(z_{c}+\frac{1}{4}z_{c}^{2}-
\frac{1}{2}z_{c}^{3}\right)\right.}\label{e77}\\
&&+\left.v_{c}^{(q)2}\left\{\sqrt{1-4z_{c}}\left(1-\frac{2}{3}z_{c}\right)+
\frac{8}{3}z_{c}+\frac{2}{3}z_{c}^{2}-\frac{4}{3}z_{c}^{3}-1\right\}\right]
e^{-i\phi_{\mbox{{\scriptsize CP}}}(B_q)},\nonumber
\end{eqnarray}
respectively, where $x_t\equiv m_t^2/M_W^2$ and $z_c\equiv m_c^2/m_b^2$. As
in subsection~\ref{subsec:BBformula} we have retained only the top
contribution to $M_{12}$. The charm contribution and the mixed top-charm
contributions are entirely negligible. The non-perturbative parameter
$B_{B_q}$ has been defined in (\ref{Def-Bpar}). QCD corrections to 
(\ref{e77}), which have been omitted in that expression and are only known 
at LO \cite{buslst,HQE}, play essentially no role for the following 
discussion of CP violation. 
In order to distinguish the CKM factors present in (\ref{e711}) and 
(\ref{e77}) from $\lambda_i=V_{is}^\ast V_{id}$ used in
$K$ decays, we have introduced the notation 
\begin{equation}\label{Viq-Def}
v_i^{(q)}\equiv V_{iq}^\ast V_{ib}
\end{equation}
with $q=d,s$. Next the phase $\phi_{\mbox{{\scriptsize CP}}}(B_q)$
parametrizing the applied CP phase convention is defined through
\begin{equation}\label{e710}
({\cal CP})\left\vert B^{0}_q\right\rangle=
e^{i\phi_{\mbox{{\scriptsize CP}}}(B_q)}
\left\vert\overline{B^{0}_q}\right\rangle.
\end{equation}

Since the expression (\ref{e77}) for the off-diagonal
element $\Gamma_{12}^{(q)}$ of the decay matrix is similarly to
$M_{12}^{(q)}$ dominated by the
term proportional to $v_t^{(q)2}$, we have
\begin{equation}\label{e712}
\frac{\Gamma_{12}^{(q)}}{M_{12}^{(q)}}\approx
-\frac{3\pi}{2S_0(x_{t})}\frac{m_b^2}{M_W^2}.
\end{equation}
Therefore, $|\Gamma_{12}^{(q)}|/|M_{12}^{(q)}|={\cal O}(m_b^2/m_t^2)\ll1$.
Expanding (\ref{e75}) in powers of this small quantity gives
\begin{equation}\label{e713}
\alpha_q=\left[1+\frac{|\Gamma_{12}^{(q)}|}{2|M_{12}^{(q)}|}\sin\delta
\Theta_{M/\Gamma}^{(q)}\right]e^{-i\left(\Theta_{M_{12}}^{(q)}+n'\pi\right)}
+{\cal O}\left(\left(\frac{|\Gamma_{12}^{(q)}|}{|M_{12}^{(q)}|}
\right)^2\right).
\end{equation}
The deviation of $|\alpha_q|$ from 1 describes CP-violating effects in
$B^0_q-\overline{B^0_q}$ oscillations. This type of CP violation is
probed by rate asymmetries in semileptonic decays of neutral
$B_q$-mesons into ``wrong charge'' leptons, i.e.\ by comparing the rate of
an initially pure $B_q^0$-meson decaying into $l^-\overline{\nu}_l X$
with that of an initially pure $\overline{B^0_q}$ decaying into
$l^+\nu_l X$:
\begin{equation}\label{e714}
{\cal A}^{(q)}_{\mbox{{\scriptsize SL}}}\equiv
\frac{\Gamma(B^0_q(t)\to l^-\overline{\nu}_l X)-\Gamma(\overline{B^0_q}(t)\to
l^+\nu_l X)}{\Gamma(B^0_q(t)\to l^-\overline{\nu}_l X)+
\Gamma(\overline{B^0_q}(t)\to l^+\nu_l X)}=
\frac{|\alpha_q|^4-1}{|\alpha_q|^4+1}\approx\frac{|\Gamma_{12}^{(q)}|}
{|M_{12}^{(q)}|}\sin\delta\Theta^{(q)}_{M/\Gamma}.
\end{equation}
Note that the time dependences cancel in (\ref{e714}). Because of
$|\Gamma_{12}^{(q)}|/|M_{12}^{(q)}|\propto
m_b^2/m_t^2$ and $\sin\delta\Theta^{(q)}_{M/\Gamma}\propto
m_c^2/m_b^2$, the asymmetry (\ref{e714}) is suppressed by
a factor $m_c^2/m_t^2={\cal O}(10^{-4})$ and is hence expected to
be very small within the Standard Model. At present there exists an
experimental upper bound $|\mbox{Re}(\varepsilon_{B_d})|\equiv |{\cal
A}^{(d)}_{\mbox{{\scriptsize SL}}}/4|<45\cdot10^{-3}$ (90\% C.L.) from
the CLEO collaboration \cite{cleo} which is about two orders of
magnitudes above the Standard Model prediction.

The time-evolution of initially, i.e.\ at $t=0$, pure $\left|B^0_q\right
\rangle$ and $\left|\overline{B^0_q}\right\rangle$ meson states is given by
\begin{eqnarray}
\left|B^0_q(t)\right\rangle&=&f_+^{(q)}(t)\left|B^{0}_q\right\rangle
+\alpha_qf_-^{(q)}(t)\left|\overline{B^{0}_q}\right\rangle\label{e715}\\
\left|\overline{B^0_q}(t)\right\rangle&=&\frac{1}{\alpha_q}f_-^{(q)}(t)
\left|B^{0}_q\right\rangle+f_+^{(q)}(t)\left|\overline{B^{0}_q}\right
\rangle,\label{e716}
\end{eqnarray}
where
\begin{equation}
f_{\pm}^{(q)}(t)=\frac{1}{2}\left(e^{-i\lambda_+^{(q)}t}\pm
e^{-i\lambda_-^{(q)}t}\right).
\end{equation}
Using these time-dependent state vectors and neglecting the very
small CP-violating effects in $B^0_q-\overline{B^0_q}$ mixing that
are described by $|\alpha_q|\not=1$ (see (\ref{e713})), a straightforward
calculation yields \cite{time-evol}
\begin{eqnarray}
\Gamma(B^0_q(t)\to f)&=&\left[\left|g_+^{(q)}(t)\right|^2+\left|\xi_f^{(q)}
\right|^2\left|g_-^{(q)}(t)\right|^2-
2\mbox{\,Re}\left\{\xi_f^{(q)}g_-^{(q)}(t)g_+^{(q)}(t)^\ast\right\}
\right]\tilde\Gamma\label{ratebf}\\
\Gamma(\overline{B^0_q}(t)\to f)&=&\left[\left|g_-^{(q)}(t)\right|^2+
\left|\xi_f^{(q)}\right|^2\left|g_+^{(q)}(t)\right|^2-
2\mbox{\,Re}\left\{\xi_f^{(q)}g_+^{(q)}(t)g_-^{(q)}(t)^\ast\right\}
\right]\tilde\Gamma\quad\label{ratebbf}\\
\Gamma(B^0_q(t)\to\overline{f})&=&\left[\left|g_+^{(q)}(t)\right|^2+
\left|\xi_{\overline{f}}^{(q)}\right|^2\left|g_-^{(q)}(t)\right|^2-
2\mbox{\,Re}\left\{\xi_{\overline{f}}^{(q)}g_-^{(q)}(t)g_+^{(q)}(t)^\ast
\right\}\right]\tilde{\overline{\Gamma}}\label{ratebfb}\\
\Gamma(\overline{B^0_q}(t)\to\overline{f})&=&\left[\left|g_-^{(q)}(t)
\right|^2+\left|\xi_{\overline{f}}^{(q)}\right|^2\left|g_+^{(q)}(t)\right|^2-
2\mbox{\,Re}\left\{\xi_{\overline{f}}^{(q)}g_+^{(q)}(t)g_-^{(q)}(t)^\ast
\right\}\right]\tilde{\overline{\Gamma}},\label{ratebbfb}
\end{eqnarray}
where
\begin{equation}
\left|g^{(q)}_{\pm}(t)\right|^2=\frac{1}{4}\left[e^{-\Gamma_L^{(q)}t}+
e^{-\Gamma_H^{(q)}t}\pm2\,e^{-\Gamma_q t}\cos(\Delta M_qt)\right]
\end{equation}
\begin{equation}
g_-^{(q)}(t)\,g_+^{(q)}(t)^\ast=\frac{1}{4}\left[e^{-\Gamma_L^{(q)}t}-
e^{-\Gamma_H^{(q)}t}+2\,i\,e^{-\Gamma_q t}\sin(\Delta M_qt)\right]
\end{equation}
and
\begin{equation}\label{e731}
\xi_f^{(q)}=e^{-i\Theta_{M_{12}}^{(q)}}
\frac{A(\overline{B_q^0}\to f)}{A(B_q^0\to f)},\quad
\xi_{\overline{f}}^{(q)}=e^{-i\Theta_{M_{12}}^{(q)}}
\frac{A(\overline{B_q^0}\to \overline{f})}{A(B_q^0\to \overline{f})}.
\end{equation}
In the time-dependent rates (\ref{ratebf})-(\ref{ratebbfb}), the
time-independent transition rates $\tilde \Gamma$ and
$\tilde{\overline{\Gamma}}$ correspond to the ``unevolved'' decay
amplitudes $A(B^0_q\to f)$ and $A(B^0_q\to\overline{f})$, respectively,
and can be calculated by performing
the usual phase space integrations. The functions $g_{\pm}^{(q)}(t)$
are related to $f_{\pm}^{(q)}(t)$. However, whereas the latter functions
depend through $\alpha_q$ on the quantity $n'$ parametrizing the
sign of the square root appearing in (\ref{e75}), $g_{\pm}^{(q)}(t)$
and the rates (\ref{ratebf})-(\ref{ratebbfb}) do not depend on that
parameter. The $n'$-dependence is cancelled by introducing the
{\it positive} mass difference
\begin{equation}
\Delta M_q\equiv M_H^{(q)}-M_L^{(q)}=2\left|M_{12}^{(q)}\right|>0
\end{equation}
of the $B_q$ mass eigenstates, where $H$ and $L$ refer to ``heavy''
and ``light'', respectively. The quantities $\Gamma_H^{(q)}$ and
$\Gamma_L^{(q)}$ denote the corresponding decay widths. Their difference
can be expressed as
\begin{equation}\label{deltagamma}
\Delta\Gamma_q\equiv\Gamma_H^{(q)}-\Gamma_L^{(q)}=\frac{4\mbox{\,Re}
\left[M_{12}^{(q)}\Gamma_{12}^{(q)\ast}\right]}{\Delta M_q},
\end{equation}
while the average decay width of the $B_q$ mass eigenstates is given by
\begin{equation}
\Gamma_q\equiv\frac{\Gamma^{(q)}_H+\Gamma^{(q)}_L}{2}=\Gamma^{(q)}_0.
\end{equation}
Whereas both the mixing phase $\Theta_{M_{12}}^{(q)}$ and
the amplitude ratios appearing in (\ref{e731})
depend on the chosen CP phase convention parametrized through
$\phi_{\mbox{{\scriptsize CP}}}(B_q)$, the quantities $\xi_f^{(q)}$
and $\xi_{\overline{f}}^{(q)}$ are {\it convention independent observables}.
We shall see the cancellation of $\phi_{\mbox{{\scriptsize CP}}}(B_q)$
explicitly in a moment.

The $B^0_q-\overline{B^0_q}$ mixing phase $\Theta_{M_{12}}^{(q)}$
appearing in the equations given above is essential for the later
discussion of ``mixing-induced'' CP violation. As can be read off
from the expression (\ref{e711}) for the off-diagonal element
$M_{12}^{(q)}$ of the mass matrix, $\Theta_{M_{12}}^{(q)}$ is related
to complex phases of CKM matrix elements through
\begin{equation}\label{e717}
\Theta_{M_{12}}^{(q)}=\pi+2\,\mbox{arg}\left(V_{tq}^\ast V_{tb}\right)-
\phi_{\mbox{{\scriptsize CP}}}(B_q).
\end{equation}
Note that the perturbative QCD corrections to $B^0_q-\overline{B^0_q}$
mixing represented by $\eta_B$ in (\ref{e711}) do not affect the 
mixing phase $\Theta_{M_{12}}^{(q)}$ and have therefore no significance 
for mixing-induced CP violation.

A measure of the strength of the $B^0_q-\overline{B^0_q}$ oscillations
is provided by the ``mixing parameter''
\begin{equation}\label{e718}
x_q\equiv\frac{\Delta M_q}{\Gamma_q}.
\end{equation}
As discussed in section~\ref{sec:epsBBUT}, the present ranges for $x_d$ 
and $x_s$ can be summarized as \cite{Gibbons}
\begin{equation}\label{e719}
x_q=\left\{
\begin{array}{ll}
0.72\pm0.03 & \mbox{\quad for $q=d$}\\
\quad{\cal O}(20) & \mbox{\quad for $q=s$}
\end{array}
\right.
\end{equation}
with $x_s={\cal O}(20)$ being the Standard Model expectation. 

The mixing parameters listed in (\ref{e719}) have interesting
phenomenological consequences for the width differences
$\Delta\Gamma_{d,s}$ defined by (\ref{deltagamma}). Using this
expression we obtain
\begin{equation}\label{dgam}
\frac{\Delta\Gamma_q}{\Gamma_q}\approx-\frac{3\pi}{2S(x_t)}\frac{m_b^2}
{M_W^2}\,x_q.
\end{equation}
Consequently $\Delta\Gamma_q$ is negative so that the decay width
$\Gamma_H^{(q)}$ of the ``heavy'' mixing eigenstate is smaller than that
of the ``light'' eigenstate. Since the numerical factor in
(\ref{dgam}) multiplying the mixing parameter $x_q$ is ${\cal O}(10^{-2})$,
the width difference $\Delta\Gamma_d$ is very small within the Standard Model.
On the other hand, the expected large value of $x_s$ implies a sizable
$\Delta\Gamma_s$ which may be as large as ${\cal O}(20\%)$. The dynamical
origin of this width difference is related to CKM favored $\bar b\to\bar
cc\bar s$ quark-level transitions into final states that are common 
to $B^0_s$ and $\overline{B^0_s}$ mesons. Theoretical analyses of
$\Delta\Gamma_s/\Gamma_s$ indicate that it may indeed be as large as
${\cal O}(20\%)$. These studies are based on box diagram
calculations \cite{boxes}, on a complementary approach where one sums over 
many exclusive $\bar b\to\bar cc\bar s$ modes \cite{excl}, and
on the Heavy Quark Expansion yielding the most recent result \cite{HQE}
\begin{equation}
\frac{\Delta\Gamma_s}{\Gamma_s}=0.16^{+0.11}_{-0.09}\,.
\end{equation}
This width difference can be determined experimentally e.g.\ from angular 
correlations in $B_s\to J/\psi\,\phi$ decays \cite{ddlr}. One expects 
$10^3-10^4$ reconstructed
$B_s\to J/\psi\,\phi$ events both at Tevatron Run II and at HERA-B which
may allow a precise measurement of $\Delta\Gamma_s$. As was pointed out
by Dunietz \cite{dunietz}, $\Delta\Gamma_s$ may lead to interesting
CP-violating effects in {\it untagged} data samples of time-evolved
$B_s$ decays where one does not distinguish between initially
present $B^0_s$ and $\overline{B^0_s}$ mesons. Before we
shall turn to detailed discussions of CP-violating asymmetries in the
$B_d$ system and of the $B_s$ system in light of $\Delta\Gamma_s$, let us
focus on $B_q$ decays ($q\in\{d,s\}$) into final CP eigenstates first. For
an analysis of $B_d$ transitions into non CP eigenstates the reader is 
referred to \cite{non-CP}, $B_s$ decays of this kind will be discussed in
detail in \ref{nonCP}.

\subsection{$B_q$ Decays into CP Eigenstates}\label{CP-Eigen}
A very promising special case in respect of extracting CKM phases from
CP-violating effects in neutral $B_q$ decays are transitions into final
states $|f\rangle$ that are eigenstates of the CP operator and hence
satisfy 
\begin{equation}\label{e733a}
({\cal CP})|f\rangle=\pm|f\rangle. 
\end{equation}
Consequently we have $\xi_f^{(q)}=\xi_{\overline{f}}^{(q)}$ in that case
(see (\ref{e731})) and have to deal only with a single observable 
$\xi_f^{(q)}$ containing essentially all the information that is needed
to evaluate the time-dependent decay rates (\ref{ratebf})-(\ref{ratebbfb}).
Decays into final states that are not eigenstates of the CP operator
play an important role in the case of the $B_s$ system to extract the UT
angle $\gamma$ and are discussed in \ref{nonCP}. 

\subsubsection{Calculation of $\xi_f^{(q)}$}
Whereas the $B^0_q-\overline{B^0_q}$ mixing phase $\Theta_{M_{12}}^{(q)}$
entering the expression (\ref{e731}) for $\xi_f^{(q)}$ is simply given 
as a function of complex phases of certain CKM matrix elements 
(see (\ref{e717})), 
the amplitude ratio $A(\overline{B^0_q}\to f)/A(B^0_q\to f)$ requires the 
calculation of hadronic matrix elements which are poorly known at present. 
In order to investigate this amplitude ratio, we shall employ the low
energy effective Hamiltonian for $|\Delta B|=1$, $\Delta C=\Delta U=0$
transitions discussed in section~\ref{Class-Ham}. Using (\ref{LEham}) we 
get
\begin{eqnarray}
\lefteqn{A\left(\overline{B^0_q}\to f\right)=\Bigl\langle f\Bigl\vert
{\cal H}_{\mbox{{\scriptsize eff}}}(\Delta B=-1)\Bigr\vert\overline{B^0_q}
\Bigr\rangle}\label{e735a}\\
&&=\Biggl\langle f\left|
\frac{G_{\mbox{{\scriptsize F}}}}{\sqrt{2}}\left[
\sum\limits_{j=u,c}V_{jr}^\ast V_{jb}\left\{\sum\limits_{k=1}^2
Q_{k}^{jr}(\mu)C_{k}(\mu)
+\sum\limits_{k=3}^{10}Q_{k}^r(\mu)C_{k}(\mu)\right\}\right]\right|
\overline{B^0_q}\Biggr\rangle,\nonumber
\end{eqnarray}
where the flavour label $r\in\{d,s\}$ distinguishes -- as in the whole
subsection -- between $b\to d$ and $b\to s$ transitions. 
On the other hand, the transition amplitude $A\left(B^0_q\to f\right)$ 
is given by
\begin{eqnarray}
\lefteqn{A\left(B^0_q\to f\right)=\left\langle f\left|
{\cal H}_{\mbox{{\scriptsize 
eff}}}(\Delta B=-1)^\dagger\right|B^0_q\right\rangle}\label{e736a}\\
&&=\Biggl\langle f\left|\frac{G_{\mbox{{\scriptsize F}}}}{\sqrt{2}}
\left[\sum\limits_{j=u,c}V_{jr}V_{jb}^\ast \left\{\sum\limits_{k=1}^2
Q_{k}^{jr\dagger}(\mu)C_{k}(\mu)+\sum\limits_{k=3}^{10}
Q_k^{r\dagger}(\mu)C_{k}(\mu)\right\}\right]\right|B^0_q
\Biggr\rangle.\nonumber
\end{eqnarray}
Performing appropriate CP transformations in this equation, i.e.\
inserting the operator $({\cal CP})^\dagger({\cal CP})=\hat 1$ both
after the bra $\langle f|$ and in front of the ket $|B^0_q\rangle$,
yields
\begin{eqnarray}
\lefteqn{A\left(B^0_q\to f\right)=\pm e^{i\phi_{\mbox{{\scriptsize CP}}}
(B_q)}}\label{e737}\\
&&\times\Biggl\langle f\left|
\frac{G_{\mbox{{\scriptsize F}}}}{\sqrt{2}}\left[\sum\limits_{j=u,c}
V_{jr}V_{jb}^\ast\left\{\sum\limits_{k=1}^2
Q_{k}^{jr}(\mu)C_{k}(\mu)+\sum\limits_{k=3}^{10}
Q_{k}^r(\mu)C_{k}(\mu)\right\}\right]\right|\overline{B^0_q}
\Biggr\rangle,\nonumber
\end{eqnarray}
where we have applied the relation
\begin{equation}\label{e738}
({\cal CP})Q_k^{jr\dagger}({\cal CP})^\dagger=Q_k^{jr}
\end{equation}
and have furthermore taken into account (\ref{e710}) and
(\ref{e733a}). Consequently we obtain
\begin{equation}\label{e739}
\frac{A(\overline{B^0_q}\to f)}{A(B^0_q\to f)}=\pm\,
e^{-i\phi_{\mbox{{\scriptsize CP}}}(B_q)}\,\frac{\sum\limits_{j=u,c} 
v_j^{(r)}\Bigl\langle f\Bigl|{\cal Q}^{jr}\Bigr|\overline{B^0_q}
\Bigr\rangle}{\sum
\limits_{j=u,c}v_j^{(r)\ast}\Bigl\langle f\Bigl|{\cal Q}^{jr}\Bigr|
\overline{B^0_q}\Bigr\rangle},
\end{equation}
where $v_j^{(r)}\equiv V_{jr}^\ast V_{jb}$ and the operators 
${\cal Q}^{jr}$ are defined by 
\begin{equation}\label{e740}
{\cal Q}^{jr}\equiv
\sum\limits_{k=1}^2Q_k^{jr}C_k(\mu)+\sum\limits_{k=3}^{10}Q_k^rC_k(\mu).
\end{equation}
Inserting (\ref{e717}) and (\ref{e739}) into the expression
(\ref{e731}) for $\xi_f^{(q)}$, we observe explicitly that the convention
dependent phases $\phi_{\mbox{{\scriptsize CP}}}(B_q)$ appearing in the 
former two equations cancel each other and arrive at 
the {\it convention independent} result
\begin{equation}\label{e741}
\xi_f^{(q)}=\mp\,e^{-i\phi_{\mbox{{\scriptsize M}}}^{(q)}}
\frac{\sum\limits_{j=u,c} v_j^{(r)}
\Bigl\langle f\Bigl|{\cal Q}^{jr}\Bigr|\overline{B^0_q}\Bigr\rangle}{\sum
\limits_{j=u,c}v_j^{(r)\ast}\Bigl\langle f\Bigl|{\cal Q}^{jr}\Bigr|
\overline{B^0_q}\Bigr\rangle}.
\end{equation}
Here the phase $\phi_{\mbox{{\scriptsize M}}}^{(q)}\equiv2\,
\mbox{arg}(V_{tq}^\ast V_{tb})$ arises from the 
$B^0_q-\overline{B^0_q}$ mixing phase $\Theta_{M_{12}}^{(q)}$. 
Applying the modified Wolfenstein 
parametrization (\ref{wolf2}), $\phi_{\mbox{{\scriptsize M}}}^{(q)}$ can be 
related to angles of the unitarity triangle as follows:
\begin{equation}\label{e742}
\phi_{\mbox{{\scriptsize M}}}^{(q)}=\left\{\begin{array}{cl}
2\beta & \mbox{for $q=d$}\\
0 & \mbox{for $q=s$.}
\end{array}\right.
\end{equation}
Consequently a non-trivial mixing phase arises only in the $B_d$ system.

In general the observable $\xi_f^{(q)}$ suffers from large hadronic 
uncertainties that are introduced through the hadronic matrix elements 
appearing in (\ref{e741}). However, there is a very important special case 
where these uncertainties cancel and theoretical clean predictions of 
$\xi_f^{(q)}$ are possible.

\subsubsection{Dominance of a Single CKM Amplitude}\label{domCKM}
If the transition matrix elements appearing in (\ref{e741}) are
dominated by a single CKM amplitude, the observable $\xi_f^{(q)}$ 
takes the very simple form
\begin{equation}\label{e743}
\xi_f^{(q)}=\mp\exp\left[-i\left\{\phi_{\mbox{{\scriptsize M}}}^{(q)}-
\phi_{\mbox{{\scriptsize D}}}^{(f)}\right\}\right],
\end{equation}
where the characteristic ``decay'' phase 
$\phi_{\mbox{{\scriptsize D}}}^{(f)}$ can be expressed in terms of angles 
of the unitarity triangle as follows:
\begin{equation}\label{e746}
\phi_{\mbox{{\scriptsize D}}}^{(f)}=\left\{\begin{array}{cl}
-2\gamma & \mbox{for dominant $\bar b\to\bar uu\bar r$ CKM amplitudes 
in $B_q^0\to f$}\\
0 & \mbox{for dominant $\bar b\to\bar cc\bar r$\,\, CKM amplitudes 
in $B_q^0\to f$.}
\end{array}\right.
\end{equation}
The validity of dominance of a single CKM amplitude 
and important phenomenological applications of (\ref{e743})
will be discussed in the following subsections. 

\subsection{The $B_d$ System}
In contrast to the $B_s$ system, the width difference is negligibly small
in the $B_d$ system. Consequently the expressions for the decay rates 
(\ref{ratebf})-(\ref{ratebbfb}) simplify considerably in that case. 

\subsubsection{CP Asymmetries in $B_d$ Decays}
Restricting ourselves, as in the previous subsection, to decays into 
final CP eigenstates $|f\rangle$ satisfying (\ref{e733a}), we obtain 
the following expressions for the time-dependent and time-integrated 
CP asymmetries:
\begin{eqnarray}
\lefteqn{a_{\mbox{{\scriptsize CP}}}(B_d\to f;t)\equiv\frac{\Gamma(B_d^0(t)
\to f)-\Gamma(\overline{B_d^0}(t)\to f)}{\Gamma(B_d^0(t)\to f)+
\Gamma(\overline{B_d^0}(t)\to f)}}\nonumber\\
&&={\cal A}_{\mbox{{\scriptsize CP}}}^{\mbox{{\scriptsize dir}}}(B_d\to f)
\cos(\Delta M_d t)+{\cal A}_{\mbox{{\scriptsize CP}}}^{\mbox{{\scriptsize 
mix-ind}}}(B_d\to f)\sin(\Delta M_d t)\label{acptimedep}
\end{eqnarray}
\begin{eqnarray}
\lefteqn{a_{\mbox{{\scriptsize CP}}}(B_d\to f)\equiv
\frac{\int\limits_0^\infty
\mbox{d}t\left[\Gamma(B_d^0(t)\to f)-
\Gamma(\overline{B_d^0}(t)\to f)\right]}
{\int\limits_0^\infty \mbox{d}t \left[\Gamma(B_d^0(t)\to f)+
\Gamma(\overline{B_d^0}(t)\to f)\right]}}\nonumber\\
&&=\frac{1}{1+x_d^2}\left[{\cal A}_{\mbox{{\scriptsize 
CP}}}^{\mbox{{\scriptsize dir}}}(B_d\to f)+x_d\,
{\cal A}_{\mbox{{\scriptsize CP}}}^{\mbox{{\scriptsize mix-ind}}}
(B_d\to f)\right],\label{acptimeint}
\end{eqnarray}
where the {\it direct} CP-violating contributions 
\begin{equation}\label{acpdir}
{\cal A}_{\mbox{{\scriptsize CP}}}^{\mbox{{\scriptsize dir}}}
(B_d\to f)\equiv
\frac{1-\left|\xi_f^{(d)}\right|^2}{1+\left|\xi_f^{(d)}\right|^2}
\end{equation}
have been separated from the {\it mixing-induced} CP-violating contributions
\begin{equation}\label{acpmi}
{\cal A}_{\mbox{{\scriptsize CP}}}^{\mbox{{\scriptsize mix-ind}}}(B_d\to 
f)\equiv\frac{2\,\mbox{Im}\,\xi_f^{(d)}}{1+\left|\xi_f^{(d)}\right|^2}.
\end{equation}
Whereas the former observables describe CP violation arising directly in
the corresponding decay amplitudes, the latter ones are due to interference
between $B^0_d-\overline{B^0_d}$ mixing- and decay-processes. Needless to
say, the expressions (\ref{acptimedep}) and (\ref{acptimeint}) have to be 
modified appropriately for the $B_s$ system because of $\Delta\Gamma_s/
\Gamma_s={\cal O}(20\%)$. In the case of the time-dependent CP asymmetry 
(\ref{acptimedep}) these effects start to become important for 
$t\,\mbox{{\scriptsize $\stackrel{>}{\sim}$}}\,2/\Delta\Gamma_s$. 

\subsubsection{CP Violation in $B_d\to J/\psi\, K_{\mbox{{\scriptsize S}}}$:
the ``Gold-plated'' Way to Extract $\beta$}\label{Beta-Ext}
The channel $B_d\to J/\psi\, K_{\mbox{{\scriptsize S}}}$ is a transition into
a CP eigenstate with eigenvalue $-1$ and originates from a $\bar b\to\bar 
cc\bar s$ quark-level decay \cite{csbs}. Consequently the corresponding 
observable $\xi^{(d)}_{\psi K_{\mbox{{\scriptsize S}}}}$ can be 
expressed as
\begin{equation}\label{xipsiks}
\xi^{(d)}_{\psi K_{\mbox{{\scriptsize S}}}}=+e^{-2i\beta}\left[
\frac{v_u^{(s)}A^{ut'}_{\mbox{{\scriptsize pen}}}+v_c^{(s)}\left(
A_{\mbox{{\scriptsize cc}}}^{c'}+A^{ct'}_{\mbox{{\scriptsize pen}}}\right)}
{v_u^{(s)\ast}A^{ut'}_{\mbox{{\scriptsize pen}}}+v_c^{(s)\ast}\left(
A_{\mbox{{\scriptsize cc}}}^{c'}+A^{ct'}_{\mbox{{\scriptsize pen}}}\right)}
\right],
\end{equation}
where $A_{\mbox{{\scriptsize cc}}}^{c'}$ denotes the $Q_{1,2}^{cs}$ 
current-current operator amplitude and $A^{ut'}_{\mbox{{\scriptsize pen}}}$ 
$(A^{ct'}_{\mbox{{\scriptsize pen}}})$ corresponds to contributions of
the penguin-type with up- and top-quarks (charm- and top-quarks) running
as virtual particles in the loops. Note that within this notation 
penguin-like matrix elements of the $Q_{1,2}^{cs}$ current-current 
operators are included by definition in the 
$A^{ct'}_{\mbox{{\scriptsize pen}}}$ amplitude, whereas those of 
$Q_{1,2}^{us}$ show up in $A^{ut'}_{\mbox{{\scriptsize pen}}}$.
The primes in (\ref{xipsiks}) 
have been introduced to remind us that we are dealing with a 
$\bar b\to\bar s$ mode. Using the modified Wolfenstein parametrization 
(\ref{wolf2}), the relevant CKM factors take the form 
\begin{equation}
v_u^{(s)}=A\lambda^4R_b\,e^{-i\gamma},\quad
v_c^{(s)}=A\lambda^2\left(1-\lambda^2/2\right)
\end{equation}
and imply that the $A^{ut'}_{\mbox{{\scriptsize pen}}}$ contribution is 
highly CKM suppressed with respect to the part containing the current-current 
amplitude. The suppression factor is given by  
\begin{equation}
\left|v_u^{(s)}/v_c^{(s)}\right|=\lambda^2R_b\approx0.02.
\end{equation}
An additional suppression arises from the fact that 
$A^{ut'}_{\mbox{{\scriptsize pen}}}$ is related to 
loop processes that are governed by Wilson coefficients 
of ${\cal O}(10^{-2})$. Moreover the colour-structure of 
$B_d\to J/\psi\, K_{\mbox{{\scriptsize S}}}$ leads to further suppression!
The point is that the $\bar c$- and $c$-quarks emerging from the gluons of 
the usual QCD penguin diagrams form a colour-octet state and consequently
cannot build up the $J/\psi$ which is a $\bar cc$ colour-singlet state. 
Therefore additional gluons are needed. Their contributions are unfortunately
very hard to estimate. However, the former colour-argument does not hold 
for EW penguins which may hence be the most important penguin
contributions to $B_d\to J/\psi\, K_{\mbox{{\scriptsize S}}}$. 
The suppression of $v_u^{(s)}A_{\mbox{{\scriptsize pen}}}^{ut'}$
relative to $v_c^{(s)}(A_{\mbox{{\scriptsize cc}}}^{c'}+
A_{\mbox{{\scriptsize pen}}}^{ct'})$ is compensated slightly since 
the dominant $Q_{1,2}^{cs}$ current-current amplitude 
$A_{\mbox{{\scriptsize cc}}}^{c'}$ 
is colour-suppressed by a phenomenological colour-suppression 
factor $a_2\approx0.2$ \cite{BSW}-\cite{a1a2-exp}. However, since 
$v_u^{(s)}A_{\mbox{{\scriptsize pen}}}^{ut'}$ is 
suppressed by {\it three} sources (CKM-structure, 
loop effects, colour-structure), we conclude that $\xi^{(d)}_{\psi 
K_{\mbox{{\scriptsize S}}}}$ is nevertheless given to an excellent 
approximation by
\begin{equation}
\xi^{(d)}_{\psi K_{\mbox{{\scriptsize S}}}}=e^{-2i\beta}
\left[\frac{v_c^{(s)}\left(A_{\mbox{{\scriptsize cc}}}^{c'}+
A^{ct'}_{\mbox{{\scriptsize pen}}}\right)}{v_c^{(s)\ast}\left(
A_{\mbox{{\scriptsize cc}}}^{c'}+A^{ct'}_{\mbox{{\scriptsize pen}}}
\right)}\right]=e^{-2i\beta}
\end{equation}
yielding 
\begin{equation}
{\cal A}_{\mbox{{\scriptsize CP}}}^{\mbox{{\scriptsize dir}}}
(B_d\to J/\psi\, K_{\mbox{{\scriptsize S}}})=0,\quad 
{\cal A}_{\mbox{{\scriptsize CP}}}^{\mbox{{\scriptsize mix-ind}}}
(B_d\to J/\psi\, K_{\mbox{{\scriptsize S}}})=-\sin(2\beta)
\end{equation}
and thus 
\begin{equation}\label{Gold-time}
a_{\mbox{{\scriptsize CP}}}(B_d\to J/\psi\, K_{\mbox{{\scriptsize S}}};t)=
-\sin(2\beta)\,\sin(\Delta M_dt)
\end{equation}
\begin{equation}\label{Gold-int}
a_{\mbox{{\scriptsize CP}}}(B_d\to J/\psi\, K_{\mbox{{\scriptsize S}}})=
-\frac{x_d}{1+x_d^2}\,\sin(2\beta)
\end{equation}
for the time-dependent and time-integrated CP asymmetries (\ref{acptimedep})
and (\ref{acptimeint}), respectively. 
Consequently these observables measure $\sin(2\beta)$ to excellent accuracy.
Therefore $B_d\to J/\psi\, K_{\mbox{{\scriptsize S}}}$ is usually referred 
to as the ``gold-plated'' mode to determine the UT angle $\beta$. The
presently expected ranges for $\sin(2\beta)$ can be read off from table 5
and imply {\it non zero} values for the CP-violating asymmetries 
(\ref{Gold-time}) and (\ref{Gold-int}). The latter asymmetry is expected to
be of the order $-30\%$ within the Standard Model. Other methods for 
extracting $\beta$ can be found e.g.\ in \cite{non-CP,beta-refs}.

\subsubsection{CP Violation in $B_d\to \pi^+\pi^-$ and Extractions of
$\alpha$}\label{Bdpipi-Alpha}
In the case of $B_d\to\pi^+\pi^-$ we have to deal with the decay of a 
$B_d$-meson into a final CP eigenstate with eigenvalue $+1$ that is caused 
by the quark-level process $\bar b\to\bar uu\bar d$. Therefore we may write
\begin{equation}\label{xipipi}
\xi^{(d)}_{\pi^+\pi^-}=-e^{-2i\beta}\left[
\frac{v_u^{(d)}\left(A_{\mbox{{\scriptsize cc}}}^{u}+
A^{ut}_{\mbox{{\scriptsize pen}}}\right)+v_c^{(d)}
A^{ct}_{\mbox{{\scriptsize pen}}}}
{v_u^{(d)\ast}\left(A_{\mbox{{\scriptsize cc}}}^{u}+
A^{ut}_{\mbox{{\scriptsize pen}}}\right)+v_c^{(d)\ast}
A^{ct}_{\mbox{{\scriptsize pen}}}}
\right],
\end{equation}
where the notation of decay amplitudes is as in the previous discussion 
of $B_d\to J/\psi\, K_{\mbox{{\scriptsize S}}}$. Using again (\ref{wolf2}),
the CKM factors are given by 
\begin{equation}\label{CKMbd}
v_u^{(d)}=A\lambda^3R_b\,e^{-i\gamma},\quad
v_c^{(d)}=-A\lambda^3.
\end{equation}
The CKM structure of (\ref{xipipi}) is very different form 
$\xi^{(d)}_{\psi K_{\mbox{{\scriptsize S}}}}$. In particular the pieces 
containing the dominant $Q_{1,2}^{ud}$ current-current contributions 
$A_{\mbox{{\scriptsize cc}}}^{u}$ are CKM suppressed with respect to 
the penguin contributions $A^{ct}_{\mbox{{\scriptsize pen}}}$ by 
\begin{equation}\label{bdckm}
\left|v_u^{(d)}/v_c^{(d)}\right| = R_b\approx0.36\,.
\end{equation}
In contrast to $B_d\to J/\psi\, K_{\mbox{{\scriptsize S}}}$, in the 
$B_d\to\pi^+\pi^-$ case the penguin amplitudes are only suppressed by the 
corresponding Wilson coefficients ${\cal O}(10^{-2})$ and not additionally 
by the colour-structure of that decay. Taking into account that the 
current-current amplitude $A_{\mbox{{\scriptsize cc}}}^{u}$ is colour-allowed 
and using both (\ref{bdckm}) and characteristic values of the Wilson 
coefficient functions, one obtains
\begin{equation}
\left|\frac{v_c^{(d)}A^{ct}_{\mbox{{\scriptsize pen}}}}
{v_u^{(d)}\left(A_{\mbox{{\scriptsize cc}}}^{u}+
A^{ut}_{\mbox{{\scriptsize pen}}}\right)}\right|={\cal O}(0.15)
\end{equation}
and concludes that
\begin{equation}
\xi^{(d)}_{\pi^+\pi^-}\approx-e^{-2i\beta}\left[\frac{v_u^{(d)}\left(
A_{\mbox{{\scriptsize cc}}}^{u}+A^{ut}_{\mbox{{\scriptsize pen}}}\right)}
{v_u^{(d)\ast}\left(A_{\mbox{{\scriptsize cc}}}^{u}+
A^{ut}_{\mbox{{\scriptsize pen}}}\right)}\right]=-e^{2i\alpha}
\end{equation}
may be a reasonable approximation to obtain an estimate for the UT angle
$\alpha$ from the CP-violating observables 
\begin{equation}\label{BdpipiCP}
{\cal A}_{\mbox{{\scriptsize CP}}}^{\mbox{{\scriptsize dir}}}
(B_d\to\pi^+\pi^-)\approx0,\quad
{\cal A}_{\mbox{{\scriptsize CP}}}^{\mbox{{\scriptsize mix-ind}}}
(B_d\to\pi^+\pi^-)\approx-\sin(2\alpha)
\end{equation}
implying the following time-dependent and time-integrated CP asymmetries 
(\ref{acptimedep}) and (\ref{acptimeint}):
\begin{equation}\label{Bpipi-time}
a_{\mbox{{\scriptsize CP}}}(B_d\to\pi^+\pi^-;t)\approx
-\sin(2\alpha)\,\sin(\Delta M_dt)
\end{equation}
\begin{equation}\label{Bpipi-int}
a_{\mbox{{\scriptsize CP}}}(B_d\to\pi^+\pi^-)\approx
-\frac{x_d}{1+x_d^2}\,\sin(2\alpha)\,.
\end{equation}
Note that a measurement of ${\cal A}_{\mbox{{\scriptsize 
CP}}}^{\mbox{{\scriptsize dir}}}(B_d\to\pi^+\pi^-)\not=0$, i.e.\ of 
a contribution to (\ref{Bpipi-time}) evolving with $\cos(\Delta M_dt)$, 
would signal the presence of penguins. We shall come back to this feature 
later. 

The formalism discussed above can also be applied straightforwardly
to $B_d\to D^+D^-$ (caused by $\bar b\to\bar cc\bar d$) 
that is dominated by the contributions proportional
to $v_c^{(d)}$. In that case (\ref{bdckm}) leads to an additional {\it 
suppression} of the $v_u^{(d)}$ amplitude originating essentially from
penguin contributions. Consequently the relations 
\begin{equation}\label{BDDCP}
{\cal A}_{\mbox{{\scriptsize CP}}}^{\mbox{{\scriptsize dir}}}
(B_d\to D^+D^-)=0,\quad
{\cal A}_{\mbox{{\scriptsize CP}}}^{\mbox{{\scriptsize mix-ind}}}
(B_d\to D^+D^-)=\sin(2\beta)
\end{equation}
are expected to be satisfied to higher accuracy than (\ref{BdpipiCP}).
However, in respect of theoretical cleanliness to extract $\beta$, the
decay $B_d\to D^+D^-$ cannot compete with the ``gold-plated'' mode
$B_d\to J/\psi\, K_{\mbox{{\scriptsize S}}}$.

The hadronic uncertainties affecting the extraction of $\alpha$ from 
CP violation in $B_d\to\pi^+\pi^-$ were analyzed by many authors in 
the literature. A selection of papers is given in 
\cite{gro-pen,uncert-alpha}. As was pointed out by Gronau and 
London \cite{gl}, the uncertainties related to QCD penguins \cite{pens} 
can be eliminated with the help of isospin relations
involving in addition to $B_d\to\pi^+\pi^-$ also the modes 
$B_d\to\pi^0\pi^0$ 
and $B^{\pm}\to\pi^{\pm}\pi^0$. The isospin relations among the 
corresponding decay amplitudes are given by
\begin{eqnarray}
A(B^0_d\to\pi^+\pi^-)+\sqrt{2}A(B^0_d\to\pi^0\pi^0)&=&\sqrt{2}A(B^+\to
\pi^+\pi^0)\label{isospin1}\\
A(\overline{B^0_d}\to\pi^+\pi^-)+\sqrt{2}A(\overline{B^0_d}\to\pi^0
\pi^0)&=&\sqrt{2}A(B^-\to\pi^-\pi^0)\label{isospin2}
\end{eqnarray}
and can be represented as two triangles in the complex plane that allow
the extraction of a value of $\alpha$ that does not suffer from QCD
penguin uncertainties. It is, however, not possible to control also the
EW penguin uncertainties using that isospin approach. The point is that
up- and down-quarks are coupled differently in EW penguin diagrams
because of their different electrical charges (see (\ref{ew-def})).
Hence one has also to think about the role of these contributions. 
We shall come back to that issue in subsection~\ref{RoEWPs}, where a more 
detailed discussion of the GL method \cite{gl} in light of EW penguin 
effects will be given. 

An experimental problem of the GL method is related to the fact that it
requires a measurement of $Br(B_d\to\pi^0\pi^0)$ which may be smaller
than ${\cal O}(10^{-6})$ because of colour-suppression effects \cite{kp}.
Therefore, despite of its attractiveness, that approach may be quite
difficult from an experimental point of view and it is important to have 
alternatives available to determine $\alpha$. Needless to say, that is 
also required in order to over-constrain the UT angle $\alpha$ as much 
as possible at future $B$-physics experiments. Fortunately such methods 
are already on the market. For example, Snyder and Quinn suggested 
to use $B\to\rho\,\pi$ modes to extract $\alpha$ \cite{sq}. 
Another method was proposed in \cite{PAPII}.
It requires a simultaneous measurement of ${\cal A}_{\mbox{{\scriptsize 
CP}}}^{\mbox{{\scriptsize mix-ind}}}(B_d\to\pi^+\pi^-)$ and ${\cal 
A}_{\mbox{{\scriptsize CP}}}^{\mbox{{\scriptsize mix-ind}}}(B_d\to K^0
\overline{K^0})$ and determines $\alpha$ with the help of a geometrical
triangle construction using the $SU(3)$ flavour symmetry of strong 
interactions. The accuracy of that approach is limited by $SU(3)$-breaking 
corrections which cannot be estimated reliably at present. 
Interestingly the penguin-induced decay $B_d\to K^0
\overline{K^0}$ may exhibit CP asymmetries as large as ${\cal O}(30\%)$ 
within the Standard Model \cite{rf-k0k0.bar}. This feature is due 
to interference 
between QCD penguins with internal up- and charm-quark exchanges \cite{bf1}. 
In the absence of these contributions, the CP-violating asymmetries of 
$B_d\to K^0\overline{K^0}$ would vanish and ``New Physics'' would be 
required (see e.g.\ \cite{mw}) to induce CP violation in that decay. 
An upper bound $Br(B_d\to K^0\overline{K^0})<1.7\cdot10^{-5}$ has been 
presented very recently by the CLEO collaboration \cite{NEW-CLEO}. 

Before discussing other methods to deal with the penguin uncertainties 
affecting the extraction of $\alpha$ from the CP-violating observables of 
$B_d\to\pi^+\pi^-$, let us next have a closer look at the above
mentioned QCD penguins with up- and charm-quarks running as virtual 
particles in the loops.

\subsubsection{Penguin Zoology}\label{Zoo}
The general structure of a generic $\bar b\to\bar q$ ($q\in\{d,s\}$) penguin
amplitude is given by
\begin{equation}\label{pen-amp}
P^{(q)}=V_{uq}V_{ub}^\ast \,P_u^{(q)}+V_{cq}V_{cb}^\ast \,P_c^{(q)}+
V_{tq}V_{tb}^\ast \,P_t^{(q)},
\end{equation} 
where $P_u^{(q)}$, $P_c^{(q)}$ and $P_t^{(q)}$ are the amplitudes of penguin 
processes with internal up-, charm- and top-quark exchanges, respectively, 
omitting CKM factors. The penguin amplitudes introduced in (\ref{xipsiks})
and (\ref{xipipi}) are related to these quantities through
\begin{equation}
\begin{array}{rclcrcl}
A^{ut}_{\mbox{{\scriptsize pen}}}&=&P_u^{(d)}-P_t^{(d)},&\quad&
A^{ct}_{\mbox{{\scriptsize pen}}}&=&P_c^{(d)}-P_t^{(d)}\\
A^{ut'}_{\mbox{{\scriptsize pen}}}&=&P_u^{(s)}-P_t^{(s)},&\quad&
A^{ct'}_{\mbox{{\scriptsize pen}}}&=&P_c^{(s)}-P_t^{(s)}.
\end{array}
\end{equation}
Using unitarity of the CKM matrix yields
\begin{equation}\label{pen-unit}
P^{(q)}=V_{cq} V_{cb}^\ast\left[P_c^{(q)}-P_u^{(q)}\right]+
V_{tq} V_{tb}^\ast\left[P_t^{(q)}-P_u^{(q)}\right],
\end{equation}
where the CKM factors can be expressed with the help of 
the Wolfenstein parametrization as follows:
\begin{equation}\label{CKMd}
V_{cd}V_{cb}^\ast=-\lambda|V_{cb}|\left(1+{\cal O}\left(\lambda^4\right)
\right),\quad V_{td}V_{tb}^\ast=|V_{td}|e^{-i\beta},
\end{equation}
\begin{equation}\label{CKMs}
V_{cs}V_{cb}^\ast=|V_{cb}|\left(1+{\cal O}\left(\lambda^2\right)\right),\quad
V_{ts}V_{tb}^\ast=-|V_{cb}|\left(1+{\cal O}\left(\lambda^2\right)\right).\quad
\end{equation}
The estimate of the non-leading terms in $\lambda$ follows 
Subsection~\ref{Wolf-Par}.
Omitting these terms and combining (\ref{pen-unit}) with (\ref{CKMd}) 
and (\ref{CKMs}), the $\bar b\to\bar d$ and $\bar b\to\bar s$ penguin 
amplitudes take the form
\begin{equation}\label{bdpenamp}
P^{(d)}=\left[e^{-i\beta}-\frac{1}{R_t}\Delta P^{(d)}\right]
\left|V_{td}\right|\left|P_{tu}^{(d)}\right|
e^{i\delta_{tu}^{(d)}}
\end{equation}
\begin{equation}\label{bspenamp}
P^{(s)}=\left[1-\Delta P^{(s)}\right]e^{-i\pi}\left|V_{cb}\right|
\left|P_{tu}^{(s)}\right|e^{i\delta_{tu}^{(s)}},
\end{equation}
where the notation
\begin{equation}
P_{q_1q_2}^{(q)}\equiv P_{q_1}^{(q)}-P_{q_2}^{(q)}
\end{equation}
has been introduced and
\begin{equation}\label{DelP}
\Delta P^{(q)}\equiv\frac{P^{(q)}_{cu}}{P^{(q)}_{tu}}
\end{equation}
describes the contributions of ``subdominant'' penguins with up- and 
charm-quarks running as virtual particles in the loops. In the limit 
of degenerate up- and charm-quark masses, $\Delta P^{(q)}$ would vanish 
because of the GIM mechanism \cite{GIM1}. However, since $m_u\approx4.5$~MeV, 
whereas $m_c\approx1.4$~GeV, this GIM cancellation is incomplete and in 
principle sizable effects arising from $\Delta P^{(q)}$ could be expected.

\begin{figure}[t]
\centerline{
\rotate[r]{
\epsfysize=9truecm
\epsffile{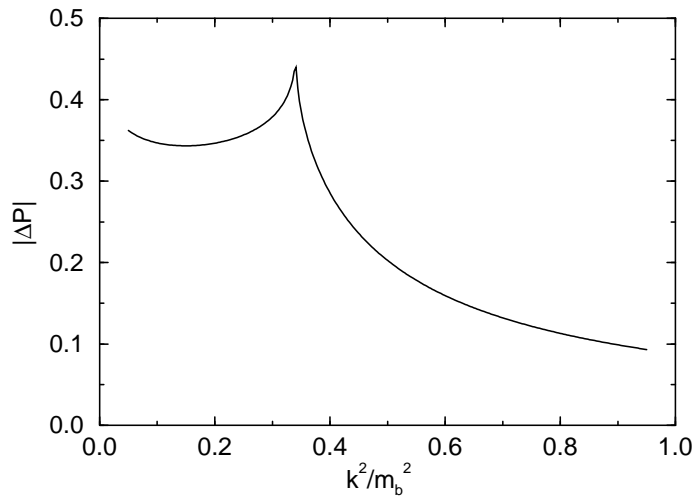}
}}
\caption{The dependence of $\left|\Delta P^{(q)}\right|$ on 
$k^2/m_b^2$.}\label{delp-fig}
\end{figure}

Usually it is assumed that the penguin amplitudes (\ref{bdpenamp}) 
and (\ref{bspenamp}) are dominated by internal top-quark exchanges, 
i.e.\ $\Delta P^{(q)}\approx0$. That is an excellent approximation for
EW penguin contributions which play an important role in certain
$B$ decays only because of the large top-quark mass as we will see in  
section~\ref{Sec-EWP}. However, QCD penguins with internal 
up- and charm-quarks may become important as is indicated by 
model calculations at the perturbative quark-level \cite{bf1}. 
Following the pioneering approach of Bander, Silverman and Soni \cite{bss},
the strong phase of $\Delta P^{(q)}$ is generated exclusively through 
absorptive parts of time-like penguin diagrams with internal up- and 
charm-quark exchanges. These estimates of $\Delta P^{(q)}$ depend strongly 
on the value of $k^2$ denoting the four-momentum of the gluon appearing 
in the time-like QCD penguin diagrams. This feature can be seen nicely in 
fig.~\ref{delp-fig}, where that dependence is shown. Simple kinematical 
considerations at the quark-level imply that $k^2$ should lie within 
the ``physical'' range \cite{rf1,detr,gh}
\begin{equation}\label{phys-range}
\frac{1}{4}\,\mbox{{\scriptsize $\stackrel{<}{\sim}$}}\,\frac{k^2}{m_b^2}
\,\mbox{{\scriptsize $\stackrel{<}{\sim}$}}\,\frac{1}{2}.
\end{equation}
A detailed discussion of the $k^2$-dependence can be found in \cite{gh}.

Looking at fig.~\ref{delp-fig}, we observe that $\Delta P^{(q)}$ may lead 
to sizable effects for such values of $k^2$. Moreover QCD penguin topologies 
with internal up- and charm-quarks contain also long-distance 
contributions, like the rescattering process 
$B^0_d\to\{D^+D^-\}\to\pi^+\pi^-$ (see e.g.\ \cite{kamal}), 
which are very hard to estimate. This feature can be seen easily by 
drawing the corresponding Feynman diagrams.  
Such long-distance contributions were 
discussed in the context of extracting $V_{td}$ from radiative $B$ decays
in \cite{abs} and are potentially very serious. 
Consequently it may not be justified to neglect the $\Delta P^{(q)}$ terms 
in (\ref{bdpenamp}) and (\ref{bspenamp}) \cite{bf1}. Recently, in a different
context, the importance of the $\Delta P^{(q)}$ contributions has been 
stressed in \cite{CFMS}. 
 
An important difference arises, however, between (\ref{bdpenamp}) and 
(\ref{bspenamp}). While
the UT angle $\beta$ shows up in the $\bar b\to\bar d$ case, there is 
only a trivial CP-violating weak phase present in the $\bar b\to\bar s$ 
case. Consequently $\Delta P^{(s)}$ cannot change the general phase 
structure of the $\bar b\to\bar s$ penguin amplitude 
$P^{(s)}$. On the other hand, if one takes into account also QCD penguins 
with internal up- and charm-quarks, the $\bar b\to \bar d$ penguin amplitude 
$P^{(d)}$ is no longer related in a simple and ``clean'' way through
\begin{equation}\label{bdpenapprox}
P^{(d)}=e^{-i\beta}e^{i\delta_P^{(d)}}\left|P^{(d)}\right|
\end{equation}
to $\beta$, where $\delta_P^{(d)}$ is a CP-conserving strong phase. 
As we pointed out in \cite{bf1}, this feature may affect some of the 
strategies to extract CKM phases with the help of $SU(3)$ amplitude 
relations that will be discussed later in this review. 

An interesting consequence of (\ref{bspenamp}) is the relation 
$P^{(s)}=\overline{P^{(s)}}$ between the $\bar b\to\bar s$ QCD penguin
amplitude and its charge-conjugate implying that penguin-induced modes of 
this type, e.g.\ the decay $B_d\to\phi\, K_{\mbox{{\scriptsize S}}}$, 
should exhibit no direct CP violation. Applying the formalism developed in
Subsection~\ref{CP-Eigen}, one finds that 
\begin{equation}\label{acpphiks}
{\cal A}_{\mbox{{\scriptsize CP}}}^{\mbox{{\scriptsize mix-ind}}}(B_d\to\phi\, 
K_{\mbox{{\scriptsize S}}})=-\sin(2\beta)
\end{equation}
measures the angle $\beta$. Within the Standard Model, small 
direct CP violation --
model calculations (see e.g.\ \cite{rfewp1,kps,gh}) indicate 
asymmetries at the ${\cal O}(1\%)$ level -- may arise from the neglected 
${\cal O}(\lambda^2)$ terms in (\ref{CKMs}) which also
limit the theoretical accuracy of (\ref{acpphiks}). An experimental
comparison between the mixing-induced CP asymmetries of $B_d\to J/\psi\, 
K_{\mbox{{\scriptsize S}}}$ and $B_d\to\phi\,K_{\mbox{{\scriptsize S}}}$,
which should be equal to very good accuracy within the Standard Model, 
would be extremely interesting since the latter decay is a ``rare'' FCNC 
process and may hence be very sensitive to physics beyond the Standard 
Model \cite{F97}. Recently this point has been discussed in more detail
by London and Soni \cite{LoSo}. The branching
ratio for $B_d\to\phi\,K_{\mbox{{\scriptsize S}}}$ is expected to be of
${\cal O}(10^{-5})$ and may be large enough to perform these measurements
at future $B$ physics facilities. 

\subsubsection{Another Look at $B_d\to\pi^+\pi^-$ and the Extraction of
$\alpha$}
The discussion presented above implies that it is important to 
reanalyze the decay $B_d\to\pi^+\pi^-$ without assuming dominance of QCD 
penguins with internal top-quark exchanges. Such a study was performed in
\cite{fm1} (see also \cite{CFMS}). To this end it is useful to introduce 
\begin{equation}
T\equiv V_{ud}V_{ub}^\ast\, A_{\mbox{{\scriptsize cc}}}^{u}
\end{equation}
and to expand the CP-violating observables (\ref{acpdir}) and (\ref{acpmi}) 
corresponding to $B_d\to\pi^+\pi^-$ in powers of $\overline{P^{(d)}}/T$ and 
$P^{(d)}/T$, which we expect to satisfy the estimate \cite{fm1}
\begin{equation}
\left|\frac{\overline{P^{(d)}}}{T}\right|
\approx\left|\frac{P^{(d)}}{T}\right|
\approx0.07-0.23,
\end{equation}
and to keep only the leading terms in that expansion:
\begin{eqnarray}
&& {\cal A}_{\mbox{{\scriptsize CP}}}^{\mbox{{\scriptsize dir}}}
(B_d\to\pi^+\pi^-) = 2 \lambda R_t\frac{|\tilde P|}{|T|}  
\sin \delta \sin \alpha + {\cal O}\left((P^{(d)}/T)^2\right)\label{cpasym}\\
&& {\cal A}_{\mbox{{\scriptsize CP}}}^{\mbox{{\scriptsize mix-ind}}}
(B_d\to\pi^+\pi^-) \nonumber\\
&&\qquad=  - \sin 2 \alpha - 2 \lambda R_t 
\frac{|\tilde P|}{|T|}\cos \delta\, \cos 2 
\alpha\,\sin \alpha+{\cal O}\left((P^{(d)}/T)^2
\right).\label{cpasym1}
\end{eqnarray}
Similar expressions were also derived by Gronau in \cite{gro-pen}.
However, it has {\it not} been assumed in (\ref{cpasym}) and (\ref{cpasym1}) 
that QCD penguins are dominated by internal top-quark exchanges and the
physical interpretation of the amplitude $\tilde P$ is quite different from
\cite{gro-pen}. This quantity is given by
\begin{equation}\label{Pprime}
\tilde P\equiv\left[1-\Delta P^{(d)}\right]|V_{cb}|\left|P_{tu}^{(d)}\right|
e^{i\delta_{tu}^{(d)}}\,,
\end{equation}
and $\delta$ appearing in (\ref{cpasym}) and (\ref{cpasym1}) is simply the 
CP-conserving strong phase of $\tilde P/T$.
If we compare (\ref{Pprime}) with (\ref{bdpenamp}) and (\ref{bspenamp}),
we observe that it is not equal to the amplitude $P^{(d)}$ -- as one would
expect naively -- but that its phase structure corresponds exactly
to the $\bar b\to\bar s$ QCD penguin amplitude $e^{i\pi}P^{(s)}$. 

The two CP-violating observables (\ref{cpasym}) and (\ref{cpasym1}) 
depend on the three ``unknowns'' $\alpha$, $\delta$ and $|\tilde 
P|/|T|$ (strategies to extract the CKM factor $R_t$ are discussed in previous
sections and $\lambda$ is the usual Wolfenstein parameter). 
Consequently an additional input is needed to determine $\alpha$ from 
(\ref{cpasym}) and (\ref{cpasym1}). Taking into account the discussion given 
in the previous paragraph, it is very natural to use the $SU(3)$ flavour 
symmetry of strong interactions to accomplish this task. In the strict 
$SU(3)$ limit one does not distinguish between down- and strange-quarks 
and $|\tilde P|$ corresponds simply to the magnitude of the decay amplitude 
of a penguin-induced $\bar b\to\bar s$ transition such as $B^+\to\pi^+ K^0$ 
with an expected branching ratio of ${\cal O}(10^{-5})$ \cite{kp}. That
decay has been measured very recently by the CLEO collaboration 
\cite{NEW-CLEO} with the branching ratio 
$Br(B^+\to\pi^+ K^0)=\left(2.3^{+1.1+0.2}_{-1.0-0.2}\pm0.2\right)
\cdot 10^{-5}$. 
On the other hand, $|T|$ can be estimated from the rate of 
$B^+\to\pi^+\pi^0$ by neglecting colour-suppressed current-current operator 
contributions. Presently only the upper bound $Br(B^+\to\pi^+\pi^0)<
2.0\cdot10^{-5}$ is available for that mode \cite{NEW-CLEO}. 

Following these lines one obtains
\begin{equation}\label{BR-rat}
\frac{|\tilde P|}{|T|}\approx\frac{F_\pi}{F_K}\sqrt{\frac{1}{2}
\frac{Br(B^+\to\pi^+K^0)}{Br(B^+\to\pi^+\pi^0)}},
\end{equation}
where $F_\pi$ and $F_K$ are the $\pi$- and $K$-meson decay constants,
respectively, taking into account factorizable $SU(3)$-breaking.
That relation allows the extraction both of $\alpha$ and $\delta$ from 
the measured CP-violating observables (\ref{cpasym}) and (\ref{cpasym1}). 
Problems of this approach arise if $\alpha$ is close to $45^\circ$ or 
$135^\circ$, where the expansion (\ref{cpasym1}) for 
${\cal A}_{\mbox{{\scriptsize CP}}}^{\mbox{{\scriptsize mix-ind}}}
(B_d\to\pi^+\pi^-)$ breaks down.
Assuming a total theoretical uncertainty of 30\% in the quantity 
\begin{equation}\label{DET-a}
a\equiv2\lambda R_t\frac{|\tilde P|}{|T|}\approx 2\lambda R_t
\frac{F_\pi}{F_K}\sqrt{\frac{1}{2}
\frac{Br(B^+\to\pi^+K^0)}{Br(B^+\to\pi^+\pi^0)}}
\end{equation}
governing (\ref{cpasym}) and 
(\ref{cpasym1}), an uncertainty of $\pm\,3^\circ$ in the extracted 
value of $\alpha$ is expected if $\alpha$ is not too close to these 
singular points \cite{fm1}. For values of $\alpha$ far away from $45^\circ$ 
and $135^\circ$, one may even have an uncertainty of only $\pm\,1^\circ$
as is indicated by the following example: Let us assume that the 
CP asymmetries are measured to be ${\cal A}_{\mbox{{\scriptsize 
CP}}}^{\mbox{{\scriptsize dir}}}(B_d\to\pi^+\pi^-)=+\,0.1$ and  
${\cal A}_{\mbox{{\scriptsize CP}}}^{\mbox{{\scriptsize mix-ind}}}
(B_d\to\pi^+\pi^-)=-\,0.25$ and that (\ref{DET-a}) gives $a=0.26$. 
Assuming a theoretical uncertainty of 30\% in $a$, i.e.\ $\Delta a=\pm0.04$, 
and inserting these numbers into (\ref{cpasym}) and (\ref{cpasym1}) gives
$\alpha=(76\pm1)^\circ$ and $\delta=(24\pm4)^\circ$. On the other hand, 
a naive analysis 
using (\ref{BdpipiCP}) where the penguin contributions are neglected would 
yield $\alpha=83^\circ$. Consequently the theoretical uncertainty of the 
extracted value of $\alpha$ is expected to be significantly smaller than
the shift through the penguin contributions. Since this method of 
extracting $\alpha$ requires neither difficult measurements of very small 
branching ratios nor complicated geometrical constructions it may turn 
out to be very useful for the early days of the $B$-factory era beginning at
the end of this millennium.

\subsubsection{A Simultaneous Extraction of $\alpha$ and 
$\gamma$}\label{SIMalp-gam}
Recently it has been pointed out by Dighe, Gronau and Rosner that a 
time-dependent measurement of $B_d\to\pi^+\pi^-$ in combination with the 
branching ratios for $B^0_d\to\pi^- K^+$, $B^+\to\pi^+ K^0$ and their
charge-conjugates may allow a simultaneous determination of the angles
$\alpha$ and $\gamma$ \cite{dgr}. These decays provide the following six 
observables
$A_1,\ldots,A_6$:
\begin{eqnarray}
\Gamma(B^0_d(t)\to\pi^+\pi^-)+\Gamma(\overline{B^0_d}(t)\to\pi^+\pi^-)&=&
e^{-\Gamma_dt}\,A_1\\
\Gamma(B^0_d(t)\to\pi^+\pi^-)-\Gamma(\overline{B^0_d}(t)\to\pi^+\pi^-)
\nonumber&=&e^{-\Gamma_dt}\left[A_2\cos(\Delta M_dt)\right.\nonumber\\
&&\quad+\left.A_3\sin(\Delta M_dt)\right]\\
\Gamma(B^0_d\to\pi^-K^+)+\Gamma(\overline{B^0_d}\to\pi^+K^-)&=&
A_4\\
\Gamma(B^0_d\to\pi^-K^+)-\Gamma(\overline{B^0_d}\to\pi^+K^-)&=&
A_5\\
\Gamma(B^+\to\pi^+K^0)+\Gamma(B^-\to\pi^-\overline{K^0})&=&
A_6.
\end{eqnarray}
Using $SU(3)$ flavour symmetry of strong interactions, neglecting
annihilation amplitudes, which should be suppressed by ${\cal O}(F_{B_d}/
m_{B_d})$ with $F_{B_d}\approx180\,\mbox{MeV}$, and assuming moreover
that the $\bar b\to\bar d$ QCD penguin amplitude is related in a simple
way to $\beta$ through (\ref{bdpenapprox}), i.e.\ assuming top-quark
dominance, the observables $A_1,\ldots,A_6$ can be expressed in terms
of six ``unknowns'' including $\alpha$ and $\gamma$. However, as we have 
outlined above, it is questionable whether the last assumption is justified 
since (\ref{bdpenapprox}) may be affected by QCD penguins 
with internal up- and charm-quark exchanges \cite{bf1}. Consequently the 
method proposed in \cite{dgr} suffers from theoretical limitations. 
Nevertheless it is an interesting approach, probably mainly in view of 
constraining $\gamma$ which is the angle of the unitarity triangle that 
is most difficult 
to measure. In order to extract that angle, $B_s$ decays play an 
important role as we will see in the following subsection. 

\subsection{The $B_s$ System}
The major phenomenological differences between the $B_d$ and $B_s$
systems arise from their mixing parameters (\ref{e719}) and from the 
fact that at leading order in the Wolfenstein expansion only a trivial 
weak mixing phase (\ref{e742}) is present in the $B_s$ case. 

\subsubsection{CP Violation in $B_s\to\rho^0K_{\mbox{{\scriptsize S}}}$:
the ``Wrong'' Way to Extract $\gamma$}\label{Bsrk-Gamma}
\noindent
Let us begin our discussion of the $B_s$ system by having a closer look
at the transition $B_s\to\rho^0K_{\mbox{{\scriptsize S}}}$ which appears
frequently in the literature as a tool to extract $\gamma$. It is a $B_s$
decay into a final CP eigenstate with eigenvalue $-1$ that is (similarly 
as the $B_d\to\pi^+\pi^-$ mode) caused by the quark-level process 
$\bar b\to\bar uu\bar d$. Hence the corresponding observable 
$\xi^{(s)}_{\rho^0K_{\mbox{{\scriptsize S}}}}$ can be expressed as
\begin{equation}\label{xirhoks}
\xi^{(s)}_{\rho^0K_{\mbox{{\scriptsize S}}}}=+e^{-i0}\left[
\frac{v_u^{(d)}\left({\cal A}_{\mbox{{\scriptsize cc}}}^{u}+
{\cal A}^{ut}_{\mbox{{\scriptsize pen}}}\right)+v_c^{(d)}
{\cal A}^{ct}_{\mbox{{\scriptsize pen}}}}
{v_u^{(d)\ast}\left({\cal A}_{\mbox{{\scriptsize cc}}}^{u}+
{\cal A}^{ut}_{\mbox{{\scriptsize pen}}}\right)+v_c^{(d)\ast}
{\cal A}^{ct}_{\mbox{{\scriptsize pen}}}}
\right],
\end{equation}
where the notation is as in \ref{Bdpipi-Alpha}. 
The structure of (\ref{xirhoks}) is 
very similar to that of the observable $\xi^{(d)}_{\pi^+\pi^-}$ given in 
(\ref{xipipi}). However, an important difference arises between 
$B_d\to\pi^+\pi^-$ and $B_s\to \rho^0K_{\mbox{{\scriptsize S}}}$:
although the penguin contributions are expected to be of equal order 
of magnitude in (\ref{xipipi}) and (\ref{xirhoks}), their importance 
is enhanced in the latter case since the current-current amplitude 
${\cal A}_{\mbox{{\scriptsize cc}}}^{u}$ is colour-suppressed
by a phenomenological colour-suppression factor $a_2\approx0.2$
\cite{BSW}-\cite{a1a2-exp}. Consequently, using in addition to that value of 
$a_2$ characteristic Wilson coefficient functions for the penguin operators 
and (\ref{bdckm}) for the ratio of CKM factors, one obtains
\begin{equation}
\left|\frac{v_c^{(d)}{\cal A}^{ct}_{\mbox{{\scriptsize pen}}}}
{v_u^{(d)}\left({\cal A}_{\mbox{{\scriptsize cc}}}^{u}+
{\cal A}^{ut}_{\mbox{{\scriptsize pen}}}\right)}\right|={\cal O}(0.5).
\end{equation}
This estimate implies that
\begin{equation}
\xi^{(s)}_{\rho^0K_{\mbox{{\scriptsize S}}}}\approx+e^{-i0}
\left[\frac{v_u^{(d)}\left({\cal A}_{\mbox{{\scriptsize cc}}}^{u}+
{\cal A}^{ut}_{\mbox{{\scriptsize pen}}}\right)}
{v_u^{(d)\ast}\left({\cal A}_{\mbox{{\scriptsize cc}}}^{u}+
{\cal A}^{ut}_{\mbox{{\scriptsize pen}}}\right)}\right]=e^{-2i\gamma}
\end{equation}
is a {\it very bad} approximation which should {\it not} allow a meaningful 
determination of $\gamma$ from the mixing-induced CP-violating asymmetry
arising in $B_s\to\rho^0K_{\mbox{{\scriptsize S}}}$. Needless to note, 
the branching ratio of that decay is expected to be of ${\cal O}(10^{-7})$
which makes its experimental investigation very difficult. Interestingly
there are other $B_s$ decays -- some of them receive also penguin 
contributions -- which {\it do} allow extractions of $\gamma$. Some of 
these strategies are even theoretically clean and suffer from no hadronic 
uncertainties. Before focussing on these modes, let us discuss an 
experimental problem of $B_s$ decays that is related to time-dependent 
measurements. 

\subsubsection{The $B_s$ System in Light of $\Delta\Gamma_s$}\label{BsiLoDG}
The large mixing parameter $x_s={\cal O}(20)$ that is expected 
within the Standard Model implies very rapid $B^0_s-\overline{B^0_s}$ 
oscillations requiring an excellent vertex resolution system to 
keep track of the $\Delta M_s t$ terms. That is obviously a formidable 
experimental task. It may, however, not be necessary to trace the rapid 
$\Delta M_s t$ oscillations in order to shed light on the mechanism of 
CP violation \cite{dunietz}. This remarkable feature is due 
to the expected sizable width difference $\Delta\Gamma_s$ which has
been discussed at the end of subsection~\ref{B0B0bar-Mix}. 
Because of that width 
difference already {\it untagged} $B_s$ rates, which are defined by 
\begin{equation}\label{untagged}
\Gamma[f(t)]\equiv\Gamma(B_s^0(t)\to f)+\Gamma(\overline{B^0_s}(t)\to f),
\end{equation}
may provide valuable information about the phase structure of the observable
$\xi_f^{(s)}$. This can be seen nicely by rewriting (\ref{untagged}) 
with the help of (\ref{ratebf}) and (\ref{ratebbf}) in a more explicit way 
as follows:
\begin{equation}\label{EE3}
\Gamma[f(t)]\propto\left[\left(1+\left|\xi_f^{(s)}
\right|^2\right)\left(e^{-\Gamma_L^{(s)} t}+e^{-\Gamma_H^{(s)} t}\right)
-2\mbox{\,Re\,}\xi_f^{(s)}\left(e^{-\Gamma_L^{(s)} t}-
e^{-\Gamma_H^{(s)} t}\right)\right].
\end{equation}
In this expression the rapid oscillatory $\Delta M_s t$ terms, which
show up in the {\it tagged} rates (\ref{ratebf}) and (\ref{ratebbf}), 
cancel \cite{dunietz}. Therefore it depends only on the two exponents 
$e^{-\Gamma_L^{(s)} t}$ and $e^{-\Gamma_H^{(s)} t}$. From an experimental 
point of view, such untagged analyses are clearly much more promising than 
tagged ones in respect of efficiency, acceptance and purity. 

In order to illustrate these untagged rates in more detail, let us 
consider an estimate of $\gamma$ using untagged $B_s\to K^+K^-$ and 
$B_s\to K^0\overline{K^0}$ decays that has been proposed 
recently in \cite{fd1}. Using the $SU(2)$ isospin
symmetry of strong interactions to relate the QCD penguin contributions to
these decays (EW penguins are colour-suppressed in these modes
and should therefore play a minor role as we will see in 
section~\ref{Sec-EWP}), 
we obtain
\begin{equation}\label{EE4}
\Gamma[K^+K^-(t)]\propto |P'|^2\Bigl[\bigl(1-2\,|r|\cos\varrho\,
\cos\gamma+|r|^2\cos^2\gamma\bigr)e^{-\Gamma_L^{(s)} t}+|r|^2\sin^2\gamma\, 
e^{-\Gamma_H^{(s)} t}\Bigr]
\end{equation}
and
\begin{equation}\label{EE5}
\Gamma[K^0\overline{K^0}(t)]\propto |P'|^2\,e^{-\Gamma_L^{(s)} t},
\end{equation}
where 
\begin{equation}\label{EE6}
r\equiv|r|e^{i\varrho}=\frac{|T'|}{|P'|}e^{i(\delta_{T'}-\delta_{P'})}.
\end{equation}
Here we have used the same notation as Gronau et al.\ in \cite{grl-gam}
which will turn out to be very useful for later discussions: $P'$ 
denotes the $\bar b\to\bar s$ QCD penguin amplitude corresponding to
(\ref{bspenamp}), $T'$ is the colour-allowed $\bar b\to\bar uu\bar s$ 
current-current amplitude, and $\delta_{P'}$ and $\delta_{T'}$ denote the 
corresponding CP-conserving strong phases. The primes remind us that we are 
dealing with $\bar b\to\bar s$ amplitudes. In order to determine $\gamma$ 
from the untagged rates (\ref{EE4}) and (\ref{EE5}), we need an additional 
input that is provided by the $SU(3)$ flavour symmetry of strong
interactions. Using that symmetry and neglecting as in (\ref{BR-rat})
the colour-suppressed current-current contributions to $B^+\to\pi^+\pi^0$, 
one finds \cite{grl-gam}
\begin{equation}\label{EE7}
|T'|\approx\lambda\,\frac{F_K}{F_\pi}\,\sqrt{2}\,|A(B^+\to\pi^+\pi^0)|,
\end{equation}
where $\lambda$ is the usual Wolfenstein parameter, $F_K/F_\pi$ 
takes into account factorizable $SU(3)$-breaking, and $A(B^+\to\pi^+\pi^0)$ 
denotes the appropriately normalized decay amplitude of $B^+\to\pi^+\pi^0$. 
Since $|P'|$ is known from the untagged $B_s\to K^0\,\overline{K^0}$ rate
(\ref{EE5}), the quantity $|r|=|T'|/|P'|$ can be estimated 
with the help of (\ref{EE7}) and allows the extraction of $\gamma$ from 
the part of (\ref{EE4}) evolving with exponent $e^{-\Gamma_H^{(s)} t}$. 
As we will see in a moment, one can even do better, i.e.\ without using an
$SU(3)$-based estimate like (\ref{EE7}), by considering the decays 
corresponding to $B_s\to K \overline{K}$ where two vector mesons 
or appropriate higher resonances are present in the final states \cite{fd1}.  

\subsubsection{$\gamma$ from $B_s\to K^{\ast+}
K^{\ast-}$ and $B_s\to K^{\ast0}\overline{K^{\ast0}}$}\label{kkbar}
\noindent
The untagged angular distributions of these decays, which take the general 
form
\begin{equation}\label{EE8}
[f(\theta,\phi,\psi;t)]=\sum_k\left[\overline{b^{(k)}}(t)+b^{(k)}(t)\right]
g^{(k)}(\theta,\phi,\psi),
\end{equation}
provide many more observables than the untagged modes
$B_s\to K^+K^-$ and $B_s\to K^0\overline{K^0}$ discussed in \ref{BsiLoDG}. 
Here $\theta$, $\phi$ and $\psi$ are generic decay angles describing 
the kinematics of the decay products arising in the
decay chain $B_s\to K^\ast(\to\pi K)\,\overline{K^\ast}(\to \pi\overline{K})$.
The observables $\left[\overline{b^{(k)}}(t)+b^{(k)}(t)\right]$ governing the
time-evolution of the untagged angular distribution (\ref{EE8}) are given
by real or imaginary parts of bilinear combinations of decay amplitudes
that are of the following structure:
\begin{eqnarray}
\lefteqn{\left[A_{\tilde f}^\ast(t)\,A_f(t)\right]\equiv\left\langle
\left(K^\ast\overline{K^\ast}\right)_{\tilde f}\left|{\cal 
H}_{\mbox{{\scriptsize eff}}}\right|\overline{B_s^0}(t)\right\rangle^\ast
\left\langle\left(K^\ast\overline{K^\ast}\right)_{f}\left|{\cal 
H}_{\mbox{{\scriptsize eff}}}\right|\overline{B_s^0}(t)\right\rangle}
\nonumber\\
&&+\left\langle\left(K^\ast\overline{K^\ast}\right)_{\tilde f}\left|{\cal 
H}_{\mbox{{\scriptsize eff}}}\right|B_s^0(t)\right\rangle^\ast
\left\langle\left(K^\ast\overline{K^\ast}\right)_{f}\left|{\cal 
H}_{\mbox{{\scriptsize eff}}}\right|B_s^0(t)\right\rangle.\label{EE9}
\end{eqnarray}
In this expression, $f$ and $\tilde f$ are labels that define the relative
polarizations of $K^\ast$ and $\overline{K^\ast}$ in final state 
configurations $\left(K^\ast\overline{K^\ast}\right)_f$ (e.g.\ linear 
polarization states \cite{rosner} $\{0,\parallel,\perp\}$) with CP 
eigenvalues $\eta_{\mbox{{\scriptsize CP}}}^f$:
\begin{equation}\label{EE10}
({\cal CP})\left|\left(K^\ast\overline{K^\ast}\right)_f\right\rangle
=\eta_{\mbox{{\scriptsize CP}}}^f\left|
\left(K^\ast\overline{K^\ast}\right)_f\right\rangle.
\end{equation}
An analogous relation holds for $\tilde f$. The observables of the 
angular distributions for $B_s\to K^{\ast+}
K^{\ast-}$ and $B_s\to K^{\ast0}\overline{K^{\ast0}}$
are given explicitly in \cite{fd1}. In the case of the latter decay
the formulae simplify considerably since it is a penguin-induced 
$\bar b\to\bar sd\bar d$ mode and receives therefore no tree contributions.
Using, as in (\ref{EE4}) and (\ref{EE5}), the $SU(2)$ isospin symmetry of 
strong interactions, the QCD penguin contributions to $B_s\to K^{\ast+}
K^{\ast-}$ and $B_s\to K^{\ast0}\overline{K^{\ast0}}$ can be 
related to each other. If one takes into account these relations and goes
very carefully through the observables of the corresponding untagged angular 
distributions, one finds that they allow the extraction of 
$\gamma$ without any additional theoretical input \cite{fd1}. 
In particular no $SU(3)$ symmetry arguments are needed and the $SU(2)$ 
isospin symmetry suffices to accomplish this task. The angular distributions 
provide moreover information about the 
hadronization dynamics of the corresponding decays, and the 
formalism developed for $B_s\to K^{\ast+}K^{\ast-}$ applies 
also to $B_s\to\rho^0\phi$ if one performs a suitable replacement of 
variables \cite{fd1}. Since that channel is expected to be dominated 
by EW penguins as discussed in \ref{Bspi0phi}, it may allow 
interesting insights into the physics of these operators.

\subsubsection{$B_s\to D_s^{\ast+}D_s^{\ast-}$ and $B_s\to J/\psi\,
\phi$: ``Gold-plated'' Transitions to Extract $\eta$}\label{gold}
The following discussion is devoted to an analysis \cite{fd1} of 
the decays $B_s\to D_s^{\ast+}(\to D_s^+\gamma)\,D_s^{\ast-}
(\to D_s^-\gamma)$ and $B_s\to J/\psi(\to l^+l^-)\,\phi(\to K^+K^-)$, 
which is the counterpart of the ``gold-plated'' mode $B_d\to J/\psi\,
K_{\mbox{{\scriptsize S}}}$ to measure $\beta$. Since these decays are 
dominated by a single CKM amplitude, the hadronic uncertainties cancel in 
$\xi_f^{(s)}$ (see \ref{domCKM}) taking in that particular case the 
following form:
\begin{equation}\label{EE11}
\xi_f^{(s)}=-\eta_{\mbox{{\scriptsize CP}}}^f
\,e^{i\,\phi_{\mbox{{\scriptsize CKM}}}}\,.
\end{equation}
Consequently the observables of the untagged angular distributions, which 
have the same general structure as (\ref{EE8}), simplify 
considerably \cite{fd1}. In (\ref{EE11}), $f$ is -- as in (\ref{EE9}) and 
(\ref{EE10}) -- a label defining the 
relative polarizations of $X_1$ and $X_2$ in final state configurations 
$\left(X_1\,X_2\right)_f$ with CP eigenvalue 
$\eta_{\mbox{{\scriptsize CP}}}^f$, where $(X_1,X_2)\in
\left\{(D_s^{\ast+},D_s^{\ast-}),(J/\psi,\phi)\right\}$. Applying 
(\ref{e743}) in combination with (\ref{e742}) and (\ref{e746}), the 
CP-violating weak phase $\phi_{\mbox{{\scriptsize CKM}}}$ would vanish. 
In order to obtain a non-vanishing result for that phase, its exact 
definition is
\begin{equation}\label{phickm}
\phi_{\mbox{{\scriptsize 
CKM}}}\equiv-2\left[\mbox{arg}(V_{ts}^\ast V_{tb})-\mbox{arg}(V_{cs}^\ast 
V_{cb})\right],
\end{equation} 
we have to take into account higher order terms in the Wolfenstein 
expansion of the CKM matrix yielding $\phi_{\mbox{{\scriptsize CKM}}}=
2\lambda^2\eta={\cal O}(0.03)$. Consequently the small weak phase 
$\phi_{\mbox{{\scriptsize CKM}}}$ measures simply $\eta$ which fixes the 
height of the UT. Another interesting interpretation of (\ref{phickm}) is 
the fact that it is related to an angle in a rather squashed and 
therefore ``unpopular'' unitarity triangle \cite{akl}. Other useful 
expressions for (\ref{phickm}) can be found in \cite{snowmass}. Let us
note that the weak phase (\ref{phickm}) is also probed by the decay $B_s\to
J/\psi\, K_{\mbox{{\scriptsize S}}}$ which is the counterpart of the mode
$B_d\to D^+D^-$ discussed briefly in \ref{Bdpipi-Alpha}. Here penguin 
contributions may lead to potential problems. 

A characteristic feature of the angular distributions for 
$B_s\to D_s^{\ast+}D_s^{\ast-}$ and $B_s\to J/\psi\,\phi$ is 
interference between CP-even and CP-odd final state configurations leading 
to untagged observables that are proportional to  
\begin{equation}\label{EE12}
\left(e^{-\Gamma_L^{(s)}t}-e^{-\Gamma_H^{(s)}t}
\right)\sin\phi_{\mbox{{\scriptsize CKM}}}.
\end{equation}
As was shown in \cite{fd1}, the angular distributions for both the 
colour-allowed channel $B_s\to D_s^{\ast+} D_s^{\ast-}$ and the 
colour-suppressed transition $B_s\to J/\psi\,\phi$ each provide separately
sufficient information to determine $\phi_{\mbox{{\scriptsize CKM}}}$ from 
their untagged data samples. The extraction of 
$\phi_{\mbox{{\scriptsize CKM}}}$ is, however, not as clean 
as that of $\beta$ from $B_d\to J/\psi\,K_{\mbox{{\scriptsize S}}}$. 
Although the unmixed amplitudes proportional to the CKM factor 
$V_{us}^\ast V_{ub}$ are similarly suppressed in both cases, the 
smallness of $\phi_{\mbox{{\scriptsize CKM}}}$ with respect to $\beta$ 
enhances the importance of this contribution for extracting 
$\phi_{\mbox{{\scriptsize CKM}}}$.

Within the Standard Model one expects a very small value of 
$\phi_{\mbox{{\scriptsize 
CKM}}}$ and $\Gamma_H^{(s)}<\Gamma_L^{(s)}$. However, that need not to
be the case in many scenarios for ``New Physics'' (see e.g.\ \cite{yg}).
An experimental study of the decays $B_s\to D_s^{\ast+}D_s^{\ast-}$ 
and $B_s\to J/\psi\,\phi$ may shed light on this issue \cite{fd1}, and an
extracted value of $\phi_{\mbox{{\scriptsize CKM}}}$ that is much larger than
${\cal O}(0.03)$ would most probably signal physics beyond the Standard Model.

\subsubsection{Clean Extractions of $\gamma$ using $B_s$ Decays caused by 
$\bar b\to\bar uc\bar s$ ($b\to c\bar us$)}\label{nonCP}
Exclusive $B_s$ decays caused by $\bar b\to\bar uc\bar s$ ($b\to c\bar u 
s$) quark-level transitions belong to decay class iii) introduced in 
Subsection~\ref{Class-Ham}, i.e.\ are pure tree decays receiving 
{\it no} penguin contributions, and probe the angle $\gamma$ \cite{gam}. 
Their transition amplitudes can be expressed as hadronic 
matrix elements of low energy effective Hamiltonians having the 
following structures \cite{fd2}:
\begin{eqnarray}
{\cal H}_{\mbox{{\scriptsize eff}}}(\overline{B^0_s}\to f)&=&
\frac{G_{\mbox{{\scriptsize F}}}}{\sqrt{2}}\,\overline{v}
\left[\overline{O}_1\,{\cal C}_1(\mu)+
\overline{O}_2\,{\cal C}_2(\mu)\right]\\
{\cal H}_{\mbox{{\scriptsize eff}}}(B^0_s\to f)&=&
\frac{G_{\mbox{{\scriptsize F}}}}{\sqrt{2}}\,v^\ast
\left[O_1^\dagger\,{\cal C}_1(\mu)+O_2^\dagger\,{\cal C}_2(\mu)\right].
\end{eqnarray}
Here $f$ denotes a final state with valence-quark content 
$s\bar u\, c\bar s$, the relevant CKM factors take the form
\begin{equation}
\overline{v}\equiv V_{us}^\ast V_{cb}=A\lambda^3,\quad
v\equiv V_{cs}^\ast V_{ub}=A\lambda^3R_b\,e^{-i\gamma},
\end{equation} 
where the modified Wolfenstein parametrization (\ref{wolf2}) has been used, 
and $\overline{O}_k$ and $O_k$ denote current-current operators (see 
(\ref{O1})) that are given by
\begin{equation}
\begin{array}{rclrcl}
\overline{O}_1&=&\left(\bar s_\alpha u_\beta\right)_{\mbox{{\scriptsize 
V--A}}}\left(\bar c_\beta b_\alpha\right)_{\mbox{{\scriptsize V--A}}},
\quad&\overline{O}_2&=&\left(\bar s_\alpha u_\alpha\right)_{\mbox{{\scriptsize 
V--A}}}\left(\bar c_\beta b_\beta\right)_{\mbox{{\scriptsize V--A}}},
\label{O-cc}\\
O_1&=&\left(\bar s_\alpha c_\beta\right)_{\mbox{{\scriptsize V--A}}}
\left(\bar u_\beta b_\alpha\right)_{\mbox{{\scriptsize V--A}}},\quad&
O_2&=&\left(\bar s_\alpha c_\alpha\right)_{\mbox{{\scriptsize V--A}}}
\left(\bar u_\beta b_\beta\right)_{\mbox{{\scriptsize V--A}}}\label{Obar-cc}.
\end{array}
\end{equation}
Nowadays the Wilson coefficient functions ${\cal C}_1(\mu)$ and 
${\cal C}_2(\mu)$ are available at NLO and the corresponding results can 
be found in \cite{BBL,ALTA,BW} and in table~\ref{tab:XXX}.

Performing appropriate CP transformations in the matrix element
\begin{eqnarray}
\lefteqn{\left\langle f\left|O_1^\dagger(\mu){\cal C}_1(\mu)+
O_2^\dagger(\mu){\cal C}_2(\mu)\right|B_s^0\right\rangle}
\nonumber\\
&&=\left\langle f\left|({\cal CP})^\dagger({\cal CP})\left[O_1^\dagger(\mu)
{\cal C}_1(\mu)+O_2^\dagger(\mu){\cal C}_2(\mu)\right]({\cal CP})^\dagger
({\cal CP})\right|B_s^0\right\rangle\\
&&=e^{i\phi_{\mbox{{\scriptsize CP}}}(B_s)}\,\Bigl\langle\overline{f}\Bigl|
O_1(\mu){\cal C}_1(\mu)+O_2(\mu){\cal C}_2(\mu)\Bigr|
\overline{B^0_s}\Bigr\rangle,
\nonumber
\end{eqnarray}
where (\ref{e710}) and the analogue of (\ref{e738}) have been taken into
account, gives
\begin{eqnarray}
A(\overline{B^0_s}\to f)&=&\left\langle f\left|{\cal 
H}_{\mbox{{\scriptsize eff}}}(\overline{B^0_s}\to f)\right|
\overline{B^0_s}\right\rangle=\frac{G_{\mbox{{\scriptsize F}}}}{\sqrt{2}}
\,\overline{v}\,\,\overline{M}_f\\
A(B^0_s\to f)&=&\left\langle f\left|{\cal H}_{\mbox{{\scriptsize eff}}}
(B^0_s\to f)\right|B^0_s\right\rangle\,=\,e^{i\phi_{\mbox{{\scriptsize 
CP}}}(B_s)}\frac{G_{\mbox{{\scriptsize F}}}}{\sqrt{2}}\,v^\ast
M_{\overline{f}}
\end{eqnarray}
with the strong hadronic matrix elements
\begin{eqnarray}
\overline{M}_f&\equiv&\Bigl\langle f\Bigl|\overline{O}_1(\mu){\cal C}_1(\mu)+
\overline{O}_2(\mu){\cal C}_2(\mu)\Bigr|\overline{B^0_s}\Bigr\rangle\\
M_{\overline{f}}&\equiv&\Bigl\langle\overline{f}\Bigl|O_1(\mu){\cal C}_1(\mu)+
O_2(\mu){\cal C}_2(\mu)\Bigr|\overline{B^0_s}\Bigr\rangle.
\end{eqnarray}
Consequently, using in addition (\ref{e717}) and (\ref{e742}), the 
observable $\xi_f^{(s)}$ defined in (\ref{e731}) is given by
\begin{equation}\label{EXIF}
\xi_f^{(s)}=-e^{-i\phi_{\mbox{{\scriptsize M}}}^{(s)}}
\frac{\overline{v}}{v^\ast}\frac{\overline{M}_f}{M_{\overline f}}
=-e^{-i\gamma}\frac{1}{R_b}\frac{\overline{M}_f}{M_{\overline f}}\,.
\end{equation}
Note that $\phi_{\mbox{{\scriptsize CP}}}(B_s)$ cancels in (\ref{EXIF})
which is a nice check. An analogous calculation yields
\begin{equation}
\xi_{\overline{f}}^{(s)}=-e^{-i\phi_{\mbox{{\scriptsize M}}}^{(s)}}
\frac{v}{\overline{v}^\ast}\frac{M_{\overline{f}}}{\overline{M}_f}
=-e^{-i\gamma}R_b\,\frac{M_{\overline f}}{\overline{M}_f}.
\end{equation}
If one measures the tagged time-dependent decay rates 
(\ref{ratebf})-(\ref{ratebbfb}), both $\xi_f^{(s)}$ and 
$\xi_{\overline{f}}^{(s)}$ can be determined and allow a {\it theoretically 
clean} determination of $\gamma$ since
\begin{equation}
\xi_f^{(s)}\cdot\xi_{\overline{f}}^{(s)}=e^{-2i\gamma}.
\end{equation}

There are by now well-known strategies on the market using time-evolutions 
of $B_s$ modes originating from $\bar b\to\bar uc\bar s$ ($b\to c\bar us$) 
quark-level transitions, e.g.\ $\stackrel{{\mbox{\tiny (---)}}}{B_s}
\to\stackrel{{\mbox{\tiny (---)}}}{D^0}\phi$ \cite{gam,glgam} and 
$\stackrel{{\mbox{\tiny (---)}}}{B_s}\to D_s^\pm K^\mp$ \cite{adk}, 
to extract $\gamma$. However, as we have noted already, in these methods 
tagging is essential and the rapid $\Delta M_s t$ oscillations have to be 
resolved which is an experimental challenge. The question what can be 
learned from {\it untagged} data samples of these decays, where the 
$\Delta M_s t$ terms cancel, has been investigated by Dunietz \cite{dunietz}. 
In the untagged case the determination of $\gamma$ requires additional 
inputs: 
\begin{itemize}
\item Colour-suppressed modes $\stackrel{{\mbox{\tiny (---)}}}{B_s}\to
\stackrel{{\mbox{\tiny (---)}}}{D^0}\phi$: a measurement of the 
untagged $B_s\to D^0_{\pm} \phi$ rate is needed, where 
$D^0_{\pm}$ is a CP eigenstate of the neutral $D$ system.
\item Colour-allowed modes $\stackrel{{\mbox{\tiny (---)}}}{B_s}\to 
D_s^\pm K^\mp$: a theoretical input corresponding to the ratio of the 
unmixed rates $\Gamma(B^0_s\to D_s^-K^+)/\Gamma(B^0_s\to D_s^-\pi^+)$ is
needed. This ratio can be estimated with the help of the ``factorization'' 
hypothesis \cite{fact1,fact2} which may work reasonably 
well for these colour-allowed channels \cite{bjorken}.
\end{itemize}
Interestingly the untagged data samples may exhibit CP-violating 
effects that are described by observables of the form
\begin{equation}\label{EE13}
\Gamma[f(t)]-\Gamma[\overline{f}(t)]\propto\left(e^{-\Gamma_L^{(s)}t}-
e^{-\Gamma_H^{(s)}t}\right)\sin\varrho_f\,\sin\gamma.
\end{equation}
Here $\varrho_f$ is a CP-conserving strong phase. Because of the 
$\sin\varrho_f$ factor, the CP-violating observables (\ref{EE13}) 
vanish within the factorization approximation predicting 
$\varrho_f\in\{0,\pi\}$. 
Since factorization may be a reasonable working assumption 
for the colour-allowed modes $\stackrel{{\mbox{\tiny 
(---)}}}{B_s}\to D_s^\pm K^\mp$, the CP-violating effects in their
untagged data samples are expected to be tiny. On the other hand,
the factorization hypothesis is very questionable for 
the colour-suppressed decays $\stackrel{{\mbox{\tiny (---)}}}{B_s}\to
\stackrel{{\mbox{\tiny (---)}}}{D^0}\phi$ and sizable CP violation may 
show up in the corresponding untagged rates \cite{dunietz}. 

Concerning such CP-violating effects and the extraction of $\gamma$ from
untagged rates, the decays $\stackrel{{\mbox{\tiny (---)}}}{B_s}\to 
D_s^{\ast\pm} K^{\ast\mp}$ and $\stackrel{{\mbox{\tiny (---)}}}{B_s}\to
\stackrel{{\mbox{\tiny(---)}}}{D^{\ast0}}\phi$ are expected to be 
more promising than the transitions discussed above. As was
shown in \cite{fd2}, the time-dependences of their untagged angular
distributions allow a clean extraction of $\gamma$ without any additional 
input. The final state configurations of these decays are not admixtures of 
CP eigenstates as in the case of the decays discussed in \ref{kkbar} and 
\ref{gold}. They can, however, be classified by their parity eigenvalues. 
A characteristic feature of the corresponding angular distributions 
is interference between parity-even and parity-odd configurations that may
lead to potentially large CP-violating effects in the untagged data samples
even when all strong phase shifts vanish. An example of such an
untagged CP-violating observable is the following quantity \cite{fd2}:
\begin{eqnarray}
\lefteqn{\mbox{Im}\left\{\left[A_f^\ast(t)\,A_\perp(t)\right]\right\}+
\mbox{Im}\left\{\left[A_f^{\mbox{{\scriptsize C}}\ast}(t)\,
A_\perp^{\mbox{{\scriptsize C}}}(t)\right]\right\}}\nonumber\\
&&\propto\left(e^{-\Gamma_L^{(s)}t}-e^{-\Gamma_H^{(s)}t}\right)
\bigl\{|R_f|\cos(\delta_f-\vartheta_\perp)+|R_\perp|
\cos(\delta_\perp-\vartheta_f)\bigr\}\,\sin\gamma.\label{EE14}
\end{eqnarray}
In that expression bilinear combinations of certain decay amplitudes 
(see (\ref{EE9})) show up, $f\in\{0,\parallel\}$ denotes a 
linear polarization state \cite{rosner} and $\delta_f$, $\vartheta_f$
are CP-conserving phases that are induced through strong final 
state interaction effects. For the details concerning the 
observable (\ref{EE14}) -- in particular the definition of the 
relevant charge-conjugate amplitudes $A_f^{\mbox{{\scriptsize C}}}$ and the 
quantities $|R_f|$ -- the reader is referred to \cite{fd2}. 
Here we would like to emphasize only that the strong phases
enter in the form of {\it cosine} terms. Therefore non-trivial 
strong phases are -- in contrast to (\ref{EE13}) -- not essential for 
CP violation in the corresponding untagged data samples and one
expects, even within the factorization approximation, which may apply to
the colour-allowed modes $\stackrel{{\mbox{\tiny (---)}}}{B_s}\to 
D_s^{\ast\pm} K^{\ast\mp}$, potentially large effects. 
 
Since the soft photons in the decays $D_s^\ast\to D_s\gamma$, $D^{\ast0}
\to D^0\gamma$ are difficult to detect, certain higher resonances exhibiting
significant all-charged final states, e.g.\ $D_{s1}(2536)^+\to
D^{\ast+}K^0$, $D_1(2420)^0\to D^{\ast+}\pi^-$ with $D^{\ast+}\to 
D^0\pi^+$, may be more promising for certain detector configurations. 
A similar comment applies also to the mode $B_s\to D_s^{\ast+}D_s^{\ast-}$
discussed in \ref{gold}.

To finish the presentation of the $B_s$ system, let us stress once
again that the untagged measurements discussed in this subsection are 
much more promising in view of efficiency, acceptance and purity than 
tagged analyses. Moreover the oscillatory $\Delta M_st$ terms, which
may be too rapid to be resolved with present vertex technology, cancel
in untagged $B_s$ data samples. However, a lot of statistics is required 
and the natural place for these experiments seems to be a hadron collider
(note that the formulae given above have to be modified appropriately for
$e^+-e^-$ machines to take into account coherence of the 
$B^0_s-\overline{B^0_s}$ pair at $\Upsilon(5\mbox{S})$). 
Obviously the feasibility of untagged strategies to extract CKM phases
depends crucially on a sizable width difference $\Delta\Gamma_s$. Even 
if it should turn out to be too small for such untagged analyses, 
once $\Delta\Gamma_s\not=0$ has been established experimentally, 
the formulae developed in \cite{fd1,fd2} have also 
to be used to determine CKM phases correctly from tagged measurements. 
Clearly time will tell and experimentalists will certainly find out which 
method is most promising from an experimental point of view. 

\subsubsection{Inclusive Decays}
So far we have considered only {\it exclusive} neutral $B_q$-meson decays. 
However, also {\it inclusive} decay processes with specific quark-flavours, 
e.g.\ $\bar b\to\bar uu\bar d$ or $\bar b\to\bar cc\bar s$, may exhibit 
mixing-induced CP-violating asymmetries \cite{ds}. Recently the determination 
of $\sin(2\alpha)$ from the CP asymmetry arising in inclusive $B_d$ decays 
into charmless final states has been analyzed by assuming local quark-hadron
duality \cite{bbd-inclCP}. Compared to exclusive 
transitions, inclusive decay processes have of course rates that are 
larger by orders of magnitudes. However, due to the summation over processes 
with asymmetries of alternating signs, the inclusive CP asymmetries are 
unfortunately diluted with respect to the exclusive case. The calculation
of the dilution factor suffers in general from large hadronic uncertainties.
Progress has been made in \cite{bbd-inclCP}, where local quark-hadron 
duality has been used to evaluate this quantity. From an experimental point 
of view, inclusive measurements, e.g.\ of inclusive $B_d^0$ decays caused by 
$\bar b\to\bar uu\bar d$, are unfortunately very difficult.

\subsection{The Charged $B$ System}
Since mixing-effects are absent in the charged $B$-meson system, 
non-vanishing CP-violating asymmetries of charged $B$ decays would 
give unambiguous evidence for direct CP violation. Due to the unitarity of 
the CKM matrix, the transition amplitude of a charged $B$ decay can be 
written in the following general form:
\begin{equation}\label{ED81}
A(B^-\to f)=v_{1}\,A_{1}\,e^{i\alpha_{1}}+v_{2}\,A_{2}\,e^{i\alpha_{2}},
\end{equation}
where $v_{1}$, $v_{2}$ are CKM factors, $A_{1}$, $A_{2}$ are ``reduced'',
i.e.\ real, hadronic matrix elements of weak transition operators and
$\alpha_{1}$, $\alpha_{2}$ denote CP-conserving phases generated 
through strong final state interaction effects. On the other hand, the 
transition amplitude of the CP-conjugate decay $B^+\to\overline{f}$ is 
given by
\begin{equation}\label{ED82}
A(B^+\to\overline{f})=v_{1}^{\ast}A_{1}\,e^{i\alpha_{1}}+v_{2}^{\ast}\,
A_{2}\,e^{i\alpha_{2}}.
\end{equation}
If the CP-violating asymmetry of the decay $B\to f$ is defined through
\begin{equation}\label{ED83}
{\cal A}_{\mbox{{\scriptsize CP}}}\equiv\frac{\Gamma(B^+\to\overline{f})-
\Gamma(B^-\to f)}{\Gamma(B^+\to\overline{f})+\Gamma(B^-\to f)},
\end{equation}
the transition amplitudes (\ref{ED81}) and (\ref{ED82}) yield
\begin{equation}\label{ED84}
{\cal A}_{\mbox{{\scriptsize CP}}}=\frac{2\,\mbox{Im}(v_{1}v_{2}^\ast)
\sin(\alpha_{1}-\alpha_{2})A_{1}A_{2}}{|v_{1}|^{2}A_{1}^{2}
+|v_{2}|^{2}A_{2}^{2}+2\,\mbox{Re}(v_{1}
v_{2}^\ast)\cos(\alpha_{1}-\alpha_{2})A_{1}A_{2}}.
\end{equation}
Consequently there are two conditions that have to be met
simultaneously in order to get a non-zero CP asymmetry 
${\cal A}_{\mbox{{\scriptsize CP}}}$:
\begin{itemize}
\item[i)] There has to be a relative {\it CP-violating} weak phase, i.e.\
$\mbox{Im}(v_1 v_2^\ast)\not=0$, between the two amplitudes contributing 
to $B\to f$. This phase difference can be expressed in terms of
complex phases of CKM matrix elements and is thus calculable.
\item[ii)] There has to be a relative {\it CP-conserving} strong phase, 
i.e.\ $\sin(\alpha_{1}-\alpha_{2})\not=0$, generated by strong final 
state interaction effects. In contrast to the CP-violating weak phase 
difference, the calculation of $\alpha_{1}-\alpha_{2}$ is very involved 
and suffers in general from large theoretical uncertainties.
\end{itemize}
These general requirements for the appearance of direct CP violation apply 
of course also to neutral $B_q$ decays, where direct CP
violation shows up as ${\cal A}_{\mbox{{\scriptsize 
CP}}}^{\mbox{{\scriptsize dir}}}\not=0$ (see (\ref{acpdir})).

Semileptonic decays of charged $B$-mesons obviously do not fulfil 
point i) and exhibit therefore no CP violation within the Standard Model. 
However, there are non-leptonic modes of charged $B$-mesons 
corresponding to decay classes i) and ii) introduced in 
Subsection~\ref{Class-Ham} 
that are very promising in respect of direct CP violation. In decays 
belonging to class i), e.g.\ in $B^+\to\pi^0K^+$,
non-zero CP asymmetries (\ref{ED83}) may arise 
from interference between current-current and penguin operator contributions, 
while non-vanishing CP-violating effects may be generated in the pure 
penguin-induced decays of class ii), e.g.\ in $B^+\to K^+\overline{K^0}$, 
through interference between penguins with internal up- and charm-quark 
exchanges (see \ref{Zoo}).

In the case of $\bar b\to \bar cc\bar s$ modes, e.g.\ $B^+\to J/\psi\,
K^+$, {\it vanishing} CP violation can be predicted to excellent 
accuracy within the Standard Model because of the arguments given 
in \ref{Beta-Ext}, 
where the ``gold-plated'' mode $B_d\to J/\psi\,K_{\mbox{{\scriptsize S}}}$ 
has been discussed exhibiting the same decay structure. In general, however, 
the CP-violating asymmetries (\ref{ED84}) suffer from large theoretical 
uncertainties arising in particular from the strong final state 
interaction phases $\alpha_1$ and $\alpha_2$. Therefore in general 
CP violation
in charged $B$ decays does not allow a clean determination
of CKM phases. The theoretical situation is a bit similar to
Re$(\varepsilon'/\varepsilon)$ discussed in section~\ref{EpsilonPrime}, 
and the major goal of a possible future measurement of non-zero CP 
asymmetries in charged $B$ decays is related to the fact that these effects 
would immediately rule out ``superweak'' models of CP violation 
\cite{wolfenstein:64}.
A detailed discussion of the corresponding calculations, which are rather 
technical, is beyond the scope of this review and the interested reader
is referred to \cite{rf1}-\cite{kps}, \cite{gh}, \cite{siwy}-\cite{remt} 
where further references can be found. 

Concerning theoretical cleanliness, there is, however, an important 
exception. In respect of extracting $\gamma$, charged $B$ decays belonging 
to decay class iii), i.e.\ pure tree decays, play an outstanding role. 
Using certain triangle relations among their decay amplitudes, a 
theoretical clean determination of this angle is possible. Let us discuss
this in explicit terms. 

\subsection{Relations among Non-leptonic $B$ Decay Amplitudes}
During recent years, relations among amplitudes of non-leptonic $B$ decays
have been very popular to develop strategies for extracting UT angles,
in particular for the ``hard'' angle $\gamma$. There are both {\it exact} 
relations and {\it approximate} relations which are based on the $SU(3)$ 
flavour symmetry of strong interactions and certain plausible dynamical 
assumptions. Let us turn to the ``prototype'' of this approach first. 

\subsubsection{$B\to DK$ Triangles}\label{BDKtri}
Applying an appropriate CP phase convention to simplify the following
discussion, the CP eigenstates $|D^0_\pm\rangle$ of the neutral $D$-meson 
system with CP eigenvalues $\pm1$ are given by
\begin{equation}\label{ED85}
\left|D^0_\pm\right\rangle=\frac{1}{\sqrt{2}}\left(\left|D^0\right
\rangle\pm\left|\overline{D^0}\right\rangle\right),
\end{equation}
so that the $B^\pm\to D^0_+K^\pm$ transition amplitudes can be expressed
as \cite{gw}
\begin{eqnarray}
\sqrt{2}A(B^+\to D^0_+K^+)&=&A(B^+\to D^0 K^+)+A(B^+\to\overline{D^0}K^+)
\label{ED86}\\
\sqrt{2}A(B^-\to D^0_+K^-)&=&A(B^-\to\overline{D^0} K^-)+A(B^-\to D^0K^-).
\label{ED87}
\end{eqnarray}
These relations, which are valid {\it exactly}, can be represented as two 
triangles in the complex plane. Taking into account that the $B^+\to DK^+$ 
decays originate from $\bar b\to\bar uc\bar s,\, \bar c u\bar s$ quark-level 
transitions yields
\begin{eqnarray}
A(B^+\to D^0K^+)&=&e^{i\gamma}\lambda\,|V_{cb}|R_b\,|a|\,e^{i\Delta_a}\,=\,
e^{2i\gamma}\,A(B^-\to\overline{D^0}K^-)\label{ED88}\\
A(B^+\to\overline{D^0}K^+)&=&\lambda\,|V_{cb}||A|\,e^{i\Delta_A}\,=\,A(B^-\to 
D^0K^-)\label{ED89},
\end{eqnarray}
where $|a|$, $|A|$ are magnitudes of hadronic matrix elements of the 
current-current operators (\ref{O-cc}) and $\Delta_a$, $\Delta_A$ 
denote the corresponding CP-conserving strong phases. Consequently 
the modes $B^+\to D^0 K^+$ and $B^+\to\overline{D^0}K^+$ exhibit no 
CP-violating effects. However, since the 
requirements for direct CP violation discussed in the previous 
subsection are fulfilled in the $B^\pm\to D^0_+K^\pm$ case because 
of (\ref{ED86}), (\ref{ED87}) and (\ref{ED88}), (\ref{ED89}), we expect
\begin{equation}\label{ED810}
|A(B^+\to D^0_+K^+)|\not=|A(B^-\to D^0_+K^-)|,
\end{equation}
i.e.\ non-vanishing CP violation in that charged $B$ decay. 

\begin{figure}[t]
\centerline{
\epsfysize=4.8truecm
\epsffile{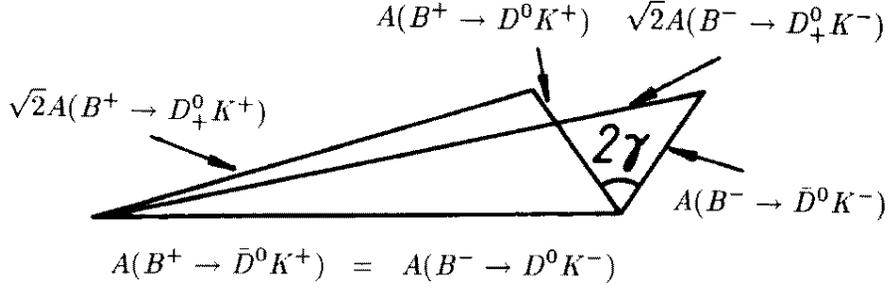}
}
\caption{Triangle relations among $B^\pm\to DK^\pm$ decay 
amplitudes.}\label{BDK-triangles}
\end{figure}

Combining all these considerations, we conclude that the triangle
relations (\ref{ED86}) and (\ref{ED87}), which are depicted in 
fig.~\ref{BDK-triangles}, can be used to extract $\gamma$ by measuring 
only the rates of the corresponding six processes. This approach was
proposed by Gronau and Wyler in \cite{gw}. It is theoretically clean 
and suffers from no hadronic uncertainties. Unfortunately the 
triangles are expected to be very squashed ones since $B^+\to D^0 K^+$
is both colour- and CKM-suppressed with respect to $B^+\to\overline{D^0}K^+$:
\begin{equation}
\frac{|A(B^+\to D^0K^+)|}{|A(B^+\to\overline{D^0}K^+)|}=
R_b\,\frac{|a|}{|A|}\approx0.36\,\frac{a_2}{a_1}\approx0.08.
\end{equation}
Here $a_1$, $a_2$ are the usual phenomenological 
colour-factors \cite{BSW,nrsx} satisfying $a_2/a_1=0.26\pm0.05\pm0.09$
\cite{a1a2-exp}. Using the $SU(3)$ flavour symmetry, 
the corresponding branching ratios can be estimated from the measured 
value $(5.3\pm0.5)\cdot10^{-3}$ \cite{PDG} of 
$Br(B^+\to\overline{D^0}\pi^+)$ to be
$Br(B^+\to\overline{D^0}K^+)\approx4\cdot10^{-4}$ and 
$Br(B^+\to D^0 K^+)\approx2\cdot10^{-6}$. While the former branching ratio
can be measured using conventional methods, the latter one suffers from
considerable experimental problems. The point is that if $Br(B^+\to D^0 
K^+)$ is measured using hadronic tags of the $D^0$, e.g.\ $D^0\to K^-\pi^+$,
one has to deal with large interference effects of ${\cal O}(1)$ with
the $\overline{D^0}$ channel, e.g.\ $B^+\to K^+\overline{D^0}[\to K^-\pi^+]$,
as has been pointed out recently \cite{atduso}. That problem is not present
in the semi-leptonic decay $D^0\to l^+\nu_l X_s$. However, here one has
to deal with huge backgrounds, e.g.\ from $B^+\to l^+\nu_lX_c$, which are
${\cal O}(10^{-6})$ larger and may be difficult to become under control 
\cite{atduso}.
Another problem is related to the CP eigenstate of the neutral $D$ system.
It is detected through $D^0_+\to\pi^+\pi^-,K^+K^-,\ldots$ and is 
experimentally challenging since the corresponding $Br\times$(detection 
efficiency) is expected to be at most of ${\cal O}(1\%)$. Therefore the 
Gronau-Wyler method \cite{gw} will unfortunately be very difficult from 
the experimental point of view. 
  
A variant of the clean determination of $\gamma$ discussed above was 
proposed by Dunietz in \cite{isi} and uses the decays $B^0_d\to D^0_+ 
K^{\ast0}$, $B^0_d\to\overline{D^0} K^{\ast0}$, $B^0_d\to D^0 
K^{\ast0}$ and their charge-conjugates. Since these modes 
are ``self-tagging'' through $K^{\ast0}\to K^+\pi^-$, no time-dependent 
measurements are needed in this method although neutral $B_d$ decays
are involved. Compared to the Gronau-Wyler approach \cite{gw}, both 
$B^0_d\to\overline{D^0} K^{\ast0}$ and $B^0_d\to D^0 K^{\ast0}$ are 
colour-suppressed, i.e.\
\begin{equation}
\frac{|A(B^0_d\to D^0K^{\ast0})|}{|A(B^0_d\to\overline{D^0}K^{\ast0})|}
\approx R_b\,\frac{a_2}{a_2}\approx0.36.
\end{equation}
Consequently the amplitude triangles are probably not as as squashed as 
in the $B^\pm\to D K^\pm$ case. The corresponding branching ratios are 
expected to be of ${\cal O}(10^{-5})$. Unfortunately one has also to 
deal with the difficulties of detecting the neutral $D$-meson 
CP eigenstate $D^0_+$.

That problem is not present in the approach to extract $\gamma$ proposed
in an interesting recent paper \cite{atduso}, where the decay chains
$B^-\to K^-D^0\,[\to f]$ and $B^-\to K^-\overline{D^0}\,[\to f]$ with
$f$ denoting a doubly Cabibbo-suppressed (Cabibbo-favoured) non-CP mode 
of $D^0$ ($\overline{D^0}$) were considered. Examples of such decays are 
$f\in\{K^+\pi^-,K\pi\pi\}$. In contrast to $B^-\to D^0_+K^-$ discussed
above, here both contributing decay amplitudes should be of comparable
size and potentially large CP-violating asymmetries proportional to the 
rate difference $Br(B^+\to K^+[\overline{f}])-Br(B^-\to K^-[f])$ are 
expected. Since several hadronic final states $f$ of neutral $D$ mesons
with different strong phases can be considered, the difficult to measure
branching ratio $Br(B^+\to K^+D^0)$ is not required in order
to extract $\gamma$. Rather both $Br(B^+\to K^+D^0)/
Br(B^+\to K^+\overline{D^0})$ and $\gamma$ can in principle be determined. 
To this end an accurate measurement of the relevant $D$ branching ratios is 
also very desirable. For the details of this approach the reader is 
referred to \cite{atduso}. 

\subsubsection{$SU(3)$ Amplitude Relations}\label{SU3REL}
In a series of interesting papers \cite{grl-gam,ghlr}, Gronau, Hern{\'a}ndez, 
London and Rosner (GHLR) pointed out that the $SU(3)$ flavour symmetry of 
strong interactions \cite{su3-sym} -- which appeared already several times
in this review -- can be combined with certain plausible dynamical 
assumptions, e.g.\ neglect of annihilation topologies, to derive 
amplitude relations among $B$ decays into $\pi\pi$, $\pi K$ and 
$K\overline{K}$ final states. These relations may allow determinations
both of weak phases of the CKM matrix and of strong final state interaction
phases by measuring {\it only} the corresponding branching ratios. 

In order to illustrate this approach, let us describe briefly 
the ``state of the art'' one had about 3 years ago. At that time it 
was assumed that EW penguins should play a very minor role in non-leptonic 
$B$ decays and consequently their contributions were not taken into account.  
Within that approximation, which will be analyzed very carefully in 
section~\ref{Sec-EWP}, the decay 
amplitudes for $B\to\{\pi\pi,\pi K,K\overline{K}\}$ transitions can be 
represented in the limit of an exact $SU(3)$ flavour symmetry in terms 
of five reduced matrix elements. This decomposition can also be performed 
in terms of diagrams. At the quark-level one finds six different topologies 
of Feynman diagrams contributing to $B\to\{\pi\pi,\pi K,K\overline{K}\}$ 
that show up in the corresponding decay amplitudes only as five independent 
linear combinations \cite{grl-gam,ghlr}. In contrast to the 
classification of non-leptonic $B$ decays performed in 
Subsection~\ref{Class-Ham}, these 
six topologies of Feynman diagrams include also three non-spectator
diagrams, i.e.\ annihilation processes, where the decaying $b$-quark
interacts with its partner anti-quark in the $B$-meson. However, due
to dynamical reasons, these three contributions are expected to be
suppressed relative to the others and hence should play a very minor role. 
Consequently, neglecting these diagrams, $6-3=3$ topologies of 
Feynman diagrams suffice to represent the transition amplitudes of $B$ 
decays into $\pi\pi$, $\pi K$ and $K\overline{K}$ final states. To be 
specific, these diagrams describe ``colour-allowed'' and ``colour-suppressed'' 
current-current processes $T$ $(T')$ and $C$ $(C')$, respectively, and QCD 
penguins $P$ $(P')$. As in \cite{grl-gam,ghlr} and in \ref{BsiLoDG}, an 
unprimed amplitude denotes strangeness-preserving decays, whereas a primed 
amplitude stands for strangeness-changing transitions. Note that the 
colour-suppressed topologies $C$ and $C'$ involve the colour-suppression 
factor $a_2\approx0.2$ \cite{BSW}-\cite{a1a2-exp}.

\begin{figure}[t]
\centerline{
\epsfxsize=12.4truecm
\epsffile{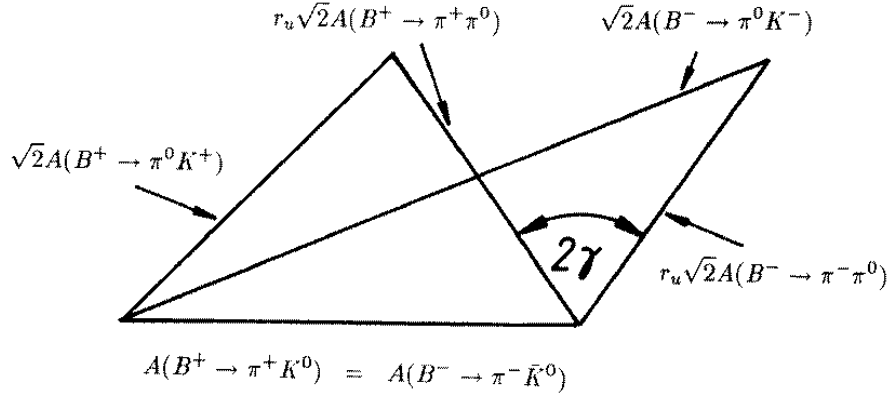}
}
\caption{Naive $SU(3)$ triangle relations among 
$B^+\to\{\pi^+\pi^0,\pi^+K^0,\pi^0K^+\}$ and charge-conjugate 
decay amplitudes {\it neglecting} EW penguin contributions.}\label{grl-const}
\end{figure}

Let us consider the decays $B^+\to\{\pi^+\pi^0,\pi^+ K^0,\pi^0 K^+\}$, 
i.e.\ the ``original'' GRL method \cite{grl-gam}, as an example.
Neglecting both EW penguins, which will be discussed later, and the 
dynamically suppressed non-spectator contributions mentioned above, the 
decay amplitudes of these modes can be expressed as
\begin{equation}\label{ED1201}
\begin{array}{rcl}
\sqrt{2}\,A(B^+\to\pi^+\pi^0)&=&-\left(T+C\right)\\
A(B^+\to\pi^+ K^0)&=&P'\\
\sqrt{2}\,A(B^+\to\pi^0 K^+)&=&-\left(T'+C'+P'\right)
\end{array}
\end{equation}
with
\begin{equation}\label{ED1202}
T=|T|\,e^{i\gamma}\,e^{i\delta_T},\quad
C=|C|\,e^{i\gamma}\,e^{i\delta_C}.
\end{equation}
Here $\delta_T$ and $\delta_C$ denote CP-conserving strong phases. 
Using the $SU(3)$ flavour symmetry, the strangeness-changing amplitudes 
$T'$ and $C'$ can be obtained easily from the strangeness-preserving ones 
through
\begin{equation}\label{ED1203}
\frac{T'}{T}\approx\frac{C'}{C}\approx\lambda\frac{F_K}{F_\pi}\equiv r_u,
\end{equation}
where $F_K$ and $F_\pi$ take into account factorizable $SU(3)$-breaking
corrections as in (\ref{EE7}). The structures of the $\bar b\to\bar d$
and $\bar b\to\bar s$ QCD penguin amplitudes $P$ and $P'$ corresponding 
to $P^{(d)}$ and $P^{(s)}$ (see (\ref{bdpenamp}) and (\ref{bspenamp})),
respectively, have been discussed in \ref{Zoo}. It is an easy exercise 
to combine the decay amplitudes given in (\ref{ED1201}) appropriately to 
derive the relations
\begin{eqnarray}
\sqrt{2}\,A(B^+\to\pi^0 K^+)+A(B^+\to\pi^+
K^0)&=&r_u\sqrt{2}\,A(B^+\to\pi^+\pi^0)\label{ED1204a}\\
\sqrt{2}\,A(B^-\to\pi^0 K^-)+A(B^-\to\pi^-
\overline{K^0})&=&r_u\sqrt{2}\,A(B^-\to\pi^-\pi^0),\label{ED1204b}
\end{eqnarray}
which can be represented as two triangles in the complex plane. If one
measures the rates of the corresponding six decays, these triangles can
easily be constructed. Their relative orientation is fixed through
$A(B^+\to\pi^+ K^0)=A(B^-\to\pi^-\overline{K^0})$, which is due to the fact 
that there is no non-trivial CP-violating weak phase present in the 
$\bar b\to\bar s$ QCD penguin amplitude governing $B^+\to\pi^+ K^0$
as we have seen in \ref{Zoo}. Taking into account moreover (\ref{ED1202}), 
we conclude that these triangles should allow a determination of $\gamma$ 
as can be seen in fig.~\ref{grl-const}. From the geometrical point of view, 
that GRL approach \cite{grl-gam} is very similar to the $B^\pm\to DK^\pm$ 
construction \cite{gw} shown in fig.~\ref{BDK-triangles}.
Furthermore it involves also only charged $B$ decays and therefore neither 
time-dependent measurements nor tagging are required. In comparison 
with the Gronau-Wyler method \cite{gw}, at first sight the major 
advantage of the GRL strategy seems to be that all branching 
ratios are expected to be of the same order of magnitude ${\cal O}(10^{-5})$, 
i.e.\ the corresponding triangles are not squashed ones, and that the 
difficult to measure CP eigenstate $D^0_+$ and $Br(B^+\to D^0K^+)$ are
not required. The decay $B^+\to\pi^+K^0$ has been measured already by the
CLEO collaboration with a branching ratio $\left(2.3^{+1.1+0.2}_{-1.0-0.2}
\pm0.2\right)\cdot10^{-5}$, while presently only the upper bounds
$Br(B^+\to K^+\pi^0)<1.6\cdot10^{-5}$ and $Br(B^+\to\pi^+\pi^0)<2.0\cdot
10^{-5}$ are available for the other decays \cite{NEW-CLEO}.

However, things are unfortunately not that simple and -- despite of its 
attractiveness -- the general GHLR approach \cite{grl-gam,ghlr} to extract 
CKM phases from $SU(3)$ amplitude relations suffers from theoretical 
limitations. The most obvious limitation is of course related to the fact 
that the relations are not, as e.g.\ (\ref{ED86}) or (\ref{ED87}), valid 
exactly but suffer from $SU(3)$-breaking corrections \cite{ghlrsu3}. 
While factorizable $SU(3)$-breaking can be included straightforwardly
through certain meson decay constants or form factors, non-factorizable
$SU(3)$-breaking corrections cannot be described in a reliable quantitative
way at present. Another limitation is related to $\bar b\to\bar d$ QCD 
penguin topologies with internal up- and charm-quark exchanges which may 
affect the simple relation (\ref{bdpenapprox}) between $\beta$ and the 
$\bar b\to\bar d$ QCD penguin amplitude $P$ significantly as we have seen 
in \ref{Zoo}. Consequently these contributions may preclude reliable 
extractions of $\beta$ using $SU(3)$ 
amplitude relations and the assumption that $\bar b\to\bar d$ QCD penguin
amplitudes are dominated by internal top-quark exchanges (see also 
\ref{SIMalp-gam}) \cite{bf1}. Remarkably also EW penguins
\cite{rfewp1,rfewp2,rfewp3}, which 
we have neglected in our discussion of $SU(3)$ amplitude relations so far, 
have a very important impact on some $SU(3)$ constructions, in particular 
on the GRL method \cite{grl-gam} of determining $\gamma$. As we will see in 
subsection~\ref{RoEWPs}, this approach is even {\it spoiled} by these 
contributions \cite{dh,ghlrewp}. However, there are other -- generally more 
involved -- $SU(3)$ methods that are not affected by EW penguins 
\cite{ghlrewp}-\cite{PAPIII}. Interestingly it is in principle also possible 
to shed light on the physics of these operators by using $SU(3)$ amplitude 
relations \cite{PAPIII,PAPI}. This issue has been one of the ``hot topics'' 
in $B$ physics over the last few years and will be the subject of the 
following section. 

\subsection{Summary and Outlook}
The $B$-meson system provides a very fertile ground for studying CP violation 
and extracting CKM phases. In this respect neutral $B_q$ decays 
($q\in\{d,s\}$) are particularly promising. The point is 
that ``mixing-induced'' 
CP-violating asymmetries are closely related to angles of the 
unitarity triangle in some
cases. For example, the ``gold-plated'' decay $B_d\to J/\psi\,
K_{\mbox{{\scriptsize S}}}$ allows an extraction of $\sin(2\beta)$ to 
excellent accuracy because of its particular decay structure, and 
$B_d\to\pi^+\pi^-$ probes $\sin(2\alpha)$. However, hadronic uncertainties
arising from QCD penguins preclude a theoretical clean determination of
$\sin(2\alpha)$ by measuring only ${\cal A}_{\mbox{{\scriptsize 
CP}}}^{\mbox{{\scriptsize mix-ind}}}(B_d\to\pi^+\pi^-)$. Consequently more 
involved strategies are required to extract $\alpha$. Such methods are 
fortunately already available and certainly time will tell which of them 
is most promising from an experimental point of view. 

In the case of $B_s\to\rho^0K_{\mbox{{\scriptsize S}}}$, which appeared
frequently in the literature as a tool to determine $\gamma$, penguin 
contributions are expected to lead to serious problems so that a meaningful
extraction of $\gamma$ from this mode should not be possible. There
are, however, other $B_s$ decays that may allow determinations of this 
angle, in some cases even in a clean way. Unfortunately 
$B^0_s-\overline{B^0_s}$ oscillations may be too fast to be resolved with 
present vertex technology so that these strategies are experimentally very 
challenging. 

An alternative route to extract CKM phases from $B_s$ decays and to explore 
CP violation in these modes may be provided by the width difference 
of the $B_s$ system that is expected to be sizable. Interestingly the 
rapid oscillatory $\Delta M_st$ terms cancel in untagged $B_s$ data 
samples that depend therefore only on two different exponents 
$e^{-\Gamma_L^{(s)}t}$ and $e^{-\Gamma_H^{(s)}t}$. Several strategies to 
extract $\gamma$ and the Wolfenstein parameter $\eta$ from untagged 
$B_s$ decays have been proposed recently. Here time-dependent angular 
distributions for $B_s$ decays into admixtures of CP eigenstates and 
exclusive channels that are caused by $\bar b\to\bar uc\bar s$ 
($b\to c\bar us$) quark-level transitions play a key role. Such untagged 
methods are obviously much more promising in respect of efficiency, 
acceptance and purity than tagged ones. However, their feasibility depends 
crucially on $\Delta\Gamma_s$ and it is not clear at present whether it 
will turn out to be large enough.

Theoretical analyses of CP violation in charged $B$ decays are usually very
technical and suffer in general from large hadronic uncertainties. 
Consequently CP-violating asymmetries in charged $B$ decays are mainly
interesting in view of excluding ``superweak'' models of CP violation
in an unambiguous way. Nevertheless, if one combines branching ratios of
charged $B$ decays in a clever way, they may allow determinations of 
angles of the unitarity triangle, 
in some cases even without hadronic uncertainties. 
To this end certain relations among decay amplitudes are used. The 
prototype of this approach are $B\to DK$ amplitude triangles that allow 
a clean determination of $\gamma$. Unfortunately one has to deal with 
experimental problems in that strategy of fixing this angle. Whereas 
the $B\to DK$ triangle relations are valid exactly, one may also use the 
$SU(3)$ flavour symmetry of strong interactions with certain plausible 
dynamical assumptions to derive approximate relations among non-leptonic 
$B\to\{\pi\pi,\pi K,K\overline{K}\}$ decay amplitudes which may
allow extractions of CKM phases and strong final state interaction phases
by measuring only the corresponding branching ratios. This approach
has been very popular over the recent years. It suffers, however, from
limitations due to non-factorizable $SU(3)$-breaking, QCD penguins with
internal up- and charm-quark exchanges and also EW penguins. A detailed
discussion of the effects introduced through the latter operators is 
the subject of the following section.

\begin{figure}
\centerline{
\rotate[r]{
\epsfysize=4.5in
\epsffile{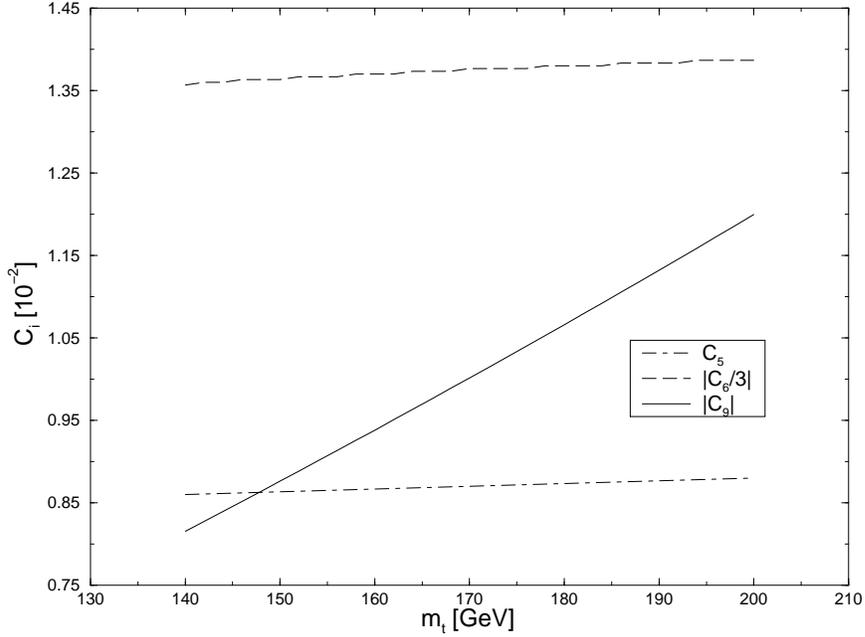}
}}
\caption{Dependence of Wilson coefficients $C_5$, $C_6$ and $C_9$ 
on the top-quark mass $\mt$.}\label{fig-clomt}
\end{figure}

\section{The Role of EW Penguins in Non-leptonic $B$ 
Decays and Strategies for Extracting CKM Phases}\label{Sec-EWP}
\setcounter{equation}{0}
\subsection{Preliminary Remarks}
Since the ratio $\alpha/\alpha_s={\cal O}(10^{-2})$ of the QED and QCD
couplings is very small, one would expect that EW penguins should only
play a minor role in comparison with QCD penguins. That would indeed be
the case if the top-quark was not ``heavy''. However, the Wilson coefficient 
of one EW penguin operator -- the operator $Q_9$ specified in 
(\ref{ew-def}) -- increases strongly with the top-quark mass and becomes
comparable in magnitude to Wilson coefficients of QCD penguin operators 
as can be seen in fig.~\ref{fig-clomt}.  
Consequently interesting EW penguin effects may arise from this feature 
in certain non-leptonic $B$ decays because of the large top-quark 
mass. As we have stressed in \ref{Mess1}, the parameter $m_t$ used in 
NLO analyses of non-leptonic weak decays is not equal to the measured ``pole'' 
mass but refers to the running top-quark current-mass normalized at the 
scale $\mu=m_t$, i.e.\ $\overline{m}_t(m_t)$. Before we shall investigate 
the role of EW penguins in methods for extracting angles of the unitarity 
triangle in 
subsection~\ref{RoEWPs}, let us have a closer 
look at a few non-leptonic $B$ decays that are affected significantly by 
EW penguin operators. These modes are listed in table~\ref{EWP-1}.

\begin{table}
\begin{displaymath}
\begin{array}{|c|c|c|c|}
\hline
\mbox{Quark-Decay}&\mbox{Exclusive Decay}&\mbox{Discussed in}&
\mbox{EW Penguin Contributions}\\
\hline
\hline
\bar b\to\bar sd\bar d&
B^+\to\pi^+K^{\ast0}&
\ref{BKphi}&
\mbox{negligible}\\
\hline
\bar b\to\bar ss\bar s&
B^+\to K^+\phi&
\ref{BKphi}&
\mbox{sizable}\\
\hline
\bar b\to\bar ds\bar s&
B^+\to\pi^+\phi&
\ref{Bpiphi}&
\mbox{dominant}\\
\hline
\bar b\to\bar s(u\bar u,d\bar d)&
B_s\to\pi^0\phi&
\ref{Bspi0phi}&
\mbox{dominant}\\
\hline
\end{array}
\end{displaymath}
\caption{EWP effects in some non-leptonic $B$ decays. The theoretical
most reliable analysis is possible in $B_s\to\pi^0\phi$ because of the
isospin symmetry of strong interactions.}\label{EWP-1}
\end{table}

\subsection{EW Penguin Effects in Non-leptonic $B$ Decays}
The EW penguin effects discussed in this subsection were 
pointed out first in \cite{rfewp1,rfewp2,rfewp3}. Meanwhile they were 
confirmed by several other authors \cite{kp,ghlrewp}, 
\cite{dh-ewp}-\cite{dy}.

\begin{figure}
\centerline{
\rotate[r]{
\epsfysize=9.8truecm
\epsffile{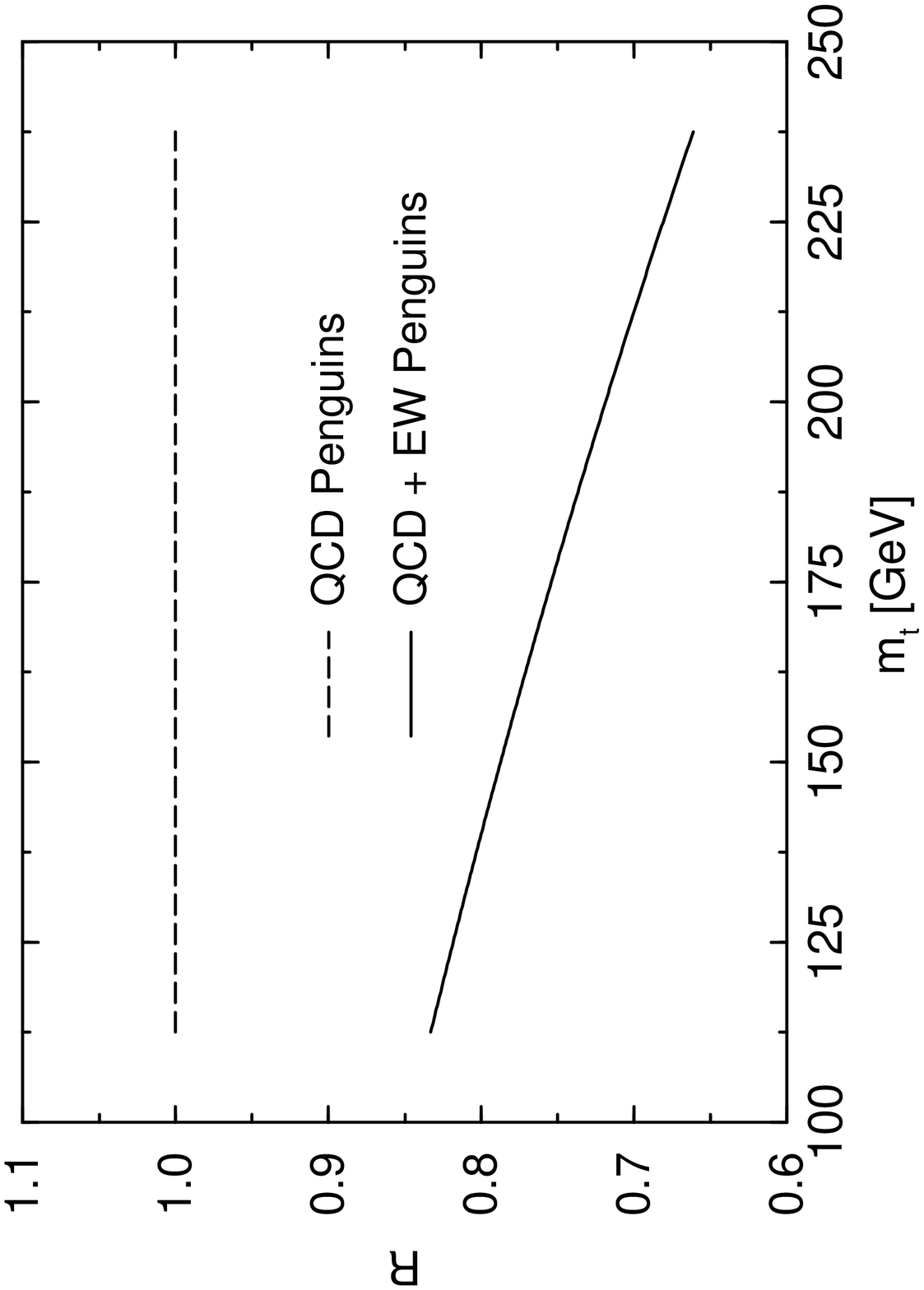}
}}
\caption{The dependence of the ratio 
$R\propto Br(B^+\to K^+\phi)/Br(B^+\to\pi^+K^{\ast0})$ 
on $m_t$.}\label{fig-Rmt}
\end{figure}

\subsubsection{EW Penguin Effects in $B^+\to K^+\phi$ and $B^+\to\pi^+
K^{\ast0}$}\label{BKphi}
The channels $B^+\to K^+\phi$ and $B^+\to\pi^+K^{\ast0}$ originating from 
the penguin-induced $\bar b$-quark decays $\bar b\to \bar ss\bar s$ 
and $\bar b\to\bar s d\bar d$, respectively, are very similar from a 
QCD point of view, i.e.\ as far as their QCD penguin contributions are 
concerned. This feature is obvious if one draws the corresponding 
Feynman diagrams which is an easy exercise. However, an important 
difference arises in respect of EW penguin contributions. We have to deal 
both with small colour-suppressed and sizable colour-allowed EW penguin 
diagrams. Whereas the former contributions are again very similar for 
$B^+\to K^+\phi$ and $B^+\to\pi^+K^{\ast0}$, the colour-allowed EW penguin 
contributions are absent in the $B^+\to\pi^+K^{\ast0}$ case and contribute 
only to $B^+\to K^+\phi$. Consequently significant EW penguin 
effects are expected in the mode $B^+\to K^+\phi$, 
while these effects should be negligible in the decay 
$B^+\to\pi^+K^{\ast0}$ \cite{rfewp1}. 

This rather qualitative kind of reasoning is in agreement with the results
of certain model calculations \cite{rfewp1,dh-ewp}, where appropriate
NLO low energy effective Hamiltonians were applied in combination with 
the ``factorization'' hypothesis \cite{fact1,fact2}. By factorization one 
means in this context that the hadronic matrix elements of the relevant
four-quark operators appearing in (\ref{LEham}) are factorized into the
product of hadronic matrix elements of two quark-currents that are 
described by a set of form factors. Usually the model proposed by 
Bauer, Stech and Wirbel (BSW) \cite{BSW} is used for these form factors and
was also applied in \cite{rfewp1,dh-ewp}. In contrast to colour-allowed
current-current processes, where ``factorization'' may work reasonably 
well \cite{bjorken}, this assumption is questionable for penguin processes 
which are classical examples of non-factorizable diagrams. Nevertheless this 
approach may give us a feeling for the expected orders of magnitudes. 
Unfortunately a more reliable analytical way of dealing with non-leptonic 
$B$ decays is not available at present. 

The corresponding calculations are quite complicated and a discussion of
their technicalities is beyond this review. Let us therefore just briefly 
discuss the main results. The model 
calculations indicate that EW penguins lead to a reduction of 
$Br(B^+\to K^+\phi)$ by ${\cal O}(30\%)$ for $m_t={\cal O}(170\,
\mbox{GeV})$, while these effects are below $2\%$ in the case of 
$Br(B^+\to\pi^+K^{\ast0})$. As in fig.~\ref{delp-fig}, the 
branching ratios, which are both of ${\cal O}(10^{-5})$, depend strongly 
on $k^2$, the four-momentum of the gluons and photons appearing in the 
corresponding time-like penguin diagrams. This ``unphysical'' 
$k^2$-dependence is due to the use of the above mentioned model \cite{gh}.
In order to reduce this dependence as well as other hadronic uncertainties, 
the ratio \cite{rfewp1}
\begin{eqnarray}
\lefteqn{R\equiv\left[\frac{F_{K^{\ast}}F_{B\pi}(M_{K^\ast}^{2};1^{-})}
{F_{\phi}F_{BK}(M_{\phi}^{2};1^{-})}\right]^{2}
\left[\frac{\Phi(M_{\pi}/M_{B},M_{K^{\ast}}/M_{B})}
{\Phi(M_{K}/M_{B},M_{\phi}/M_{B})}
\right]^{3}}\nonumber\\
&&\times\left[\frac{Br(B^{+}\to K^{+}\phi)}{Br(B^{+}\to 
\pi^{+}K^{\ast0})}\right]\approx0.5\times
\left[\frac{Br(B^{+}\to K^{+}\phi)}{Br(B^{+}\to \pi^{+}
K^{\ast0})}\right]\label{e919}
\end{eqnarray}
turns out to be very useful. 
Here $F_V$ are meson decay constants, $F_{PP'}$ are quark-current form
factors and $\Phi(x,y)$ is the usual two-body phase space function.
Although $R$ is affected in almost the same 
way by EW penguins as the branching ratio $Br(B^{+}\to K^{+}\phi)$, it 
suffers much less from hadronic uncertainties, is very stable against 
variations both of the momentum transfer $k^{2}/m_{b}^{2}$ and of the QCD 
scale parameter $\Lambda_{\overline{\mbox{{\scriptsize MS}}}}$, and
does not depend on CKM factors if the ${\cal O}(\lambda^2)$ terms in 
(\ref{CKMs}) are neglected. These terms play a minor role and may lead to 
tiny direct CP-violating asymmetries of ${\cal O}(1\%)$. One should keep
in mind, however, that 
$Br(B^{+}\to K^{+}\phi)$ and $Br(B^{+}\to\pi^{+} K^{\ast0})$ could 
receive quite different contributions in principle if ``factorization'' does 
not hold. Therefore $R$ could be affected by such unknown corrections.

The effects of EW penguins can be seen nicely in fig.~\ref{fig-Rmt}, 
where the top-quark mass dependence of $R$ is shown. For details of
the calculation of these curves the reader is referred to \cite{F97,rfewp1}. 
Whereas the dashed line corresponds to the case where only QCD
penguins are included, the solid line describes the calculation taking
into account both QCD and EW penguin operators. 

There are not only some non-leptonic $B$ decays that are affected 
significantly by EW penguins. There are even a few channels where the 
corresponding operators may play the {\it dominant} role as we will 
see in the remainder of this subsection. 

\subsubsection{EW Penguin Effects in $B^+\to\pi^+\phi$}\label{Bpiphi}
In respect of EW penguin effects, the mode $B^+\to\pi^+\phi$ is also quite
interesting \cite{rfewp2}. Within the spectator model, it originates from the 
penguin-induced $\bar b$-quark decay $\bar b\to\bar d s\bar s$, where the 
$s\bar s$ pair hadronizes into the $\phi$-meson which is present in a 
colour-singlet state. The $s$- and $\bar s$-quarks emerging from the gluons 
of the usual QCD penguin diagrams form, however, a colour-octet state and 
consequently cannot build up that $\phi$-meson (see also \ref{Beta-Ext}). 
Thus, using both an appropriate NLO low energy effective Hamiltonian and the 
BSW model in combination with the factorization assumption to estimate the 
relevant hadronic matrix elements of the QCD penguin operators, 
one finds a very small branching ratio 
$Br\left.(B^+\to\pi^+\phi)\right|_{\mbox{{\scriptsize QCD}}}={\cal 
O}(10^{-10})$. The non-vanishing result is due to the renormalization group 
evolution from $\mu={\cal O}(M_W)$ down to $\mu={\cal O}(m_b)$. Neglecting 
this evolution would give a vanishing branching ratio because of the 
colour-arguments given above. Since these arguments do not apply to EW 
penguins, their contributions are expected to become important \cite{rfewp2}. 
In fact, taking into account also these operators gives a branching ratio 
$Br\left.(B^+\to\pi^+\phi)\right|_{\mbox{{\scriptsize QCD+EW}}}={\cal 
O}(10^{-8})$ for $m_t={\cal O}(170\,\mbox{GeV})$ that increases strongly 
with the top-quark mass. Unfortunately the enhancement by a factor of 
${\cal O}(10^2)$ through EW penguins is not strong enough to make the 
decay $B^+\to\pi^+\phi$ measurable in the foreseeable future.

The colour-arguments for the QCD penguins may be affected by additional 
soft gluon exchanges which are not under quantitative 
control at present. These contributions would show up as non-factorizable 
contributions to the hadronic matrix elements of the penguin operators 
which were neglected in \cite{rfewp2}. Nevertheless there is no 
doubt that EW penguins play a very important -- probably even dominant -- 
role in the decay $B^+\to\pi^+\phi$ and related modes like $B^+\to\rho^+\phi$.

\subsubsection{EW Penguin Effects in $B_s\to\pi^0\phi$}\label{Bspi0phi}
The theoretical situation arising in the decay $B_s^0\to\pi^0\phi$ caused by
$\bar b\to\bar s\,(u\bar u,d\bar d)$ quark-level transitions is much more 
favourable than in the channels discussed previously 
because of the $SU(2)$ 
isospin symmetry of strong interactions. Let us therefore be more detailed 
in the presentation of that transition which is expected to be dominated 
by EW penguins \cite{rfewp3}. In contrast to the decays discussed in 
\ref{BKphi} and \ref{Bpiphi}, it receives not only penguin but also 
current-current operator contributions at the tree level. The final state 
is an eigenstate of the CP operator with eigenvalue +1 and has strong 
isospin quantum numbers $(I,I_3)=(1,0)$, whereas the initial state is an 
isospin singlet. Thus we have to deal with a $\Delta I=1$ transition. 

Looking at the operator basis given in (\ref{cc-def})-(\ref{ew-def}),
we observe that the current-current operators $Q^{us}_{1,2}$ and the
EW penguin operators can lead to final states both with isospin $I=0$ 
and $I=1$, whereas the QCD penguin operators give only final states 
with $I=0$. Therefore the $\Delta I=1$ transition $B_s\to\pi^0\phi$ 
receives no QCD penguin contributions and arises purely from the 
current-current operators $Q^{us}_{1,2}$  and the EW penguin operators. 
For the same reason, QCD penguin matrix elements of the current-current 
operators $Q_{2}^{us}$ and $Q_{2}^{cs}$ with up- and 
charm-quarks running as virtual particles in the loops, respectively, 
do not contribute to that decay.
Consequently, using in addition the unitarity of the CKM matrix and
applying the modified Wolfenstein parametrization (\ref{wolf2}) yielding
\begin{equation}\label{ED1105}
V_{us}^{\ast}V_{ub}=\lambda|V_{ub}|\,e^{-i\gamma},\quad
V_{ts}^{\ast}V_{tb}=-|V_{ts}|=-|V_{cb}|(1+{\cal O}(\lambda^2)),
\end{equation}
the hadronic matrix element of the Hamiltonian (\ref{LEham}) can be
expressed as
\begin{eqnarray}
\lefteqn{\Bigl\langle\pi^0\phi\Bigl|{\cal H}_{\mbox{{\scriptsize eff}}}
(\Delta B=-1)\Bigr|\overline{B^0_s}\Bigr\rangle=
\frac{G_{\mbox{{\scriptsize F}}}}{\sqrt{2}}\,|V_{ts}|}\label{ED1106}\\
&&\times\left[\lambda^2R_b\,e^{-i\gamma}
\sum\limits_{k=1}^2\Bigl\langle\pi^{0}\phi\Bigl|
Q_k^{us}(\mu)\Bigr|\overline{B^0_s}\Bigr\rangle C_k(\mu)
+\sum\limits_{k=7}^{10}\Bigl\langle\pi^0\phi\Bigl|
Q_k^s(\mu)\Bigr|\overline{B^0_s}\Bigr\rangle C_k(\mu)\right],\nonumber
\end{eqnarray}
where the correction of  ${\cal O}(\lambda^{2})$ in (\ref{ED1105}) has
been omitted.

Neglecting EW penguin operators for a moment and applying the formalism 
developed in \ref{domCKM}, we would find ${\cal A}_{\mbox{{\scriptsize 
CP}}}^{\mbox{{\scriptsize mix-ind}}}(B_s\to\pi^0\phi)=\sin(2\gamma)$.
The approximation of neglecting EW penguin operator contributions to 
$B_s\to\pi^0\phi$ is, however, very bad since the current-current 
amplitude $A_{\mbox{{\scriptsize CC}}}$ is suppressed relative to the
EW penguin part $A_{\mbox{{\scriptsize EW}}}$ by the CKM factor
$\lambda^2R_b\approx0.02$. Moreover the current-current operator
contribution is colour-suppressed by $a_2\approx0.2$. On the other hand, 
in the presence of a heavy top-quark, the Wilson coefficient of the dominant
EW penguin operator $Q_9^s$ contributing to $B_s\to\pi^0\phi$
in colour-allowed form is of ${\cal O}(10^{-2})$ (see fig.~\ref{fig-clomt}). 
Therefore we expect 
$|A_{\mbox{{\scriptsize EW}}}|/|A_{\mbox{{\scriptsize CC}}}|
={\cal O}(10^{-2}/(0.02\cdot0.2))={\cal O}(2.5)$ and conclude that EW penguins
have not only to be taken into account in an analysis of $B_s\to\pi^0\phi$
but should even give the {\it dominant} contribution to that channel.

In order to simplify the following discussion, let us neglect the 
influence of QCD corrections to EW penguins for a moment. Using isospin
symmetry and taking into account that $B_s\to\pi^0\phi$ is a $\Delta I=1$
transition, the correpsonding decay amplitude can be expressed as
\begin{equation}\label{ED1115}
\Bigl\langle\pi^0\phi\Bigl|{\cal H}_{\mbox{{\scriptsize eff}}}
(\Delta B=-1)\Bigr|\overline{B^0_s}\Bigr\rangle=\frac{G_{\mbox{{\scriptsize 
F}}}}{\sqrt{2}}\,A_{\mbox{{\scriptsize CC}}}\left(e^{-i\gamma}+x\right)
\end{equation}
with
\begin{equation}\label{ED1121}
x\equiv\frac{A_{\mbox{{\scriptsize EW}}}}{A_{\mbox{{\scriptsize CC}}}}\approx
\frac{\alpha}{2\pi\lambda^2R_b\,a_2\sin^2\Theta_{\mbox{{\scriptsize
W}}}}\bigl[5B_0(x_t)-2C_0(x_t)\bigr],
\end{equation}
where the Inami-Lim functions \cite{IL} $B_0(x_t)$ and $C_0(x_t)$ are
given in (\ref{BF}) and (\ref{C0xt}) and describe box diagrams and $Z$ 
penguins, respectively. The phenomenological colour-suppression 
factor $a_2$ takes into account that $A_{\mbox{{\scriptsize CC}}}$
is colour-suppressed. For the details of this calculation leading to
(\ref{ED1115}) and (\ref{ED1121}), the reader is
referred to \cite{rfewp3}.

\begin{figure}[t]
\centerline{
\rotate[r]{
\epsfysize=10truecm
\epsffile{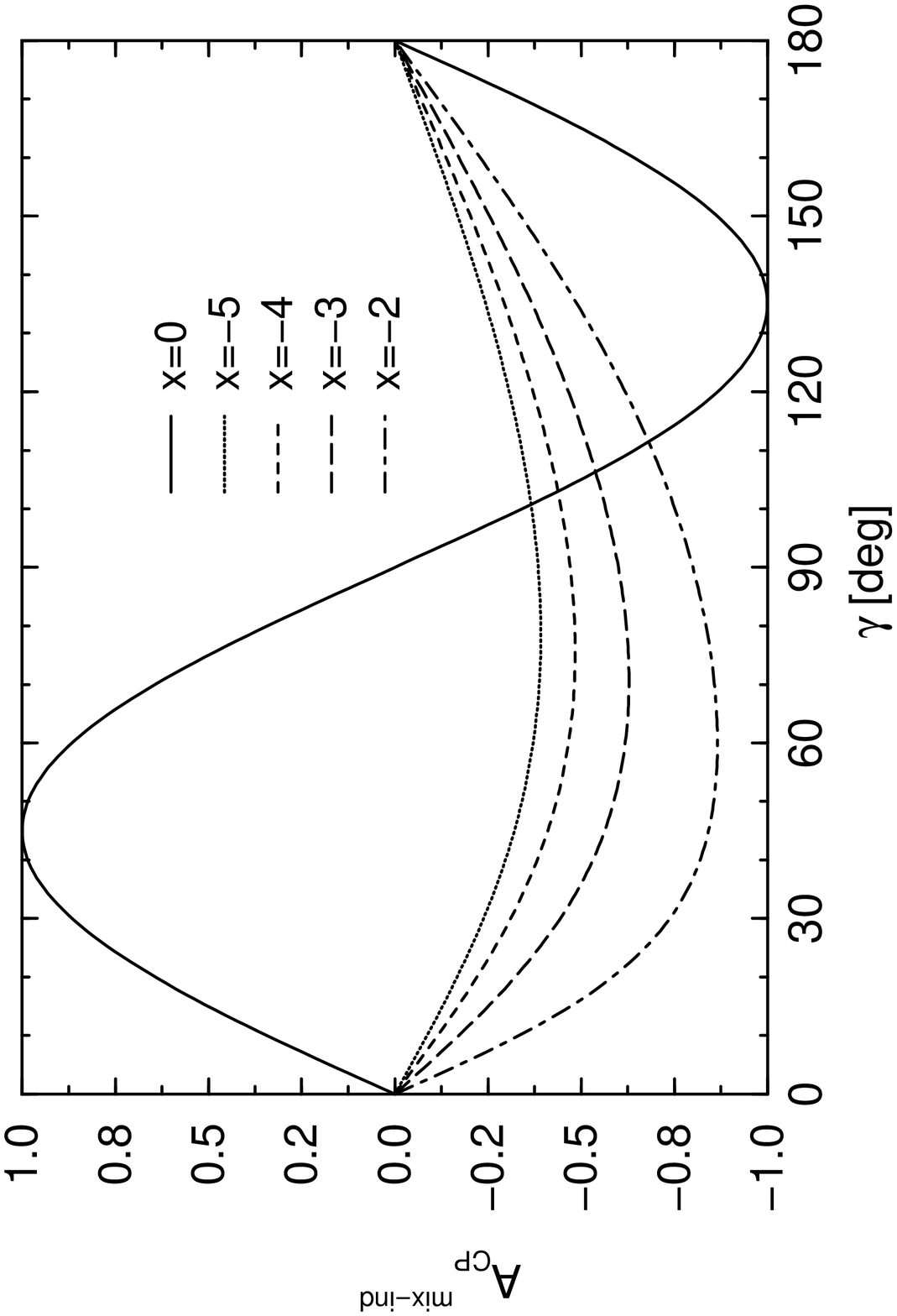}}}
\caption{Dependence of ${\cal A}_{\mbox{{\scriptsize 
CP}}}^{\mbox{{\scriptsize mix-ind}}}(B_s\to\pi^0\phi)$ on 
$\gamma$ for various values of $x$.}\label{fig-Acp}
\end{figure}

With the help of (\ref{ED1115}), the CP-violating observables of 
$B_s\to\pi^0\phi$ can be expressed as 
\begin{equation}\label{ED1116}
{\cal A}_{\mbox{{\scriptsize CP}}}^{\mbox{{\scriptsize dir}}}
(B_s\to\pi^0\phi)=0,\quad
{\cal A}_{\mbox{{\scriptsize CP}}}^{\mbox{{\scriptsize mix-ind}}}
(B_s\to\pi^0\phi)=\frac{2\,(x+\cos\gamma)\sin\gamma}{x^2+2\,x\cos\gamma+1},
\end{equation}
while the branching ratio $Br(B_s\to\pi^0\phi)$ takes the form 
\begin{equation}\label{ED1117}
{\cal R}\equiv\frac{Br(B_s\to\pi^0\phi)}{Br_{\mbox{{\scriptsize
CC}}}(B_s\to\pi^0\phi)}=x^2+2\,x\cos\gamma+1,
\end{equation}
where $Br_{\mbox{{\scriptsize CC}}}(B_s\to\pi^0\phi)={\cal O}
(10^{-8})$ denotes the current-current branching ratio.  

\begin{figure}
\centerline{
\rotate[r]{
\epsfysize=10truecm
\epsffile{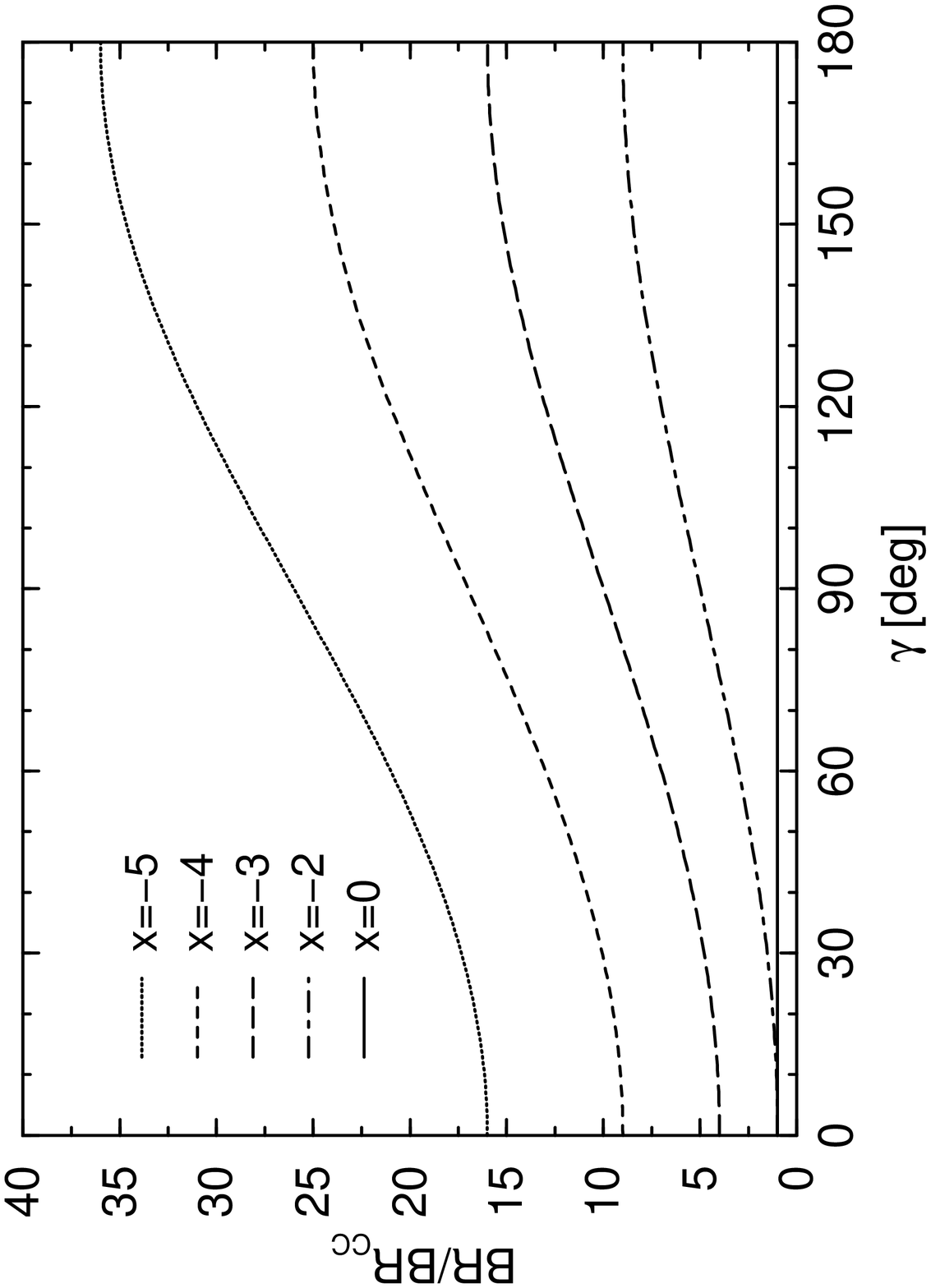}}}
\caption{Dependence of $Br(B_s\to\pi^0\phi)/
Br_{\mbox{{\scriptsize CC}}}(B_s\to\pi^0\phi)$ on $\gamma$ for
various values of $x$.}\label{fig-br}
\end{figure}

Note that (\ref{ED1121}) is rather clean concerning hadronic uncertainties. 
This nice feature is due to the fact that we are in a position to absorb 
all non-perturbative $B$-parameters related to deviations from naive 
factorization of the hadronic matrix elements by introducing the 
phenomenological colour-suppression factor $a_2$. Concerning short-distance 
QCD corrections, which have been neglected so far, we have to consider only
those affecting the box diagrams and $Z$ penguins contributing to $x$,
since the QCD corrections to the current-current operators are
incorporated effectively in $a_2$. The corresponding short-distance 
QCD corrections are small if we use $\overline{m}_t(m_t)$ 
(see section 7).
On the other hand, the QCD corrections to EW penguin operators arising from 
the renormalization group evolution from $\mu={\cal O}(M_W)$ down 
to $\mu={\cal O}(m_b)$ modify $x$ by only a few percent and are hence
also negligibly small.

Using as an example $a_2=0.25$, $R_b=0.36$ and $m_t=170\,
\mbox{GeV}$ yields $x\approx-3$ and confirms nicely our qualitative
expectation that EW penguins should play the dominant role in 
$B_s\to\pi^0\phi$. Varying $a_2$ within $0.2\,\,\mbox{{\scriptsize 
$\stackrel{<}{\sim}$}}\,\,a_2\,\,\mbox{{\scriptsize $\stackrel{<}{\sim}$}}\,\,
0.3$ and $R_b$ and $m_t$ within their presently allowed experimental ranges 
gives $-5\,\,\mbox{{\scriptsize $\stackrel{<}{\sim}$}}\,\,x\,\,
\mbox{{\scriptsize $\stackrel{<}{\sim}$}}-2$. The EW penguin contributions 
lead to dramatic effects in the mixing-induced CP asymmetry as well as in 
the branching ratio as can be seen in Figs.~\ref{fig-Acp} and 
\ref{fig-br}, where the dependences of 
${\cal A}_{\mbox{{\scriptsize CP}}}^{\mbox{{\scriptsize mix-ind}}}
(B_s\to\pi^0\phi)$ and of the ratio ${\cal R}$ on $\gamma$ are shown for
various values of $x$. The solid lines in these figures correspond to 
the case where EW penguins are neglected completely. In the case of 
${\cal A}_{\mbox{{\scriptsize CP}}}^{\mbox{{\scriptsize mix-ind}}}
(B_s\to\pi^0\phi)$ even the sign is changed through the EW penguin 
contributions for $\gamma<90^\circ$, whereas the branching ratio is 
enhanced by a factor of ${\cal O}(10)$ with respect to the pure 
current-current case. The resulting $Br(B_s\to\pi^0\phi)$ is of 
${\cal O}(10^{-7})$, so that an experimental investigation of that 
decay -- which would be interesting to explore EW penguins -- 
will unfortunately be very difficult. Needless to say, the modes 
$B_s\to\rho^0\phi, \pi^0\eta, \rho^0\eta$ exhibiting a very similar 
dynamics should also be dominated by their EW penguin 
contributions \cite{ghlrewp,dht-ewp}. 

\subsection{EW Penguin Effects in Strategies for Extracting CKM 
Phases}\label{RoEWPs}
In the strategies for extracting CKM phases reviewed in 
section~\ref{CP-SEC}, EW penguins do not lead to problems wherever it 
has not been emphasized 
explicitly. That is in fact the case for most of these methods. However,
the GL approach \cite{gl} to eliminate the penguin uncertainties affecting
the determination of $\alpha$ from ${\cal A}_{\mbox{{\scriptsize 
CP}}}^{\mbox{{\scriptsize mix-ind}}}(B_d\to\pi^+\pi^-)$ with the help
of isospin relations among $B\to\pi\pi$ decays (see \ref{Bdpipi-Alpha}), 
as well as the GRL method \cite{grl-gam} to determine $\gamma$ from $SU(3)$ 
amplitude relations involving $B^+\to\{\pi^+\pi^0,\pi^+K^0,\pi^0K^+\}$ and 
their charge-conjugates (see \ref{SU3REL}) require a careful 
investigation \cite{dh,ghlrewp,PAPI}.

\begin{figure}[t]
\centerline{
\epsfxsize=13.5truecm
\epsffile{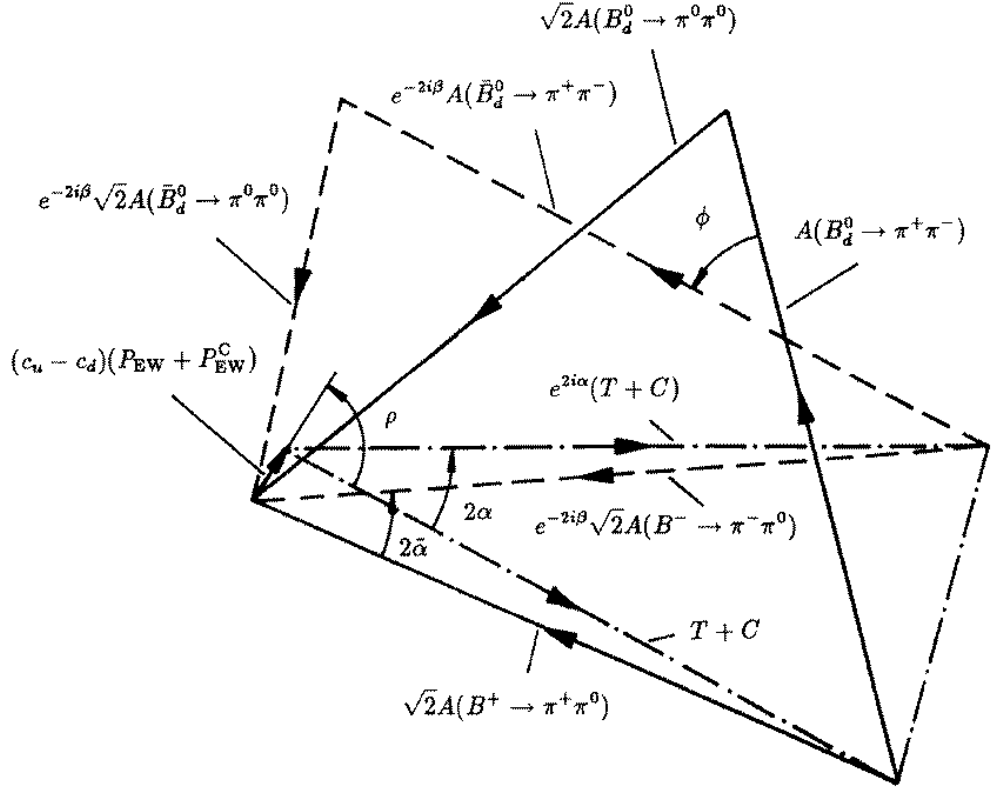}}
\caption{The determination of $\alpha$ from $B\to\pi\pi$ isospin 
triangles in the presence of EW penguins.}\label{fig-Bpipi}
\end{figure}

\subsubsection{The GL Method of Extracting $\alpha$}
If one redraws the GL construction \cite{gl} to determine $\alpha$ from 
$B\to\pi\pi$ isospin triangles by taking into account EW penguin 
contributions, one obtains the situation shown in fig.~\ref{fig-Bpipi}. This 
construction \cite{PAPI} is a bit different from the original one presented 
in \cite{gl}, since the $A(\overline{B}\to\pi\pi)$ amplitudes have been 
rotated by $e^{-2i\beta}$. The angle $\phi$ fixing the relative orientation 
of the two isospin triangles, which are constructed by measuring only the
corresponding six branching ratios, is determined from the mixing-induced CP 
asymmetry of $B_d\to\pi^+\pi^-$ with the help of the relation
\begin{equation}
{\cal A}_{\mbox{{\scriptsize CP}}}^{\mbox{{\scriptsize mix-ind}}}
(B_d\to\pi^+\pi^-)=-\,\frac{2\,\bigl|A(\overline{B^0_d}\to\pi^+\pi^-)\bigr|
\bigl|A(B^0_d\to\pi^+\pi^-)\bigr|}{\bigl|A(\overline{B^0_d}\to\pi^+
\pi^-)\bigr|^2+\bigl|A(B^0_d\to\pi^+\pi^-)\bigr|^2}\,\sin\phi\,.
\end{equation}
In fig.~\ref{fig-Bpipi}, the notation of GHLR \cite{ghlrewp} has been used, 
where $P_{\mbox{{\scriptsize EW}}}$ and 
$P_{\mbox{{\scriptsize EW}}}^{\mbox{{\scriptsize C}}}$ denote colour-allowed
and colour-suppressed $\bar b\to\bar d$ EW penguin amplitudes and $c_u=+2/3$ 
and $c_d=-1/3$ are the electrical up- and down-type quark charges, 
respectively. Because of the presence of EW penguins, the construction shown 
in that figure does {\it not} allow the determination of the {\it exact} 
angle $\alpha$ of the UT. It allows only the extraction of an angle 
$\tilde\alpha$ that is related to $\alpha$ through 
\begin{equation}\label{deltaalpha}
\alpha=\tilde\alpha+\Delta\alpha,
\end{equation}
where $\Delta\alpha$ is given by
\begin{equation}
\Delta\alpha=r\,\sin\alpha\,\cos\,(\rho-\alpha)+{\cal O}(r^2)
\end{equation}
with
\begin{equation}
r\equiv\frac{\left|(c_u-c_d)(P_{\mbox{{\scriptsize EW}}}+
P_{\mbox{{\scriptsize EW}}}^{\mbox{{\scriptsize C}}})\right|}{|T+C|}\approx
\left|\frac{P_{\mbox{{\scriptsize EW}}}}{T}\right|.
\end{equation}
Since $r$ is expected to be of ${\cal O}(10^{-2})$ as can be shown by using
a plausible hierarchy of $\bar b\to\bar d$ decay amplitudes \cite{ghlrewp},
EW penguins should not lead to serious problems in the GL method. 
This statement can also be put on more quantitative ground. Unfortunately 
$\rho$ contains strong final state interaction phases and hence cannot be 
calculated at present. However, using $|\cos(\rho-\alpha)|\leq1$, one may 
estimate the following upper bound for the uncertainty 
$\Delta\alpha$ \cite{PAPIII}:
\begin{equation}\label{deltaalpha-est}
|\Delta\alpha|\,\,\mbox{{\scriptsize $\stackrel{<}{\sim}$}}\,\,
\frac{\alpha}{2\pi a_1\sin^2\Theta_{\mbox{{\scriptsize
W}}}}\bigl|5B(x_t)-2C(x_t)\bigr|\cdot\left\vert\frac{V_{td}}{V_{ub}}
\right\vert\left\vert\sin\alpha\right\vert\,.
\end{equation}
Taking into account the present status of the CKM matrix yielding
$|V_{td}|/|V_{ub}|\leq4.6$ \cite{ALUT} 
gives $|\Delta\alpha|/|\sin\alpha|\,\,
\mbox{{\scriptsize $\stackrel{<}{\sim}$}}\,4^\circ$ for a top-quark 
mass $m_t=170\,\mbox{GeV}$ and a phenomenological colour-factor 
$a_1=1$.

\begin{figure}[t]
\centerline{
\epsfxsize=14truecm
\epsffile{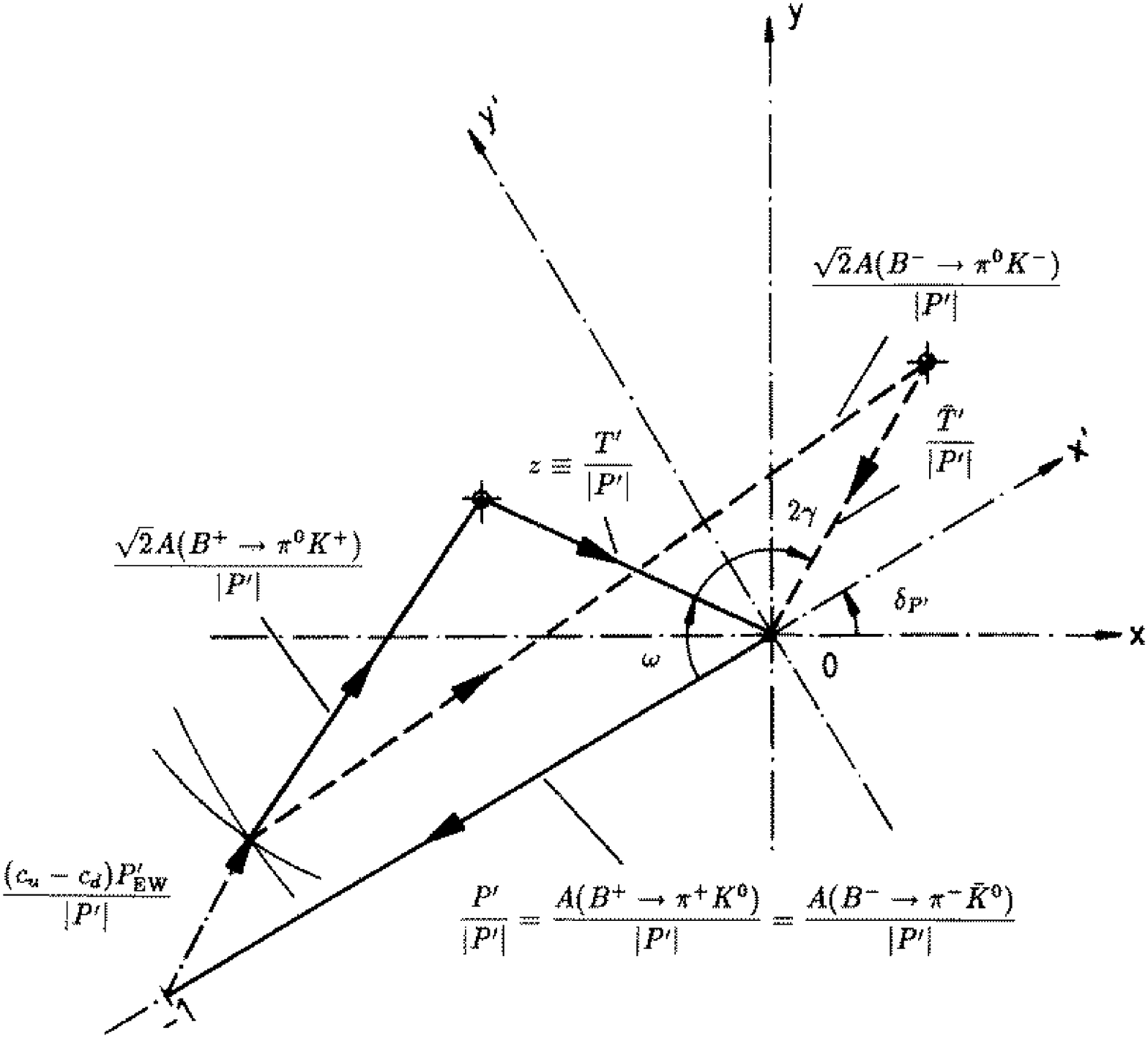}}
\caption{$SU(3)$ relations among $B^+\to\{\pi^+\pi^0,\pi^+K^0,
\pi^0K^+\}$ and charge-conjugate decay amplitudes {\it including} EW penguin
contributions.}\label{fig-GRL-EWP}
\end{figure}

\subsubsection{The GRL Method of Extracting $\gamma$}
In the case of the GRL strategy \cite{grl-gam} of extracting the  
angle $\gamma$ from the construction shown in fig.~\ref{grl-const},
we have to deal with $\bar b\to\bar s$ modes which exhibit an 
interesting hierarchy of decay amplitudes that is very
different from the $\bar b\to\bar d$ case \cite{dh,ghlrewp}. Since the
colour-allowed current-current amplitude $T'$ is highly CKM suppressed 
by $\lambda^2R_b\approx0.02$, one expects that the QCD penguin amplitude 
$P'$ plays the dominant role in this decay class and that $T'$ and 
the colour-allowed EW penguin amplitude $P'_{\mbox{{\scriptsize EW}}}$
are equally important \cite{ghlrewp}:
\begin{equation}\label{bs-hierarchy}
\left|\frac{T'}{P'}\right|={\cal O}(0.2),\quad
\left|\frac{P'_{\mbox{{\scriptsize EW}}}}{T'}\right|={\cal O}(1)\,.
\end{equation}\label{hier-est}
The last ratio can be estimated more quantitatively as \cite{PAPIII} 
\begin{equation}\label{REW-est}
\left|\frac{P'_{\mbox{{\scriptsize EW}}}}{T'}\right|\approx
\frac{\alpha}{2\pi\lambda^2 R_b\, a_1\sin^2\Theta_{\mbox{{\scriptsize
W}}}}\bigl|5B_0(x_t)-2C_0(x_t)\bigr|\,r_{SU(3)}.
\end{equation}
Here $r_{SU(3)}$ takes into account $SU(3)$-breaking corrections. 
Factorizable corrections are described by
\begin{equation}
\left.r_{SU(3)}\right\vert_{\mbox{{\scriptsize fact}}}=\frac{F_{\pi}}{F_K}
\frac{F_{BK}(0;0^+)}{F_{B\pi}(0;0^+)},
\end{equation}
where the BSW form factors \cite{BSW} parametrizing the corresponding
quark-current matrix elements yield 
$\left.r_{SU(3)}\right\vert_{\mbox{{\scriptsize fact}}}\approx1$. The
ratio (\ref{REW-est}) increases significantly with the top-quark mass. 
Using $m_t=170\,\mbox{GeV}$, $R_b=0.36$, $a_1=1$ and $r_{SU(3)}=1$ gives 
$|P'_{\mbox{{\scriptsize EW}}}|/|T'|\approx0.8$ and confirms the 
expectation (\ref{bs-hierarchy}). 

Consequently EW penguins are very important in that case and even 
{\it spoil} the GRL approach \cite{grl-gam} to determine $\gamma$ as 
was pointed out by Deshpande and He \cite{dh}. This feature can be seen 
in fig.~\ref{fig-GRL-EWP}, where colour-suppressed EW penguin and 
current-current amplitudes are neglected to simplify the 
presentation \cite{PAPI}. If the EW penguin amplitude $(c_u-c_d)\,
P'_{\mbox{{\scriptsize EW}}}$ were not there, this figure would correspond 
to fig.~\ref{grl-const} and we would simply have to deal with two triangles 
in the complex plane that could be fixed by measuring only the six branching 
ratios corresponding to 
$B^+\to\{\pi^+\pi^0,\pi^+K^0,\pi^0K^+\}$ and their charge-conjugates.
However, EW penguins do contribute and since the magnitude of their 
``unknown'' amplitude $(c_u-c_d)\,P'_{\mbox{{\scriptsize EW}}}$ is of the 
same size as $|T'|$, it is unfortunately not possible to determine $\gamma$
with the help of this construction.  This feature led to the
development of other methods using $SU(3)$ amplitude relations to extract 
$\gamma$ that require similarly as the GRL method only measurements of 
branching ratios, and 
to strategies to control EW penguins in a quantitative way to shed light
on the physics of these FCNC processes. 

\subsection{$SU(3)$ Strategies for Extracting $\gamma$ that are not
affected by EW Penguins}
In the recent literature some solutions have been proposed to solve the 
problem arising from EW penguins in the GRL approach 
\cite{ghlrewp}-\cite{PAPIII}. Let us have a closer look at them in this 
subsection. 

\begin{figure}[t]
\centerline{
\rotate[r]{
\epsfysize=8.5truecm
\epsffile{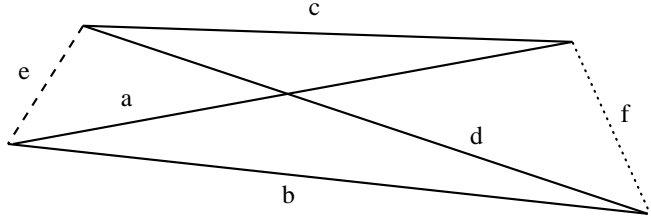}}}
\caption{Amplitude quadrangle for $B\to\pi K$ decays. The labels are 
explained in the text.}\label{fig-quadrangle}
\end{figure}

\subsubsection{Amplitude Quadrangle for $B\to\pi K$ Decays}\label{AmplQuad}
A quadrangle construction involving $B\to\pi K$ decay amplitudes was 
proposed in \cite{ghlrewp} that can be used in principle to determine
$\gamma$ irrespectively of the presence of EW penguins. This construction
is shown in fig.~\ref{fig-quadrangle}, where (a) corresponds to 
$A(B^+\to\pi^+K^0)$, (b) to $\sqrt{2}\,A(B^+\to\pi^0K^+)$,
(c) to $\sqrt{2}\,A(B^0_d\to\pi^0K^0)$, (d) to $A(B^0_d\to\pi^-K^+)$ and 
the dashed line (e) to the decay amplitude $\sqrt{3}\,A(B_s^0\to\pi^0\eta)$. 
The dotted line (f) denotes an $I=3/2$ isospin amplitude $A_{3/2}$ that is 
composed of two parts and can be written as \cite{ghlrewp}
\begin{equation}\label{A32}
A_{3/2}=\bigl|A_{\pi K}^{\mbox{{\scriptsize T}}}\bigr|\,e^{i\tilde\delta_T}
\,e^{i\gamma}-\bigl|A_{\pi K}^{\mbox{{\scriptsize EWP}}}\bigr|\,
e^{i\tilde\delta_{\mbox{{\scriptsize EWP}}}}\,.
\end{equation}
The corresponding charge-conjugate amplitude takes on the other hand the
form
\begin{equation}\label{A32CP}
\overline{A}_{3/2}=\bigl|A_{\pi K}^{\mbox{{\scriptsize T}}}\bigr|\,
e^{i\tilde\delta_T}\,e^{-i\gamma}-\bigl|A_{\pi K}^{\mbox{{\scriptsize 
EWP}}}\bigr|\,e^{i\tilde\delta_{\mbox{{\scriptsize EWP}}}}\,,
\end{equation}
so that the EW penguin contributions cancel in the difference of (\ref{A32})
and (\ref{A32CP}):
\begin{equation}\label{ampl-diff}
A_{3/2}-\overline{A}_{3/2}=2\,i\,e^{i\tilde\delta_T}\,
\bigl|A_{\pi K}^{\mbox{{\scriptsize T}}}\bigr|\,\sin\gamma\,.
\end{equation}
In order to determine this amplitude difference geometrically, both the 
quadrangle depicted in fig.~\ref{fig-quadrangle} and the one corresponding 
to the charge-conjugate processes have to be constructed by measuring the
branching ratios corresponding to (a)--(e). Moreover the relative 
orientation of these two quadrangles in the complex plane has to be fixed. 
This can be done through the side (a) as no non-trivial CP-violating 
weak phase is present in the $\bar b\to\bar s$ penguin-induced decay 
$B^+\to\pi^+K^0$, i.e.\  $A(B^-\to\pi^-\overline{K^0})=A(B^+\to\pi^+K^0)$ 
(see \ref{Zoo}). Since the quantity 
$\bigl|A_{\pi K}^{\mbox{{\scriptsize T}}}\bigr|$ 
corresponds to $|T'+C'|$, it can be determined with the help of the 
$SU(3)$ flavour symmetry (note (\ref{ED1201}) and (\ref{ED1203})) by 
measuring the branching ratio for $B^+\to\pi^+\pi^-$, i.e.\ through 
$\bigl|A_{\pi K}^{\mbox{{\scriptsize T}}}\bigr|=r_u\,\sqrt{2}\,
|A(B^+\to\pi^+\pi^0)|$, so that both $\sin\gamma$ and the strong 
phase $\tilde\delta_T$ can be extracted from the amplitude difference 
(\ref{ampl-diff}). Unfortunately the dashed line (e) corresponds to the 
decay $B^0_s\to\pi^0\eta$ that is dominated by EW 
penguins \cite{rfewp3,dht-ewp} (see \ref{Bspi0phi}) and is 
therefore expected to exhibit a branching ratio at the ${\cal O}(10^{-7})$ 
level. Consequently the amplitude quadrangles are rather squashed ones and 
this approach to determine $\gamma$ is very difficult from an experimental 
point of view. 

\begin{figure}[t]
\centerline{
\rotate[r]{
\epsfysize=9.5truecm
\epsffile{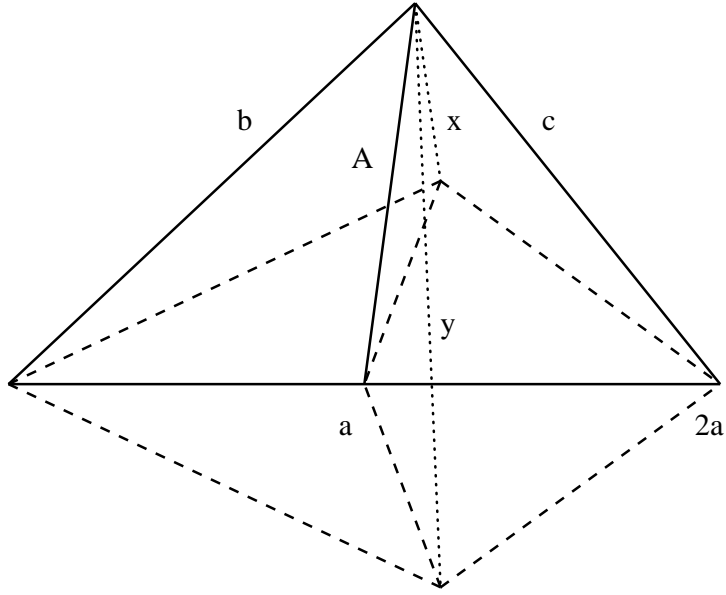}}}
\caption{$SU(3)$ amplitude relations involving $B^+\to\{\pi^+K^0,
\pi^0K^+,\eta_8\,K^+\}$ and charge-conjugates (dashed lines). The labels are 
explained in the text.}\label{fig-dehe}
\end{figure}

\subsubsection{$SU(3)$ Relations among $B^+\to\{\pi^+K^0,
\pi^0K^+,\eta_8\,K^+\}$ Decay Amplitudes}\label{DeHeApp}
Another approach to extract $\gamma$ involving the decays 
$B^+\to\{\pi^+K^0,\pi^0K^+,\eta_8\,K^+\}$ and their charge-conjugates
was proposed by Deshpande and He in \cite{dh-gam}. Using $SU(3)$ flavour 
symmetry, it is possible to derive relations among the corresponding decay 
amplitudes that can be represented in the complex plane as shown in 
fig.~\ref{fig-dehe}. Here the solid lines labelled (a), (b) and (c) 
correspond to the decay amplitudes $A(B^+\to\pi^+K^0)$, $\sqrt{2}\,
A(B^+\to\pi^0 K^+)$ and $\sqrt{6}\,A(B^+\to\eta_8\,K^+)$, respectively, and 
the dashed lines represent the corresponding charge-conjugate amplitudes. 
Note that $A(B^-\to\pi^-\overline{K^0})=A(B^+\to\pi^+K^0)$ has also been 
used in this construction. Similarly as in \ref{AmplQuad}, the determination 
of $\gamma$ can be accomplished by considering the difference of a 
particularly useful chosen combination $A$ of decay amplitudes and its
charge-conjugate $\overline{A}$, where the penguin contributions cancel:
\begin{equation}
A-\overline{A}=2\,\sqrt{2}\,i\,e^{i\tilde\delta_T}r_u\,|A(B^+\to\pi^+\pi^0)|
\,\sin\gamma\,.
\end{equation}
Here the magnitude of the $B^+\to\pi^+\pi^0$ amplitude is used --  as 
in the $B\to\pi K$ quadrangle approach \cite{ghlrewp} -- to fix $|T'+C'|$.
In fig.~\ref{fig-dehe}, the dotted lines (x) and (y) represent two possible 
solutions for this amplitude difference. The fact that this construction 
does not give a unique solution for $A-\overline{A}$ is a well-known 
characteristic feature of all geometrical constructions of this kind, i.e.\
one has in general to deal with several discrete ambiguities. 

Compared to the method using $B\to\pi K$ quadrangles discussed in 
\ref{AmplQuad}, the
advantage of this strategy is that all branching ratios are expected to
be of the same order of magnitude ${\cal O}(10^{-5})$. In particular one
has not to deal with an EW penguin dominated channel with an
expected branching ratio at the ${\cal O}(10^{-7})$ level. However, 
the accuracy of
the strategy is limited by $\eta-\eta'$ mixing, i.e.\ the $A(B^\pm\to
\eta_8\,K^\pm)$ amplitudes have to be determined through
\begin{equation}
A(B^\pm\to\eta_8\,K^\pm)=A(B^\pm\to\eta\, K^\pm)\cos\Theta+
A(B^\pm\to\eta'K^{\pm})\sin\Theta
\end{equation}
with a mixing angle $\Theta\approx20^\circ$, and by other $SU(3)$-breaking
effects which cannot be calculated at present. A similar approach to 
determine $\gamma$ was proposed by Gronau and Rosner in \cite{gr-gam},
where the amplitude construction is expressed in terms of the physical
$\eta$ and $\eta'$ states. A detailed discussion of $SU(3)$ amplitude 
relations for $B$ decays involving $\eta$ and $\eta'$ in light of 
extractions of CKM phases can be found in \cite{eta-etap}.

\subsubsection{A Simple Strategy for Fixing $\gamma$ and Obtaining Insights
into the World of EW Penguins}
\begin{figure}[t]
\centerline{
\rotate[r]{
\epsfysize=10truecm
\epsffile{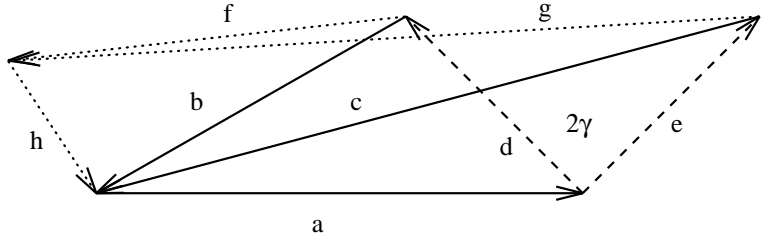}}}
\caption{$SU(2)$ isospin relations among $B^+\to\pi^+K^0$, 
$B^0_d\to\pi^-K^+$ and charge-conjugates. The labels are
explained in the text.}\label{fig-EWdet}
\end{figure}

Since the geometrical constructions discussed in \ref{AmplQuad} and 
\ref{DeHeApp} are
quite complicated and appear to be very challenging from an experimental
point of view, let us consider a much simpler approach to determine 
$\gamma$ \cite{PAPIII}. It uses the decays $B^+\to\pi^+K^0$, 
$B^0_d\to\pi^-K^+$ and their charge-conjugates. In the case of these 
transitions, EW penguins contribute only in colour-suppressed form and hence
play a minor role. Neglecting these contributions and using 
the $SU(2)$ isospin symmetry of strong interactions -- not $SU(3)$ -- 
to relate their QCD penguin contributions (note the similarity to the 
example given in \ref{BsiLoDG}), the corresponding decay amplitudes can be
written in the GHLR notation as \cite{ghlr}
\begin{eqnarray}
A(B^+\to\pi^+K^0)&=&P'\,=\,A(B^-\to\pi^-\overline{K^0})\nonumber\\
A(B^0_d\to\pi^-K^+)&=&-\,(P'+T')\\
A(\overline{B^0_d}\to\pi^+K^-)&=&-\,(P'+e^{-2i\gamma}\,T')\,.\nonumber
\end{eqnarray}
Let us note that these relations are on rather solid ground from a 
theoretical point of view. They can be represented in the complex plane 
as shown in fig.~\ref{fig-EWdet}. Here (a) corresponds to 
$A(B^+\to\pi^+K^0)=P'=A(B^-\to\pi^-\overline{K^0})$,
(b) to $A(B^0_d\to\pi^-K^+)$, (c) to $A(\overline{B^0_d}\to\pi^+K^-)$ and
the dashed lines (d) and (e) to the colour-allowed current-current 
amplitudes $T'$ and $e^{-2i\gamma}\,T'$, respectively. The dotted lines 
(f)--(h) will be discussed in a moment. Note that these $B\to\pi K$ decays
appeared already in \ref{SIMalp-gam}. Combining their branching ratios with 
the observables of a time-dependent measurement of $B_d\to\pi^+\pi^-$, a
simultaneous extraction of $\alpha$ and $\gamma$ may be possible \cite{dgr}. 
The information provided by the $B\to\pi K$ modes can, however, also be
used for a quite different approach that may finally allow the determination
of EW penguin amplitudes.

In order to determine $\gamma$ from fig.~\ref{fig-EWdet}, we have to know the 
length $|T'|$ of the dashed lines (d) and (e). In fact, the situation is 
analogous to the 
extraction of $\gamma$ from (\ref{EE4}) and (\ref{EE5}) in \ref{BsiLoDG}. 
There we 
saw that $B^+\to\pi^+\pi^0$ provides an estimate of that quantity through 
(\ref{EE7}) which is based on two assumptions: $SU(3)$ flavour symmetry and 
neglect of colour-suppressed current-current contributions to 
$B^+\to\pi^+\pi^0$. Consequently, following these lines, it is possible to
obtain an estimate of $\gamma$ by measuring only $Br(B^+\to\pi^+K^0)=
Br(B^-\to\pi^-\overline{K^0})$, $Br(B^0_d\to\pi^-K^+)$,
$Br(\overline{B^0_d}\to\pi^+K^-)$ and $Br(B^+\to\pi^+\pi^0)=Br
(B^-\to\pi^-\pi^0)$. Note that the neutral $B_d$ decays are ``self-tagging''
modes so that no time-dependent measurements are needed and that this 
estimate of $\gamma$ is very similar to the ``original'' GRL 
approach \cite{grl-gam} shown in fig.~\ref{grl-const} that is unfortunately
spoiled by EW penguins. Needless to say, this strategy is very simple from 
a geometrical point of view -- just triangle constructions -- and very 
promising from an experimental point of view since all branching ratios are 
of the same order of magnitude ${\cal O}(10^{-5})$. Moreover no 
experimentally difficult CP eigenstate of the neutral $D$ system is 
required as in \ref{BDKtri}.

Let us emphasize that the ``weak'' point of this approach -- and of 
the one using untagged $B_s$ decays discussed in \ref{BsiLoDG} -- is the 
relation (\ref{EE7}) to estimate $|T'|$. Therefore this ``estimate'' of 
$\gamma$ may well turn into a solid ``determination'' if it should become
possible to fix the magnitude of the colour-allowed current-current
amplitude contributing to $B^0_d\to\pi^-K^+$ in a more reliable way. 
Another possibility of fixing $|T'|$ is of course the factorization 
hypothesis which may work reasonably well for that colour-allowed 
amplitude \cite{bjorken} and could be used as some kind of cross-check 
for (\ref{EE7}). Maybe the ``final'' result for $|T'|$ will come from 
lattice gauge theory one day. 

Interestingly the construction shown in fig.~\ref{fig-EWdet} provides even
more information if one takes into account the amplitude relations
\begin{eqnarray}
\sqrt{2}\,A(B^+\to\pi^0K^+)&\approx&-\,\left[P'+T'+(c_u-c_d)
P_{\mbox{{\scriptsize EW}}}'\right]\\
\sqrt{2}\,A(B^-\to\pi^0K^-)&\approx&-\,\left[P'+e^{-2i\gamma}\,T'+(c_u-c_d)
P_{\mbox{{\scriptsize EW}}}'\right],
\end{eqnarray}
where colour-suppressed current-current and EW penguin amplitudes have
been neglected. Consequently, the dotted lines (f) and (g) corresponding to
$\sqrt{2}\,A(B^+\to\pi^0K^+)$ and $\sqrt{2}\,
A(B^-\to\pi^0K^-)$, respectively, allow a determination 
of the dotted line (h) denoting the colour-allowed \mbox{$\bar b\to
\bar s$} EW penguin amplitude $(c_u-c_d)P_{\mbox{{\scriptsize EW}}}'$. 
Since EW penguins are -- in contrast to QCD penguins -- dominated to 
excellent accuracy by internal top-quark exchanges, the $\bar b\to\bar d$
EW penguin amplitude $(c_u-c_d)P_{\mbox{{\scriptsize EW}}}$ is related in 
the limit of an exact $SU(3)$ flavour symmetry to the corresponding 
$\bar b\to\bar s$ amplitude through the simple relation
\begin{equation}
(c_u-c_d)P_{\mbox{{\scriptsize EW}}}=-\lambda\,R_t\,e^{-i\beta}\,(c_u-c_d)
P_{\mbox{{\scriptsize EW}}}'
\end{equation}
and may consequently be determined from the constructed 
$(c_u-c_d)P_{\mbox{{\scriptsize EW}}}'$ amplitude. 

\subsubsection{Towards Control over EW Penguins}
It would be very useful to determine the EW penguin contributions 
experimentally. That would allow several predictions, consistency checks
and tests of certain Standard Model calculations \cite{PAPIII}. For example, 
one may determine the quantity $x$ parametrizing the EW penguin effects in 
$B_s\to\pi^0\phi$ experimentally and may compare this result with the 
Standard Model expression (\ref{ED1121}). That way 
one may obtain predictions for ${\cal A}_{\mbox{{\scriptsize 
CP}}}^{\mbox{{\scriptsize mix-ind}}}(B_s\to\pi^0\phi)$ and 
$Br(B_s\to\pi^0\phi)$ long before it might be possible (if it is possible 
at all!) to measure them directly. Another interesting point is that the
$\bar b\to\bar d$ EW penguin amplitude $P_{\mbox{{\scriptsize EW}}}$ 
allows in principle to fix the uncertainty $\Delta\alpha$ (see 
(\ref{deltaalpha})) arising from EW penguins in the GL 
method \cite{gl} for extracting $\alpha$ and to check whether it is e.g.\
in agreement with (\ref{deltaalpha-est}). Since EW penguins are ``rare'' 
FCNC processes that are absent at tree level within the 
Standard Model, it may well be that ``New Physics'' contributes to 
them significantly 
through additionally present virtual particles in the loops. Consequently
EW penguins may give hints to physics beyond the Standard Model.

We have just seen an example of a simple strategy to determine EW penguin 
amplitudes experimentally. In \cite{PAPI}, where more involved methods
to accomplish this task are discussed, we pointed out that the 
central input to control EW penguins in a quantitative way is the CKM angle 
$\gamma$. Consequently determinations of this UT angle are not only 
important in respect of testing the Standard Model description of CP 
violation but also to shed light on the physics of EW penguins. 

\subsection{Summary and Outlook}
Contrary to naive expectations, EW penguins may play an important --
in some cases, e.g.\ $B_s\to\pi^0\phi$, even dominant -- role in certain
non-leptonic $B$ decays because of the large top-quark mass. The EW
penguin contributions spoil the determination of $\gamma$ using 
$B^+\to\{\pi^+\pi^0,\pi^+K^0,\pi^0K^+\}$ (and charge-conjugate)
$SU(3)$ triangle relations and require in general more involved 
geometrical constructions, e.g.\ $B\to\pi K$ quadrangles, to extract this 
UT angle which are difficult from an experimental point of view. There is, 
however, also a simple ``estimate'' of $\gamma$ using only triangles which
involve the $B^+\to\pi^+K^0$, $B^0_d\to\pi^-K^+$ and charge-conjugate
decay amplitudes. This approximate approach is more promising for 
experimentalists and may turn into a ``determination'' if the magnitude 
of the colour-allowed $\bar b\to\bar s$ current-current amplitude, which is
its major input, can be determined reliably. Measuring in addition the
branching ratios for $B^\pm\to\pi^0K^\pm$, also the EW penguin amplitudes
can be determined experimentally which should allow valuable insights into
the physics of these FCNC processes.  There are more refined strategies
to control EW penguins in a quantitative way that require $\gamma$ as
an input. These methods may allow valuable insights into the world of
EW penguins and could give indications for ``New
Physics''.

\section{Classification}
\setcounter{equation}{0}
In this review we have discussed a large number of $K$- and $B$-decays
paying attention to their theoretical cleanliness and their usefulness in
the determination of the parameters of the Standard Model.
It is probably a good idea to summarize the situation by grouping various 
decays and related quantities into four distinct classes with respect to 
theoretical uncertainties. We are aware of the fact that not everybody
would fully agree with this classification and that some decays
could be moved upwards or downwards by one class. Moreover we
expect that this classification may change in time as our
understanding of non-perturbative effects improves. 

\subsection{Gold-Plated Class}
\begin{itemize}
\item
CP asymmetry in $B_d\to J/\psi K_{\rm S}$, which
measures the angle $\beta$, and 
CP asymmetries in $B_{u,d}\to DK$, $B_s\to D_sK,\,D_s^\ast K^\ast$ and 
$B_s\to D\phi,\,D^\ast \phi$ all relevant for the angle $\gamma$. 
\item
The ratio  $Br(B\to X_d\nu\bar\nu)/ Br(B\to X_s\nu\bar\nu)$
which offers the cleanest direct determination of the ratio 
$|V_{td}/V_{ts}|$,
\item
Rare $K$-decays $K_{\rm L} \to \pi^0\nu\bar\nu$ and $K^+\to \pi^+\nu\bar\nu$
which offer very clean determinations of $\IM\lambda_t (\eta)$ and
$|V_{td}|$, respectively. In particular $\klpn$ seems to allow the
cleanest determination of $\imlt$. Taking together these two decays
offers a clean determination of $\sin 2 \beta$.
\end{itemize}
Except for the extraction of $\vtd$ from $\kpn$,
which suffers from roughly $\pm 4\%$
uncertainty due to the charm contribution, the theoretical
uncertainties in the remaining quantities in this class are 
conservatively below $\pm 2\%$.

\subsection{Class 1}
\begin{itemize}
\item
CP asymmetries in $B_d\to \pi^+\pi^-$, $B_d\to D^+D^-,\,\phi K_{\rm S}$,
$B_s\to K^+K^-,\,K^{*+}K^{*-}$ and
$B_s\to J/\psi\,\phi,\, D_s^{*+}D_s^{*-}$ 
relevant for the angles $\alpha$, $\beta$, $\gamma$ 
and the parameter $\eta$, respectively.
As discussed in previous sections, most of these CP asymmetries
require additional strategies in order to determine the CKM parameters
in question  without hadronic uncertainties.
\item
Ratios $Br(B_d\to l\bar l)/Br(B_s\to l\bar l)$ and
$(\Delta M)_d/(\Delta M)_s$
which give  good measurements of $|V_{td}/V_{ts}|$ 
provided the $SU(3)$ breaking effects in the ratios $F_{B_d}/F_{B_s}$
and $\sqrt{B_d}F_{B_d}/\sqrt{B_s}F_{B_s}$ can be brought under control.
\item
$\vcb_{{\rm excl}}$, 
$\vcb_{{\rm incl}}$,
$| V_{ub}/V_{cb}|_{{\rm incl}}$
\end{itemize}
Due to the need for various strategies, which involve
generally several channels and the need for non-perturbative
estimates in certain cases, it appears that it will be difficult to
achieve the extraction of the Standard Model parameters to better
than $(5-10)\%$ from the decays in this class.
\subsection{Class 2}
\begin{itemize}
\item
$B\to X_{s,d}\gamma$, $B\to X_{s,d} e^+ e^-$, $B\to K^*(\rho)e^+e^-$
\item
$(\Delta M)_d$, $(\Delta M)_s$ 
\item
CP asymmetries in $B_s\to J/\psi K_{\rm S}$ and $B_{u,d} \to \pi K$.
\item
$\varepsilon_K$ and $K_{\rm L}\to \pi^0 e^+e^-$
\end{itemize}
Here we group  quantities or decays with presently moderate or substantial
theoretical uncertainties which should be considerably reduced in the next
five years. In particular we assume that the uncertainties in $B_K$
and $\sqrt{B}F_B$ will be reduced below 10\% and that the question of
the importance of the CP-conserving and indirectly CP-violating
contributions to $K_{\rm L}\to \pi^0 e^+e^-$ will be answered somehow.
We also assume that the knowledge of the long distance contributions to
$B\to X_{s,d}\gamma$, $B\to X_{s,d} \mu^+ \mu^-$ and 
$B\to K^*(\rho)\mu^+\mu^-$
will be improved. In view of all these requirements it is difficult
to estimate to which level the theoretical uncertainties can be reduced,
but an estimate of $(10-15)\%$ depending on the quantity considered
appears to be a reasonable one.
\subsection{Class 3}
\begin{itemize}
\item
CP asymmetries in most $B^{\pm}$-decays
\item
$B_d\to K^*\gamma$, non-leptonic $B$-decays, $| V_{ub}/V_{cb}|_{{\rm excl}}$
\item
$\varepsilon^{\prime} /\varepsilon$, $K\to \pi\pi$, 
$\Delta M(K_{\rm L}-K_{\rm S})$,
$K_{\rm L}\to\mu\bar\mu$, hyperon decays and so on.
\end{itemize}
Here we have a list of important decays with large theoretical
uncertainties which can only be removed by a dramatic progress
in non-perturbative techniques.
It should be stressed that even in the presence of theoretical
uncertainties a measurement of a non-vanishing 
ratio $\varepsilon^{\prime}/\varepsilon$ or a non-vanishing CP asymmetry
in charged $B$-decays would signal direct CP violation excluding
superweak scenarios \cite{wolfenstein:64}. This is not guaranteed by
several clean decays of the gold-plated class or class 1 \cite{WIN} except for 
$\klpn$ and $B^{\pm}\to D_{\rm CP} K^{\pm}$.

\section{Future Visions}
\setcounter{equation}{0}
Let us next have a look in the future and ask the question how well various
parameters of the Standard Model can be determined provided the
cleanest decays of the ``gold-plated'' class and class 1 have been measured
to some respectable precision. We have made already such an exercise in
section \ref{sec:Kpnn:Triangle}
using the decays $\klpn$ and $\kpn$. Now we want to make
an analogous analysis using CP-asymmetries in $B$-decays. This way we
will be able to compare the potentials of the CP asymmetries in
determining the parameters of the Standard Model with those
of the cleanest rare $K$-decays: $K_{\rm L}\to\pi^0\nu\bar\nu$ and
$K^+\to\pi^+\nu\bar\nu$. This section is based on \cite{BLO,AJB94,BB96}.

\subsection{CP-Asymmetries in $B$-Decays versus $K \to \pi \nu\bar\nu$}

In what follows let us assume that the problems with the determination
of $\alpha$ will be solved somehow. Since in the usual rescaled 
unitarity triangle  one side is known, it suffices to measure
two angles to determine the triangle completely. This means that
the measurements of $\sin 2\alpha$ and $\sin 2\beta$ can determine
the parameters $\varrho$ and $\eta$.
As the standard analysis of the unitarity triangle of section 4
shows, $\sin 2\beta$ is expected to be large: $\sin 2\beta=0.58\pm 0.22$
implying the time-integrated CP asymmetry  
$a_{\rm CP}(B_d\to J/\psi K_{\rm S})$
as high as $(30 \pm 10)\%$.
The prediction for $\sin 2\alpha$ is very
uncertain on the other hand $(0.1\pm0.9)$ and even a rough measurement
of $\alpha$ would have a considerable impact on our knowledge of
the unitarity triangle as stressed in \cite{BLO} and recently in
\cite{BB96}.

Measuring then $\sin 2\alpha$ and $\sin 2\beta$ from CP asymmetries in
$B$ decays allows, in principle, to fix the 
parameters $\bar\eta$ and $\bar\varrho$, which can be expressed as
\cite{AJB94}
\begin{equation}\label{ersab}
\bar\eta=\frac{r_-(\sin 2\alpha)+r_+(\sin 2\beta)}{1+
r^2_+(\sin 2\beta)}\,,\qquad
\bar\varrho=1-\bar\eta r_+(\sin 2\beta)\,,
\end{equation}
where $r_\pm(z)=(1\pm\sqrt{1-z^2})/z$.
In general the calculation of $\bar\varrho$ and $\bar\eta$ from
$\sin 2\alpha$ and $\sin 2\beta$ involves discrete ambiguities.
As described in \cite{AJB94}
they can be resolved by using further information, e.g.\ bounds on
$|V_{ub}/V_{cb}|$, so that eventually the solution (\ref{ersab})
is singled out.

Let us then consider two scenarios of the measurements of CP asymmetries 
in $B_d\to\pi^+\pi^-$ and $B_d\to J/\psi K_{\rm S}$, expressed in terms 
of $\sin 2\alpha$ and
$\sin 2\beta$:
\begin{equation}\label{sin2a2bI}
\sin 2\alpha=0.40\pm 0.10\,, \qquad \sin 2\beta=0.70\pm 0.06
\qquad ({\rm scenario\ I})
\end{equation}
\begin{equation}\label{sin2a2bII}
\sin 2\alpha=0.40\pm 0.04\,, \qquad \sin 2\beta=0.70\pm 0.02
\qquad ({\rm scenario\ II})\,.
\end{equation}
Scenario I corresponds to the accuracy being aimed for at $B$-factories
and HERA-B prior to the LHC era. An improved precision can be anticipated from
LHC experiments, which we illustrate with the scenario II.

In table \ref{tabkb} this way of the determination of
the Standard Model parameters is compared with the analogous analysis
using $\klpn$ and $\kpn$ which has been presented in section 7.4. We
recall that in the latter analysis
the following input has been used:
\begin{equation}\label{vcbmt}
|V_{cb}|=0.040\pm 0.002(0.001)\,, \qquad m_t=(170\pm 3) \mbox{GeV}
\end{equation}
\begin{equation}\label{bklkp}
Br(K_{\rm L}\to\pi^0\nu\bar\nu)=(3.0\pm 0.3)\cdot 10^{-11}\,,\qquad
Br(K^+\to\pi^+\nu\bar\nu)=(1.0\pm 0.1)\cdot 10^{-10}\,.
\end{equation}

As can be seen in table \ref{tabkb}, the CKM determination
using $K\to\pi\nu\bar\nu$ is competitive with the one based
on CP violation in $B$ decays studied prior to the LHC era, except 
for $\bar\varrho$ which
is less constrained by the rare kaon processes.
The LHC-B experiment should generally give higher precision than
obtainable from $\kpn$ and $\klpn$ unless the assumed experimental
errors in (\ref{bklkp}) are lowered.
\begin{table}
\begin{center}
\begin{tabular}{|c||c|c|c|}\hline
&$K\to\pi\nu\bar\nu$&$B\to\pi\pi, J/\psi K_{\rm S}$ (I) 
&$B\to\pi\pi, J/\psi K_{\rm S}$ (II) \\
\hline
\hline
$|V_{td}|/10^{-3}$&$10.3\pm 1.1(\pm 0.9)$&$8.8\pm 0.5(\pm 0.3)$ 
&$8.8\pm 0.5(\pm 0.2)$ \\
\hline
$|V_{ub}/V_{cb}|$&$0.089\pm 0.017(\pm 0.011)$
&$0.087\pm 0.009(\pm 0.009)$&$0.087\pm 0.003(\pm 0.003)$ \\
\hline 
$\bar\varrho$&$-0.10\pm 0.16(\pm 0.12)$&$0.07\pm 0.03(\pm 0.03)$
&$0.07\pm 0.01(\pm 0.01)$ \\
\hline
$\bar\eta$&$0.38\pm 0.04(\pm 0.03)$&$0.38\pm 0.04(\pm 0.04)$
&$0.38\pm 0.01(\pm 0.01)$ \\
\hline
$\sin 2\beta$&$0.62\pm 0.05(\pm 0.05)$&$0.70\pm 0.06(\pm 0.06)$
&$0.70\pm 0.02(\pm 0.02)$ \\
\hline
${\rm Im}\lambda_t/10^{-4}$&$1.37\pm 0.07(\pm 0.07)$
&$1.37\pm 0.19(\pm 0.15)$&$1.37\pm 0.14(\pm 0.08)$ \\
\hline
\end{tabular}
\end{center}
\caption[]{Illustrative example of the determination of CKM
parameters from $K\to\pi\nu\bar\nu$ and from CP-violating
asymmetries in $B$ decays \cite{BB96}. The relevant input is as described
in the text. Shown in brackets are the errors one obtains
using $V_{cb}=0.040\pm 0.001$ instead of $V_{cb}=0.040\pm 0.002$.
\label{tabkb}}
\end{table}
On the other hand, ${\rm Im}\lambda_t$ is better determined
in the kaon scenario. It can be obtained from
$K_{\rm L}\to\pi^0\nu\bar\nu$ alone and does not require knowledge
of $V_{cb}$ which enters ${\rm Im}\lambda_t$ when derived
from $\sin 2\alpha$ and $\sin 2\beta$.
This analysis suggests that $K_{\rm L}\to\pi^0\nu\bar\nu$ should eventually 
yield the most accurate value of ${\rm Im}\lambda_t$.

There is another virtue of the comparision of the determinations
of various parameters using CP-B asymmetries with the determinations
in very clean decays $K\to\pi\nu\bar\nu$. Any substantial deviations
from these two determinations would signal new physics beyond the
Standard Model. Formula (\ref{kbcon}) is an example of such a comparision.

\subsection{Unitarity Triangle from $\klpn$ and $\sin 2\alpha$}

Next, results from CP asymmetries in $B$ decays could also be
combined with measurements of $K\to\pi\nu\bar\nu$.
As an illustration we would like to present a scenario \cite{BB96}
where
the unitarity triangle is determined by $\lambda$, $V_{cb}$,
$\sin 2\alpha$ and $Br(K_{\rm L}\to\pi^0\nu\bar\nu)$.
In this case $\bar\eta$ follows directly from 
$Br(K_{\rm L}\to\pi^0\nu\bar\nu)$ (\ref{bklpn1}) and $\bar\varrho$ is
obtained using \cite{AJB94}
\begin{equation}\label{rhoalpha}
\bar\varrho=\frac{1}{2}-\sqrt{\frac{1}{4}-\bar\eta^2+
\bar\eta r_-(\sin 2\alpha)}\,,
\end{equation}
where $r_-(z)$ is defined after eq.\ (\ref{ersab}).
The advantage of this strategy is that most CKM quantities are
not very sensitive to the precise value of $\sin 2\alpha$.
Moreover a high accuracy in 
${\rm Im}\lambda_t$ is automatically guaranteed. As shown in
table \ref{tabkl2a}, very respectable results can be expected
for other quantities as well with only modest requirements
on the accuracy of $\sin 2\alpha$. 
It is conceivable that theoretical uncertainties due to penguin
contributions could eventually be brought under control at least
to the level assumed in table \ref{tabkl2a}. 
\begin{table}
\begin{center}
\begin{tabular}{|c||c|c|c|}\hline
&&A&B \\
\hline
\hline
$\bar\eta$&$0.380$&$\pm 0.043$&$\pm 0.028$ \\
\hline
$\bar\varrho$&$0.070$&$\pm 0.058$&$\pm 0.031$ \\
\hline
$\sin 2\beta$&$0.700$&$\pm 0.077$&$\pm 0.049$ \\
\hline
$|V_{td}|/10^{-3}$&$8.84$&$\pm 0.67$&$\pm 0.34$ \\
\hline
$|V_{ub}/V_{cb}|$&$0.087$&$\pm 0.012$&$\pm 0.007$ \\
\hline 
\end{tabular}
\end{center}
\caption[]{Determination of the CKM matrix from $\lambda$, $V_{cb}$,
$K_{\rm L}\to\pi^0\nu\bar\nu$ and $\sin 2\alpha$ from the CP asymmetry
in $B_d\to\pi^+\pi^-$ \cite{BB96}. Scenario A (B) assumes
$V_{cb}=0.040\pm 0.002 (\pm 0.001)$
and $\sin 2\alpha=0.4\pm 0.2 (\pm 0.1)$. In both cases we take
$Br(K_{\rm L}\to\pi^0\nu\bar\nu)\cdot 10^{11}=3.0\pm 0.3$ and
$\mt=(170\pm 3)\gev$. 
\label{tabkl2a}}
\end{table}
As an alternative, $\sin 2\beta$ from $B_d\to J/\psi K_{\rm S}$ 
could be used as an independent input instead of $\sin 2\alpha$.
Unfortunately the combination of $K_{\rm L}\to\pi^0\nu\bar\nu$ and
$\sin 2\beta$ tends to yield somewhat less restrictive constraints
on the unitarity triangle \cite{BB96}. 
On the other hand it has of course the
advantage of being practically free of any theoretical uncertainties.   

\subsection{Unitarity Triangle and $\vcb$ from $\sin 2\alpha$,
$\sin 2\beta$ and $\klpn$}
As proposed in \cite{AJB94},
unprecedented precision for all basic CKM
parameters could be achieved by combining the cleanest $K$ and 
$B$ decays. 
While $\lambda$ is obtained as usual from
$K\to\pi e\nu$, $\bar\varrho$ and $\bar\eta$ could be determined
from $\sin 2\alpha$ and $\sin 2\beta$ as measured in CP
violating asymmetries in $B$ decays. Given $\eta$, one could
take advantage of the very clean nature of $K_{\rm L}\to\pi^0\nu\bar\nu$
to extract $A$ or, equivalently $|V_{cb}|$. As seen in (\ref{vcbklpn}),
this determination
benefits further from the very weak dependence of $|V_{cb}|$ on
the $K_{\rm L}\to\pi^0\nu\bar\nu$ branching ratio, which is only with
a power of $0.25$. Moderate accuracy in $Br(K_{\rm L}\to\pi^0\nu\bar\nu)$
would thus still give a high precision in $|V_{cb}|$.
As an example we take $\sin 2\alpha=0.40\pm 0.04$,
$\sin 2\beta=0.70\pm 0.02$ and 
$Br(K_{\rm L}\to\pi^0\nu\bar\nu)=(3.0\pm 0.3)\cdot 10^{-11}$,
$m_t=(170\pm 3)$ GeV. 
This yields \cite{BB96}:
\begin{equation}\label{rhetvcb}
\bar\varrho=0.07\pm 0.01\,,\qquad
\bar\eta=0.38\pm 0.01\,,\qquad
|V_{cb}|=0.0400\pm 0.0013\,,
\end{equation}
which would be a truly remarkable result. Again the comparision of
this determination of $|V_{cb}|$ with the usual one in tree level
$B$-decays would offer an excellent test of the standard model
and in the case of discrepancy would signal physics beyond the
standard model.  

\subsection{Unitarity Triangle from $R_t$ and $\sin 2\beta$} 
Another strategy is to use the measured value of $R_t$ together with
$\sin 2\beta$. Useful measurements of $R_t$ can be achieved
using the ratios $Br(B\to X_d \nu\bar\nu)/Br(B\to X_s \nu\bar\nu)$,
$\Delta M_d/\Delta M_s$,
$Br(B_d\to l^+l^-)/Br(B_s \to l^+l^-)$
 and $Br(\kpn)$. Then (\ref{ersab})
is replaced by \cite{B95}
\begin{equation}\label{5a}
\bar\eta=\frac{R_t}{\sqrt{2}}\sqrt{\sin 2\beta \cdot r_{-}(\sin 2\beta)}\,,
\quad\quad
\bar\varrho = 1-\bar\eta r_{+}(\sin 2\beta)\,.
\end{equation}
The numerical results of this exercise can be found in \cite{B95}.
Additional strategies involving the angle $\gamma$ 
can be found in \cite{BLO}.

\section{Summary and Outlook}
\setcounter{equation}{0}
We are approaching the end of our review. We hope, we have given here
a proper 
account of the highlights of this field and succeeded to equip
the reader with a collection of formulae for most interesting
quantities which should be useful in various phenomenological
applications. 
We also hope that we have convinced the reader about the important 
role this field
plays in the deeper understanding of the Standard Model 
and particle physics in
general. It is also evident that the experimental work to be 
done in the next ten
years at BNL, CERN, CORNELL, DA$\Phi$NE, DESY, 
FNAL, KEK  and SLAC will have considerable impact on this field.

\begin{table}[thb]
\begin{center}
\begin{tabular}{|c||c||c||c|}\hline
{\bf Quantity} & {\bf Scanning} & {\bf Gaussian} & {\bf Experiment}
 \\ \hline
$Br(K_L\to\pi^0 e^+e^-)_{\rm dir} $ &$(4.5\pm 2.6)\cdot 10^{-12}$ 
&$(4.2\pm 1.4)\cdot 10^{-12}  $ &$<  4.3\cdot 10^{-9}$\\ \hline
$Br(B \to X_s\gamma)$ &$ - $ &$ (3.28\pm 0.33)\cdot 10^{-4}$
&$ (2.32\pm 0.67)\cdot 10^{-4}$    \\ \hline
$Br(B\to X_s \mu^+\mu^-)_{\rm NR}$ &$-$ &$ (5.7\pm 0.9)\cdot 10^{-6} $
& $ < 3.6 \cdot 10^{-5}$  \\ \hline
$Br(K^+\to \pi^+\nu\bar\nu)$ &$(9.1\pm 3.2)\cdot 10^{-11} $ 
&$ (8.0\pm 1.5)\cdot 10^{-11} $ & $ <2.4\cdot 10^{-9}$ \\ \hline
$Br(K_L\to \pi^0\nu\bar\nu)$ &$(2.8\pm 1.7)\cdot 10^{-11} $ 
&$ (2.6\pm 0.9)\cdot 10^{-11} $ & $ <5.8\cdot 10^{-5}$ \\ \hline
$Br(B\to X_s\nu\bar\nu)$ &$(3.4\pm 0.7)\cdot 10^{-5} $ 
&$ (3.2\pm 0.4)\cdot 10^{-5} $ & $ <7.7\cdot 10^{-4}$ \\ \hline
$Br(B_s\to \mu^+\mu^-)$ &$(3.6\pm 1.9)\cdot 10^{-9} $ 
&$ (3.4\pm 1.2)\cdot 10^{-9} $ & $ <8.4\cdot 10^{-6}$ \\ \hline
$Br(K_L\to \mu \mu)$ &$(1.2\pm 0.6)\cdot 10^{-9}(*) $ 
&$ (1.0\pm 0.3)\cdot 10^{-9}(*) $ & $(7.2\pm 0.5)\cdot 10^{-9} $  
\\ \hline
\end{tabular}
\caption[]{Predictions for various rare decays in the Standard
Model. The * in the last
row indicates prediction for the short distance contribution
only. 
\label{TAB5}}
\end{center}
\end{table}

Indeed the field of weak decays and of CP violation is one of the least
understood sectors of the Standard Model.
Even if the Standard Model is fully consistent with the existing data for
weak decay processes (see table \ref{TAB5}), the near future could change 
this picture
dramatically through the advances in experiment and theory. Let us
then enumarate what one could expect in the coming ten years:

\begin{itemize}
\item
The error on $\vcb$ and $\vub$ could be decreased below 0.002
and 0.01, respectively. This progress should come mainly from
Cornell, $B$-factories and new theoretical efforts.
\item
The error on $\mt$ should be decreased down to $\pm 3\gev$
at Tevatron in the Main Injector era and to $\pm 1\gev$ at LHC.
\item
The improved measurements of $\epe$ at the $\pm (1-2) \cdot 10^{-4}$ 
level from CERN, FNAL and DA$\Phi$NE should give some insight into the 
physics of 
direct CP violation in spite of large theoretical uncertainties. 
Excluding confidently the superweak models would be an important result. 
In this respect measurements of CP-violating asymmetries in charged $B$
decays will also play an outstanding role. These experiments can be
performed e.g.\ at CLEO since no time-dependences
are needed. The situation concerning hadronic uncertainties is quite similar
to $\epe$. Although these CP asymmetries cannot be calculated
reliably, any measured non-vanishing values would unambiguously rule out 
superweak scenarios. Simultaneously one should hope 
that some definite progress in calculating relevant hadronic matrix elements 
will be made. 
\item
The first events for $K^+\to\pi^+\nu\bar\nu$ could in principle
be seen at BNL already this or next year. In view of the theoretical 
cleanliness of this decay an observation of events at the $2\cdot 10^{-10}$
level would signal physics beyond the Standard Model.
A detailed study of this very
important decay requires, however, new experimental ideas and
new efforts. The new efforts \cite{AGS2,Cooper} in this direction allow 
to hope that
a measurement of $Br(\kpn)$ with an accuracy of $\pm 10 \%$ should
be possible before 2005.
\item
The future improved inclusive $B \to X_{s,d} \gamma$ measurements
confronted with improved Standard Model predictions could
give the first signals of new physics. It appears that the
theoretical error on $Br(B\to X_s\gamma)$ could be decreased
confidently down to $\pm 10 \%$ in the next years. The same
accuracy in the experimental branching ratio will hopefully
come soon from CLEO II. This may, however, be insufficient to
disentangle new physics contributions although such an accuracy
should put important constraints on the physics beyond the Standard
Model. It would also be desirable to look for $B \to X_d \gamma$,
but this is clearly a much harder task.
\item
Similar comments apply to transitions $B \to X_s l^+l^-$
which appear to be even  more sensitive to new physics contributions
than $ B \to X_{s,d} \gamma$. An observation of
$B \to X_s \mu\bar\mu$ is expected from D0 and $B$-physics dedicated
experiments at the beginning of the next 
decade. The distributions of various kind when measured should
be very useful in the tests of the Standard Model and its extensions.
\item
The theoretical status of $K_{\rm L}\to \pi^0 e^+ e^-$ and of 
$K_{\rm L}\to \mu\bar\mu$ should be improved to confront future
data. Experiments at DA$\Phi$NE should be very helpful in this
respect. The first events of $K_{\rm L}\to \pi^0 e^+ e^-$ should
come in the first years of the next decade from KAMI at FNAL.
The experimental status of $K_{\rm L}\to \mu\bar\mu$, with the 
experimental error of $\pm 7\%$ to be decreased soon down to $\pm 1\%$,
is truly impressive.
\item
The newly approved experiment at BNL to
measure $Br(\klpn)$ at the $\pm 10\%$ level before 2005 may make a decisive
impact on the field of CP violation. 
In particular $\klpn$ seems to allow the
cleanest determination of $\imlt$. Taken together with $\kpn$
a very clean determination of $\sin 2 \beta$ can be obtained.
\item
The measurement of the $B^0_s-\bar B^0_s$ mixing and in particular of
$B \to X_{s,d}\nu\bar\nu$ and 
$B_{s,d}\to \mu\bar\mu$ will take most probably longer time but
as stressed in this review all efforts should be made to measure
these transitions. Considerable progress on $B^0_s-\bar B^0_s$ mixing
should be expected from HERA-B, SLAC and TEVATRON in the first years
of the next decade. LHC-B should measure it to a high precision.
With the improved calculations of $\xi$ in (\ref{107b}) this will have
important impact on the determination of $\vtd$ and on the
unitarity triangle. 
\item
Clearly future precise studies of CP violation at SLAC-B, KEK-B, 
HERA-B, CORNELL, FNAL and  LHC-B providing first
direct measurements of $\alpha$, $\beta$ and $\gamma$ may totally
revolutionize our field. In particular the first signals
of new physics could be found in the $(\bar\varrho,\bar\eta)$ plane.
During the recent years several, in some cases quite sophisticated and
involved, strategies have been developed to extract these angles with
small or even no hadronic uncertainties. Certainly the future will bring
additional methods to determine $\alpha$, $\beta$ and $\gamma$. 
Obviously it is very desirable to have as many such strategies as possible
available in order to overconstrain the unitarity triangle and to resolve 
certain discrete ambiguities which are a characteristic feature of these 
methods.
\item
The forbidden or strongly suppressed transitions such as
$D^0-\bar D^0$ mixing and $K_{\rm L}\to \mu e$ are also very
important in this respect. Considerable progress in this area
should come from the experiments at BNL, FNAL and KEK.
\item
One should hope that the non-perturbative
methods will be considerably improved. In this connection important
lessons will come from DA$\Phi$NE which is an excellent machine
for testing chiral perturbation theory and other non-perturbative
methods. Further lessons will come from $D$- and $B$-physics experiments
studying in particular non-leptonic decays.
\end{itemize}

In any case the field of weak decays and in particular of the FCNC 
transitions and of CP violation have a great future and
one should expect that they could dominate particle physics in the first 
part of the next decade. 

Clearly the next ten years should be very exciting.

\vspace{0.5cm}

{\bf Acknowledgements}

First of all we would like to thank 
Gerhard Buchalla, Isi Dunietz, Matthias Jamin, Markus Lautenbacher,
Thomas Mannel, Mikolaj Misiak, Manfred M{\"u}nz, Ulrich Nierste, 
Gaby Ostermaier, Nicolas Pott and Oliver B{\"a}r for
fruitful discussions and collaborations on several topics presented in
this review. Special thanks go to Markus Lautenbacher and Manfred M{\"u}nz
for help in some technical aspects of this work.
 
This work has been supported by the
German Bundesministerium f{\"u}r Bildung and Forschung under contract 
06 TM 743 and DFG Project Li 519/2-2.

\vfill\eject

\end{document}